\pdfoutput=1
\documentclass[11pt]{article}

\usepackage[letterpaper,margin=1in]{geometry}

\usepackage{graphicx}
\usepackage{multirow}
\usepackage{amsmath,amssymb,amsfonts}
\usepackage{amsthm}
\usepackage{mathtools}
\usepackage{mathrsfs}
\usepackage{dsfont}
\usepackage[title]{appendix}
\usepackage{xcolor}
\usepackage{textcomp}
\usepackage{booktabs}
\usepackage{algorithm}
\usepackage{algorithmic}
\usepackage{listings}
\usepackage{float}
\usepackage{url}
\usepackage[numbers,sort&compress]{natbib}
\usepackage[colorlinks=true,linkcolor={blue!50!black},citecolor={blue!50!black},urlcolor={blue!50!black},
  breaklinks=true,bookmarks=true,bookmarksnumbered=true,bookmarksopen=true,bookmarksdepth=2]{hyperref}
\hypersetup{pdftitle={The Curved Exponential Family of Cylindrical Distributions: m-Projection, Curvature and a Test of the von Mises\textendash Fisher Component},
  pdfauthor={Amine M. Aboussalah}}

\makeatletter
\newtheoremstyle{thmstyleone}
  {\topsep}{\topsep}{\itshape}{0pt}{\bfseries}{}{.5em}%
  {\thmname{#1}\thmnumber{\@ifnotempty{#1}{ }\@upn{#2}}\thmnote{ {\the\thm@notefont(#3)}}}
\newtheoremstyle{thmstyletwo}
  {\topsep}{\topsep}{\normalfont}{0pt}{\itshape}{}{.5em}%
  {\thmname{#1}\thmnumber{\@ifnotempty{#1}{ }{#2}}\thmnote{ {\the\thm@notefont(#3)}}}
\newtheoremstyle{thmstylethree}
  {\topsep}{\topsep}{\normalfont}{0pt}{\bfseries}{}{.5em}%
  {\thmname{#1}\thmnumber{\@ifnotempty{#1}{ }\@upn{#2}}\thmnote{ {\the\thm@notefont(#3)}}}
\makeatother

\theoremstyle{thmstyleone}
\newtheorem{theorem}{Theorem}
\newtheorem{proposition}[theorem]{Proposition}
\newtheorem{lemma}[theorem]{Lemma}
\newtheorem{corollary}[theorem]{Corollary}
\theoremstyle{thmstyletwo}

\theoremstyle{thmstylethree}
\newtheorem{definition}[theorem]{Definition}
\newtheorem{assumption}[theorem]{Assumption}

\DeclarePairedDelimiter\norm{\lVert}{\rVert}

\DeclareMathOperator*{\argmin}{arg\,min}
\DeclareMathOperator{\tr}{tr}
\DeclareMathOperator{\diag}{diag}
\DeclareMathOperator{\rank}{rank}
\DeclareMathOperator{\cov}{cov}

\newcommand{\E}{\mathbb{E}}
\newcommand{\V}{\mathrm{Var}}
\newcommand{\R}{\mathbb{R}}

\newcommand{\Z}{\mathbb{Z}}

\newcommand{\calN}{\mathcal{N}}
\newcommand{\calM}{\mathcal{M}}

\newcommand{\calP}{\mathcal{P}}

\newcommand{\bx}{\mathbf{x}}
\newcommand{\bmu}{\boldsymbol{\mu}}
\newcommand{\bSigma}{\boldsymbol{\Sigma}}
\newcommand{\btheta}{\boldsymbol{\theta}}
\newcommand{\bpi}{\boldsymbol{\pi}}
\newcommand{\bB}{\mathbf{B}}
\newcommand{\bS}{\mathbf{S}}
\newcommand{\bI}{\mathbf{I}}
\newcommand{\sphere}{\mathbb{S}}
\newcommand{\vMF}{\mathrm{vMF}}

\newcommand{\ARI}{\mathrm{ARI}}
\newcommand{\HLL}{\mathrm{HLL}}

\newcommand{\balpha}{\boldsymbol{\alpha}}
\newcommand{\beps}{\boldsymbol{\epsilon}}
\newcommand{\bnu}{\boldsymbol{\nu}}
\newcommand{\bxi}{\boldsymbol{\xi}}
\newcommand{\bLambda}{\boldsymbol{\Lambda}}
\newcommand{\bGamma}{\boldsymbol{\Gamma}}
\newcommand{\bPsi}{\boldsymbol{\Psi}}
\newcommand{\bOmega}{\boldsymbol{\Omega}}
\newcommand{\bDelta}{\boldsymbol{\Delta}}
\newcommand{\bvartheta}{\boldsymbol{\vartheta}}
\newcommand{\bfeta}{\boldsymbol{\eta}}
\newcommand{\bTheta}{\boldsymbol{\Theta}}
\newcommand{\bt}{\mathbf{t}}
\newcommand{\bs}{\mathbf{s}}
\newcommand{\br}{\mathbf{r}}
\newcommand{\bz}{\mathbf{z}}
\newcommand{\sct}[1]{\widetilde{\mathbf{s}}_{#1}}
\newcommand{\Sympp}[1]{\operatorname{Sym}^{#1}_{++}}
\newcommand{\Symmm}[1]{\operatorname{Sym}^{#1}_{--}}
\newcommand{\Symzero}[1]{\operatorname{Sym}^{#1}_{0}}
\newcommand{\Hnat}{\mathcal{H}}
\newcommand{\calI}{\mathcal{I}}
\newcommand{\Eamb}{\mathcal{E}_{\mathrm{amb}}}
\newcommand{\Ecyl}{\mathcal{E}_{\mathrm{cyl}}}
\newcommand{\Eprod}{\mathcal{E}_{\mathrm{G}\otimes\mathrm{vMF}}}
\newcommand{\FB}{\mathrm{FB}}
\newcommand{\Am}{A_m}
\newcommand{\Cm}{C_m}
\newcommand{\CG}{\mathrm{CG}}
\newcommand{\cyl}{\mathrm{cyl}}
\newcommand{\fcyl}{f_{\mathrm{cyl}}}
\newcommand{\fCG}{f_{\mathrm{CG}}}
\newcommand{\fG}{f_{\mathrm{G}}}
\newcommand{\fvMF}{f_{\mathrm{vMF}}}
\newcommand{\KL}{D_{\mathrm{KL}}}
\newcommand{\lhier}{\ell_{\mathrm{hier}}}
\newcommand{\liso}{\ell_{\mathrm{iso}}}
\newcommand{\Dcyl}{D_{\mathrm{cyl}}}
\newcommand{\Pperp}{\mathbf{P}^{\perp}}
\newcommand{\indep}{\perp\!\!\!\perp}
\DeclareMathOperator{\Floor}{Floor}
\DeclareMathOperator{\clip}{clip}
\DeclareMathOperator{\sym}{sym}
\DeclareMathOperator{\blockdiag}{blockdiag}
\newcommand{\Pext}{\bar{\calP}}
\newcommand{\Ieff}{\mathcal{I}_{\mathrm{eff}}}            
\newcommand{\LRstat}{\mathrm{LR}}                          
\newcommand{\Tmix}{T^{\mathrm{mix}}}                       
\newcommand{\bomega}{\boldsymbol{\omega}}                  
\newcommand{\bbeta}{\boldsymbol{\beta}}

\newcommand{\backmatter}{}
\let\botrule\bottomrule
\makeatletter
\newcommand\bmhead{\@startsection{subparagraph}{5}{\z@}{6pt \@plus1ex \@minus .2ex}{-1em}{\normalfont\normalsize\bfseries}}

\let\arxiv@appendices\appendices
\let\arxiv@endappendices\endappendices
\renewenvironment{appendices}{%
  \setcounter{figure}{0}\setcounter{table}{0}\setcounter{equation}{0}%
  \arxiv@appendices
  \renewcommand\thefigure{\@Alph\c@section\arabic{figure}}%
  \renewcommand\thetable{\@Alph\c@section\arabic{table}}%
  \renewcommand\theequation{\@Alph\c@section\arabic{equation}}%
}{\arxiv@endappendices}
\makeatother
\AtBeginDocument{\allowdisplaybreaks}
\makeatletter
\let\arxiv@table\table
\def\arxiv@tablesetup{\let\footnotesize\scriptsize}
\@ifundefined{@@startpbox}{}{
  \let\arxiv@startpbox\@@startpbox
  \def\arxiv@cmpbox#1{\in@{cm}{#1}%
    \ifin@\expandafter\@firstoftwo\else\expandafter\@secondoftwo\fi
    {\arxiv@startpbox{1.142857\dimexpr#1\relax}}{\arxiv@startpbox{#1}}}%
  \def\arxiv@tablesetup{\let\footnotesize\scriptsize\let\@@startpbox\arxiv@cmpbox}}
\renewcommand*{\table}{\arxiv@tablesetup\arxiv@table}
\makeatother

\title{The Curved Exponential Family of Cylindrical Distributions: m-Projection, Curvature and a Test of the von Mises--Fisher Component}
\author{Amine M. Aboussalah\\[4pt]
  \normalsize Department of Finance and Risk Engineering\\
  \normalsize New York University Tandon School of Engineering\\
  \normalsize Brooklyn, NY 11201, USA\\
  \normalsize \href{mailto:ama10288@nyu.edu}{\nolinkurl{ama10288@nyu.edu}}}
\date{}

\begin{document}

\maketitle

\begin{abstract}
Cylindrical data pair a vector in $\mathbb{R}^d$ with a direction in $\mathbb{S}^{m-1}$. We study a family in which the direction is von Mises--Fisher and, given the direction, the vector is Gaussian with mean affine in it; this generalizes the Mardia--Sutton model. The family is a curved exponential family inside a regular family with Fisher--Bingham directions. We state necessary and sufficient conditions for the maximum-likelihood estimator to exist and be unique. The estimator is explicit except for one scalar root and is the m-projection (Kullback--Leibler projection) of the moment-matching member. Every moment it does not match is a function of one residual, the gap between the sample and fitted second moments of the direction. The squared length of this gap in the Fisher metric is the classical score test of von Mises--Fisher against Fisher--Bingham directions, due to Cox on the circle and extended by Mardia, Holmes and Kent. Specific here are this reading and the exact invariance of its law to the regression part. The exponential curvature lies only in the directions of the regression matrix. In closed form, the information lost on this matrix against the expected information is $m-1$ observations for uniform directions and $(m+1)^2/(m-1)$ in the limit of high concentration; conditioning on the directions recovers it. Finite mixtures are identifiable. On buoy records the test rejects von Mises--Fisher directions, and generalized von Mises directions raise the held-out log-density by $0.45$ nats per observation with one component.

\end{abstract}

\section{Introduction}\label{sec:intro}

Many real-world observations pair a Euclidean vector with a direction. For example, a moored buoy records wave height, period, and wind speed alongside the direction of the waves (Section~\ref{sec:ocean}). Similarly, in drone localization, the orientations of map boundaries constrain the position of the drone, and the Fisher information measures the uncertainty associated with these directions~\cite{noumane2026when}. An observation is a pair
\begin{equation}\label{eq:sample-space}
\bx=(\bx_1,\bx_2)\in\R^d\times\sphere^{m-1},\qquad \sphere^{m-1}=\{\bx\in\R^m:\norm{\bx}=1\},
\end{equation}
consisting of a Euclidean block $\bx_1$ and a unit direction $\bx_2$; on the circle ($m=2$), this direction is given by $\bx_2=(\cos\theta,\sin\theta)^{\top}$. Such data are cylindrical on the circle and directional--linear in general~\cite{leyverdebout2017modern}, though we refer to them as cylindrical in every dimension.

We study the \emph{cylindrical family}, in which $\bx_2$ follows a von Mises--Fisher law and, given $\bx_2$, the vector $\bx_1$ is Gaussian with covariance $\bS$ and mean $\balpha+\bB\bx_2$. In this structure, the regression matrix $\bB$ captures all dependence between the two blocks. For $d=1$, $m=2$, this corresponds to the model of Mardia and Sutton~\cite{mardia1978cylindrical}. This family is a curved exponential family~\cite{efron1975curvature,amari1982curved} embedded within a regular ambient family where the directional factor is Fisher--Bingham~\cite{kent1982fisherbingham}. Moreover, $\bx_2$ is a cut of this ambient family~\cite{barndorffnielsen1978information}. This cut property implies three facts: the likelihood splits, the Fisher metric is block-diagonal, and tests of the directional law depend only on the directions. While the fact that the estimator is an m-projection follows from Amari's general results~\cite{amari1982curved,amari1985differential}, we provide results specific to this family. These include explicit conditions for the estimator, the m-projection residual interpreted as the classical score test, and the exponential curvature. This curvature lies only in the directions of $\bB$ and allows us to express the information loss in closed form, an exact Efron--Hinkley example. Finite mixtures are identifiable and are fitted by exact EM, while two-layer hierarchies, where a latent class replaces $\bB$ as the source of dependence, are treated as a secondary topic (Section~\ref{sec:hier}). Material whose section, result, equation, table and figure numbers carry the prefix S is available in Online Resource~1.

\subsection{Related work}\label{subsec:related}

Cylindrical models go back to Mardia and Sutton~\cite{mardia1978cylindrical}, who conditioned a trivariate normal vector on the circle, and to Johnson and Wehrly~\cite{johnsonwehrly1977correlation,johnsonwehrly1978angularlinear}, who developed angular--linear correlations and distributions. The likelihood gain of $\bB$ is given by the canonical correlations between a direction in any dimension and a vector, as defined by Jupp and Mardia~\cite{juppmardia1980correlation} (Corollary~\ref{cor:cyl-dependence-gain}). Kato and Shimizu~\cite{katoshimizu2008dependent} extended the Mardia--Sutton model to include the second harmonic of the angle; for $d=1$, $m=2$, our ambient family is equivalent to theirs up to reparametrization and boundary values. Abe and Ley~\cite{abe2017tractable} couple a sine-skewed von Mises angle with a Weibull variable on $[0,\infty)$. The most closely related work is by Luengo-S\'anchez, Larra\~naga and Bielza~\cite{luengo2019directional}, who consider von Mises marginals and a Gaussian mean affine in the cosines and sines of several angles, with estimation, mixtures and Kullback--Leibler computations. For a single circle, their model matches ours at $m=2$, though we do not claim their estimator and mixtures. Further constructions involve regressing a Gaussian vector on a rotation matrix with a Fisher--Bingham-type marginal~\cite{kumeprestonwood2013fisherbingham}, or using projected normal vectors~\cite{mastrantonio2015hmm,mastrantonio2018polycylindrical} and kernels~\cite{garciaportugues2013kde}. Regarding cylindrical mixtures, previous approaches have used Abe--Ley~\cite{sadeghianpourhamami2019abeley}, semi-wrapped and marginal-specified components~\cite{roy2016swgmm,roy2017jclmm}, or products of Gaussian and von Mises--Fisher laws on $\R^d\times\sphere^2$ (which corresponds to our case with $\bB=\mathbf 0$)~\cite{hasnat2016jcsd}. Directional mixtures have also employed Kent~\cite{peel2001kent} and von Mises--Fisher laws~\cite{banerjee2005clustering,sra2012vmf}. Our mixture is a linear Gaussian cluster-weighted model~\cite{gershenfeld1997cwm,ingrassia2012cwm} with a von Mises--Fisher covariate density. Finally, while sea-state models introduce serial or spatial dependence~\cite{bulla2012searegimes,lagona2015hmmcylindrical,wang2015wavesadriatic}, our model does not.

The Fisher--Bingham family~\cite{kent1982fisherbingham} encompasses both the von Mises--Fisher and Bingham~\cite{bingham1974antipodal} laws. On the circle, this corresponds to the generalized von Mises family~\cite{gattojammalamadaka2007gvm}, while on higher spheres, calculating the normalizer requires saddlepoint approximation~\cite{kumewood2005saddlepoint} or the holonomic gradient method~\cite{kumesei2018holonomic}. The Bingham law is used to describe axes or lines through the origin; higher-dimensional subspaces are points on a Grassmann manifold, compared via principal angles, which can be used to measure mode collapse in generative models of time series~\cite{abbahaddou2025geometry}. To test the von Mises law, Cox~\cite{cox1975discussion} used the second trigonometric moments about the mean: a circular kurtosis and the sine term of Pewsey's symmetry test~\cite{pewsey2002symmetry}. Kent~\cite{kent1982fisherbingham} provided score tests for the Fisher law within Kent's own five-parameter family and for the Bingham law within the Fisher--Bingham family. Mardia, Holmes and Kent~\cite{mardiaholmeskent1984gof} extended Cox's test to the von Mises--Fisher law, against Fisher--Bingham alternatives, and Rivest~\cite{rivest1986kent} adjusted Kent's statistic for concentrated samples (review:~\cite[Sect.~12.3]{mardiajupp2000directional}). Statistical curvature is Efron's~\cite{efron1975curvature,efron1978geometry}. Within Amari's geometry of curved exponential families~\cite{amari1982curved,amari1985differential}, the exponential curvature determines the information lost by the estimator, and the ancillary direction supports conditional inference~\cite{amari1982ancillarity}, as seen in Efron and Hinkley~\cite{efronhinkley1978observed}. Kumon and Amari~\cite{kumonamari1983higher} studied higher-order properties of tests (see also~\cite{amarinagaoka2000methods,kassvos1997geometrical}). The Fisher metric also preconditions gradient learning through Amari's natural gradient~\cite{amari1998natural}, and its quantum counterpart, the quantum geometric tensor, is used to stabilize the training of recurrent neural-network quantum states~\cite{attar2026quantum}. In the low-rank fine-tuning of language models, the rank of the Fisher information is interpreted as the local dimension of the parameter manifold and motivates a lower bound on the adapter rank~\cite{edib2025gelora}. The score test is Rao's~\cite{rao1948score}, and the $C(\alpha)$ test is Neyman's~\cite{neyman1959optimal}. The EM algorithm~\cite{dempster1977em,wu1983em} alternately minimizes Kullback--Leibler divergences~\cite{csiszartusnady1984alternating}, as in Amari's em algorithm~\cite{amari1995em}, and each step increases the Neal--Hinton free energy~\cite{nealhinton1998em}. Finite mixtures of Gaussian and various directional laws are identifiable~\cite{yakowitz1968identifiability,kent1983identifiability}, and unbounded mixture likelihoods are controlled by constraints~\cite{hathaway1985constrained}. In machine learning, Gaussian mixtures fitted by EM can generate training data: when fitted to the hidden representations of a graph neural network, they produce new representations that augment the training set, while influence scores based on the Fisher information rank these samples~\cite{abbahaddou2025graph}. Finally, two-layer hierarchies are mixtures of mixtures~\cite{malsinerwalli2017identifying}; their isolated estimator is, up to the weights, equivalent to the two-stage scheme of SoccerMix~\cite{decroos2021soccermix} and a stepwise latent-class estimator, which understates associations unless corrected~\cite{bolck2004threestep,vermunt2010threestep,bakkkuha2018twostep}.

\subsection{Contributions}\label{subsec:contributions}

The status of every result is summarized in Table~\ref{tab:known-new}, where we distinguish between known results, specializations of general results (with their respective sources), and new results for which no earlier statement was found.\footnote{Definition~\ref{def:cyl}, Proposition~\ref{prop:cyl-curved}(a), Proposition~S3, Theorem~\ref{thm:cyl-mle} under sufficient conditions, the estimators of Section~\ref{sec:hier}, and the experimental designs come from the M.S.\ thesis of J.~E. Garay Vergara~\cite{garayvergara2026thesis}.}\enlargethispage{2pt}
\begin{enumerate}
\item \textbf{The curved family in a cut ambient family.} The cylindrical family is curved, of codimension $m(m+1)/2-1$, because $\bB$ enters the natural parameter quadratically; each member is a Gaussian density restricted to the cylinder (Propositions~\ref{prop:cyl-restriction} and~\ref{prop:cyl-curved}).
\item \textbf{Estimation and identifiability.} Necessary and sufficient conditions for existence and uniqueness of the maximum-likelihood estimator, assembled from least squares and the von Mises--Fisher estimator through the cut (Theorem~\ref{thm:cyl-mle}); identifiability from two moments and the closed-form Kullback--Leibler divergence between components (Propositions~\ref{prop:cyl-identifiable} and~\ref{prop:cyl-kl}); identifiability of finite mixtures with a von Mises--Fisher covariate, and exact EM with closed-form M-steps (Theorem~\ref{thm:mix-identifiable}, Section~\ref{subsec:mix-em}).
\item \textbf{The m-projection residual.} Every unmatched moment is a function of $\bDelta$, the gap between the sample and fitted second moments of the direction (Proposition~\ref{prop:cyl-mprojection}). The squared length of $\bDelta$ in the Fisher metric is the classical score test of von Mises--Fisher against Fisher--Bingham directions~\cite{cox1975discussion,mardiaholmeskent1984gof}, and its law does not depend on the regression part at any $N$. We write its efficient information in closed form for every $m$ (computed here, not claimed as new) (Theorem~\ref{thm:geomtest}, Proposition~\ref{prop:geomtest-ieff}).
\item \textbf{Curvature and conditioning.} The exponential curvature lies only in the $\bB$-directions. Paired with $\bDelta$, it gives the exact defect of the Pythagorean relation and the observed-minus-expected information for $\bB$. Its square, Amari's form of Efron's statistical curvature, gives the information lost on $\bB$ against the expected information ($m-1$ observations at $\kappa=0$, $(m+1)^2/(m-1)$ as $\kappa\to\infty$); conditioning on the directions recovers it (Proposition~\ref{prop:geomtest-curv}, Theorem~\ref{thm:geomtest-loss}).
\item \textbf{For mixtures.} A $C(\alpha)$ score test calibrated by the expected information and valid under the conditions of Theorem~\ref{thm:geomtest-mix}, a missing-information account of the plug-in, and a calibration for serial dependence~\cite{kiefervogelsang2005fixedb}, justified for $K=1$ (Section~\ref{subsec:geomtest-mix}).
\item \textbf{Hierarchies (secondary).} The isolated two-stage criterion is the Neal--Hinton bound at the $\bx_1$-only posterior; its ascent can lower the joint likelihood (Proposition~\ref{prop:hier-elbo}).
\item \textbf{Illustration.} In simulations ($20$ replicates), $\bB_j$ raises the held-out log-density mainly where the dependence lies within components ($0.30$ nats, natural-log units, per observation when strong; about zero through a latent class when $K$ equals the number of classes). On hourly buoy records the test rejects von Mises--Fisher directions at $K=1$ (Kullback--Leibler effect size $0.45$ nats), and generalized von Mises components repair this out of sample ($0.454$ nats); direct dependence gains clearly for $K\le5$, at most $0.012$ nats beyond, and not at a fixed parameter budget up to $95$ parameters (Sections~\ref{sec:synth}--\ref{sec:ocean}).
\end{enumerate}

\begin{table}[!t]
\caption{Status of the results: ``known'', classical; ``specialization'', a general result applied to the cylindrical family, with what is specific to it; ``new'', no earlier statement found}\label{tab:known-new}
\scriptsize
\setlength{\tabcolsep}{3pt}
\begin{tabular}{@{}lp{0.145\textwidth}p{0.73\textwidth}@{}}
\toprule
Result & Status & Source; what is specific \\
\midrule
Prop.~\ref{prop:cyl-restriction} & specialization & Mardia--Sutton's conditioning of a normal vector on the circle~\cite{mardia1978cylindrical}, lifted to $\sphere^{m-1}$ \\
Prop.~\ref{prop:cyl-curved} & specialization & curved exponential family in the sense of Efron and Amari~\cite{efron1975curvature,amari1982curved}, with $\bx_2$ a cut~\cite{barndorffnielsen1978information}; specific: the constraint $\bOmega=\mathbf 0$, linear in the parameters of the cut factorization but quadratic in $\btheta_{12}$ in natural coordinates, its codimension, and that the family is neither e-flat nor m-flat \\
Thm.~\ref{thm:cyl-mle} & specialization & by the cut~\cite{barndorffnielsen1978information}, Gaussian least squares~\cite{anderson2003multivariate} and the von Mises--Fisher estimator~\cite{mardiajupp2000directional}; formulas known~\cite{juppmardia1980correlation}, for this model at $m=2$~\cite{mardia1978cylindrical,luengo2019directional}; specific: the necessary and sufficient conditions for the weighted estimator, with $\kappa=0$ and zero weights \\
Prop.~\ref{prop:cyl-identifiable} & specialization & Gaussian regression and minimality of the von Mises--Fisher family; specific: the explicit moment map, including $\kappa=0$ \\
Prop.~\ref{prop:cyl-mprojection} & specialization & the estimator of a curved family as an m-projection, $\bDelta$ its ancillary coordinate~\cite{amari1982curved,amari1985differential}; specific: the explicit $\bDelta$ and unmatched moments \\
Prop.~\ref{prop:cyl-fisher} & specialization & pullback of the ambient metric, block-diagonal by the cut~\cite{barndorffnielsen1978information,kassvos1997geometrical}; specific: the explicit blocks \\
Prop.~\ref{prop:cyl-kl} & specialization & chain rule of the divergence; computed at $m=2$ in~\cite{luengo2019directional}; specific: the closed form for general $m$ \\
Cor.~\ref{cor:cyl-dependence-gain} & known & Wilks' likelihood ratio in the canonical correlations of~\cite{juppmardia1980correlation}; Mardia--Sutton's test at $d=1$, $m=2$ \\
Thm.~\ref{thm:mix-identifiable} & specialization & Yakowitz--Spragins~\cite{yakowitz1968identifiability} and Kent~\cite{kent1983identifiability,fraser1981vonmises} for the two blocks; known for mixtures of regressions~\cite{hennig2000identifiability} and Gaussian-covariate cluster-weighted models~\cite{ingrassia2015glmcwm}; specific: a von Mises--Fisher covariate (no earlier statement found), shared $(\balpha,\bB,\bS)$, $\kappa=0$ \\
Thm.~\ref{thm:geomtest} & known & score test of von Mises--Fisher within Fisher--Bingham~\cite{cox1975discussion,mardiaholmeskent1984gof}, reduced to the directions by S-ancillarity~\cite{barndorffnielsen1978information}; deviance~\cite{efron1978geometry}; specific: the residual-length reading, the exact invariance (iii) \\
Prop.~\ref{prop:geomtest-ieff} & specialization & efficient information of the test of~\cite{mardiaholmeskent1984gof}; closed form for every $m$ computed here, not claimed as new; Cox's at $m=2$~\cite{cox1975discussion}, split~\cite{pewsey2002symmetry}; Kent's at $m=3$~\cite{kent1982fisherbingham}; Bingham's as $\kappa\to0$~\cite{bingham1974antipodal} \\
Prop.~\ref{prop:geomtest-curv} & specialization & e-embedding curvature~\cite{amari1982curved} and three-point identity~\cite{amarinagaoka2000methods}; specific: curvature only in the $\bB$-directions, the exact Pythagorean defect \\
Thm.~\ref{thm:geomtest-loss} & specialization & Amari's loss theorem~\cite{efron1975curvature,amari1982curved}, observed versus expected information~\cite{efronhinkley1978observed}; new: the closed-form information loss on $\bB$ for every $m$, its exact recovery by conditioning \\
Thm.~\ref{thm:geomtest-mix} & specialization & Neyman's $C(\alpha)$ test~\cite{neyman1959optimal} for the cylindrical mixture; new: the calibration study (expected versus outer-product information) and the missing-information account of the plug-in~\cite{louis1982em} (Proposition~S9) \\
Prop.~\ref{prop:hier-elbo} & known & Neal--Hinton identity~\cite{nealhinton1998em} at the $\bx_1$-only posterior; specific: an explicit example in which ascent of the isolated criterion lowers the joint likelihood \\
\botrule
\end{tabular}
\end{table}

\section{Exponential families, projections and EM}
\label{sec:background}\label{subsec:ef-duality}\label{subsec:mixtures-em}

This section establishes the necessary notation, while Section~\ref{subsec:flat-families} reviews the von Mises--Fisher family. Consider an exponential family on a space $\mathcal X$ equipped with a $\sigma$-finite measure $\nu$ and a statistic $\bt:\mathcal X\to\R^p$. The densities are given by
\begin{equation}
f(\bx;\btheta)=\exp\bigl(\langle\bt(\bx),\btheta\rangle-\psi(\btheta)\bigr),\qquad
\psi(\btheta)=\log\int_{\mathcal X}\exp\bigl(\langle\bt(\bx),\btheta\rangle\bigr)\,d\nu(\bx),
\label{eq:ef-canonical}
\end{equation}
where $\btheta\in\Hnat:=\{\psi<\infty\}$, and $\langle\cdot,\cdot\rangle$ denotes the Euclidean inner product (or the Frobenius inner product for matrix blocks). We assume the family is \emph{minimal}, meaning no nonzero affine function of $\bt$ is $\nu$-a.e.\ constant, and \emph{regular}, meaning $\Hnat$ is open. Under these conditions, $\psi$ is smooth and strictly convex, and the map $\bfeta:=\nabla\psi(\btheta)=\E_{\btheta}[\bt]$ ranges over the mean space $\calM$, which is the interior of the convex hull of the support of $\bt$. Here, $\nabla^2\psi$ represents the Fisher information. Let $\phi:=\psi^{*}$ be the dual function; then $\nabla\phi=(\nabla\psi)^{-1}$, and the Bregman divergence $D_\phi[\bfeta:\bfeta']=\KL[p_{\bfeta}:p_{\bfeta'}]$ coincides with the Kullback--Leibler divergence $\KL[p:q]:=\int p\log(p/q)\,d\nu$~\cite{barndorffnielsen1978information,amarinagaoka2000methods,amari2016information,ay2017information}. A subfamily is termed \emph{e-flat} (or \emph{m-flat}) if its natural parameters (or expectation parameters) form an open subset of an affine subspace. A \emph{curved exponential family} of dimension $k<p$ is defined as $\{f(\cdot;\btheta(\mathbf u)):\mathbf u\in U\}$, where $U\subset\R^k$ is open and $\btheta$ is a smooth embedding~\cite{efron1975curvature,amari1982curved,amari1985differential}. The \emph{m-projection} of a distribution $P$ onto a set $\mathcal E$ minimizes $\KL[P:q]$ over $q\in\mathcal E$, while the \emph{e-projection} minimizes $\KL[q:P]$. For an empirical distribution $P$, the m-projection is the maximum-likelihood estimate on a subfamily; if $\E_P[\bt]\in\calM$, it is the moment match $\nabla\psi(\hat{\btheta})=\E_P[\bt]$ on the full family. To fit a finite mixture $p(\bx;\bTheta)=\sum_j\pi_jf(\bx;\btheta_j)$ with $\bTheta:=\{\pi_j,\btheta_j\}_{j=1}^K$, the EM algorithm~\cite{dempster1977em} alternates between e- and m-projections~\cite{csiszartusnady1984alternating,amari1995em}. In the E-step, we set $r_{nj}\propto\pi_jf(\bx_n;\btheta_j)$ to maximize the Neal--Hinton free energy~\cite{nealhinton1998em}. In the M-step, we set $\pi_j=N_j/N$ with $N_j:=\sum_nr_{nj}$, and we moment-match each component via $\nabla\psi(\btheta_j)=N_j^{-1}\sum_nr_{nj}\bt(\bx_n)$, provided the right side lies in $\calM$ (Lemma~S1, after~\cite{banerjee2005bregman}; see also~\cite{hino2022geometryem}).

\subsection{The von Mises--Fisher family}
\label{subsec:flat-families}

The Gaussian density on $\R^d$ is $\fG$. We now turn to the von Mises--Fisher law $\vMF_m(\bmu,\kappa)$ on $\sphere^{m-1}$, where $\bmu$ is the mean direction and $\kappa\ge0$ is the concentration. With respect to the surface measure $\sigma$, its density is given by $\fvMF(\bx;\bmu,\kappa)=C_m(\kappa)\exp(\kappa\bmu^\top\bx)$, where $C_m(\kappa)=\kappa^{m/2-1}/\{(2\pi)^{m/2}I_{m/2-1}(\kappa)\}$ and $C_m(0)=\Gamma(m/2)/(2\pi^{m/2})$, and $I_\nu$ is the modified Bessel function of the first kind. This family is regular and minimal with $\bt=\bx$, natural parameters $\btheta=\kappa\bmu\in\R^m$, and potential function $\psi=-\log C_m(\kappa)$. The expectation parameters are $\bfeta=A_m(\kappa)\bmu$, where $A_m(\kappa):=I_{m/2}(\kappa)/I_{m/2-1}(\kappa)$ is the mean resultant length. Since $A_m(0)=0$ and $A_m(\kappa)$ increases strictly from $[0,\infty)$ onto $[0,1)$, the mean space is $\calM=\{\norm{\bfeta}<1\}$~\cite{schou1978concentration,mardiajupp2000directional}. Consequently, for $\bfeta\neq\mathbf 0$, the dual map is $\nabla\phi(\bfeta)=A_m^{-1}(\norm{\bfeta})\,\bfeta/\norm{\bfeta}$, with $\nabla\phi(\mathbf 0)=\mathbf 0$, and the dual potential is $\phi=\kappa\norm{\bfeta}+\log C_m(\kappa)$ where $\kappa=A_m^{-1}(\norm{\bfeta})$. Using Bessel recurrences~\cite{schou1978concentration}, we obtain
\begin{align}
\tfrac{d}{d\kappa}\log C_m(\kappa)&=-A_m(\kappa),
\label{eq:vmf-logpartition}\\
A_m'(\kappa)&=1-A_m(\kappa)^2-\tfrac{m-1}{\kappa}A_m(\kappa).
\label{eq:riccati}
\end{align}
The first two moments are classical~\cite{mardiajupp2000directional,schou1978concentration} (Lemma~S2): $\E\bx=A_m(\kappa)\bmu$, and the second moment $\E\bx\bx^{\top}$ has eigenvalue $A_m(\kappa)/\kappa$ on $\bmu^{\perp}$ and $1-(m-1)A_m(\kappa)/\kappa$ on $\bmu$, tending to $\bI_m/m$ as $\kappa\downarrow0$. The Fisher information in $\btheta=\kappa\bmu$ is given by the covariance $\kappa^{-1}A_m(\kappa)(\bI_m-\bmu\bmu^{\top})+A_m'(\kappa)\bmu\bmu^{\top}$. Because the second moment is uniquely determined by the mean, the cylindrical estimator leaves it unmatched. Table~\ref{tab:notation} lists recurring symbols.

\begin{table}[t]
\caption{Recurring notation, by the section that introduces it; symbols used in one result only are defined there. Bold $\bDelta$ is the residual of Section~\ref{sec:cyl}; a plain $\Delta$ denotes a contrast}
\label{tab:notation}
\scriptsize
\begin{tabular}{@{}lp{0.9\textwidth}@{}}
\toprule
Sect. & Symbols \\
\midrule
2--3 & $(\balpha,\bB,\bS)$ intercept, regression matrix, conditional covariance; $\bmu_c(\bx_2)=\balpha+\bB\bx_2$; $\bvartheta_2=\kappa\bmu_2$ (mean direction, concentration); $\calP$, $\Pext$ parameter spaces ($\kappa>0$; $\bvartheta_2\in\R^m$); $\bbeta=[\balpha\ \bB]$, $\bz=(1,\bx_2^{\top})^{\top}$; $A_m$, $C_m$ mean resultant length, normalizer; $\bt$, $\btheta$, $\bfeta$ ambient statistic, natural and expectation coordinates; $\psi_{\mathrm{amb}}$, $\phi_{\mathrm{amb}}$ ambient potentials; $\Eamb\supset\Ecyl\supset\Eprod$ ambient, cylindrical and $\bB=\mathbf 0$ families; $\FB_m(\bvartheta,\bOmega)$ Fisher--Bingham law; $\bOmega^{\circ}$, $\Symzero{m}$ traceless part, traceless symmetric matrices; $\bfeta_2$, $\mathbf V$ mean and covariance of $\bx_2$; $\bfeta_{22}(\bvartheta)=\E\bx_2\bx_2^{\top}$ under $\vMF_m(\bvartheta)$; $w_n$, $\bar{\bx}_i^{w}$, $\sct{ij}^{w}$, $\br^{w}$, $E^{w}$, (A0)--(A3): Assumption~\ref{ass:design}; $\bar{\bt}$, $\bar{\mathbf M}_{22}$ averages of $\bt$ and $\bx_2\bx_2^{\top}$; $\bDelta=\bar{\mathbf M}_{22}-\bfeta_{22}(\hat{\bvartheta})$; $\calI$, $g$ Fisher information, metric \\
4 & $r_{nj}$, $N_j$ responsibilities, masses; $q_{\cyl}$, $q_{\mathrm{ind}}$, $q_{\mathrm{hier}}$ free-parameter counts \\
5 & $q_m=m(m+1)/2-1$; $\Ieff$ efficient information, with eigenvalues $\nu_{\mathrm{rad}}$, $\nu_{\mathrm{cpl}}$, $\nu_{\mathrm{tan}}$; $T_N$, $\LRstat_N$ score and likelihood-ratio statistics; $H^{\mathrm e}$ exponential curvature; $\iota_\parallel$, $\iota_\perp$ information loss; $\mathbf s_n$, $\boldsymbol\varsigma$ scores, null parameters; $\mathcal J$, $\mathcal J_{\mathrm{eff}}$ mixture information and its efficient part; $\mathbf s^{\natural}$ effective score; $\Tmix_N$, $\Tmix_{N,j}$ mixture statistics \\
6 & $L_j$ local sizes; $f$, $h$, $h_j$ first-layer, local-component and local-mixture densities; $a_{nj}$ $\bx_1$-only posterior; $\lhier$, $\liso$, $F$ joint and isolated criteria, free energy \\
7--8 & $\HLL=\HLL_2+\HLL_{1\mid2}$ held-out log-density, directional and conditional parts; $\ARI$ adjusted Rand index; $\bar\ell_y(\hat f)$, $\Delta\bar\ell^{\mathrm{dep}}_{y,K}$ test log-density in year $y$, direct-dependence contrast; $T^{\mathrm{fb}}$ fixed-$b$ statistic; $\boldsymbol\zeta_j$, $\mathbf h_{\mathrm G}$ generalized von Mises parameter and statistic \\
\botrule
\end{tabular}
\end{table}

\section{The cylindrical family: curved exponential-family structure and maximum likelihood}\label{sec:cyl}

We augment the product of the Gaussian and von Mises--Fisher families with the simplest form of dependence: a regression of $\bx_1$ on $\bx_2$. Results that specialize the general theory are credited where they appear and in Table~\ref{tab:known-new}.

\subsection{The family}\label{subsec:cyl-family}

We consider densities on $\R^d\times\sphere^{m-1}$, $m\ge2$, defined with respect to the product measure $d\nu(\bx_1,\bx_2)=d\bx_1\,d\sigma(\bx_2)$. For symmetric matrices, the notation $\mathbf A\succ\mathbf A'$ ($\mathbf A\succeq\mathbf A'$) indicates that $\mathbf A-\mathbf A'$ is positive definite (semidefinite).

\begin{definition}[Cylindrical family]\label{def:cyl}
Let $\calP:=\R^d\times\R^{d\times m}\times\Sympp{d}\times\sphere^{m-1}\times(0,\infty)$, where $\Sympp{d}$ is the cone of symmetric positive-definite $d\times d$ matrices. For $(\balpha,\bB,\bS,\bmu_2,\kappa)\in\calP$, the conditional-Gaussian density of $\bx_1$ given $\bx_2$ is $\fCG(\bx_1\mid\bx_2;\balpha,\bB,\bS):=\fG(\bx_1;\balpha+\bB\bx_2,\bS)$, and the cylindrical density is
\begin{equation}\label{eq:cyl-density}
\fcyl(\bx_1,\bx_2;\balpha,\bB,\bS,\bmu_2,\kappa):=\fCG(\bx_1\mid\bx_2;\balpha,\bB,\bS)\,\fvMF(\bx_2;\bmu_2,\kappa).
\end{equation}
We call $\balpha$ the intercept, $\bB$ the direct-dependence parameter (the regression matrix), $\bS$ the conditional covariance, and $\bmu_c(\bx_2):=\balpha+\bB\bx_2$ the conditional mean.
\end{definition}

\begin{figure}[t]
\centering
\includegraphics[width=0.58\textwidth]{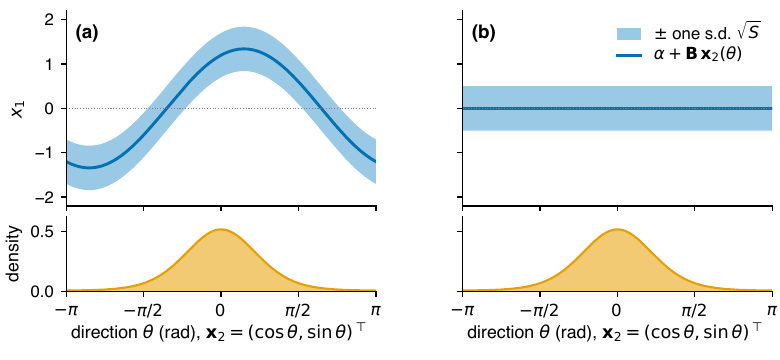}
\caption{One cylindrical component on $\R\times\sphere^{1}$ (a) and its independent restriction $\bB=\mathbf 0$ (b). Above, the conditional mean $\alpha+\bB\bx_2(\theta)$, $\bx_2(\theta)=(\cos\theta,\sin\theta)^{\top}$ (solid), its band of one conditional standard deviation $\sqrt{S}$ (shaded) and $x_1=0$ (dotted); below, the von Mises density of $\theta$, the same in both panels. Here $\alpha=0$, $\bB=(1.2,0.6)$, $S=0.25$, $\bmu_2=(1,0)$, $\kappa=2$}
\label{fig:component}
\end{figure}

Equivalently, the Euclidean block can be viewed as a linear regression on the direction (Figure~\ref{fig:component}),
\begin{equation}\label{eq:cyl-regression}
\bx_2\sim\vMF_m(\bmu_2,\kappa),\qquad \bx_1=\balpha+\bB\bx_2+\beps,\qquad \beps\sim\calN(\mathbf 0,\bS),\qquad \beps\indep\bx_2,
\end{equation}
and the case where $\bB=\mathbf 0$ corresponds to the independent restriction
\begin{equation}\label{eq:independent-restriction}
\fcyl(\bx_1,\bx_2;\balpha,\mathbf 0,\bS,\bmu_2,\kappa)=\fG(\bx_1;\balpha,\bS)\,\fvMF(\bx_2;\bmu_2,\kappa).
\end{equation}
The direction enters the model only through $\bvartheta_2:=\kappa\bmu_2$ (Section~\ref{subsec:flat-families}), where the mapping $(\bmu_2,\kappa)\mapsto\kappa\bmu_2$ is a diffeomorphism from $\sphere^{m-1}\times(0,\infty)$ onto $\R^m\setminus\{\mathbf 0\}$. By defining $\fvMF(\bx_2)=\Cm(\norm{\bvartheta_2})\exp(\bvartheta_2^{\top}\bx_2)$, the expression in \eqref{eq:cyl-density} defines $\fcyl(\cdot;\balpha,\bB,\bS,\bvartheta_2)$ on the \emph{extended parameter space} $\Pext:=\R^d\times\R^{d\times m}\times\Sympp{d}\times\R^m$. This space extends $\calP\cong\{\bvartheta_2\neq\mathbf 0\}$ by adding the uniform direction, where only the polar chart $(\bmu_2,\kappa)$ degenerates. For convenience, Section~\ref{sec:mix} writes $\bvartheta_{2,j}=\kappa_j\bmu_{2,j}$, while Section~\ref{sec:geomtest} drops the subscript $2$.

By Tonelli's theorem $\fcyl$ is a probability density, also on $\Pext$ (Proposition~S3).

For $\kappa>0$, the Mardia--Sutton parametrization~\cite{mardia1978cylindrical} of the conditional mean provides the \emph{original coordinates}, which differ from the \emph{regression coordinates} $(\balpha,\bB,\bS,\bmu_2,\kappa)$:
\begin{equation}\label{eq:cyl-original-params}
\bSigma_{12}:=\bB/\kappa,\qquad \bmu_1:=\balpha+\bB\bmu_2,\qquad\text{so that}\qquad \bmu_c(\bx_2)=\bmu_1+\kappa\bSigma_{12}(\bx_2-\bmu_2).
\end{equation}
For fixed $(\bmu_2,\kappa)$, this mapping is a bijection, although $\bSigma_{12}$ has no limit as $\kappa\downarrow0$ unless $\bB=\mathbf 0$. The following proposition extends the conditioning used by Mardia and Sutton to $\sphere^{m-1}$. In this case, a general Gaussian kernel results in a conditional Gaussian multiplied by a Fisher--Bingham law $\FB_m(\bvartheta,\bOmega)$, with density proportional to $\exp(\bvartheta^{\top}\bx_2+\bx_2^{\top}\bOmega\bx_2)$ \cite{kent1982fisherbingham}, \cite[Chapter~9]{mardiajupp2000directional}. Since $\bx_2^{\top}\bx_2=1$, the density depends on the symmetric matrix $\bOmega$ only through its traceless part $\bOmega^{\circ}:=\bOmega-\tfrac1m\tr(\bOmega)\bI_m$. Finally, we note that $\FB_m(\kappa\bmu_2,\mathbf 0)=\vMF_m(\bmu_2,\kappa)$.

\begin{proposition}[Restriction of a Gaussian kernel to the cylinder]\label{prop:cyl-restriction}
Let $\bmu_1\in\R^d$, $\bmu_2\in\sphere^{m-1}$, $\kappa>0$, and let $\bSigma\succ\mathbf 0$ be a $(d+m)\times(d+m)$ matrix with blocks $\bSigma_{11}$, $\bSigma_{12}$ and $\bSigma_{22}=\kappa^{-1}\bI_m$. The restriction of the kernel $K(\bx):=\exp\{-\tfrac12(\bx-(\bmu_1,\bmu_2))^{\top}\bSigma^{-1}(\bx-(\bmu_1,\bmu_2))\}$ to $\R^d\times\sphere^{m-1}$ is $\nu$-integrable, and its normalization is $\fcyl$ with original coordinates $(\bmu_1,\bSigma_{12},\bSigma_{11}-\kappa\bSigma_{12}\bSigma_{12}^{\top},\bmu_2,\kappa)$. Conversely, every $\fcyl$ with parameter in $\calP$ arises in this way, with $\bSigma_{11}:=\bS+\kappa\bSigma_{12}\bSigma_{12}^{\top}$, whose Schur complement $\bS$ makes $\bSigma\succ\mathbf 0$.
\end{proposition}

Here $\bSigma_{12}$ is the cross-covariance of $\bx_1$ and $\bx_2$ only if $\bB=\mathbf 0$ (Section~S7).

\subsection{Curved exponential-family structure}\label{subsec:cyl-curved}

The logarithm of a cylindrical density is a quadratic polynomial in $(\bx_1,\bx_2)$. We define the ambient family $\Eamb$ as the set of all normalizable densities of this form. Its sufficient statistic is
\begin{equation*}
\bt(\bx):=\bigl(\bx_1,\bx_2,\bx_1\bx_1^{\top},\bx_1\bx_2^{\top},\bx_2\bx_2^{\top}-\tfrac1m\bI_m\bigr),
\end{equation*}
where the final block belongs to $\Symzero{m}$, the space of traceless symmetric $m\times m$ matrices. Since $\tr(\bx_2\bx_2^{\top})=1$ on the sphere, the trace part is constant; we remove it to ensure a minimal representation. The natural parameter $\btheta=(\btheta_1,\btheta_2,\btheta_{11},\btheta_{12},\btheta_{22})$, with $\btheta_{11}$ symmetric and $\btheta_{22}\in\Symzero{m}$, pairs with $\bt(\bx)$ via the Frobenius inner product: $\langle\bt(\bx),\btheta\rangle:=\btheta_1^{\top}\bx_1+\btheta_2^{\top}\bx_2+\bx_1^{\top}\btheta_{11}\bx_1+\bx_1^{\top}\btheta_{12}\bx_2+\bx_2^{\top}\btheta_{22}\bx_2$. The members of $\Eamb$ are then the densities $p_{\btheta}:=\exp(\langle\bt,\btheta\rangle-\psi_{\mathrm{amb}}(\btheta))$, where $\psi_{\mathrm{amb}}(\btheta):=\log\int\exp\langle\bt,\btheta\rangle\,d\nu$ is the log-partition function defined on $\Hnat_{\mathrm{amb}}:=\{\psi_{\mathrm{amb}}<\infty\}$. Within this ambient space, we identify the subfamilies $\Eprod\subset\Ecyl$, which consist of the product family of independent restrictions \eqref{eq:independent-restriction} and the cylindrical densities \eqref{eq:cyl-density} with parameters in $\calP$. Finally, the expectation coordinates are given by $\bfeta=\nabla\psi_{\mathrm{amb}}(\btheta)$.

\begin{proposition}[The cylindrical family as a curved exponential family]\label{prop:cyl-curved}
The following hold.
\begin{enumerate}
\item[(a)] For $(\balpha,\bB,\bS,\bmu_2,\kappa)\in\calP$, the density \eqref{eq:cyl-density} equals $\exp(\langle\bt(\bx),\btheta\rangle-\psi_{\cyl})$ with
\begin{equation}\label{eq:amb-natural}
\begin{gathered}
\btheta_{11}=-\tfrac12\bS^{-1},\qquad \btheta_{12}=\bS^{-1}\bB,\qquad \btheta_{22}=\bigl(-\tfrac12\bB^{\top}\bS^{-1}\bB\bigr)^{\circ},\\
\btheta_1=\bS^{-1}\balpha,\qquad \btheta_2=\kappa\bmu_2-\bB^{\top}\bS^{-1}\balpha,
\end{gathered}
\end{equation}
\begin{equation}\label{eq:cyl-logpartition}
\psi_{\cyl}=\tfrac12\balpha^{\top}\bS^{-1}\balpha+\tfrac12\log|\bS|+\tfrac d2\log(2\pi)-\log\Cm(\kappa)+\tfrac{1}{2m}\tr(\bB^{\top}\bS^{-1}\bB),
\end{equation}
and these natural parameters satisfy
\begin{equation}\label{eq:cyl-constraint}
\bOmega(\btheta):=\bigl(\btheta_{22}-\tfrac14\btheta_{12}^{\top}\btheta_{11}^{-1}\btheta_{12}\bigr)^{\circ}=\mathbf 0,
\qquad
\bvartheta_2(\btheta):=\btheta_2-\tfrac12\btheta_{12}^{\top}\btheta_{11}^{-1}\btheta_1=\kappa\bmu_2\neq\mathbf 0 .
\end{equation}
\item[(b)] $\Eamb$ is regular and minimal, with $\Hnat_{\mathrm{amb}}=\R^d\times\R^m\times\Symmm{d}\times\R^{d\times m}\times\Symzero{m}$, $\Symmm{d}$ the symmetric negative-definite matrices. Each member factorizes as
\begin{equation*}
p_{\btheta}(\bx)=\fG\bigl(\bx_1;\,-\tfrac12\btheta_{11}^{-1}(\btheta_1+\btheta_{12}\bx_2),\,-\tfrac12\btheta_{11}^{-1}\bigr)\times\FB_m\bigl(\bvartheta_2(\btheta),\bOmega(\btheta)\bigr)(\bx_2),
\end{equation*}
and $\psi_{\mathrm{amb}}(\btheta)$ is the sum of a closed-form Gaussian term and the Fisher--Bingham normalizer $\log\int_{\sphere^{m-1}}\exp(\bvartheta_2^{\top}\bx_2+\bx_2^{\top}\bOmega\bx_2)\,d\sigma(\bx_2)$, which has no closed form \cite{kent1982fisherbingham,kumewood2005saddlepoint,kumeprestonwood2013fisherbingham}; on $\{\bOmega=\mathbf 0\}$ the normalizer equals $-\log\Cm(\norm{\bvartheta_2})$ and $\psi_{\mathrm{amb}}=\psi_{\cyl}$.
\item[(c)] In natural coordinates
\begin{equation*}
\Eprod=\{\btheta_{12}=\mathbf 0,\ \bOmega=\mathbf 0,\ \bvartheta_2\neq\mathbf 0\}\ \subset\ \Ecyl=\{\bOmega=\mathbf 0,\ \bvartheta_2\neq\mathbf 0\}\ \subset\ \Eamb,
\end{equation*}
of dimensions $d+m+\tfrac{d(d+1)}2$, $d+m+\tfrac{d(d+1)}2+dm$ and $d+m+\tfrac{d(d+1)}2+dm+\tfrac{m(m+1)}2-1$, and \eqref{eq:amb-natural} is a bijection of $\calP$ onto $\Ecyl$. $\Ecyl$ is an open subset of the graph of the smooth non-affine map $(\btheta_1,\btheta_2,\btheta_{11},\btheta_{12})\mapsto(\tfrac14\btheta_{12}^{\top}\btheta_{11}^{-1}\btheta_{12})^{\circ}$, hence an embedded submanifold of codimension $\tfrac{m(m+1)}2-1$. $\Ecyl$ is neither e-flat nor m-flat, whereas $\Eprod$ is e-flat, and the constraint $\bOmega=\mathbf 0$ says that the directional marginal of $p_{\btheta}$ is von Mises--Fisher rather than Fisher--Bingham.
\item[(d)] With $\bfeta_1:=\E[\bx_1]$, $\bfeta_2:=\E[\bx_2]$, $\bfeta_{11}:=\E[\bx_1\bx_1^{\top}]$, $\bfeta_{12}:=\E[\bx_1\bx_2^{\top}]$, $\bfeta_{22}:=\E[\bx_2\bx_2^{\top}]$ (the expectation coordinate of the last block of $\bt$ being $\bfeta_{22}-\tfrac1m\bI_m$), the expectation coordinates of the cylindrical family are closed-form:
\begin{equation}\label{eq:cyl-expectation}
\begin{gathered}
\bfeta_2=\Am(\kappa)\bmu_2,\qquad \bfeta_{22}=\tfrac{\Am(\kappa)}\kappa\bI_m+\bigl(1-\tfrac{m\Am(\kappa)}{\kappa}\bigr)\bmu_2\bmu_2^{\top},\qquad \bfeta_1=\balpha+\bB\bfeta_2,\\
\bfeta_{12}=\balpha\bfeta_2^{\top}+\bB\bfeta_{22},\qquad
\bfeta_{11}=\bS+\balpha\balpha^{\top}+\balpha\bfeta_2^{\top}\bB^{\top}+\bB\bfeta_2\balpha^{\top}+\bB\bfeta_{22}\bB^{\top}.
\end{gathered}
\end{equation}
\end{enumerate}
\end{proposition}

The proofs of the proposition are provided in Sections~S7 and~S8. The constraint $\bOmega=\mathbf 0$ is linear in the parameters $(\balpha,\bB,\bS,\bvartheta_2,\bOmega)$ of (b), but these are not affine in $\btheta$. Specifically, since $\btheta_{22}=\bOmega+(\tfrac14\btheta_{12}^{\top}\btheta_{11}^{-1}\btheta_{12})^{\circ}$, the condition $\bOmega=\mathbf 0$ ties $\btheta_{22}$ nonlinearly to $\btheta_{12}$. On the subfamily $\Eprod$, which also factorizes with a von Mises--Fisher marginal, this relationship becomes linear ($\btheta_{22}=\mathbf 0$), and $\Eprod$ is e-flat: the curvature arises from $\bB$, rather than from the sphere or the factorization. In the case $d=1$, $m=2$, this constraint removes $\cos2\theta$ and $\sin2\theta$. By replacing $\kappa\bmu_2$ with $\bvartheta_2$, the mapping \eqref{eq:amb-natural} maps $\Pext$ onto the closed embedded submanifold $\{\bOmega=\mathbf 0\}$. Within this manifold, $\Ecyl$ is open and dense, and the expressions in \eqref{eq:cyl-logpartition} and \eqref{eq:cyl-expectation} extend continuously, with $\bfeta_2=\mathbf 0$ and $\bfeta_{22}=\bI_m/m$ at $\bvartheta_2=\mathbf 0$.

\subsection{Maximum likelihood, identifiability and Fisher metric}\label{subsec:cyl-mle}

\begin{definition}[Cut~\cite{barndorffnielsen1978information}]\label{def:cut}
Let $\{p_{\boldsymbol\omega}\}$ be a family of densities of $\bx=(\bx_1,\bx_2)$. The statistic $\bx_2$ is a \emph{cut} if the parameter can be written $\boldsymbol\omega=(\boldsymbol\lambda,\boldsymbol\varphi)$ with $(\boldsymbol\lambda,\boldsymbol\varphi)$ ranging over a product set $\Lambda\times\Phi$ and $p_{\boldsymbol\omega}(\bx)=p(\bx_2;\boldsymbol\varphi)\,p(\bx_1\mid\bx_2;\boldsymbol\lambda)$.
\end{definition}

For the cylindrical family on $\Pext$, we observe that $\bx_2$ is a cut with parameters $\boldsymbol\lambda=(\balpha,\bB,\bS)$ and $\boldsymbol\varphi=\bvartheta_2$. This structure ensures that the likelihood splits into a conditional part and a marginal part, which can be maximized separately. Consequently, $\bx_2$ is S-sufficient for $\bvartheta_2$ and S-ancillary for $(\balpha,\bB,\bS)$~\cite{barndorffnielsen1978information}, where the S in these standard terms is unrelated to the covariance $\bS$. We state the conditions using weights, as this formulation is required for Section~\ref{sec:mix}.

\begin{assumption}[Design conditions]\label{ass:design}
Let $(\bx_{1n},\bx_{2n})\in\R^d\times\sphere^{m-1}$, $n=1,\dots,N$, be observations with weights $w_n\ge0$, $W:=\sum_nw_n>0$. Let $\bar{\bx}_i^{w}:=W^{-1}\sum_nw_n\bx_{in}$, $\sct{ij}^{w}:=\sum_nw_n(\bx_{in}-\bar{\bx}_i^{w})(\bx_{jn}-\bar{\bx}_j^{w})^{\top}$ ($i,j\in\{1,2\}$), $\br^{w}:=\sum_nw_n\bx_{2n}$, $\bz_n:=(1,\bx_{2n}^{\top})^{\top}$, and let $E^{w}:=\sum_nw_n(\bx_{1n}-\hat{\bbeta}\bz_n)(\bx_{1n}-\hat{\bbeta}\bz_n)^{\top}$ be the residual scatter of the weighted least-squares regression of $\bx_1$ on $\bz$, for any solution $\hat{\bbeta}$ of its normal equations (it does not depend on the choice). The conditions, on the observations with positive weight, are:
\begin{enumerate}
\item[(A0w)] $\norm{\br^{w}}<W$: the directions are not all equal;
\item[(A1w)] $\rank(\sct{22}^{w})=m$: the directions lie in no affine hyperplane of $\R^m$;
\item[(A2w)] $E^{w}\succ\mathbf 0$: no nonzero linear combination of $\bx_1$ is an exact affine function of $\bx_2$; under (A1w), $\sct{11}^{w}-\sct{12}^{w}(\sct{22}^{w})^{-1}(\sct{12}^{w})^{\top}\succ\mathbf 0$;
\item[(A3w)] $\br^{w}\neq\mathbf 0$.
\end{enumerate}
With unit weights we write (A0)--(A3) and drop the superscript $w$.
\end{assumption}

\begin{theorem}[Maximum-likelihood estimator of the cylindrical distribution]\label{thm:cyl-mle}
Let the observations, weights and statistics be as in Assumption~\ref{ass:design}, and $\ell_w(\btheta):=\sum_nw_n\log\fcyl(\bx_{1n},\bx_{2n};\btheta)$ for $\btheta\in\Pext$.
\begin{enumerate}
\item[(i)] \emph{Existence.} $\ell_w$ has a maximizer over $\Pext$ if and only if (A0w) and (A2w) hold; otherwise $\sup_{\Pext}\ell_w=+\infty$. Every maximizer has $\hat{\bS}=E^{w}/W$, $\hat{\bvartheta}_2=\nabla\phi(\br^{w}/W)$ (the von Mises--Fisher natural parameter with mean $\br^{w}/W$) and, as $[\hat{\balpha}\ \ \hat{\bB}]$, a weighted least-squares coefficient of $\bx_1$ on $(1,\bx_2)$.
\item[(ii)] \emph{Uniqueness.} The maximizer is unique if and only if (A1w) also holds, and it is then
\begin{equation}\label{eq:mle-formulas}
\begin{gathered}
\hat{\bB}=\sct{12}^{w}(\sct{22}^{w})^{-1},\qquad \hat{\balpha}=\bar{\bx}_1^{w}-\hat{\bB}\bar{\bx}_2^{w},\qquad \hat{\bS}=\frac1W\bigl(\sct{11}^{w}-\sct{12}^{w}(\sct{22}^{w})^{-1}(\sct{12}^{w})^{\top}\bigr),\\
\hat{\bvartheta}_2=\hat\kappa\hat{\bmu}_2,\qquad \hat{\bmu}_2=\frac{\br^{w}}{\norm{\br^{w}}},\qquad \hat\kappa=\Am^{-1}\Bigl(\frac{\norm{\br^{w}}}W\Bigr),
\end{gathered}
\end{equation}
with $\hat{\bvartheta}_2=\mathbf 0$ if $\br^{w}=\mathbf 0$.
\item[(iii)] \emph{The polar chart.} Over $\calP$, a maximizer exists if and only if (A0w), (A2w) and (A3w) hold, and the maximizers are those of (i), with $\hat{\bSigma}_{12}=\hat{\bB}/\hat\kappa$, $\hat{\bmu}_1=\hat{\balpha}+\hat{\bB}\hat{\bmu}_2$ in the original coordinates \eqref{eq:cyl-original-params}; under (A2w) with $\br^{w}=\mathbf 0$, the supremum over $\calP$ is finite and is approached, not attained, as $\kappa\downarrow0$.
\end{enumerate}
Every block is in closed form except $\hat\kappa$, the unique root of the strictly increasing scalar equation $\Am(\kappa)=\norm{\br^{w}}/W$.
\end{theorem}

The proof of the theorem is provided in Appendix~\ref{app:theory-proofs}, while Section~S7 examines the boundary cases and explains why dependence is reported through $\bB$. The formulas in \eqref{eq:mle-formulas} correspond to least squares on the embedded direction~\cite{juppmardia1980correlation}, the von Mises--Fisher estimator~\cite[Ch.~10]{mardiajupp2000directional}, and, at $m=2$, to those in~\cite{mardia1978cylindrical,luengo2019directional}.

\begin{proposition}[Identifiability from the first two moments]\label{prop:cyl-identifiable}
Let $(\balpha,\bB,\bS,\bvartheta_2)\in\Pext$, and let $\bfeta_2:=\E[\bx_2]$, $\mathbf V:=\cov(\bx_2)$, $\mathbf K_{12}:=\cov(\bx_1,\bx_2)$ and $\mathbf K_{11}:=\cov(\bx_1)$ under $\fcyl(\cdot;\balpha,\bB,\bS,\bvartheta_2)$. The map $\bvartheta_2\mapsto\bfeta_2$ is a diffeomorphism of $\R^m$ onto the open unit ball, with inverse $\nabla\phi$ of Section~\ref{subsec:flat-families}, and $\mathbf V\succ\mathbf 0$. Hence
\begin{equation*}
\bvartheta_2=\nabla\phi(\bfeta_2),\qquad \bB=\mathbf K_{12}\mathbf V^{-1},\qquad \balpha=\E[\bx_1]-\bB\bfeta_2,\qquad \bS=\mathbf K_{11}-\mathbf K_{12}\mathbf V^{-1}\mathbf K_{12}^{\top},
\end{equation*}
and $(\balpha,\bB,\bS,\bvartheta_2)\mapsto\fcyl$ is injective on $\Pext$; so is $(\balpha,\bB,\bS,\bmu_2,\kappa)\mapsto\fcyl$ on $\calP$, and likewise in the original coordinates.
\end{proposition}

Section~S8 proves the proposition from the moments of \eqref{eq:cyl-regression}.

In a curved exponential family, the maximum-likelihood estimate is obtained as the m-projection of the observed point $\bar{\bt}$ onto the model, along the ancillary family through the estimate~\cite{amari1982curved,amari1985differential}. For the cylindrical family, both the m-projection and the ancillary family are explicit.

\begin{proposition}[The estimator as an m-projection]\label{prop:cyl-mprojection}
Let $(\bx_{1n},\bx_{2n})\in\R^d\times\sphere^{m-1}$, $n=1,\dots,N$, be observations, $\bar{\bt}:=N^{-1}\sum_n\bt(\bx_n)$, with blocks $\bar{\bx}_1$, $\bar{\bx}_2$, $\bar{\bt}_{11}$, $\bar{\bt}_{12}$ and $\bar{\mathbf M}_{22}-\tfrac1m\bI_m$, where $\bar{\mathbf M}_{22}:=N^{-1}\sum_n\bx_{2n}\bx_{2n}^{\top}$, and let $\calM_{\mathrm{amb}}:=\nabla\psi_{\mathrm{amb}}(\Hnat_{\mathrm{amb}})$ and $\phi_{\mathrm{amb}}$ be the mean space and dual potential of $\Eamb$.
\begin{enumerate}
\item[(i)] \emph{m-projection.} $\bar{\bt}\in\calM_{\mathrm{amb}}$ if and only if (A1)--(A2) of Assumption~\ref{ass:design} hold. Then a unique $p_{\bar{\bt}}\in\Eamb$ has $\E_{p_{\bar{\bt}}}[\bt(\bx)]=\bar{\bt}$, with conditional block $\fCG(\cdot\mid\bx_2;\hat{\balpha},\hat{\bB},\hat{\bS})$ and a Fisher--Bingham directional marginal. For every $\btheta\in\Hnat_{\mathrm{amb}}$,
\begin{equation*}
\KL[p_{\bar{\bt}}:p_{\btheta}]=\phi_{\mathrm{amb}}(\bar{\bt})+\psi_{\mathrm{amb}}(\btheta)-\langle\btheta,\bar{\bt}\rangle,\qquad
\tfrac1N\ell(\btheta)=\langle\btheta,\bar{\bt}\rangle-\psi_{\mathrm{amb}}(\btheta)\ \text{ on }\Ecyl,
\end{equation*}
so if (A3) also holds, the natural parameter $\hat{\btheta}\in\Ecyl$, by \eqref{eq:amb-natural}, of the estimator of Theorem~\ref{thm:cyl-mle} is $\argmin_{\btheta\in\Ecyl}\KL[p_{\bar{\bt}}:p_{\btheta}]$: the estimator is the m-projection of $p_{\bar{\bt}}$ onto the curved family, and it differs from $p_{\bar{\bt}}$ only in the directional marginal, where the Fisher--Bingham law is replaced by the von Mises--Fisher law with the same first moment $\bar{\bx}_2$.
\item[(ii)] \emph{Unmatched moments.} Under (A1)--(A3), at the estimator, with $\bDelta:=\bar{\mathbf M}_{22}-\hat{\bfeta}_{22}$, traceless because both matrices have unit trace,
\begin{equation}\label{eq:mprojection-mismatch}
\begin{gathered}
\hat{\bfeta}_1=\bar{\bx}_1,\qquad \hat{\bfeta}_2=\bar{\bx}_2,\qquad
\hat{\bfeta}_{22}-\bar{\mathbf M}_{22}=-\bDelta,\\
\hat{\bfeta}_{12}-\bar{\bt}_{12}=-\hat{\bB}\bDelta,\qquad
\hat{\bfeta}_{11}-\bar{\bt}_{11}=-\hat{\bB}\bDelta\hat{\bB}^{\top},
\end{gathered}
\end{equation}
so the score of the ambient log-likelihood $\sum_n\log p_{\btheta}(\bx_n)$ at $\hat{\btheta}$ is $N(\bar{\bt}-\nabla\psi_{\mathrm{amb}}(\hat{\btheta}))=N(\mathbf 0,\mathbf 0,\hat{\bB}\bDelta\hat{\bB}^{\top},\hat{\bB}\bDelta,\bDelta)$. It annihilates the tangent space of $\Ecyl$ at $\hat{\btheta}$, because $\hat{\btheta}$ maximizes the likelihood on $\Ecyl$, and it is determined by its $\btheta_{22}$ block $N\bDelta$, which measures the departure of the directional second moment from the von Mises--Fisher constraint $\bOmega=\mathbf 0$.
\end{enumerate}
\end{proposition}

Section~S8 proves the proposition; $\bDelta$ is the ancillary coordinate of $\bar{\bt}$, which Section~\ref{sec:geomtest} uses.

\begin{proposition}[Fisher metric]\label{prop:cyl-fisher}
At $(\balpha,\bB,\bS,\bvartheta_2)\in\Pext$, with $\bz:=(1,\bx_2^{\top})^{\top}$, $\bbeta:=[\balpha\ \ \bB]$, $\mathbf V:=\cov(\bx_2)$ and $\mathbf D_d$ the duplication matrix, $\operatorname{vec}(\bS)=\mathbf D_d\operatorname{vech}(\bS)$, the Fisher information in the cut coordinates $(\operatorname{vec}\bbeta,\operatorname{vech}\bS,\bvartheta_2)$ is
\begin{equation}\label{eq:fisher-info}
\calI=\blockdiag\Bigl(\E[\bz\bz^{\top}]\otimes\bS^{-1},\ \ \tfrac12\mathbf D_d^{\top}(\bS^{-1}\otimes\bS^{-1})\mathbf D_d,\ \ \mathbf V\Bigr),
\end{equation}
positive definite on all of $\Pext$; it is the pullback $(\partial\btheta_{\cyl})^{\top}\nabla^2\psi_{\mathrm{amb}}\,\partial\btheta_{\cyl}$ of the ambient metric by the map $\btheta_{\cyl}$ of \eqref{eq:amb-natural}.
\end{proposition}

Section~S8 proves Propositions~\ref{prop:cyl-fisher}--\ref{prop:cyl-kl} and Corollary~\ref{cor:cyl-dependence-gain}, while Proposition~S4 provides the squared length of a tangent vector, the polar form and the Hessian of $\psi_{\mathrm{amb}}\circ\btheta_{\cyl}$. Within this structure, orthogonality between $(\bbeta,\bS)$ and $\bvartheta_2$ holds for every cut, and orthogonality between $\bbeta$ and $\bS$ holds for every Gaussian conditional. As in any regular curved exponential family~\cite{barndorffnielsen1978information,kassvos1997geometrical}, $\sqrt N$ times the estimation error is asymptotically $\calN(\mathbf 0,\calI^{-1})$, and the block estimates are asymptotically independent. In the coordinates $\operatorname{vech}\bS$, the middle block of~\eqref{eq:fisher-info} defines the quadratic form $\tfrac12\operatorname{tr}(\bS^{-1}\,\mathrm d\bS\,\bS^{-1}\,\mathrm d\bS)$. This is half the affine-invariant metric on symmetric positive-definite matrices~\cite{pennec2006riemannian}, a metric used by geometric Hamiltonian neural networks for inertia matrices~\cite{aboussalah2025geohnns}. Finally, the inverse Fisher information is also the asymptotic covariance of estimators fitted to time series augmented by recursive interpolation~\cite{aboussalah2023recursive}.

\begin{proposition}[Kullback--Leibler divergence between two components]\label{prop:cyl-kl}
Let $\btheta=(\balpha,\bB,\bS,\bvartheta_2)$ and $\btheta'=(\balpha',\bB',\bS',\bvartheta_2')$ lie in $\Pext$, and let $\bfeta_2$ and $\mathbf V$ be the mean and the covariance of $\bx_2$ under $\btheta$. Then
\begin{equation}\label{eq:cyl-kl}
\begin{split}
\KL\bigl[\fcyl(\cdot;\btheta):\fcyl(\cdot;\btheta')\bigr]
={}&\log\frac{\Cm(\norm{\bvartheta_2})}{\Cm(\norm{\bvartheta_2'})}+(\bvartheta_2-\bvartheta_2')^{\top}\bfeta_2+\tfrac12\Bigl[\tr(\bS'^{-1}\bS)-d+\log\frac{|\bS'|}{|\bS|}\Bigr]\\
&+\tfrac12\bigl(\balpha-\balpha'+(\bB-\bB')\bfeta_2\bigr)^{\top}\bS'^{-1}\bigl(\balpha-\balpha'+(\bB-\bB')\bfeta_2\bigr)\\
&+\tfrac12\tr\bigl(\bS'^{-1}(\bB-\bB')\mathbf V(\bB-\bB')^{\top}\bigr).
\end{split}
\end{equation}
\end{proposition}

Applying the chain rule for a cut, we find that the first two terms represent the directional divergence, while the remaining terms average the conditional Gaussian divergence over $\bx_2$.

\begin{corollary}[Likelihood gain of direct dependence]\label{cor:cyl-dependence-gain}
Let observations satisfy (A1)--(A2), let $\hat\ell_{\cyl}$ be the maximized log-likelihood of Theorem~\ref{thm:cyl-mle} and $\hat\ell_{\mathrm{ind}}$ that of the independent restriction \eqref{eq:independent-restriction}, whose maximizer is $\hat{\balpha}_{\mathrm{ind}}=\bar{\bx}_1$, $\hat{\bS}_{\mathrm{ind}}=\sct{11}/N$ and the same $\hat{\bvartheta}_2$, and let $\hat\rho_1,\dots,\hat\rho_{\min(d,m)}$ be the sample canonical correlations between $\bx_1$ and $\bx_2$. Then
\begin{equation}\label{eq:dependence-gain}
\begin{aligned}
\hat\ell_{\cyl}-\hat\ell_{\mathrm{ind}}
&=\frac N2\log\frac{|\sct{11}|}{|\sct{11}-\sct{12}\sct{22}^{-1}\sct{12}^{\top}|}
=-\frac N2\sum_{i=1}^{\min(d,m)}\log(1-\hat\rho_i^{2})\\
&=\sum_{n=1}^N\KL\bigl[\calN(\hat{\bmu}_c(\bx_{2n}),\hat{\bS}):\calN(\bar{\bx}_1,\hat{\bS}_{\mathrm{ind}})\bigr];
\end{aligned}
\end{equation}
the von Mises--Fisher term cancels.
\end{corollary}

Twice the gain equals Wilks' statistic for independence between $\bx_1$ and the embedded direction, written in the canonical correlations of Jupp and Mardia~\cite{juppmardia1980correlation}. When $\bB=\mathbf 0$, this statistic tends to a $\chi^2_{dm}$ distribution~\cite[Ch.~16]{vandervaart1998asymptotic}. Given the directions, which are S-ancillary for $(\balpha,\bB,\bS)$, the null law is exact for every $N\ge d+m+1$: specifically, $\prod_i(1-\hat\rho_i^2)$ follows Wilks' distribution $\Lambda(d,m,N-m-1)$~\cite[Ch.~8]{anderson2003multivariate}. Against this exact law, the $\chi^2_{dm}$ limit is liberal at small $N$ (Section~S8).

\section{Mixtures of cylindrical distributions: identifiability and the EM algorithm}
\label{sec:mix}

Finite mixtures preserve the closed forms described in Section~\ref{sec:cyl}: they allow for an explicit E-step and, by Theorem~\ref{thm:cyl-mle} with weights, an exact M-step.

\subsection{Model and identifiability}
\label{subsec:mix-constraints}

For parameters $\btheta_j\in\Pext$ and weights $\pi_j>0$ satisfying $\sum_{j=1}^K\pi_j=1$, the cylindrical mixture density is
\begin{equation}
\label{eq:mix-density}
p(\bx_1,\bx_2;\bTheta):=\textstyle\sum_{j=1}^K\pi_j\,\fcyl(\bx_1,\bx_2;\btheta_j),\qquad \bTheta:=\{\pi_j,\btheta_j\}_{j=1}^K.
\end{equation}
This structure represents a linear Gaussian cluster-weighted model~\cite{gershenfeld1997cwm,ingrassia2012cwm} with a von Mises--Fisher covariate density. For $m=2$, this is a special case of the mixtures described by~\cite{luengo2019directional}. If every $\bB_j=\mathbf 0$, the model reduces to the \emph{independent cylindrical mixture}. Regarding the uniqueness of these models, finite mixtures of a family are identifiable exactly when its distinct members are linearly independent~\cite{yakowitz1968identifiability}.

\begin{theorem}[Identifiability of finite cylindrical mixtures]\label{thm:mix-identifiable}
\begin{enumerate}
\item[(i)] If $\btheta_1,\dots,\btheta_J\in\Pext$ are distinct, the functions $\fcyl(\cdot;\btheta_j)$, $j=1,\dots,J$, on $\R^d\times\sphere^{m-1}$ are linearly independent.
\item[(ii)] If $\sum_{j=1}^K\pi_j\fcyl(\cdot;\btheta_j)=\sum_{k=1}^{K'}\pi'_k\fcyl(\cdot;\btheta'_k)$ $\nu$-almost everywhere, with positive weights summing to one, $\btheta_1,\dots,\btheta_K$ distinct and $\btheta'_1,\dots,\btheta'_{K'}$ distinct in $\Pext$, then $K'=K$ and there is a permutation $\sigma$ of $\{1,\dots,K\}$ with $\pi'_{\sigma(j)}=\pi_j$ and $\btheta'_{\sigma(j)}=\btheta_j$ for every $j$.
\end{enumerate}
This includes components that share their regression part $(\balpha,\bB,\bS)$ or their direction, components with uniform directions, and the independent mixture; on $\calP$ it holds equally in polar and original coordinates.
\end{theorem}

The proof of identifiability (Appendix~\ref{app:theory-proofs}) begins by fixing the direction, which ensures that $\bx_1$ is a finite Gaussian mixture~\cite{yakowitz1968identifiability}. Off a null set of directions, components that share the same conditional Gaussian distribution must also share $(\balpha,\bB,\bS)$; within such a class, we identify the von Mises--Fisher mixture directly (cf.~\cite{fraser1981vonmises,kent1983identifiability}). While identifiability is already known for mixtures of linear regressions~\cite{hennig2000identifiability} and Gaussian-covariate cluster-weighted models~\cite{ingrassia2015glmcwm,ingrassia2014linearcwm}, Theorem~\ref{thm:mix-identifiable} adds a von Mises--Fisher covariate, components sharing $(\balpha,\bB,\bS)$, and the case where $\kappa=0$.

The log-likelihood $\ell(\bTheta):=\sum_n\log p(\bx_{1n},\bx_{2n};\bTheta)$ is unbounded for $K\ge2$~\cite{hathaway1985constrained}. This occurs because a component can collapse, $\bS_j\to\mathbf 0$, onto $m+1$ observations in general position, or concentrate, $\kappa_j\to\infty$, on tied directions, which the size rule described in Section~\ref{subsec:mix-em} does not detect. While constraints can bound the likelihood (Lemma~S8), the fits reported in this paper use none.

\subsection{The EM algorithm}
\label{subsec:mix-em}

At the current iterate, the E-step computes the responsibilities
\begin{equation}
r_{nj}:=\pi_j\,\fcyl(\bx_{1n},\bx_{2n};\btheta_j)\big/p(\bx_{1n},\bx_{2n};\bTheta), \label{eq:mix-resp}
\end{equation}
through the e-projection described in Section~\ref{subsec:mixtures-em}. In the M-step, we set $\pi_j=N_j/N$, where $N_j:=\sum_nr_{nj}$, and maximize the weighted log-likelihood for each component using weights $w_n=r_{nj}$. Let $\bar{\bx}_{i,j}$, $\mathbf C_{ik,j}:=\sct{ik}^{w}/N_j$, $\bs_{2,j}:=\br^{w}$ and $\bar R_j:=\norm{\bs_{2,j}}/N_j$ be the weighted statistics under Assumption~\ref{ass:design}. Theorem~\ref{thm:cyl-mle} then provides the exact M-step: since every $r_{nj}>0$, a maximizer exists if and only if (A0) and (A2) hold, and it is unique if and only if (A1) also holds (Lemma~S7). The maximizer is then given by
\begin{gather*}
\hat{\bB}_j=\mathbf C_{12,j}\mathbf C_{22,j}^{-1},\quad
\hat{\balpha}_j=\bar{\bx}_{1,j}-\hat{\bB}_j\bar{\bx}_{2,j},\\
\hat{\bS}_j=\mathbf C_{11,j}-\mathbf C_{12,j}\mathbf C_{22,j}^{-1}\mathbf C_{12,j}^{\top},\quad
\hat{\bvartheta}_{2,j}=\hat\kappa_j\,\frac{\bs_{2,j}}{\norm{\bs_{2,j}}},
\end{gather*}
where $\hat\kappa_j$ is the root of $\Am(\kappa)=\bar R_j$ (with $\hat{\bvartheta}_{2,j}=\mathbf 0$ if $\bs_{2,j}=\mathbf 0$). This corresponds to the cluster-weighted M-step~\cite{ingrassia2012cwm} with a von Mises--Fisher block; for the independent mixture, $\hat{\bB}_j=\mathbf 0$, $\hat{\balpha}_j=\bar{\bx}_{1,j}$, and $\hat{\bS}_j=\mathbf C_{11,j}$. The algorithm monitors the average log-likelihood and the effective component sizes,
\begin{gather}
\bar\ell^{(t)}:=\ell(\bTheta^{(t)})/N,\label{eq:mix-avg-ll}\\
N_{j,\mathrm{eff}}:=\Bigl(\textstyle\sum_{n=1}^N r_{nj}\Bigr)^2\Big/\textstyle\sum_{n=1}^N r_{nj}^2,\label{eq:mix-ess}
\end{gather}
discarding any start as degenerate if some $N_{j,\mathrm{eff}}<N_{\min}:=d+m+1$, which is the smallest size for which (A1)--(A2) can hold. Algorithm~S1 executes ten $k$-means starts (Section~\ref{subsec:mix-init}) until the increase in $\bar\ell^{(t)}$ is at most $10^{-8}\max\{1,|\bar\ell^{(t)}|\}$ (up to $10^{4}$ iterations), selecting the non-degenerate fit with the largest final $\bar\ell$.

Since every iteration is an exact EM step, the average log-likelihood $\bar\ell^{(t)}$ is nondecreasing~\cite{dempster1977em}. If the iterates remain within a compact set, every limit point is stationary~\cite{wu1983em}, although different starts may converge to different stationary points. We provide further technical details in Section~S9, including the root-finding procedure, a split start method based on a $(K-1)$-component fit, and optional safeguards that were not used in the reported fits.

\subsection{Initialization and parameter counts}
\label{subsec:mix-init}\label{subsec:mix-selection}

Each start is initialized using $k$-means labels based on a dissimilarity measure. This measure sums the scaled Mahalanobis and cosine dissimilarities of the Euclidean and directional blocks (Section~S9). Because this process ignores $\bB_j$, no start targets components that differ only in their $\bB_j$ values. We consider several model structures: cylindrical mixtures, independent mixtures, and the Euclidean-first and directional-first hierarchies described in Section~\ref{sec:hier} (consisting of $K$ classes with $L$ local components). The number of free parameters for these models is given by
\begin{align}
\begin{split}
q_{\cyl}(K,d,m)&=(K-1)+K\bigl[d+dm+\tfrac{d(d+1)}2+m\bigr],\qquad q_{\mathrm{ind}}=q_{\cyl}-Kdm,\\
q_{\mathrm{hier}}(K,L,d,m)&=(K-1)+K\bigl[d+\tfrac{d(d+1)}2+(L-1)+Lm\bigr],
\end{split}\label{eq:param-counts}\\
q^{\mathrm{dir}}_{\mathrm{hier}}(K,L,d,m)&=(K-1)+K\bigl[m+(L-1)+L\bigl(d+\tfrac{d(d+1)}2\bigr)\bigr].\label{eq:param-count-dir}
\end{align}
Section~S9 evaluates these counts, with the information criteria reported in Online Resource~1.

\section{The m-projection residual: a test and the inferential cost of curvature}
\label{sec:geomtest}

The unmatched moment $\bDelta$ from Proposition~\ref{prop:cyl-mprojection} is the score of the von Mises--Fisher assumption. This expression allows us to develop a test in Section~\ref{subsec:geomtest-one}. Next, when paired with the exponential curvature, $\bDelta$ gives the gap between the observed and expected information for $\bB$, which measures the information lost by $\hat{\bB}$ (Section~\ref{subsec:geomtest-curv}). Finally, Section~\ref{subsec:geomtest-mix} treats mixtures.

\subsection{The cut, and a test of the von Mises--Fisher component}
\label{subsec:geomtest-one}

Several facts below hold for any model with a cut~\cite{barndorffnielsen1978information}; we state them here without claiming them. By Proposition~\ref{prop:cyl-curved}(b), the map $\btheta\mapsto\bxi:=(\balpha,\bB,\bS,\bvartheta,\bOmega)$ is a diffeomorphism of $\Hnat_{\mathrm{amb}}$ onto $\R^d\times\R^{d\times m}\times\Sympp{d}\times\R^m\times\Symzero{m}$, where $\bS=-\tfrac12\btheta_{11}^{-1}$, $\bB=\bS\btheta_{12}$, $\balpha=\bS\btheta_1$ and $(\bvartheta,\bOmega)=(\bvartheta_2(\btheta),\bOmega(\btheta))$ as defined in \eqref{eq:cyl-constraint}. In these \emph{cut coordinates}, the density is $p_{\btheta}=\fCG(\cdot\mid\bx_2;\balpha,\bB,\bS)\,\FB_m(\bvartheta,\bOmega)$ with variation-independent blocks. Consequently, $\bx_2$ is S-sufficient for $(\bvartheta,\bOmega)$ and S-ancillary for $(\balpha,\bB,\bS)$. This structure ensures that the Fisher metric $g$ of $\Eamb$ is block-diagonal in $\bxi$, and likelihood tests on $(\bvartheta,\bOmega)$ depend only on the directions. In this setting, the likelihood ratio is the deviance~\cite{efron1978geometry}, the score statistic is a squared normal length~\cite{amari1985differential}, and these two are connected by a Bregman remainder. The cylindrical components ($\kappa\ge0$, Theorem~\ref{thm:cyl-mle}) form $\{\bOmega=\mathbf 0\}$. Thus, the component assumption is $H_0:\bOmega=\mathbf 0$, which imposes $q_m:=\tfrac{m(m+1)}2-1$ restrictions.

To analyze these statistics, we first define the moments of the directional data. For $\bvartheta\in\R^m$, let $\bfeta_{22}(\bvartheta):=\E\bx_2\bx_2^{\top}$ and $\mathbf V(\bvartheta):=\cov(\bx_2)$ under $\vMF_m(\bvartheta)$, noting that $\bfeta_{22}(\mathbf 0)=\bI_m/m$. We define the \emph{efficient information} as a positive-definite form on $\Symzero{m}$, with $\langle\mathbf A,\mathbf A'\rangle:=\tr(\mathbf A\mathbf A')$ the inner product,
\begin{equation}\label{eq:geomtest-ieff}
\Ieff(\bvartheta)[\mathbf A,\mathbf A]:=\min_{\mathbf b\in\R^m}\V\bigl(\bx_2^{\top}\mathbf A\bx_2-\mathbf b^{\top}\bx_2\bigr),\qquad\bx_2\sim\vMF_m(\bvartheta),
\end{equation}
which we identify with a self-adjoint operator on $\Symzero{m}$. We now introduce the necessary estimators and statistics. Let $\hat{\btheta}$ be the maximum-likelihood estimate from Theorem~\ref{thm:cyl-mle} in natural coordinates, with $\hat{\bvartheta}:=\hat\kappa\hat{\bmu}_2$. We let $\hat{\bfeta}:=\nabla\psi_{\mathrm{amb}}(\hat{\btheta})$ be the corresponding mean, $\bar{\bt}$ be the average statistic, and $p_{\bar{\bt}}$ be the ambient member with mean $\bar{\bt}$ (Proposition~\ref{prop:cyl-mprojection}). Next, we define the residual $\bDelta:=\bar{\mathbf M}_{22}-\bfeta_{22}(\hat{\bvartheta})$, where $\bar{\mathbf M}_{22}:=N^{-1}\sum_n\bx_{2n}\bx_{2n}^{\top}$, and let $\gamma(s):=p_{\hat{\bfeta}+s(\bar{\bt}-\hat{\bfeta})}$ for $s\in[0,1]$ be the m-geodesic from $p_{\hat{\btheta}}$ to $p_{\bar{\bt}}$. Finally, we define the score statistic~\cite{rao1948score} as $T_N:=N(\bar{\bt}-\hat{\bfeta})^{\top}\nabla^2\psi_{\mathrm{amb}}(\hat{\btheta})^{-1}(\bar{\bt}-\hat{\bfeta})$ and the likelihood ratio as $\LRstat_N:=2\sup\ell_{\mathrm{amb}}-2\sup_{\bOmega=\mathbf 0}\ell_{\mathrm{amb}}$, where $\ell_{\mathrm{amb}}(\btheta):=\sum_n\log p_{\btheta}(\bx_n)$.

\begin{theorem}[Classical tests of the von Mises--Fisher law~\cite{cox1975discussion,mardiaholmeskent1984gof}, in cut coordinates]\label{thm:geomtest}
Let the observations satisfy (A1)--(A2).
\begin{enumerate}
\item[(i)] $\dot\gamma(0)$ is $g$-normal to $\{\bOmega=\mathbf 0\}$ at $\hat{\btheta}$, has no $(\balpha,\bB,\bS)$-component in cut coordinates, and $T_N=Ng(\dot\gamma(0),\dot\gamma(0))=N\langle\bDelta,\Ieff(\hat{\bvartheta})^{-1}\bDelta\rangle$.
\item[(ii)] $\LRstat_N=2N\KL[\FB^{*}:\vMF_m(\hat{\bvartheta})]=2N\int_0^1(1-s)\,g(\dot\gamma(s),\dot\gamma(s))\,ds$, with $\FB^{*}$ the directional marginal of $p_{\bar{\bt}}$; hence
\[
|\LRstat_N/T_N-1|\le\sup_{s\le1}|g(\dot\gamma(s),\dot\gamma(s))/g(\dot\gamma(0),\dot\gamma(0))-1|.
\]
\item[(iii)] \emph{Exact invariance.} For independent draws from any ambient member $\fCG(\cdot\mid\bx_2;\balpha,\bB,\bS)\FB_m(\bvartheta,\bOmega)$, $T_N$ and $\LRstat_N$ are functions of $N$ independent $\FB_m(\bvartheta,\bOmega)$ directions, so for every $N\ge d+m+1$ their law, and the size and power of the test, do not depend on $(\balpha,\bB,\bS)$.
\item[(iv)] Under $\bOmega=\mathbf 0$ (any $\kappa\ge0$), (A1)--(A2) hold almost surely for $N\ge d+m+1$, and $T_N$, $\LRstat_N\to\chi^2_{q_m}$ with $\LRstat_N-T_N\to0$ in probability. Under $\bOmega=N^{-1/2}\boldsymbol\Upsilon$, $\boldsymbol\Upsilon\in\Symzero{m}$, both converge to $\chi^2_{q_m}(\Ieff(\bvartheta)[\boldsymbol\Upsilon,\boldsymbol\Upsilon])$, and $\Ieff(\bvartheta)[\boldsymbol\Upsilon,\boldsymbol\Upsilon]=\lim_N2N\inf_p\KL[p_N:p]$ over cylindrical $p$, $p_N$ the law of the observations.
\end{enumerate}
\end{theorem}

The proof of the theorem is provided in Section~S14.1. According to (ii), the two statistics agree to relative order $O_p(N^{-1/2})$. We use the score form because it is closer to nominal at the 5\% level for small $N$, although its far tail is heavier than $\chi^2_{q_m}$ at high $\kappa$ (Tables~S10--S11). To report the effect size, we use $\LRstat_N/(2N)=\KL[\FB^{*}:\vMF_m(\hat{\bvartheta})]$, expressed in nats per observation. Finally, by (iii), the curvature of $\Ecyl$ (Proposition~\ref{prop:geomtest-curv}) cannot affect the test at any order, as its null family $\vMF_m\subset\FB_m$ is e-flat.

\begin{proposition}[Closed form and classical special cases]\label{prop:geomtest-ieff}
Let $\kappa>0$, $t:=\bmu_2^{\top}\bx_2$ with $\bx_2\sim\vMF_m(\bmu_2,\kappa)$, $\varrho_j(\kappa):=I_{m/2-1+j}(\kappa)/I_{m/2-1}(\kappa)$, $\Pperp:=\bI_m-\bmu_2\bmu_2^{\top}$ and $\mathbf R:=\bmu_2\bmu_2^{\top}-\Pperp/(m-1)$. On the orthogonal decomposition of $\Symzero{m}$ into the radial line $\R\mathbf R$, the coupling space $\{\bmu_2\mathbf b^{\top}+\mathbf b\bmu_2^{\top}:\mathbf b\perp\bmu_2\}$ and the tangential space $\{\mathbf C=\Pperp\mathbf C\Pperp:\tr\mathbf C=0\}$, of dimensions $1$, $m-1$, $\tfrac{m(m-1)}2-1$, $\Ieff(\kappa\bmu_2)$ is multiplication by
\begin{equation}\label{eq:geomtest-blocks}
\nu_{\mathrm{rad}}=\frac{m}{m-1}\,\frac{\det[\E t^{i+j}]_{i,j=0}^{2}}{\Am'(\kappa)},\qquad
\nu_{\mathrm{cpl}}=\frac{2\Am(\kappa)A_{m+2}'(\kappa)}{\kappa},\qquad
\nu_{\mathrm{tan}}=\frac{2\varrho_2(\kappa)}{\kappa^2},
\end{equation}
with $\E t^k$ polynomials in $\varrho_1,\dots,\varrho_4$ and $1/\kappa$ (Appendix~\ref{app:geomtest}); all three tend to $2/(m(m+2))$ as $\kappa\downarrow0$. Hence $T_N$ is the sum of a radial, a coupling and a tangential part, asymptotically independent $\chi^2_1$, $\chi^2_{m-1}$ and $\chi^2_{m(m-1)/2-1}$ under $\bOmega=\mathbf 0$ (written out in Section~S14).
\end{proposition}

Parts (i), (ii) and the null limit in (iv) are classical results. Mardia, Holmes and Kent~\cite{mardiaholmeskent1984gof} provided the score test of the von Mises--Fisher model within the Fisher--Bingham family, which extended the test by Cox~\cite{cox1975discussion}; Jupp~\cite{jupp2005sobolev} and~\cite[Sect.~12.3]{mardiajupp2000directional} identify it as this score test. For $m=2$, the statistic and its two-part split correspond to those of Cox: Cox's second trigonometric moments about $\hat{\bmu}_2$, standardized by their efficient variances at $\hat\kappa$~\cite[pp.~142--143]{mardiajupp2000directional}, yield $T_N$ and its components, which are the centred circular kurtosis and the skewness~\cite{pewsey2002symmetry}. For $m=3$, the tangential part is Kent's statistic~\cite{kent1982fisherbingham,rivest1986kent}, and as $\hat\kappa\downarrow0$, the statistic tends to Bingham's~\cite{bingham1974antipodal}. Specific to this family are the reading of $T_N$ as the squared length of the m-projection residual and the exact invariance in (iii); the closed forms \eqref{eq:geomtest-blocks} are computed here, but not claimed as new.

\subsection{Curvature, observed information and the information lost by \texorpdfstring{$\hat{\bB}$}{the estimate of B}}
\label{subsec:geomtest-curv}

\begin{proposition}[Exponential curvature of the cylindrical family]\label{prop:geomtest-curv}
Let $\btheta\in\{\bOmega=\mathbf 0\}$ have regression block $(\balpha,\bB,\bS)$ and directional parameter $\bvartheta$, and for $\mathbf A\in\Symzero{m}$ let $\mathbf c(\mathbf A):=(\mathbf 0,\mathbf 0,\bB\mathbf A\bB^{\top},\bB\mathbf A,\mathbf A)$ in the block order of $\bt$.
\begin{enumerate}
\item[(i)] $\{\mathbf c(\mathbf A)\}$ is the annihilator of the tangent space of $\{\bOmega=\mathbf 0\}$ at $\btheta$; at the estimate, $\bar{\bt}-\hat{\bfeta}=\mathbf c(\bDelta)$, so all unmatched moments are functions of $\bDelta$: those of $\bx_2\bx_2^{\top}$, $\bx_1\bx_2^{\top}$ and $\bx_1\bx_1^{\top}$ are $\bDelta$, $\hat{\bB}\bDelta$ and $\hat{\bB}\bDelta\hat{\bB}^{\top}$.
\item[(ii)] For tangent vectors $X$, $Y$ with $\bB$-components $\dot{\bB}_X$, $\dot{\bB}_Y$, the second fundamental form for the exponential connection paired with $\mathbf c(\mathbf A)$ is $H^{\mathrm e}(X,Y;\mathbf A)=-\tr(\mathbf A\dot{\bB}_X^{\top}\bS^{-1}\dot{\bB}_Y)$; each leaf $\{\bB=\bB_0\}$ is e-flat, $\btheta$ being affine in $(\bS^{-1},\bS^{-1}\balpha,\bvartheta)$ on it.
\item[(iii)] Under (A1)--(A2), for every $p_{\btheta}$ with $\bOmega(\btheta)=\mathbf 0$, regression matrix $\bB$ and covariance $\bS$,
\begin{equation}\label{eq:geomtest-pyth}
\KL[p_{\bar{\bt}}:p_{\btheta}]=\KL[p_{\bar{\bt}}:p_{\hat{\btheta}}]+\KL[p_{\hat{\btheta}}:p_{\btheta}]+\tfrac12\tr\bigl\{(\hat{\bB}-\bB)^{\top}\bS^{-1}(\hat{\bB}-\bB)\bDelta\bigr\}.
\end{equation}
\end{enumerate}
\end{proposition}

Part (iii) evaluates the final term $\langle\hat{\btheta}-\btheta,\bar{\bt}-\hat{\bfeta}\rangle$ of the three-point identity~\cite{amarinagaoka2000methods}. This term has no definite sign and equals $-\tfrac12H^{\mathrm e}(\hat{\bB}-\bB,\hat{\bB}-\bB;\bDelta)$, where $\bS$ is the covariance of $\btheta$; this provides an exact identity for a finite displacement rather than the curvature at a single point. Consequently, the Pythagorean relation holds on the e-flat leaf $\{\bB=\hat{\bB}\}$ and fails off it only through the curvature, a fact that has an inferential consequence.

\begin{theorem}[Information lost by the estimate of $\bB$]\label{thm:geomtest-loss}
Let $\btheta\in\{\bOmega=\mathbf 0\}$ have regression coordinates $(\balpha,\bB,\bS,\bvartheta)$, $\bvartheta=\kappa\bmu_2$, $\kappa\ge0$; let $\mathbf V:=\mathbf V(\bvartheta)$, $\sigma_\parallel^2:=\Am'(\kappa)$ and $\sigma_\perp^2:=\Am(\kappa)/\kappa$ be its radial and tangential variances, $\mathbf Z\sim\calN(\mathbf 0,\Ieff(\bvartheta))$ a Gaussian random element of $\Symzero{m}$, and $g$ the Fisher metric in regression coordinates.
\begin{enumerate}
\item[(i)] \emph{Observed versus expected information.} Under (A1)--(A2), the observed information $g^{\mathrm{obs}}_N:=-N^{-1}\nabla^2\ell$ at the estimate satisfies $g^{\mathrm{obs}}_N(X,X)=g_{\hat{\btheta}}(X,X)-H^{\mathrm e}_{\hat{\btheta}}(X,X;\bDelta)=g_{\hat{\btheta}}(X,X)+\tr(\bDelta\dot{\bB}_X^{\top}\hat{\bS}^{-1}\dot{\bB}_X)$ for every tangent vector $X$ and every $N$. With $\balpha$ profiled, the $\bB$-block of $g^{\mathrm{obs}}_N$ is $\bar{\mathbf V}\otimes\hat{\bS}^{-1}$, with $\bar{\mathbf V}=\mathbf V(\hat{\bvartheta})+\bDelta$ the sample covariance of the directions, and that of $g_{\hat{\btheta}}$ is $\mathbf V(\hat{\bvartheta})\otimes\hat{\bS}^{-1}$. Given the directions, $\operatorname{vec}[\hat{\balpha}\ \hat{\bB}]$ is Gaussian with covariance $(\sum_n\bz_n\bz_n^{\top})^{-1}\otimes\bS$, $\bz_n:=(1,\bx_{2n}^{\top})^{\top}$.
\item[(ii)] \emph{Squared curvature.} $\sqrt N\bDelta\to\mathbf Z$ in distribution, and the squared exponential curvature~\cite{amari1982curved} is $(H^{\mathrm e})^2(X,X)=\E\tr(\mathbf Z\dot{\bB}_X^{\top}\bS^{-1}\dot{\bB}_X\mathbf Z\mathbf V^{-1})$.
\item[(iii)] \emph{Information loss.} The nonzero eigenvalues of $g^{-1}(H^{\mathrm e})^2$ are those of $\mathbf V^{-1/2}\E[\mathbf Z\mathbf V^{-1}\mathbf Z]\mathbf V^{-1/2}$, each repeated $d$ times: $\iota_\parallel(\kappa)$, with eigenvector $\bmu_2$, and $\iota_\perp(\kappa)$, of multiplicity $m-1$, where
\begin{equation}\label{eq:geomtest-loss}
\begin{gathered}
\iota_\parallel=\frac{1}{\sigma_\parallel^2}\Bigl[\frac{(m-1)\nu_{\mathrm{rad}}}{m\sigma_\parallel^2}+\frac{(m-1)\nu_{\mathrm{cpl}}}{2\sigma_\perp^2}\Bigr],\\
\iota_\perp=\frac{1}{\sigma_\perp^2}\Bigl[\frac{\nu_{\mathrm{rad}}}{m(m-1)\sigma_\perp^2}+\frac{\nu_{\mathrm{cpl}}}{2\sigma_\parallel^2}+\frac{(m+1)(m-2)\nu_{\mathrm{tan}}}{2(m-1)\sigma_\perp^2}\Bigr],
\end{gathered}
\end{equation}
with \eqref{eq:geomtest-blocks}. They do not depend on $(\balpha,\bB,\bS,\bmu_2)$; $\iota_\parallel=\iota_\perp=m-1$ at $\kappa=0$, and $\iota_\parallel\to(m+1)^2/(m-1)$, $\iota_\perp\to m+1$ as $\kappa\to\infty$. Consequently~\cite{efron1975curvature,amari1982curved}, the Fisher information carried by the maximum-likelihood estimator, $G_N$, satisfies $Ng-G_N\to(H^{\mathrm e})^2$: the estimate of $\bB$ loses the information of about $\iota_\parallel$ observations in the direction of $\bmu_2$ and $\iota_\perp$ in each tangential direction (eigenvectors $(\dot{\balpha},\dot{\bB})$ with $\dot{\balpha}=-\dot{\bB}\bfeta_2$).
\end{enumerate}
\end{theorem}

\begin{figure}[t]
\centering
\includegraphics[width=\textwidth]{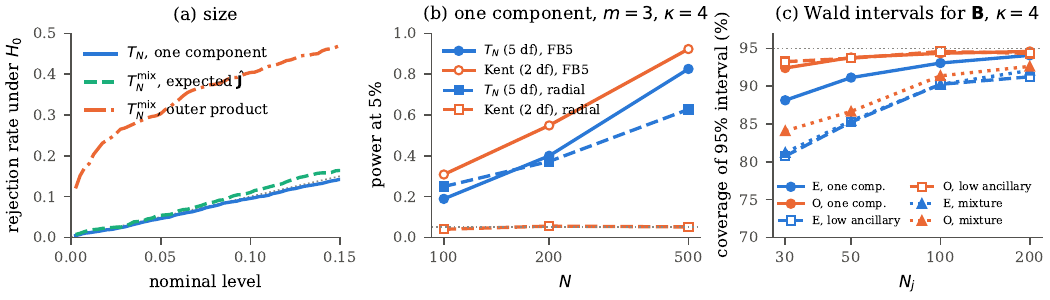}
\caption{The test in simulation (designs, seeds and tables in Section~S14). (a) Rejection rate under $H_0$ against the nominal level (dotted, the diagonal): one component ($m=3$, $\kappa=4$, $N=100$) and a three-component mixture ($d=3$, $m=2$, $N=500$) with the expected and the outer-product information. (b) Power at 5\% on $\sphere^2$ ($\kappa=4$) of $T_N$ and of Kent's block against a five-parameter Fisher--Bingham law ($\beta=0.5$) and a rotationally symmetric one ($\bOmega=\mathbf R$). (c) Coverage of 95\% Wald intervals for the radial entry of $\bB$ with the expected (E) and observed (O) information: one component ($m=2$, $\kappa=4$), the fifth of the samples with the smallest ancillary, and each component of a two-component mixture on $\R\times\sphere^1$; dotted, nominal}
\label{fig:geomtest}
\end{figure}

Theorem~\ref{thm:geomtest-loss} provides an exact Efron--Hinkley example~\cite{efronhinkley1978observed}, where information loss is measured against the expected information. Because the loss matrix $(H^{\mathrm e})^2$ vanishes outside the $\bB$-block, no information is lost on $(\bS,\bvartheta)$, nor on $\balpha$ when $\bB$ is known. However, when $\bB$ is estimated, the estimator $\hat{\balpha}=\bar{\bx}_1-\hat{\bB}\bar{\bx}_2$ loses $\Am^2\iota_\parallel/(\Am'+\Am^2)$ observations, which equals $7.45$ for $m=2$ and $\kappa=4$. The eigenvalues $\iota$ of $g^{-1}(H^{\mathrm e})^2$ correspond to Amari's multiparameter form of Efron's statistical curvature $\gamma^2$~\cite{efron1975curvature,amari1982curved}, and $g^{-1}(H^{\mathrm e})^2$ reduces to $\gamma^2$ for a single parameter. Beyond the proved values at $\kappa=0$ and $\kappa\to\infty$, numerically, tests ($m=2,\dots,8$, $\kappa\le10^3$) show that $\iota_\parallel\in[m-1,(m+1)^2/(m-1))$ and $\iota_\perp\in[m-1,m+1)$, both of which increase except for $\iota_\parallel$ at $m=2$ (Figure~S3). This loss is negligible for large samples, but it becomes relevant for small mixture components. The design $\bx_2$ is an exact S-ancillary for $(\balpha,\bB,\bS)$, and conditioning on it recovers the lost information exactly. Given $\bx_2$, the covariance of $\hat{\bB}$ is $(N\bar{\mathbf V})^{-1}\otimes\bS$, which is the inverse observed information with $\bS$ for $\hat{\bS}$. The departure of this result from the expected information depends only on $\bDelta$, and $\sqrt N\Ieff^{-1/2}\bDelta$ is the standardized ancillary coordinate of the m-projection~\cite{amari1982ancillarity}. This coordinate is ancillary to first order only, as its $O(N^{-1/2})$ mean and skewness depend on $\kappa$ (higher-order theory:~\cite{kumonamari1983higher,kassvos1997geometrical}). The loss $\iota$ represents only part of the second-order variance of $\hat{\bB}$, and the coverage of intervals based on expected information depends on the full second-order ratio, which is about $10$ at $m=2$, $\kappa=4$ (Section~S14.4). Figure~\ref{fig:geomtest}(c) and Table~S14 show the effect on 95\% Wald intervals for $\bB$: in the radial direction, the expected information undercovers, especially when directions are less spread than predicted by the model. In contrast, the observed information, which is exact given the design, does not undercover. In mixtures, both methods undercover (81--92\% for $N_j\le200$), mainly through uncertain memberships.

\subsection{Mixtures: responsibility-weighted residuals}
\label{subsec:geomtest-mix}

We now consider a mixture \eqref{eq:mix-density} where the alternative replaces each $\vMF_m(\bvartheta_j)$ with a Fisher--Bingham distribution $\FB_m(\bvartheta_j,\bOmega_j)$. The null hypothesis $H_0:\bOmega_1=\dots=\bOmega_K=\mathbf 0$ imposes $Kq_m$ restrictions and concerns the entire fitted $K$-component mixture, including the Gaussian conditionals. At a null parameter with responsibilities $r_{nj}$ and masses $N_j$, the score for $\bOmega_j$ is given by $N_j\bDelta_j$, where $\bDelta_j:=N_j^{-1}\sum_nr_{nj}\bx_{2n}\bx_{2n}^{\top}-\bfeta_{22}(\bvartheta_j)$. This term $\bDelta_j$ is the residual of the weighted m-projection at a fixed point of the exact M-step. According to Proposition~\ref{prop:geomtest-curv}(iii) with weights, this step increases the expected complete-data log-likelihood of component $j$ by $N_j$ times a divergence plus the curvature defect (Section~S14).

Because the responsibilities couple $\bOmega_j$ with every other parameter, the information matrix is no longer block-diagonal. Consequently, the efficient information must be estimated. Theorem~\ref{thm:geomtest-mix} is Neyman's C$(\alpha)$ test~\cite{neyman1959optimal} specialized to this model; what is new is the calibration study presented below and the characterization of the plug-in statistic.

\begin{theorem}[Score test for a fitted mixture]\label{thm:geomtest-mix}
Let the observations be independent draws from \eqref{eq:mix-density} with parameter $\bTheta_0$ ($\kappa_j\ge0$ allowed). For a fitted null parameter $\tilde{\bTheta}$ with responsibilities $r_{nj}$, let $\mathbf s_n=(\mathbf s_n^{\varsigma},\mathbf s_n^{\Omega})$ be the per-observation scores of the ambient mixture at $(\tilde{\bTheta},\mathbf 0)$, with $\boldsymbol\varsigma$ the null parameters in any smooth coordinates and $\mathbf s_n^{\Omega_j}=r_{nj}(\bx_{2n}\bx_{2n}^{\top}-\bfeta_{22}(\bvartheta_j))$; let $\mathcal J$ be the per-observation Fisher information of the ambient mixture at $(\bTheta_0,\mathbf 0)$, $\hat{\mathcal J}$ an estimate, $\hat{\mathcal J}_{\mathrm{eff}}:=\hat{\mathcal J}_{\Omega\Omega}-\hat{\mathcal J}_{\Omega\varsigma}\hat{\mathcal J}_{\varsigma\varsigma}^{-1}\hat{\mathcal J}_{\varsigma\Omega}$, $\mathbf s^{\natural}:=\sum_n(\mathbf s_n^{\Omega}-\hat{\mathcal J}_{\Omega\varsigma}\hat{\mathcal J}_{\varsigma\varsigma}^{-1}\mathbf s_n^{\varsigma})$, which is $(N_j\bDelta_j)_j$ at a stationary point, $\Tmix_N:=N^{-1}\mathbf s^{\natural\top}\hat{\mathcal J}_{\mathrm{eff}}^{-1}\mathbf s^{\natural}$, and $\Tmix_{N,j}$ the same statistic from the $\bOmega_j$-scores alone, the other $\bOmega_k$ fixed at $\mathbf 0$ and not profiled. Assume (M1) every $\pi_j>0$, the components pairwise distinct; (M2) $\sqrt N(\tilde{\bTheta}-\bTheta_0)=O_p(1)$ up to relabelling, as for a consistent root of the likelihood equations~\cite{rednerwalker1984mixture}, the mixture being identifiable by Theorem~\ref{thm:mix-identifiable}; (M3) $\mathcal J$ nonsingular; (M4) $\hat{\mathcal J}\to\mathcal J$ in probability. Then $\Tmix_N\to\chi^2_{Kq_m}$, $\Tmix_{N,j}\to\chi^2_{q_m}$, and under $\bOmega_j=N^{-1/2}\boldsymbol\Upsilon_j$, $\Tmix_N\to\chi^2_{Kq_m}(\boldsymbol\upsilon^{\top}\mathcal J_{\mathrm{eff}}\boldsymbol\upsilon)$, with $\mathcal J_{\mathrm{eff}}$ the efficient $\bOmega$-block of $\mathcal J$ and $\boldsymbol\upsilon$ the coordinates of $(\boldsymbol\Upsilon_j)_j$. (M4) holds for the outer-product estimate $N^{-1}\sum_n\mathbf s_n\mathbf s_n^{\top}$, for $\mathcal J(\tilde{\bTheta})$, and for the average of $\mathbf s\mathbf s^{\top}$ over $n_{\mathrm{mc},N}\to\infty$ draws from the fitted model.
\end{theorem}

Conditions (M1)--(M3) exclude the case of an overfitted $K$, where the information becomes singular~\cite{watanabe2013wbic}; we assume, but do not show, that a fit refined by exact EM is the consistent root of (M2). In practice, we select the best fit among several local maxima, and $\hat{\mathcal J}$ is computed by averaging over a fixed $n_{\mathrm{mc}}=2\times10^4$ draws, although Theorem~\ref{thm:geomtest-mix} covers this only as $n_{\mathrm{mc}}\to\infty$. For the size results, we use fits started at the truth; $k$-means++ and split starts were tested only in design M1 with 200 samples (Section~S14.4). Since the hypothesis concerns the entire fitted mixture, a rejection at $K>1$ may stem from the conditional Gaussian part ($t_5$ errors raised the rejection rate of $\Tmix_N$ to 12.0--27.8\%, Section~S14.4). Consequently, a directional interpretation rests on $K=1$, where $T_N$ depends only on the directions, or on the per-component parts. Note that $\Tmix_{N,j}$ has its limit only under the global null; if another component departs, the fitted null parameters are pseudo-true values and the level is not guaranteed. The choice of $\hat{\mathcal J}$ also matters: for $N\le500$, the outer-product form rejected in 24--30\% of the samples, while the expected information rejected in 3.5--6.1\%. Following the missing-information principle~\cite{louis1982em}, $\mathcal J_{\mathrm{eff}}\preceq\blockdiag_j(\pi_j\Ieff(\bvartheta_j))$, with the two coinciding in the limit of separated components (Proposition~S9(iii)); thus, the plug-in $\sum_jN_j\langle\bDelta_j,\Ieff(\hat{\bvartheta}_j)^{-1}\bDelta_j\rangle$ is undersized (0.5--1.5\%).

We now consider a stationary ergodic sequence with the null law as marginal. Assuming a $\sqrt N$-consistent fit and a central limit theorem for the effective scores with a consistent long-run covariance (all assumed, not derived), the studentized statistic is asymptotically $\chi^2_{Kq_m}$ (Proposition~S11). This result is delicate when applied to the persistence of wave records (Section~S14.4). In these tests, calibrations using Newey--West~\cite{neweywest1987hac} and the moving-block bootstrap~\cite{kunsch1989block} rejected in up to 18.7\% and 17\% of cases at the 5\% level, respectively. In contrast, the fixed-$b$ calibration of Kiefer and Vogelsang~\cite{kiefervogelsang2005fixedb}, which fixes the Bartlett bandwidth at a fraction $b$ of $N$ and uses the resulting nonstandard limit, rejected in 4.7--6.3\% for $b=0.1$; the self-normalized calibration~\cite{shao2010selfnormalized} rejected in 5.3--6.7\%. We therefore use the fixed-$b$ calibration. For one component, its limit is proved (Proposition~S12), given stationarity, ergodicity, a $\sqrt N$ rate for $\bar{\bx}_2$, and a functional central limit theorem; for $K>1$, this result is a heuristic. Consequently, dependence-robust $p$-values are indicative, and the conclusion rests on the effect size $\LRstat_N/(2N)$, whose null scale is about $\lambda q_m/(2N)$, where $\lambda$ is the ratio of the long-run to the sample variance of the scores.

\section{Two-layer feature-block hierarchies}
\label{sec:hier}\label{subsec:hier-model}\label{subsec:hier-em}\label{subsec:hier-iso}\label{rem:hier-ordering}

A hierarchy links the two blocks without $\bB$: they are independent given a latent class.
\begin{definition}[Two-layer feature-block hierarchy]
\label{def:hier}
Let $f(\cdot;\btheta_j)$ and $h(\cdot;\btheta_{jk})$ be densities on $\mathcal X_1$ and $\mathcal X_2$ with respect to reference measures $\nu_1$ and $\nu_2$. The two-layer hierarchy with first-layer size $K$ and local sizes $L_1,\dots,L_K$ is the family of densities on $\mathcal X_1\times\mathcal X_2$
\begin{equation}
\label{eq:hier-density}
p(\bx_1,\bx_2;\bTheta)=\sum_{j=1}^K\pi_j\,f(\bx_1;\btheta_j)\,h_j(\bx_2)
=\sum_{j=1}^K\sum_{k=1}^{L_j}\pi_j\pi_{jk}\,f(\bx_1;\btheta_j)\,h(\bx_2;\btheta_{jk}),
\end{equation}
where $h_j:=\sum_k\pi_{jk}h(\cdot;\btheta_{jk})$ is the local mixture of class $j$, $\bTheta^{(1)}=\{\pi_j,\btheta_j\}_j$, $\bTheta^{(2)}_j=\{\pi_{jk},\btheta_{jk}\}_k$ and $\bTheta=(\bTheta^{(1)},\bTheta^{(2)})$.
\end{definition}

We consider two ways of estimating the parameters. The \emph{full estimator} maximizes $\lhier(\bTheta)=\sum_n\log p(\bx_{1n},\bx_{2n};\bTheta)$ using the exact EM algorithm. In contrast, the \emph{isolated} estimator~\cite{garayvergara2026thesis} maximizes $\ell_1(\bTheta^{(1)}):=\sum_n\log\sum_j\pi_jf(\bx_{1n};\btheta_j)$, freezes the \emph{$\bx_1$-only posterior}, and fits each local mixture to a weighted criterion; this process does not correspond to an EM step for $\lhier$. Let the posteriors be normalized over $j$, and let $a$ be a responsibility matrix. Then the posteriors, the criteria and the free energy~\cite{nealhinton1998em} ($0\log0=0$) are given by
\begin{align}
r^{(1)}_{nj}&:=p(Y_1=j\mid\bx_{1n},\bx_{2n};\bTheta)\propto\pi_j\,f(\bx_{1n};\btheta_j)\,h_j(\bx_{2n}),\label{eq:hier-resp1}\\
a_{nj}&:=p(Y_1=j\mid\bx_{1n};\bTheta^{(1)})\propto\pi_j\,f(\bx_{1n};\btheta_j),\label{eq:hier-iso-resp}\\
\ell_{2,j}(\bTheta^{(2)}_j\mid\mathbf a_j)&:=\textstyle\sum_{n=1}^N a_{nj}\log h_j(\bx_{2n}),\label{eq:hier-iso-criterion}\\
F(a,\bTheta)&:=\textstyle\sum_{n=1}^N\sum_{j=1}^K a_{nj}\bigl[\log\pi_j+\log f(\bx_{1n};\btheta_j)+\log h_j(\bx_{2n})-\log a_{nj}\bigr],\label{eq:hier-elbo}
\end{align}
and the \emph{isolated criterion} becomes $\liso(\bTheta):=\ell_1(\bTheta^{(1)})+\sum_j\ell_{2,j}(\bTheta^{(2)}_j\mid\mathbf a_j(\bTheta^{(1)}))$. Proposition~\ref{prop:hier-elbo} is the Neal--Hinton identity, which is the Pythagorean relation of the em algorithm~\cite{csiszartusnady1984alternating,amari1995em} evaluated at the $\bx_1$-only posterior.

\begin{proposition}[The isolated criterion as the Neal--Hinton bound at the $\bx_1$-only posterior]
\label{prop:hier-elbo}
Let $\bTheta$ be a parameter of the hierarchy of Definition~\ref{def:hier} at which $\lhier(\bTheta)$ is finite, and let $r^{(1)}_{n\cdot}(\bTheta)$ be the joint first-layer posterior \eqref{eq:hier-resp1}.
\begin{enumerate}
\item[(i)] For every responsibility matrix $a$,
\begin{equation}
\label{eq:hier-elbo-gap}
\lhier(\bTheta)-F(a,\bTheta)=\sum_{n=1}^N\KL\bigl[a_{n\cdot}:r^{(1)}_{n\cdot}(\bTheta)\bigr]\ \ge\ 0,
\end{equation}
with equality if and only if $a_{n\cdot}=r^{(1)}_{n\cdot}(\bTheta)$ for every $n$.
\item[(ii)] Let $a=a(\bTheta^{(1)})$ be the $\bx_1$-only posterior \eqref{eq:hier-iso-resp} at the first-layer part of the same $\bTheta$. The joint posterior is this posterior reweighted by the local mixtures, $r^{(1)}_{nj}(\bTheta)=a_{nj}h_j(\bx_{2n})/\sum_{j'}a_{nj'}h_{j'}(\bx_{2n})$, and $F(a,\bTheta)=\liso(\bTheta)$. Hence, for every $\bTheta$, in particular at $\hat{\bTheta}_{\mathrm{iso}}$,
\begin{equation}
\label{eq:hier-gap}
\begin{split}
\lhier(\bTheta)-\liso(\bTheta)&=\sum_{n=1}^N\KL\bigl[a_{n\cdot}:r^{(1)}_{n\cdot}(\bTheta)\bigr]\\
&=\sum_{n=1}^N\Bigl[\log\sum_{j}a_{nj}\,h_j(\bx_{2n})-\sum_ja_{nj}\log h_j(\bx_{2n})\Bigr]\ \ge\ 0.
\end{split}
\end{equation}
The $n$th term is the Jensen gap of the logarithm under $a_{n\cdot}$; it is zero if and only if $h_j(\bx_{2n})$ takes the same value for all $j$ with $a_{nj}>0$.
\item[(iii)] At fixed $a$ and $\bTheta^{(1)}$, $F(a,\bTheta)$ depends on $\bTheta^{(2)}$ only through $\sum_{j}\ell_{2,j}(\bTheta^{(2)}_j\mid\mathbf a_j)$, so its maximization over $\bTheta^{(2)}$ is the collection of $K$ decoupled problems \eqref{eq:hier-iso-criterion} that the second stage targets.
\end{enumerate}
\end{proposition}

The proof of this identity is provided in Section~S10.3, while Section~S10 details the orderings, identifiability, and estimators in full. Although the exact EM algorithm never lowers $\lhier$, the second stage of the isolated estimator can do so: it raises $\liso$ for a frozen responsibility matrix $a$. We illustrate this with an example where $K=2$, $L_j=1$, $\pi_j=\tfrac12$, and the components are Gaussians $\calN(\pm\delta,1)$ with $\delta>0$. If the data are $x_{1n}=0$, then $a_{nj}=\tfrac12$. For directions $\pm\mathbf e_1$ and local components $\vMF_2(\pm\mathbf e_1,\kappa)$, the weighted resultant of each class is zero. In this case, the exact second stage moves both local components to the uniform law; this raises $\liso$ by $2\log I_0(\kappa)>0$ but changes $\lhier$ by $2\log\{I_0(\kappa)/\cosh\kappa\}<0$, since $1<I_0(\kappa)<\cosh\kappa$ (Section~S10.6).

In Section~\ref{sec:synth}, we compare the estimators and models on data held out. Joint estimation outperforms isolated estimation by $0.038$ nats per observation, a margin that grows as the dependence rises ($0.007$, $0.016$, $0.090$). Along with this improvement, the gap \eqref{eq:hier-gap} at the isolated fits also increases ($0.0003$, $0.005$, $0.033$). While the matched hierarchy beats the three-component cylindrical mixture by $0.017$ nats, it is closely matched by the independent mixture with as many directional components; this latter model, of which the matched hierarchy is a parameter-sharing submodel, comes within $0.001$ nats.

\section{Simulation studies: direct versus hierarchy-mediated dependence}\label{sec:synth}

We examine two projected-normal generators~\cite{presnell1998projected}, each corresponding to a different mechanism of cross-block dependence. Both generators lie outside every fitted family, providing a basis to show when $\bB_j$ earns its parameters (Question~1) and what an explicit hierarchy adds to the model (Question~2).

\subsection{Design}\label{subsec:synth-design}

We define two specific generators to test our models. \emph{Generator~1 (within-component dependence)} assumes $d=m\in\{2,3\}$, where a label $z_n\in\{1,2,3\}$ follows probabilities $\bpi=(0.36,0.34,0.30)$. Given $z_n=j$, we let $(\mathbf u_{1n},\mathbf u_{2n})\sim\calN_{2d}(\bnu_j,\bGamma_j)$, setting $\bx_{1n}=\mathbf u_{1n}$ and $\bx_{2n}=\mathbf u_{2n}/\norm{\mathbf u_{2n}}$. Here, the cross-covariance $\bGamma_{12,j}$ controls the within-component dependence, which we set to be strong, weak ($\tfrac13$ of strong), or zero. In the case of zero cross-covariance, the blocks are conditionally independent and associated only through $z_n$. \emph{Generator~2 (dependence through a latent class)} maintains the Euclidean marginal $\bx_{1n}\mid Y_{1n}=j\sim\calN_d(\bnu_{1j},\bGamma_{11,j})$ with $Y_{1n}\sim\operatorname{Categorical}(\bpi)$, but introduces a subtype $Y_{2n}\in\{1,2\}$ within each class. Given $(Y_{1n},Y_{2n})=(j,k)$, we set $\bx_{2n}=\mathbf u_{2n}/\norm{\mathbf u_{2n}}$ with $\mathbf u_{2n}\sim\calN_m(\mathbf c_{jk},\bLambda_k)$. This setup follows the structure of Definition~\ref{def:hier}, where blocks are associated only through $Y_1$, which rotates the directional means strongly, weakly, or not at all (\emph{marginal independence}). At high separability, we spread the Euclidean means by a factor of $6$ and the directional angles by $2.25$. The full arrays are listed in Online Resource~1 (Section~S1), with further details provided in Section~S11.

Each data set contains $N=8{,}000$ observations, split into $6{,}400$ for training and $1{,}600$ for testing, and the seeds are the independent replicates. We fit every model, including the hierarchies, using exact EM from ten starts until a stopping rule is met (Algorithm~S1 of Online Resource~1). Performance is measured by the held-out log-density $\HLL$ in nats per observation. This score admits an exact decomposition, $\HLL=\HLL_2+\HLL_{1\mid2}$, into the test averages of $\log p(\bx_2)$ and $\log p(\bx_1\mid\bx_2)$; both terms are available in closed form for every fitted family (Online Resource~1, Section~S11). To compare models, we compute the contrast $\Delta M=M_A-M_B$ on the same data set, averaging within each seed and reporting 95\% Student-$t$ intervals across seeds. For clustering recovery, we use the adjusted Rand index (ARI)~\cite{hubertarabie1985ari} on the test sample, expressed in percentage points.

\subsection{Question 1: direct dependence versus its restriction}\label{subsec:synth-q1}

\begin{table}[t]
\caption{Converged paired contrasts of the simulation studies: held-out log-density $\HLL$ and its parts $\HLL_{1\mid2}$ ($\log p(\bx_1\mid\bx_2)$) and $\HLL_2$ ($\log p(\bx_2)$), in nats per observation, and first-layer $\ARI$, in percentage points. A: cylindrical minus independent mixture at $K=3$, pooled over dimension and separability ($4$ data sets per seed; 20 seeds). B: generator-matched full $(3,2)$ hierarchy minus three alternatives under Generator~2 at high separability, pooled over dimension and dependence ($6$ data sets per seed; 20 seeds). CI: two-sided 95\% Student-$t$ interval across seeds}\label{tab:cyl-vs-hier}
\centering\footnotesize
\setlength{\tabcolsep}{2.3pt}
\begin{tabular}{@{}llrcrrrc@{}}
\toprule
& & $\Delta\HLL$ & 95\% CI & $\Delta\HLL_{1\mid2}$ & $\Delta\HLL_2$ & $\Delta\ARI$ & 95\% CI\\
\midrule
\multicolumn{8}{@{}l}{\textit{A. Cylindrical minus independent mixture, $K=3$}}\\
Gen.~1 & Cond.\ independent & $0.000$ & $[-0.001,\,0.001]$ & $-0.001$ & $0.001$ & $-2.7$ & $[-3.7,\,-1.8]$\\
 & Weak & $0.016$ & $[0.014,\,0.017]$ & $0.014$ & $0.002$ & $1.7$ & $[1.1,\,2.4]$\\
 & Strong & $0.301$ & $[0.294,\,0.307]$ & $0.278$ & $0.023$ & $43.6$ & $[41.7,\,45.6]$\\
Gen.~2 & Marg.\ independent & $-0.001$ & $[-0.002,\,-0.0003]$ & $-0.001$ & $-0.0004$ & $0.9$ & $[-0.3,\,2.1]$\\
 & Weak & $0.0002$ & $[-0.001,\,0.001]$ & $0.001$ & $-0.001$ & $0.2$ & $[-0.5,\,0.9]$\\
 & Strong & $-0.0002$ & $[-0.001,\,0.0005]$ & $-0.001$ & $0.001$ & $-4.0$ & $[-5.2,\,-2.7]$\\
\midrule
\multicolumn{8}{@{}l}{\textit{B. Full $(3,2)$ hierarchy minus the model of the row, Generator~2, high separability}}\\
& isolated $(3,2)$ hierarchy & $0.038$ & $[0.035,\,0.040]$ & $0.022$ & $0.015$ & $8.8$ & $[7.5,\,10.2]$\\
& independent mixture, $K=6$ & $0.001$ & $[0.001,\,0.002]$ & $0.002$ & $-0.001$ & $15.6$ & $[14.7,\,16.5]$\\
& cylindrical mixture, $K=3$ & $0.017$ & $[0.015,\,0.018]$ & $0.007$ & $0.010$ & $10.1$ & $[9.0,\,11.1]$\\
\bottomrule
\end{tabular}
\end{table}

Table~\ref{tab:cyl-vs-hier}(A) compares the cylindrical mixture with its independent restriction at $K=3$, where 3 is the number of generating classes; these two models differ only in whether $\bB_j$ is estimated. Under Generator~1, the gain follows the within-component dependence: there is no gain under conditional independence ($0.000$ nats, interval $[-0.001,0.001]$), a gain of $0.016$ under weak dependence, and $0.301$ under strong dependence. Under strong dependence, the cylindrical mixture also recovers labels better by $43.6$ percentage points. Most of this gain occurs in the conditional part $\HLL_{1\mid2}$ ($0.278$ of the $0.301$ nats), while the remainder comes from better-placed components. Under Generator~2, the same parameter $\bB_j$ changes the held-out density by at most $0.001$ nats at every dependence level; under marginal independence, it costs $0.001$ nats (an interval that just excludes zero). This result is expected, as $\bx_1$ does not depend on $\bx_2$ within a class and $K$ equals the number of classes. Thus, estimating $\bB_j$ provides a benefit mainly where dependence lives inside the components. It also absorbs some misspecification of the components, by up to $0.003$ nats in a cell (Online Resource~1, Section~S2.1).

\subsection{Question 2: an explicit hierarchy}\label{subsec:synth-q2}

Table~\ref{tab:cyl-vs-hier}(B) compares the full $(3,2)$ hierarchy, which follows the structure of Generator~2, against three alternatives under conditions of high separability. Joint estimation outperforms isolated estimation by $0.038$ nats, increasing from $0.007$ under marginal independence to $0.090$ under strong dependence; concurrently, the variational gap of Proposition~\ref{prop:hier-elbo} at the isolated fits grows in a similar fashion ($0.0003$, $0.005$ and $0.033$ nats). The hierarchy also beats the cylindrical mixture with $K=3$ by $0.017$ nats, although the independent mixture with $K=6$, which contains the hierarchy as the submodel whose six Gaussian components are shared in pairs, comes within $0.001$ nats of the hierarchy. This advantage over the cylindrical mixture stems from the number of directional components; when counts are matched, the two-layer structure provides parsimony (with $15$ to $27$ fewer parameters) and better recovery of $Y_1$ ($15.6$ percentage points), rather than improved density (Online Resource~1, Sections~S2.3 and~S3).

\emph{Exact specification.} We further evaluate the model on data drawn from a three-component cylindrical mixture (Online Resource~1, Section~S4). When the components are separated and strongly dependent, the parameters are recovered with normalized errors of $0.06$ or less; however, location and cross-dependence degrade as the components overlap. The score test from Section~\ref{sec:geomtest} rejects in $13$ of $240$ fits at the 5\% level, as it should under its hypothesis. These conclusions hold for the factor levels used here, where $K$ equals the number of generating classes and no fixed-budget comparison is performed, for approximations under misspecified generators.

\section{Illustration on buoy sea-state records}\label{sec:ocean}

We use hourly buoy records, which consist of a circular direction ($m=2$) and serially dependent rows, to illustrate mainly the test described in Section~\ref{sec:geomtest} and its ambient alternative. In this example, the focus is the test itself; direct dependence is secondary.

\subsection{Data and protocol}\label{subsec:ocean-data}\label{subsec:ocean-protocol}

The data consist of hourly standard-meteorological records from the National Data Buoy Center (NDBC) station 46042, located west of Monterey, California, from 2000 to 2022~\cite{ndbc46042}. We define the linear component $\bx_{1i}^{\mathrm{raw}}\in\R^3$ using significant wave height (WVHT), average wave period (APD), and wind speed (WSPD). The circular component $\bx_{2i}\in\sphere^1$ is the mean wave direction (MWD), which represents the direction from which the waves at the dominant period arrive; thus, $d=3$ and $m=2$. Out of the $335{,}789$ archive rows, we retain the $187{,}238$ complete hourly records (Online Resource~1, Section~S13, with the missing years). While gaps in NOAA station records can be filled by sampling from a fitted model of day-to-day dependence, as seen for temperature records in~\cite{aboussalah2016forecasting}, we keep only the complete cases.

We evaluate the models using a sliding window approach for each test year $y\in\{2004,\dots,2022\}$. For a given year $y$, the models are fitted on data from years $y-4,\dots,y-1$ and then scored on year $y$; this pair is a \emph{window}, and the $19$ test years hold $152{,}784$ observations. We apply a Yeo--Johnson transformation~\cite{yeojohnson2000power} to the Euclidean coordinates and standardize them within each training block, so the Jacobian cancels from every contrast. For each $K\in\{1,\dots,8\}$, we fit three different families: the cylindrical mixture, its independent restriction, and the \emph{generalized von Mises (GvM) mixture}. The GvM mixture maintains the Gaussian regression of $\bx_1$ on $\bx_2$ but replaces the von Mises--Fisher factor with the generalized von Mises law~\cite{gattojammalamadaka2007gvm}, which is the directional factor of the ambient family at $m=2$ and is the alternative of the test, adding two parameters per component. Both the cylindrical and independent mixtures are initialized with an equal budget of $k$-means++ starts and run to convergence via Algorithm~S1, while the GvM mixture is initialized from the distinct cylindrical optima (Online Resource~1, Section~S5.2). Let $\bar\ell_y(\hat f)$ denote the average test log-density of a fit in year $y$. The annual direct-dependence contrast for a fixed $K$ is
\begin{equation}\label{eq:ocean-contrast}
\Delta\bar\ell^{\mathrm{dep}}_{y,K}=\bar\ell_y\bigl(\hat f^{\,\cyl}_{y,K}\bigr)-\bar\ell_y\bigl(\hat f^{\,\mathrm{ind}}_{y,K}\bigr),
\end{equation}
and the macro contrast is defined as its average over the $19$ years. Because adjacent windows share training data, no interval procedure is exact; we therefore report Student-$t$ intervals over the $19$ annual contrasts in Table~\ref{tab:ocean-q1}, while Online Resource~1, Section~S5.3, compares four other procedures.

\subsection{A worked example: the 2022 window with one component}\label{subsec:ocean-example}

\begin{table}[t]
\caption{The 2022 window with one component ($N=34{,}545$ training hours, 2018--2021); $\bB$ in standardized units, gains in nats per observation}\label{tab:ocean-example}
\centering\footnotesize
\setlength{\tabcolsep}{4pt}
\begin{tabular}{@{}lll@{}}
\toprule
Result & Quantity & Value\\
\midrule
Thm.~\ref{thm:cyl-mle} & $\hat\kappa$; bearing of $\hat{\bmu}_2$ & $3.04$; $294^\circ$\\
Thm.~\ref{thm:cyl-mle} & $\hat{\bB}\hat{\bmu}_2$; $\hat{\bB}\hat{\bmu}_2^{\perp}$ (WVHT, APD, WSPD) & $(0.41,1.12,-0.54)$; $(0.41,-0.81,0.67)$\\
Prop.~\ref{prop:cyl-mprojection} & $\Delta_\parallel=\hat{\bmu}_2^{\top}\bDelta\hat{\bmu}_2$; $\hat{\bmu}_2^{\perp\top}\bDelta\hat{\bmu}_2$ & $0.082$; $0.125$\\
Prop.~\ref{prop:geomtest-curv}(i) & $(\hat{\bfeta}_{12}-\bar{\bt}_{12})\hat{\bmu}_2=-\hat{\bB}\bDelta\hat{\bmu}_2$, WVHT & $-0.085$\\
Prop.~\ref{prop:geomtest-ieff} & $T_N$ (Thm.~\ref{thm:geomtest}(i)): radial $+$ coupling part & $10{,}616+18{,}864=29{,}480$\\
Prop.~S12 & $T^{\mathrm{fb}}$ ($b=0.1$); $p$-value & $494$; $\le5\times10^{-5}$\\
Thm.~\ref{thm:geomtest-loss}(iii) & $\iota_\parallel$; $\iota_\perp$ (each $d$ times) & $5.75$; $1.81$\\
Thm.~\ref{thm:geomtest-loss}(i) & $(\hat{\bB}\hat{\bmu}_2)_{\mathrm{WVHT}}=0.413$, 95\% CI from $\mathbf V(\hat{\bvartheta})$; $\bar{\mathbf V}$ & $[0.375,0.450]$; $[0.375,0.451]$\\
Thm.~\ref{thm:geomtest-loss}(i) & $(\hat{\bB}\hat{\bmu}_2^{\perp})_{\mathrm{WVHT}}=0.411$, the same & $[0.392,0.431]$; $[0.377,0.446]$\\
Thm.~\ref{thm:geomtest}(ii) & Gain, GvM $-$ vMF: held out in 2022 (training) & $0.423$ ($0.478$)\\
Cor.~\ref{cor:cyl-dependence-gain} & Gain, cylindrical $-$ independent: the same & $0.168$ ($0.123$)\\
\bottomrule
\end{tabular}
\end{table}

Table~\ref{tab:ocean-example} illustrates the application of our theory to a single window. Using the closed form (Theorem~\ref{thm:cyl-mle}), we find the mean wave direction $\hat{\bmu}_2$ at bearing $294^\circ$. We then regress the transformed $\bx_1$ on the cosine and sine of the angle relative to this mean, given by $\hat{\bmu}_2^{\top}\bx_2$ and $\hat{\bmu}_2^{\perp\top}\bx_2$, where $\hat{\bmu}_2^{\perp}$ is a quarter turn clockwise from $\hat{\bmu}_2$. While the fit matches the first moments, it fails to match the second moments of the direction. Specifically, the radial entry of $\bDelta$ shows a sample variance of the cosine of $0.153$, whereas the fit assumes $\sigma_\parallel^2=0.072$; meanwhile, the coupling entry captures a cosine--sine covariance that the symmetric fit sets to zero. These two entries determine every other unmatched moment and, when standardized by $\nu_{\mathrm{rad}}=0.044$ and $\nu_{\mathrm{cpl}}=0.057$, they define the kurtosis and skewness parts of $T_N$. The $\chi^2_2$ calibration for $T_N$ assumes independent hours and is not accurate in the far tail. In contrast, the fixed-$b$ statistic $T^{\mathrm{fb}}$, which accounts for the dependence between hours, exceeds all $20{,}000$ simulated null draws. Finally, $\LRstat_N/(2N)$ is exactly the training gain of the GvM component.

The same residual $\bDelta$ separates the observed from the expected information for $\hat{\bB}$, $\bar{\mathbf V}\otimes\hat{\bS}^{-1}$ and $\mathbf V(\hat{\bvartheta})\otimes\hat{\bS}^{-1}$ ($\balpha$ profiled), since $\bar{\mathbf V}-\mathbf V(\hat{\bvartheta})=\bDelta$. If the model were correct, $\bDelta$ would be a fluctuation of order $N^{-1/2}$, costing $\hat{\bB}$, against the expected information, about $\iota_\parallel\approx6$ hours of information radially and $\iota_\perp\approx2$ tangentially, under $2\times10^{-4}$ of the sample. Here, $\bDelta$ is a misfit that does not shrink with $N$: the observed-information intervals for wave height are $1.02$ times as wide as the expected ones radially and $1.79$ times tangentially. The observed intervals would be exact given the directions if the hours were independent and $\bx_1\mid\bx_2$ Gaussian; however, neither holds here (a Newey--West standard error with lag $168$ is about six times larger), so both intervals illustrate the algebra only. Both gains persist on the 2022 test year (Table~\ref{tab:ocean-example}; Figure~\ref{fig:ocean-q1}(a), ringed).

\subsection{Diagnosis and repair of the directional law}\label{subsec:ocean-q2}

\begin{table}[t]
\caption{Fixed-$K$ contrasts on the buoy records, in nats per observation, averaged over the $19$ test years. CI: 95\% Student-$t$ interval over the annual contrasts; $>0$: years with a positive contrast. Ambient components: GvM minus cylindrical mixture, in total and in $\log p(\bx_2)$ (Dir.). Direct dependence: cylindrical minus independent mixture. Test: medians over the windows of the fixed-$b$ statistic ($2K$ restrictions) and of the KL effect size for the fitted cylindrical mixture; for $K>1$ the fixed-$b$ statistic is descriptive (Section~\ref{subsec:ocean-q2})}\label{tab:ocean-q1}
\centering\footnotesize
\setlength{\tabcolsep}{3.2pt}
\begin{tabular}{@{}r rr rcrr rcr@{}}
\toprule
& \multicolumn{2}{c}{Test} & \multicolumn{4}{c}{Ambient components} & \multicolumn{3}{c@{}}{Direct dependence}\\
\cmidrule(r){2-3}\cmidrule(lr){4-7}\cmidrule(l){8-10}
$K$ & $T^{\mathrm{fb}}$ & KL & Macro & 95\% CI & $>0$ & Dir. & Macro & 95\% CI & $>0$\\
\midrule
1 & $238$ & $0.449$ & $0.454$ & $[0.416,\,0.492]$ & $19$ & $0.454$ & $0.121$ & $[0.097,\,0.145]$ & $18$\\
2 & $132$ & $0.050$ & $0.191$ & $[0.170,\,0.212]$ & $19$ & $0.034$ & $0.080$ & $[0.068,\,0.092]$ & $19$\\
3 & $162$ & $0.027$ & $0.057$ & $[0.043,\,0.070]$ & $18$ & $0.016$ & $0.054$ & $[0.045,\,0.064]$ & $19$\\
4 & $193$ & $0.010$ & $0.047$ & $[0.035,\,0.059]$ & $19$ & $0.002$ & $0.029$ & $[0.018,\,0.040]$ & $17$\\
5 & $218$ & $0.010$ & $0.029$ & $[0.017,\,0.040]$ & $17$ & $0.001$ & $0.022$ & $[0.014,\,0.030]$ & $18$\\
6 & $243$ & $0.006$ & $0.021$ & $[0.010,\,0.032]$ & $17$ & $0.001$ & $0.012$ & $[0.005,\,0.019]$ & $14$\\
7 & $262$ & $0.004$ & $0.021$ & $[0.011,\,0.031]$ & $17$ & $-0.0001$ & $0.012$ & $[0.005,\,0.019]$ & $17$\\
8 & $313$ & $0.005$ & $0.018$ & $[0.010,\,0.027]$ & $18$ & $-0.0003$ & $0.006$ & $[-0.001,\,0.013]$ & $15$\\
\bottomrule
\end{tabular}
\end{table}

\begin{figure}[t]
\centering
\includegraphics[width=\textwidth]{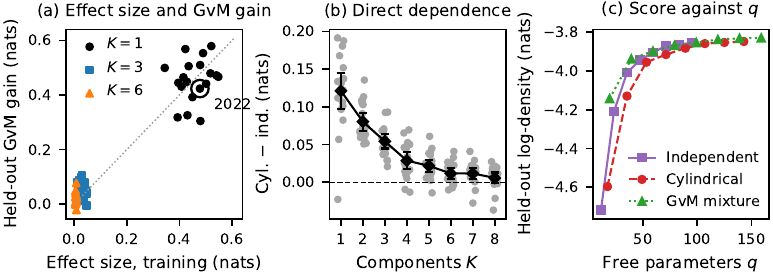}
\caption{Buoy records, $19$ rolling windows, in nats per observation. (a) Per window, the KL effect size of the test on the training block against the held-out gain of the GvM mixture with the same $K$ (ringed: the example of Table~\ref{tab:ocean-example}); dotted, the diagonal. (b) Annual contrasts $\Delta\bar\ell^{\mathrm{dep}}_{y,K}$ \eqref{eq:ocean-contrast} (grey) and their macro average with its 95\% Student-$t$ interval (black). (c) Macro held-out log-density (transformed scale) of the three families against the number of free parameters $q$, for $K=1,\dots,8$ from left to right}
\label{fig:ocean-q1}
\end{figure}

The test in Section~\ref{sec:geomtest} examines whether the directional factor of each fitted component follows a von Mises--Fisher distribution. For $K=1$, the statistic depends only on the directions (Theorem~\ref{thm:geomtest}(iii)), and the misfit observed in Section~\ref{subsec:ocean-example} recurs across every window: $\LRstat_N/(2N)$ ranges from $0.344$ to $0.548$ nats (median $0.449$), while the fixed-$b$ statistic ranges from $117$ to $494$, with $p\le5\times10^{-5}$, the smallest attainable value. These $p$-values are indicative: although the long-run variance of the hourly scores (Newey--West, lag $168$) is $32$--$46$ times their sample variance, a ratio that falls within the range of the null simulations in Section~\ref{subsec:geomtest-mix}, the real records contain gaps and an annual cycle that those simulations lack. The conclusion instead follows from the effect size: under dependence, the null scale is about $\lambda q_m/(2N)$, where $\lambda$ is that variance ratio. This scale is at most $0.007$ nats, even with the fixed-$b$ bandwidth $0.1N$ ($\lambda\le233$), which is two orders of magnitude below the observed values. For $K>1$, the null hypothesis includes the Gaussian part $\bx_1\mid\bx_2$ and the calibration is heuristic; consequently, a mixture statistic is descriptive and need not be directional. In this case, the per-component directional effect sizes have medians of $0.026$ nats at $K=3$ and $0.005$ at $K=6$.

The alternative of the test is also a repair. Replacing each von Mises--Fisher factor with the generalized von Mises law increases the next-year log-density for every $K$ (Table~\ref{tab:ocean-q1}; Online Resource~1, Figure~S2). At $K=1$, the gain is $0.454$ nats, which is entirely directional and close to the average in-sample effect size of $0.457$ nats observed in the test. For higher values of $K$, the gain is $0.191$ nats at $K=2$ and $0.021$ nats at $K=6$. Starting from $K=2$, most of this gain occurs in the conditional part; from $K=4$ on, nearly all of it does, where the directional gain (column Dir.) is at most $0.002$ nats with intervals including zero. While the effect size predicts the level of the gain at $K=1$, it does not predict its variation across years (Figure~\ref{fig:ocean-q1}(a); correlation $0.20$, and between $-0.49$ and $-0.07$ for $K>1$).

\subsection{Direct dependence at fixed \texorpdfstring{$K$}{K}}\label{subsec:ocean-q1}

The cylindrical mixture predicts the next year better than its independent restriction by $0.121$ nats per observation at $K=1$, $0.054$ at $K=3$, and $0.022$ at $K=5$. These gains are separated from zero by every interval procedure (Table~\ref{tab:ocean-q1}, Figure~\ref{fig:ocean-q1}(b)). Beyond $K=5$, the gain is at most $0.012$ nats, which falls within the start-to-start variation described in Section~\ref{subsec:ocean-audit}; here, whether an interval excludes zero depends on the procedure (for instance, at $K=6$ the sign test gives $p=0.064$). The gain is located in $\log p(\bx_1\mid\bx_2)$ and shrinks as the independent mixture captures the association with components. When we compare performance against the number of parameters (Figure~\ref{fig:ocean-q1}(c)), the independent mixture with more components scores higher than the cylindrical mixture at every $q$ up to $95$, which is the largest independent fit. This difference is $0.121$ $[0.101,0.141]$ nats at $q=35$ and $0.045$ $[0.031,0.059]$ at $q=71$, the budgets both families attain. The GvM mixture clearly leads both models up to $q\approx40$ (by $0.071$ $[0.055,0.086]$ nats over the independent mixture with $K=3$); from there to $q\approx100$, it and independent mixtures of similar size differ by less than $0.007$ nats, with intervals that include zero.

\subsection{Fits and scope}\label{subsec:ocean-audit}

All $5{,}031$ runs converged, and none was degenerate. Since MWD is recorded in whole degrees ($271$ distinct values), a component could potentially concentrate on one recorded direction with $\kappa_j\to\infty$ and unbounded likelihood; however, no reported fit does this (the largest fitted concentration is $89$, and the smallest component consists of $206$ hours; Online Resource~1, Section~S13). From $K=4$ onward, and in some windows at $K=3$, different starts reach distinct optima. At $K=6$, the held-out scores of these optima span $0.028$ nats for the cylindrical mixture and $0.031$ nats for the independent mixture on average. Increasing the number of starts does not reduce this spread. Moreover, using a second-best optimum can change the contrasts at $K\ge6$ by up to $0.015$ nats and can even change the sign at $K=6$; thus, these results describe the retained optima (Online Resource~1, Section~S5.2). For the illustration, we use one station and fit the hourly rows as if independent. In this specific case, we select no model, fit no hierarchies (Section~\ref{subsec:synth-q2}), and do not read components as sea states.

\section{Discussion}\label{sec:conclusion}

Most of the first-order geometry follows from the cut and the e-flat von Mises--Fisher factor. What is specific is explicit: these include the conditions for the estimator and the residual $\bDelta$, which carries every unmatched moment and provides the classical score test. We find that curvature exists only in the $\bB$-directions, a property that leaves the test untouched. The square of this curvature provides the information lost on $\bB$ relative to the expected information, expressed in closed form. This is the one new geometric result (Table~\ref{tab:known-new}): the information loss lies between $m-1$ and $(m+1)^2/(m-1)$ observations. This result is proved at the two ends and holds numerically in between; the loss matters only for small samples and components.

Empirically, $\bB_j$ earns its parameters mainly where the dependence lives inside the components. The gain is $0.30$ nats held out when dependence is strong and about zero through a latent class when $K$ equals the number of classes (Table~\ref{tab:cyl-vs-hier}). Using buoy records, the test diagnoses a non-von Mises--Fisher direction with a Kullback--Leibler effect size of $0.45$ nats at $K=1$, a misfit that the ambient alternative repairs out of sample with $0.454$ nats (Table~\ref{tab:ocean-q1}). The gain from direct dependence is clear only for $K\le5$, reaching at most $0.012$ nats beyond that point, within the variation between local optima, and absent at a fixed parameter budget up to $95$ parameters (Figure~\ref{fig:ocean-q1}); beyond about $40$ parameters, the gain from the generalized von Mises model is also absent at a fixed parameter budget.

\textbf{Limitations.} The theory assumes independent observations from the stated models, and the test is conducted against Fisher--Bingham alternatives only. For finite mixtures, Theorem~\ref{thm:geomtest-mix} assumes a $\sqrt N$-consistent fit at an interior parameter with the correct $K$, which exact EM does not guarantee. It also assumes a Gaussian $\bx_1$ given $\bx_2$; consequently, for $K>1$, a rejection may also arise from the conditional part. The fixed-$b$ calibration is justified for $K=1$ (Proposition~S12) but remains heuristic for $K>1$. Identifiability does not select $K$. In the provided illustration, we use one station with $m=2$, treat the hourly rows as independent, and check its conclusions across interval procedures (Table~S5). The synthetic designs share a projected-normal misspecification with $d=m\le3$.

\textbf{Outlook.} Future work may consider Bartlett corrections and intervals conditional on $\bDelta$~\cite{kumonamari1983higher}, as well as normalizers on higher spheres~\cite{kumewood2005saddlepoint,kumesei2018holonomic}. We could also examine components outside the ambient family, such as those described by Abe and Ley~\cite{abe2017tractable}. Note that at $d=1$ and $m=2$, the ambient family is equivalent to the model of Kato and Shimizu~\cite{katoshimizu2008dependent}, up to reparametrization. To narrow the gap described in Proposition~\ref{prop:hier-elbo}, corrected stepwise estimators~\cite{vermunt2010threestep,bakkkuha2018twostep} may be useful. Finally, cylindrical hidden Markov models~\cite{lagona2015hmmcylindrical} could be used to carry serial dependence.

\backmatter

\bmhead{Supplementary information}
Online Resource 1, appended after the References of this article, contains further experimental results (Sections~S1--S5), background, derivations and proofs (Sections~S6--S10 and~S14) and design and preprocessing details (Sections~S11--S13).

\bmhead{Acknowledgements}
Parts of this article build on the M.S.\ thesis of Joaquin E. Garay Vergara, ``Geometry-Aware Latent Variable Models for Cylindrical Data''~\cite{garayvergara2026thesis}, supervised by the author. The author thanks him for this work, which first set out the model and the experimental designs used here. The author also thanks the National Data Buoy Center (NOAA) for the station 46042 archive.


\begin{appendices}
\section{Proofs for Sections~\ref{sec:cyl} and~\ref{sec:mix}}\label{app:theory-proofs}

Throughout, $\psi(\bvartheta):=-\log\Cm(\norm{\bvartheta})$ is the von Mises--Fisher log-partition function on $\R^m$, $\phi$ its dual, $\bz_n:=(1,\bx_{2n}^{\top})^{\top}$ and $\bbeta:=[\balpha\ \ \bB]$.

\begin{proof}[Proof of Theorem~\ref{thm:cyl-mle}]
Observations with $w_n=0$ do not enter $\ell_w$. Put $\mathbf M^{w}:=\sum_nw_n\bz_n\bz_n^{\top}$, $E_w(\bbeta):=\sum_nw_n(\bx_{1n}-\bbeta\bz_n)(\bx_{1n}-\bbeta\bz_n)^{\top}$ and $\bar{\br}:=\br^{w}/W$.

\emph{Splitting.} By \eqref{eq:cyl-density}, $\ell_w=\ell_{\CG}(\bbeta,\bS)+\ell_{\vMF}(\bvartheta_2)$, with $\ell_{\CG}=-\tfrac{Wd}2\log(2\pi)-\tfrac W2\log|\bS|-\tfrac12\tr(\bS^{-1}E_w(\bbeta))$ and $\ell_{\vMF}=W\{\langle\bvartheta_2,\bar{\br}\rangle-\psi(\bvartheta_2)\}$. As $\bx_2$ is a cut, the blocks range over the product $(\R^{d\times(m+1)}\times\Sympp{d})\times\R^m$, so a maximizer exists if and only if both parts have one, and the maximizers are the pairs of maximizers.

\emph{Regression.} With $\mathbf Z_w$, $\mathbf X_w$ the matrices of rows $\sqrt{w_n}\bz_n^{\top}$, $\sqrt{w_n}\bx_{1n}^{\top}$, the normal equations $\mathbf Z_w^{\top}\mathbf Z_w\bbeta^{\top}=\mathbf Z_w^{\top}\mathbf X_w$ are solvable, because the columns of $\mathbf Z_w^{\top}\mathbf X_w$ lie in the range of $\mathbf Z_w^{\top}$, which is that of $\mathbf Z_w^{\top}\mathbf Z_w$. For a solution $\hat{\bbeta}$ the cross terms vanish and $E_w(\bbeta)=E^{w}+(\bbeta-\hat{\bbeta})\mathbf M^{w}(\bbeta-\hat{\bbeta})^{\top}$; so, for every $\bS\succ\mathbf 0$, $\tr(\bS^{-1}E_w(\bbeta))\ge\tr(\bS^{-1}E^{w})$, with equality if and only if $(\bbeta-\hat{\bbeta})\mathbf M^{w}=\mathbf 0$, that is, if $\bbeta$ solves the normal equations. The solution is unique if and only if $\mathbf M^{w}\succ\mathbf 0$; as $\mathbf M^{w}$ has leading entry $W>0$ and Schur complement $\sct{22}^{w}$, this is (A1w), and eliminating $\hat{\balpha}$ from the normal equations then gives \eqref{eq:mle-formulas}.

\emph{Covariance.} The profile is $G(\bS):=-\tfrac W2\log|\bS|-\tfrac12\tr(\bS^{-1}E^{w})$. If $E^{w}\succ\mathbf 0$, then $G(\bS)-G(E^{w}/W)=-W\,\KL[\calN(\mathbf 0,E^{w}/W):\calN(\mathbf 0,\bS)]\le0$, with equality only at $\bS=E^{w}/W$; under (A1w), expanding the residuals with $\hat{\bB}\sct{22}^{w}=\sct{12}^{w}$ gives $E^{w}=\sct{11}^{w}-\sct{12}^{w}(\sct{22}^{w})^{-1}(\sct{12}^{w})^{\top}$. If $E^{w}\mathbf v=\mathbf 0$ for a unit $\mathbf v$, then $\bS_t:=t\mathbf v\mathbf v^{\top}+\bI_d-\mathbf v\mathbf v^{\top}$ gives $G(\bS_t)=-\tfrac W2\log t-\tfrac12\tr\{(\bI_d-\mathbf v\mathbf v^{\top})E^{w}\}\to+\infty$ as $t\downarrow0$.

\emph{Direction.} $\ell_{\vMF}$ is strictly concave, as $\psi$ is strictly convex, with gradient $W(\bar{\br}-\nabla\psi(\bvartheta_2))$. If $\norm{\bar{\br}}<1$, $\bar{\br}$ lies in the von Mises--Fisher mean space and $\nabla\phi(\bar{\br})$, which is $\Am^{-1}(\norm{\bar{\br}})\bar{\br}/\norm{\bar{\br}}$ or $\mathbf 0$, is the unique stationary point and maximizer; $\Am$ is strictly increasing, so $\hat\kappa$ is the unique root. By the triangle inequality $\norm{\br^{w}}\le W$, with equality if and only if all directions with positive weight equal one $\mathbf v$; then $\ell_{\vMF}(t\mathbf v)=-W\log\int\exp\{t(\mathbf v^{\top}\mathbf u-1)\}\,d\sigma(\mathbf u)\to+\infty$, by dominated convergence. This proves (i) and (ii).

\emph{Polar chart.} $\calP\cong\{\bvartheta_2\neq\mathbf 0\}$. Under (A3w) the maximizers lie in $\calP$. If (A0w) or (A2w) fails, the unbounded sequences above can be taken in $\calP$. If (A2w) holds and $\br^{w}=\mathbf 0$, the unique directional maximizer is $\mathbf 0\notin\calP$, while $\ell_w(\hat{\bbeta},\hat{\bS},\kappa\bmu_2)\to\max_{\Pext}\ell_w$ as $\kappa\downarrow0$; so the supremum over $\calP$ is finite and not attained. The original coordinates follow from the bijection \eqref{eq:cyl-original-params}.
\end{proof}

\begin{proof}[Proof of Theorem~\ref{thm:mix-identifiable}]
\emph{(i) implies (ii).} Moving all terms to one side and merging equal parameters gives a vanishing linear combination of distinct members, whose coefficients vanish by (i). As the weights are positive, every $\btheta_j$ equals some $\btheta'_k$ with $\pi'_k=\pi_j$ and conversely, and distinctness on each side makes this a bijection.

\emph{(i).} Let $\btheta_j=(\balpha_j,\bB_j,\bS_j,\bvartheta_j)$, $j=1,\dots,J$, be distinct with $\sum_jc_j\fcyl(\cdot;\btheta_j)=0$ $\nu$-almost everywhere. The densities are continuous and $\nu$ has full support, so the identity holds everywhere. Write $j\sim k$ when $(\balpha_j,\bB_j,\bS_j)=(\balpha_k,\bB_k,\bS_k)$. For $j\not\sim k$, the set of $\mathbf u\in\sphere^{m-1}$ with $(\balpha_j+\bB_j\mathbf u,\bS_j)=(\balpha_k+\bB_k\mathbf u,\bS_k)$ is empty if $\bS_j\neq\bS_k$ or $\bB_j=\bB_k$, and otherwise lies in the proper affine subspace $\{(\bB_j-\bB_k)\mathbf u=\balpha_k-\balpha_j\}$, whose intersection with the sphere is $\sigma$-null. For $\mathbf u$ outside the union $\mathcal N$ of these null sets,
\begin{equation*}
\textstyle\sum_jc_j\,\fvMF(\mathbf u;\bvartheta_j)\,\fG(\bx_1;\balpha_j+\bB_j\mathbf u,\bS_j)=0\qquad\text{for all }\bx_1,
\end{equation*}
and the Gaussian densities that occur are distinct for distinct classes of $\sim$. Finitely many distinct Gaussian densities are linearly independent~\cite{yakowitz1968identifiability} (Section~S8), so $\sum_{j\in R}c_j\fvMF(\mathbf u;\bvartheta_j)=0$ for each class $R$ and every $\mathbf u\notin\mathcal N$, hence, by continuity, for every $\mathbf u\in\sphere^{m-1}$. Within a class the $\bvartheta_j$ are distinct, so it remains to show that $\sum_{l=1}^Lb_l\exp(\bvartheta_l^{\top}\mathbf u)=0$ on the sphere, with distinct $\bvartheta_l$, forces $b=\mathbf 0$. Choose orthonormal $\mathbf v$, $\mathbf w$ such that the numbers $a_l:=\mathbf v^{\top}\bvartheta_l$ are distinct, which excludes $\mathbf v$ only from finitely many hyperplanes, put $e_l:=\mathbf w^{\top}\bvartheta_l$, and evaluate on the great circle $\mathbf u(t)=\cos t\,\mathbf v+\sin t\,\mathbf w$. The function $F(t):=\sum_lb_l\exp(a_l\cos t+e_l\sin t)$ is entire and vanishes on $\R$, so it vanishes on $\mathbb C$. At $t=-\mathrm i\tau$, $\tau>0$, $\cos t=\cosh\tau$ and $\sin t=-\mathrm i\sinh\tau$, so $F(-\mathrm i\tau)=\sum_lb_l\exp(a_l\cosh\tau-\mathrm ie_l\sinh\tau)$. Let $a_{l^{*}}$ be the largest $a_l$. Dividing by $\exp(a_{l^{*}}\cosh\tau)$ and letting $\tau\to\infty$ gives $b_{l^{*}}\exp(-\mathrm ie_{l^{*}}\sinh\tau)\to0$, so $b_{l^{*}}=0$, and induction on $L$ gives $b=\mathbf 0$. Hence $c_j\Cm(\norm{\bvartheta_j})=0$ and $c_j=0$ for every $j$. On $\calP$, the polar and original coordinates are bijective images of $\Pext\setminus\{\bvartheta_2=\mathbf 0\}$, so identifiability carries over.
\end{proof}

\section{Proofs for Section~\ref{sec:geomtest}}\label{app:geomtest}

\emph{Notation.} $\FB_m(\bvartheta,\bOmega)$ has statistic $\boldsymbol\tau(\bx_2):=(\bx_2,\bx_2\bx_2^{\top}-\bI_m/m)$, log-partition function $\psi_{\FB}$, dual potential $\phi_{\FB}$ and information $\mathbf W:=\nabla^2\psi_{\FB}=\cov\boldsymbol\tau$, with blocks $\mathbf W_{11}=\cov(\bx_2)$, $\mathbf W_{21}[\mathbf A]=\cov(\bx_2^{\top}\mathbf A\bx_2,\bx_2)$, $\mathbf W_{22}[\mathbf A,\mathbf A]=\V(\bx_2^{\top}\mathbf A\bx_2)$; it is regular on $\R^m\times\Symzero{m}$ (the sphere is compact) and minimal (Section~S8), so $\mathbf W\succ\mathbf 0$. The minimum in \eqref{eq:geomtest-ieff} is a regression residual variance, so $\Ieff=\mathbf W_{22}-\mathbf W_{21}\mathbf W_{11}^{-1}\mathbf W_{12}$ at $(\bvartheta,\mathbf 0)$ and $[\mathbf W^{-1}]_{\Omega\Omega}=\Ieff^{-1}$. Let $h:=(\nabla\psi_{\vMF})^{-1}$ on the open unit ball; $D\bfeta_{22}=\mathbf W_{21}$. In cut coordinates the inverse map is $\btheta_{11}=-\tfrac12\bS^{-1}$, $\btheta_{12}=\bS^{-1}\bB$, $\btheta_1=\bS^{-1}\balpha$, $\btheta_2=\bvartheta-\bB^{\top}\bS^{-1}\balpha$, $\btheta_{22}=\bOmega-\tfrac12(\bB^{\top}\bS^{-1}\bB)^{\circ}$, and the score is $(\mathbf s_{\CG},\boldsymbol\tau-\nabla\psi_{\FB})$ with $\E[\mathbf s_{\CG}\mid\bx_2]=\mathbf 0$, so $g=\blockdiag(g_{\CG},\mathbf W)$.

\begin{proof}[Proof of Proposition~\ref{prop:geomtest-ieff}]
Under $\vMF_m(\bmu_2,\kappa)$, $\bx_2=t\bmu_2+\sqrt{1-t^2}\,\mathbf w$ with $t$ and $\mathbf w$ independent, $\mathbf w$ uniform on the unit sphere of $\bmu_2^{\perp}$ and $t$ with density proportional to $e^{\kappa t}(1-t^2)^{(m-3)/2}$ on $[-1,1]$~\cite{mardiajupp2000directional}. For $\mathbf A=a\mathbf R+\bmu_2\mathbf b^{\top}+\mathbf b\bmu_2^{\top}+\mathbf C$,
\[
\bx_2^{\top}\mathbf A\bx_2=\underbrace{\tfrac{a}{m-1}(mt^2-1)}_{f_{\mathrm r}}+\underbrace{2t\sqrt{1-t^2}\,\mathbf b^{\top}\mathbf w}_{f_{\mathrm c}}+\underbrace{(1-t^2)\,\mathbf w^{\top}\mathbf C\mathbf w}_{f_{\mathrm t}},
\]
and the regressors are $1$, $t$ and the coordinates of $\sqrt{1-t^2}\,\mathbf w$. The odd moments of $\mathbf w$ vanish, $\E\mathbf w\mathbf w^{\top}=\Pperp/(m-1)$ and $\E[\mathbf w^{\top}\mathbf C\mathbf w]=0$. Hence $f_{\mathrm c}$ and $\sqrt{1-t^2}\,\mathbf w$ are orthogonal in $L^2$ to every function of $t$, and $f_{\mathrm t}$ is orthogonal to $f_{\mathrm r}$, $f_{\mathrm c}$ and all regressors. The regression therefore splits: $f_{\mathrm r}$ is regressed on $(1,t)$, $f_{\mathrm c}$ on $\sqrt{1-t^2}\,\mathbf w$, $f_{\mathrm t}$ on nothing, and the three residuals are orthogonal. So $\Ieff[\mathbf A,\mathbf A]$ is the sum of the three residual variances, the subspaces are $\Ieff$-orthogonal, and on each the form is $\lambda\norm{\cdot}_F^2$ with
\begin{itemize}
\item radial: $(am/(m-1))^2v_{\mathrm r}/\norm{a\mathbf R}_F^2=\tfrac m{m-1}v_{\mathrm r}$, with $v_{\mathrm r}$ the residual variance of $t^2$ on $(1,t)$, which is the ratio of the Gram determinants $\det[\E t^{i+j}]_{i,j\le2}/\det[\E t^{i+j}]_{i,j\le1}$, and $\det[\E t^{i+j}]_{i,j\le1}=\V t=\Am'(\kappa)$ (Lemma~S2);
\item coupling: the best coefficient of $\sqrt{1-t^2}\,\mathbf w$ is $2\mathbf b\,\E[t(1-t^2)]/\E(1-t^2)$, which leaves $\tfrac{4\norm{\mathbf b}^2}{m-1}\E(1-t^2)\V_w(t)$, with $\V_w$ the variance under the law proportional to $(1-t^2)\,dP$; with $\norm{\bmu_2\mathbf b^{\top}+\mathbf b\bmu_2^{\top}}_F^2=2\norm{\mathbf b}^2$, $\nu_{\mathrm{cpl}}=\tfrac2{m-1}\E(1-t^2)\V_w(t)$;
\item tangential: $\E(\mathbf w^{\top}\mathbf C\mathbf w)^2=2\tr(\mathbf C^2)/((m-1)(m+1))$, so $\nu_{\mathrm{tan}}=2\E(1-t^2)^2/(m^2-1)$.
\end{itemize}
The integral $\int_{-1}^1e^{\kappa t}(1-t^2)^{\nu-1/2}dt=\sqrt\pi\,\Gamma(\nu+\tfrac12)(2/\kappa)^{\nu}I_\nu(\kappa)$ at $\nu=\tfrac m2-1+j$ gives $\E(1-t^2)^j=\{\Gamma(\tfrac{m-1}2+j)/\Gamma(\tfrac{m-1}2)\}(2/\kappa)^j\varrho_j(\kappa)$, so $\E(1-t^2)=(m-1)\Am/\kappa$ and $\E(1-t^2)^2=(m^2-1)\varrho_2/\kappa^2$, which gives $\nu_{\mathrm{tan}}$. The law proportional to $(1-t^2)\,dP$ has density proportional to $e^{\kappa t}(1-t^2)^{((m+2)-3)/2}$, that of $t$ under $\vMF_{m+2}$, so $\V_w(t)=A_{m+2}'(\kappa)$ by Lemma~S2, which gives $\nu_{\mathrm{cpl}}$. The moments $\E t^k$ and the limits at $\kappa=0$ are computed at the end of this appendix. The projections of $\bDelta$ on the three subspaces are $\delta_0\mathbf R$, $\bmu_2\boldsymbol\delta_{\mathrm c}^{\top}+\boldsymbol\delta_{\mathrm c}\bmu_2^{\top}$ and $\bar{\mathbf T}-\tr(\bar{\mathbf T})\hat{\mathbf P}/(m-1)$ (with $\hat{\bmu}_2$ for $\bmu_2$), because $\bfeta_{22}(\hat{\bvartheta})$ has eigenvector $\hat{\bmu}_2$ and is a multiple of $\hat{\mathbf P}$ on $\hat{\bmu}_2^{\perp}$; with $\norm{\mathbf R}_F^2=m/(m-1)$ this gives the decomposition (S66). For $\kappa>0$, $\hat{\bmu}_2\to\bmu_2$ almost surely, so the projections of $\sqrt N\bDelta$ converge jointly to independent Gaussian vectors with covariances $\nu_{\mathrm{rad}}$, $\nu_{\mathrm{cpl}}$, $\nu_{\mathrm{tan}}$ times the identity, and the three terms to independent $\chi^2$ variables. As $\kappa\downarrow0$, $\bfeta_{22}\to\bI_m/m$ and $\Ieff\to\tfrac2{m(m+2)}\mathrm{Id}$, uniformly in $\bmu_2$.
\end{proof}

\begin{proof}[Proof of Proposition~\ref{prop:geomtest-curv}]
\emph{(i)} In the graph chart $\mathbf u:=(\btheta_1,\btheta_2,\btheta_{11},\btheta_{12})$ of Proposition~\ref{prop:cyl-curved}(c), $\{\bOmega=\mathbf 0\}$ is $\btheta_{22}=G(\mathbf u):=\tfrac14(\btheta_{12}^{\top}\btheta_{11}^{-1}\btheta_{12})^{\circ}$, and $\btheta_{11}^{-1}\btheta_{12}=-2\bB$ there. So $dG=(-\tfrac12(d\btheta_{12}^{\top}\bB+\bB^{\top}d\btheta_{12})-\bB^{\top}d\btheta_{11}\bB)^{\circ}$ and, for traceless symmetric $\mathbf A$, $\langle\mathbf c(\mathbf A),d\btheta\rangle=\tr(\bB\mathbf A\bB^{\top}d\btheta_{11})+\tr(\mathbf A\bB^{\top}d\btheta_{12})+\tr(\mathbf A\,dG)=0$; $\mathbf A\mapsto\mathbf c(\mathbf A)$ is injective and the annihilator has dimension $q_m$. At the estimate, $\bar{\bt}-\hat{\bfeta}=\mathbf c(\bDelta)$ is \eqref{eq:mprojection-mismatch}.

\emph{(ii)} The exponential connection is flat in $\btheta$, so for $X$, $Y$ tangent, $\nabla^{\mathrm e}_XY$ has natural expression $X(Y^{\btheta})$, and the second fundamental form is its normal part~\cite{amari1982curved,amarinagaoka2000methods}. Its pairing $H^{\mathrm e}(X,Y;\mathbf A):=\langle\mathbf c(\mathbf A),\nabla^{\mathrm e}_XY\rangle$ is tensorial, so coordinate fields of $\mathbf u$ may be used, for which $\nabla^{\mathrm e}_XY=(\mathbf 0,\mathbf 0,\mathbf 0,\mathbf 0,D^2G[X,Y])$. With $\mathbf M:=\btheta_{11}^{-1}$, $\mathbf K:=\mathbf M\btheta_{12}$ and $(\mathbf H_X,\mathbf C_X)$ the $(\btheta_{11},\btheta_{12})$-components of $X$, $D^2(\btheta_{12}^{\top}\mathbf M\btheta_{12})[X,Y]=(\mathbf C_X-\mathbf H_X\mathbf K)^{\top}\mathbf M(\mathbf C_Y-\mathbf H_Y\mathbf K)+(X\leftrightarrow Y)$ (end of this appendix), and on $\{\bOmega=\mathbf 0\}$, $\mathbf M=-2\bS$, $\mathbf K=-2\bB$, $\mathbf C_X-\mathbf H_X\mathbf K=d\btheta_{12}+2d\btheta_{11}\bB=\bS^{-1}\dot{\bB}_X$. Hence $D^2G[X,Y]=-\tfrac12(\dot{\bB}_X^{\top}\bS^{-1}\dot{\bB}_Y+\dot{\bB}_Y^{\top}\bS^{-1}\dot{\bB}_X)^{\circ}$, which gives $H^{\mathrm e}$. On $\{\bB=\bB_0\}$, $\btheta=(\mathbf L\balpha,\bvartheta-\bB_0^{\top}\mathbf L\balpha,-\tfrac12\mathbf L,\mathbf L\bB_0,-\tfrac12(\bB_0^{\top}\mathbf L\bB_0)^{\circ})$, $\mathbf L:=\bS^{-1}$, is linear in $(\mathbf L\balpha,\mathbf L,\bvartheta)$.

\emph{(iii)} Write $p_{\btheta}=\fCG(\cdot\mid\bx_2;\boldsymbol\gamma)\fvMF(\cdot;\bvartheta)$ and $k(\bx_2)$ for the divergence between the conditional Gaussians with blocks $\hat{\boldsymbol\gamma}$ and $\boldsymbol\gamma$. By the chain rule $\KL[p_{\bar{\bt}}:p_{\btheta}]=\KL[\FB^{*}:\vMF(\bvartheta)]+\E_{\FB^{*}}k$, $\KL[p_{\hat{\btheta}}:p_{\btheta}]=\KL[\vMF(\hat{\bvartheta}):\vMF(\bvartheta)]+\E_{\hat{\bvartheta}}k$ and $\KL[p_{\bar{\bt}}:p_{\hat{\btheta}}]=\KL[\FB^{*}:\vMF(\hat{\bvartheta})]$. As $\FB^{*}$ and $\vMF(\hat{\bvartheta})$ have mean $\bar{\bx}_2$ and the log-ratio of two von Mises--Fisher densities is affine, $\KL[\FB^{*}:\vMF(\bvartheta)]=\KL[\FB^{*}:\vMF(\hat{\bvartheta})]+\KL[\vMF(\hat{\bvartheta}):\vMF(\bvartheta)]$. Finally $k(\bx_2)=\tfrac12\{\tr(\bS^{-1}\hat{\bS})-d+\log(|\bS|/|\hat{\bS}|)+(\mathbf a+\mathbf D\bx_2)^{\top}\bS^{-1}(\mathbf a+\mathbf D\bx_2)\}$, $\mathbf a:=\hat{\balpha}-\balpha$, $\mathbf D:=\hat{\bB}-\bB$, and the second moments of the two laws differ by $\bDelta$, so $\E_{\FB^{*}}k-\E_{\hat{\bvartheta}}k=\tfrac12\tr(\mathbf D^{\top}\bS^{-1}\mathbf D\bDelta)$.
\end{proof}

\begin{proof}[Proof of Theorem~\ref{thm:geomtest-loss}]
\emph{(i)} In any coordinates $\boldsymbol\zeta$ on $\{\bOmega=\mathbf 0\}$, $\ell(\boldsymbol\zeta)/N=\langle\btheta(\boldsymbol\zeta),\bar{\bt}\rangle-\psi_{\mathrm{amb}}(\btheta(\bxi))$, so $-N^{-1}\partial_a\partial_b\ell=g_{ab}-\langle\partial_a\partial_b\btheta,\bar{\bt}-\nabla\psi_{\mathrm{amb}}\rangle$. At $\hat{\boldsymbol\zeta}$, $\bar{\bt}-\hat{\bfeta}=\mathbf c(\bDelta)$ by Proposition~\ref{prop:geomtest-curv}(i), and $\langle\mathbf c(\bDelta),\partial_a\partial_b\btheta\rangle=H^{\mathrm e}(\partial_a,\partial_b;\bDelta)$, since $\partial_a\partial_b\btheta=\nabla^{\mathrm e}_{\partial_a}\partial_b$; Proposition~\ref{prop:geomtest-curv}(ii) gives the trace. Directly, in $\bGamma:=[\balpha\ \bB]$ the observed information of the regression is $\bar{\mathbf M}_z\otimes\hat{\bS}^{-1}$, $\bar{\mathbf M}_z:=N^{-1}\sum_n\bz_n\bz_n^{\top}$, and the expected one $\mathbf M_z(\hat{\bvartheta})\otimes\hat{\bS}^{-1}$; their $\bB$-Schur complements are $\bar{\mathbf V}$ and $\mathbf V(\hat{\bvartheta})$ because $\bar{\bx}_2=\nabla\psi_{\vMF}(\hat{\bvartheta})$, and $\bar{\mathbf V}-\mathbf V(\hat{\bvartheta})=\bar{\mathbf M}_{22}-\bfeta_{22}(\hat{\bvartheta})=\bDelta$. The blocks pairing $\bS$ or $\bvartheta$ with the regression vanish at the estimate (normal equations and the cut). Given the directions, $\bx_{1n}=\bGamma\bz_n+\beps_n$ is a Gaussian linear model with fixed design.

\emph{(ii)} The limit of $\sqrt N\bDelta$ is Theorem~\ref{thm:geomtest}(iv) with $\boldsymbol\Upsilon=\mathbf 0$. Take a basis $\{\mathbf A_r\}$ of $\Symzero{m}$ and the normal vectors $\mathbf n_r:=g^{-1}\mathbf c(\mathbf A_r)$; in cut coordinates $\mathbf c(\mathbf A)$ is $(\mathbf 0,\mathbf 0,\mathbf A)$, so $g(\mathbf n_r,\mathbf n_s)=\langle\mathbf A_r,\Ieff^{-1}\mathbf A_s\rangle$, and $H^{\mathrm e}_{abr}=\langle\mathbf A_r,\mathbf h_{ab}\rangle$ with $\mathbf h_{ab}:=-(\dot{\bB}_a^{\top}\bS^{-1}\dot{\bB}_b)^{\circ}_{\mathrm{sym}}$. Then $\sum_{rs}g^{rs}H^{\mathrm e}_{acr}H^{\mathrm e}_{bds}=\langle\mathbf h_{ac},\Ieff\mathbf h_{bd}\rangle=\E\langle\mathbf h_{ac},\mathbf Z\rangle\langle\mathbf h_{bd},\mathbf Z\rangle$. Contracting with $g^{cd}$, the covector $c\mapsto\langle\mathbf h_{Xc},\mathbf Z\rangle$ is $\dot{\bB}\mapsto-\langle\bS^{-1}\dot{\bB}_X\mathbf Z,\dot{\bB}\rangle_F$, supported on $\bB$, where the inverse metric of the regression block $\mathbf M_z\otimes\bS^{-1}$ is $\mathbf V^{-1}\otimes\bS$ ($[\mathbf M_z^{-1}]_{BB}=\mathbf V^{-1}$). This gives $\E\tr(\bS\,\bS^{-1}\dot{\bB}_X\mathbf Z\mathbf V^{-1}\mathbf Z\dot{\bB}_X^{\top}\bS^{-1})$, the stated form.

\emph{(iii)} By (ii), $(H^{\mathrm e})^2$ vanishes off the $\bB$-block and equals $\mathbf Q\otimes\bS^{-1}$ there, $\mathbf Q:=\E[\mathbf Z\mathbf V^{-1}\mathbf Z]$; as $[g^{-1}]_{BB}=\mathbf V^{-1}\otimes\bS$, the nonzero eigenvalues of $g^{-1}(H^{\mathrm e})^2$ are those of $\mathbf V^{-1}\mathbf Q\otimes\bI_d$. Write $\mathbf Z=\sum_iz_i\lambda_i^{1/2}\mathbf E_i$ in a Frobenius-orthonormal basis adapted to the decomposition of Proposition~\ref{prop:geomtest-ieff}, $z_i$ independent standard normal, so $\mathbf Q=\sum_i\lambda_i\mathbf E_i\mathbf V^{-1}\mathbf E_i$, with $\mathbf V^{-1}=\sigma_\parallel^{-2}\bmu_2\bmu_2^{\top}+\sigma_\perp^{-2}\Pperp$. For $\mathbf E=\mathbf R/\norm{\mathbf R}$, $\mathbf E\mathbf V^{-1}\mathbf E=\tfrac{m-1}m\{\sigma_\parallel^{-2}\bmu_2\bmu_2^{\top}+(m-1)^{-2}\sigma_\perp^{-2}\Pperp\}$; summed over the coupling basis $(\bmu_2\mathbf b^{\top}+\mathbf b\bmu_2^{\top})/\sqrt2$, $\tfrac12\{(m-1)\sigma_\perp^{-2}\bmu_2\bmu_2^{\top}+\sigma_\parallel^{-2}\Pperp\}$; and over the tangential basis, $\sigma_\perp^{-2}\sum_i\mathbf E_i^2=\sigma_\perp^{-2}\tfrac{(m+1)(m-2)}{2(m-1)}\Pperp$. So $\mathbf Q$ has eigenvector $\bmu_2$ and eigenspace $\bmu_2^{\perp}$, and dividing by $\sigma_\parallel^2$, $\sigma_\perp^2$ gives \eqref{eq:geomtest-loss}. The limits follow from the values $2/(m(m+2))$ and $\sigma_\parallel^2=\sigma_\perp^2=1/m$ at $\kappa=0$, and from the Gamma limit of $1-t$ as $\kappa\to\infty$ (end of this appendix). The loss statement is the theorem of Efron~\cite{efron1975curvature} and Amari~\cite{amari1982curved} for the maximum-likelihood estimator of a curved exponential family, whose loss is the squared exponential curvature.
\end{proof}

\begin{proof}[Proof of Theorem~\ref{thm:geomtest-mix}]
Let $\ell_n(\boldsymbol\varsigma,\bomega)$ be the ambient-mixture log-density at $\bx_n$, $\boldsymbol\varsigma$ the null parameters in cut coordinates ($\bvartheta_j\in\R^m$, so $\kappa_j=0$ is interior), $\bomega:=(\bOmega_j)_j$, with the labelling of (M2). On a compact neighbourhood $U$ of $(\boldsymbol\varsigma_0,\mathbf 0)$ with weights bounded below and covariances with eigenvalues in a compact subset of $(0,\infty)$, derivatives of $\ell_n$ up to order three are polynomials in responsibilities in $[0,1]$ and in derivatives of quadratic component log-densities with smooth coefficients, hence bounded by $C(1+\norm{\bx_1}^6)$, integrable uniformly on $U$. So the model is differentiable in quadratic mean, $\E\mathbf s\mathbf s^{\top}=-\E\nabla^2\ell_n=\mathcal J$, and averages of $\nabla^2\ell_n$ and $\mathbf s_n\mathbf s_n^{\top}$ converge uniformly on $U$~\cite[Chapters~5 and~7]{vandervaart1998asymptotic}. By the integral form of the mean-value theorem, on $\{\tilde{\boldsymbol\varsigma}\in U\}$, $N^{-1/2}\mathbf s(\tilde{\boldsymbol\varsigma})=N^{-1/2}\mathbf s(\boldsymbol\varsigma_0)+\{\int_0^1N^{-1}\nabla_\varsigma\mathbf s(\boldsymbol\varsigma_0+u(\tilde{\boldsymbol\varsigma}-\boldsymbol\varsigma_0))du\}\sqrt N(\tilde{\boldsymbol\varsigma}-\boldsymbol\varsigma_0)$, and the braces tend to $-(\mathcal J_{\varsigma\varsigma},\mathcal J_{\Omega\varsigma})$. With (M2), (M4), the terms in $\sqrt N(\tilde{\boldsymbol\varsigma}-\boldsymbol\varsigma_0)$ cancel in $\mathbf s^{\natural}$, and $N^{-1/2}\mathbf s^{\natural}=N^{-1/2}\sum_n(\mathbf s_n^{\Omega}-\mathcal J_{\Omega\varsigma}\mathcal J_{\varsigma\varsigma}^{-1}\mathbf s_n^{\varsigma})(\boldsymbol\varsigma_0,\mathbf 0)+o_p(1)\to\calN(\mathbf 0,\mathcal J_{\mathrm{eff}})$, which gives the null limits, with $\bOmega_j$ alone for $\Tmix_{N,j}$. The local alternatives are contiguous by local asymptotic normality, and Le Cam's third lemma~\cite[Chapter~6]{vandervaart1998asymptotic} shifts the limit by $\mathcal J_{\mathrm{eff}}\boldsymbol\upsilon$. The outer-product estimate satisfies (M4) by uniform convergence and the information identity, $\mathcal J(\tilde{\bTheta})$ by continuity of $\mathcal J$ on $U$, and a Monte Carlo average over $n_{\mathrm{mc},N}\to\infty$ draws by Chebyshev's inequality, its conditional variance being at most $C/n_{\mathrm{mc},N}$.
\end{proof}

\paragraph{Computations used above.} \emph{Moments of $t$.} Let $t=\bmu_2^{\top}\bx_2$ under $\vMF_m(\bmu_2,\kappa)$, $\nu_0:=\tfrac m2-1$, and $c(\kappa):=\int_{-1}^1e^{\kappa t}(1-t^2)^{\nu_0-1/2}dt$, which is proportional to $\kappa^{-\nu_0}I_{\nu_0}(\kappa)$. Then $\E t^k=c^{(k)}(\kappa)/c(\kappa)$. With $g_j:=\kappa^{-\nu_0}I_{\nu_0+j}=\kappa^j\cdot\kappa^{-(\nu_0+j)}I_{\nu_0+j}$ and $\tfrac{d}{d\kappa}\{\kappa^{-\mu}I_\mu(\kappa)\}=\kappa^{-\mu}I_{\mu+1}(\kappa)$, $g_j'=(j/\kappa)g_j+g_{j+1}$. Starting from $c\propto g_0$ and differentiating four times,
\[
c'\propto g_1,\quad c''\propto g_1/\kappa+g_2,\quad c'''\propto 3g_2/\kappa+g_3,\quad c''''\propto 3g_2/\kappa^2+6g_3/\kappa+g_4,
\]
which, divided by $g_0$, are the four moments of Proposition~\ref{prop:geomtest-ieff} in terms of $\varrho_j=g_j/g_0$. For example, $(g_1/\kappa+g_2)'=-g_1/\kappa^2+(g_1/\kappa+g_2)/\kappa+2g_2/\kappa+g_3=3g_2/\kappa+g_3$. The Bessel recurrence $\varrho_{j+1}=\varrho_{j-1}-\tfrac{2(\nu_0+j)}{\kappa}\varrho_j$ gives $\varrho_2=1-m\Am/\kappa$, so $\E t^2=1-(m-1)\Am/\kappa$, in agreement with Lemma~S2. At $\kappa=0$, $t$ has density proportional to $(1-t^2)^{(m-3)/2}$, with $\E t^2=1/m$, $\E t^4=3/(m(m+2))$ and vanishing odd moments. Then $v_{\mathrm r}=\V(t^2)=2(m-1)/(m^2(m+2))$, $\E(1-t^2)\V_w(t)=\E t^2-\E t^4=(m-1)/(m(m+2))$ and $\E(1-t^2)^2=(m^2-1)/(m(m+2))$, so the three eigenvalues of the proof of Proposition~\ref{prop:geomtest-ieff} all equal $2/(m(m+2))$. The moments are continuous in $\kappa\ge0$ by dominated convergence, and so are the three eigenvalues. Near $\kappa=0$ the Bessel forms of Proposition~\ref{prop:geomtest-ieff} are evaluated as ratios of exponentially scaled Bessel functions; for large $\kappa$, where $\nu_{\mathrm{rad}}\asymp\kappa^{-4}$ is a difference of terms of order $\kappa^{-2}$, the code computes the residual variances by one-dimensional quadrature in $s=1-t$ with centred integrands, which agrees with the Bessel forms to $10^{-9}$ for $\kappa\le50$ (Section~S14.5).

\emph{Limits of the information loss.} As $\kappa\to\infty$, $s:=1-t$ satisfies $\kappa s\to\mathrm{Gamma}(a,1)$ in law with all moments, $a:=(m-1)/2$, because the density of $t$ is proportional to $e^{-\kappa s}\{s(2-s)\}^{a-1}$. With $\theta:=1/\kappa$ and $\E s^k\sim\theta^k\Gamma(a+k)/\Gamma(a)$: $\sigma_\parallel^2=\V s\sim a\theta^2$; $\sigma_\perp^2=\E s(2-s)/(m-1)\sim\theta$; the residual variance of $t^2$ on $(1,t)$, that of $s^2$ on $(1,s)$, is $\sim2a(a+1)\theta^4$, so $\nu_{\mathrm{rad}}\sim m(m+1)\theta^4/2$; $\E(1-t^2)\V_w(t)\sim2a\theta\cdot(a+1)\theta^2$, so $\nu_{\mathrm{cpl}}\sim(m+1)\theta^3$; and $\nu_{\mathrm{tan}}=2\varrho_2/\kappa^2\sim2\theta^2$. Inserting these in (\ref{eq:geomtest-loss}) gives $\iota_\parallel\to(m+1)^2/(m-1)$ and $\iota_\perp\to m+1$ (in $\iota_\perp$ the coupling and tangential terms are of order $\theta$ and the radial term of order $\theta^3$). At $\kappa=0$ the three eigenvalues equal $\lambda:=2/(m(m+2))$ and $\sigma_\parallel^2=\sigma_\perp^2=1/m$; both brackets in (\ref{eq:geomtest-loss}) then equal $\lambda(m-1)(m+2)/2$, using $2+m(m-1)+m(m+1)(m-2)=(m-1)^2(m+2)$ for $\iota_\perp$, so $\iota_\parallel=\iota_\perp=m\lambda(m-1)(m+2)/2=m-1$.

\emph{The second derivative in Proposition~\ref{prop:geomtest-curv}(ii).} Let $\mathcal Q(\mathbf H,\mathbf C):=\mathbf C^{\top}\mathbf H^{-1}\mathbf C$ for symmetric invertible $\mathbf H$, $\mathbf M:=\mathbf H^{-1}$ and $\mathbf K:=\mathbf M\mathbf C$. Since $d\mathbf M=-\mathbf M\,d\mathbf H\,\mathbf M$,
\[
D\mathcal Q[X]=\mathbf C_X^{\top}\mathbf M\mathbf C+\mathbf C^{\top}\mathbf M\mathbf C_X-\mathbf C^{\top}\mathbf M\mathbf H_X\mathbf M\mathbf C,
\]
and differentiating again along $Y$,
\begin{align*}
D^2\mathcal Q[X,Y]={}&\mathbf C_X^{\top}\mathbf M\mathbf C_Y+\mathbf C_Y^{\top}\mathbf M\mathbf C_X-\mathbf C_X^{\top}\mathbf M\mathbf H_Y\mathbf K-\mathbf K^{\top}\mathbf H_Y\mathbf M\mathbf C_X\\
&-\mathbf C_Y^{\top}\mathbf M\mathbf H_X\mathbf K-\mathbf K^{\top}\mathbf H_X\mathbf M\mathbf C_Y\\
&+\mathbf K^{\top}\mathbf H_X\mathbf M\mathbf H_Y\mathbf K+\mathbf K^{\top}\mathbf H_Y\mathbf M\mathbf H_X\mathbf K.
\end{align*}
Expanding $(\mathbf C_X-\mathbf H_X\mathbf K)^{\top}\mathbf M(\mathbf C_Y-\mathbf H_Y\mathbf K)$ and its transpose, the version with $X$ and $Y$ exchanged, returns these eight terms, which is the identity used in the proof of Proposition~\ref{prop:geomtest-curv}. At a cylindrical point, $\mathbf H=\btheta_{11}=-\tfrac12\bS^{-1}$ and $\mathbf C=\btheta_{12}=\bS^{-1}\bB$, so $\mathbf M=-2\bS$ and $\mathbf K=-2\bB$.

\end{appendices}

\bibliographystyle{sn-mathphys-num}
\bibliography{reference}


\begin{thebibliography}{106}
\ifx \bisbn   \undefined \def \bisbn  #1{ISBN #1}\fi
\ifx \binits  \undefined \def \binits#1{#1}\fi
\ifx \bauthor  \undefined \def \bauthor#1{#1}\fi
\ifx \batitle  \undefined \def \batitle#1{#1}\fi
\ifx \bjtitle  \undefined \def \bjtitle#1{#1}\fi
\ifx \bvolume  \undefined \def \bvolume#1{\textbf{#1}}\fi
\ifx \byear  \undefined \def \byear#1{#1}\fi
\ifx \bissue  \undefined \def \bissue#1{#1}\fi
\ifx \bfpage  \undefined \def \bfpage#1{#1}\fi
\ifx \blpage  \undefined \def \blpage #1{#1}\fi
\ifx \burl  \undefined \def \burl#1{\textsf{#1}}\fi
\ifx \doiurl  \undefined \def \doiurl#1{\url{https://doi.org/#1}}\fi
\ifx \betal  \undefined \def \betal{\textit{et al.}}\fi
\ifx \binstitute  \undefined \def \binstitute#1{#1}\fi
\ifx \binstitutionaled  \undefined \def \binstitutionaled#1{#1}\fi
\ifx \bctitle  \undefined \def \bctitle#1{#1}\fi
\ifx \beditor  \undefined \def \beditor#1{#1}\fi
\ifx \bpublisher  \undefined \def \bpublisher#1{#1}\fi
\ifx \bbtitle  \undefined \def \bbtitle#1{#1}\fi
\ifx \bedition  \undefined \def \bedition#1{#1}\fi
\ifx \bseriesno  \undefined \def \bseriesno#1{#1}\fi
\ifx \blocation  \undefined \def \blocation#1{#1}\fi
\ifx \bsertitle  \undefined \def \bsertitle#1{#1}\fi
\ifx \bsnm \undefined \def \bsnm#1{#1}\fi
\ifx \bsuffix \undefined \def \bsuffix#1{#1}\fi
\ifx \bparticle \undefined \def \bparticle#1{#1}\fi
\ifx \barticle \undefined \def \barticle#1{#1}\fi
\bibcommenthead
\ifx \bconfdate \undefined \def \bconfdate #1{#1}\fi
\ifx \botherref \undefined \def \botherref #1{#1}\fi
\ifx \url \undefined \def \url#1{\textsf{#1}}\fi
\ifx \bchapter \undefined \def \bchapter#1{#1}\fi
\ifx \bbook \undefined \def \bbook#1{#1}\fi
\ifx \bcomment \undefined \def \bcomment#1{#1}\fi
\ifx \oauthor \undefined \def \oauthor#1{#1}\fi
\ifx \citeauthoryear \undefined \def \citeauthoryear#1{#1}\fi
\ifx \endbibitem  \undefined \def \endbibitem {}\fi
\ifx \bconflocation  \undefined \def \bconflocation#1{#1}\fi
\ifx \arxivurl  \undefined \def \arxivurl#1{\textsf{#1}}\fi
\csname PreBibitemsHook\endcsname

\bibitem[\protect\citeauthoryear{Noumane and
  Aboussalah}{2026}]{noumane2026when}
\begin{barticle}
\bauthor{\bsnm{Noumane}, \binits{O.}},
\bauthor{\bsnm{Aboussalah}, \binits{A.M.}}:
\batitle{When structure is not enough: Geometry and reliability in {UAV}-to-map
  localization}.
\bjtitle{SSRN Electronic Journal}
(\byear{2026})
\doiurl{10.2139/ssrn.7549358}
\end{barticle}
\endbibitem

\bibitem[\protect\citeauthoryear{Ley and
  Verdebout}{2017}]{leyverdebout2017modern}
\begin{bbook}
\bauthor{\bsnm{Ley}, \binits{C.}},
\bauthor{\bsnm{Verdebout}, \binits{T.}}:
\bbtitle{Modern Directional Statistics}.
\bpublisher{Chapman and Hall/CRC},
\blocation{Boca Raton, FL}
(\byear{2017}).
\doiurl{10.1201/9781315119472}
\end{bbook}
\endbibitem

\bibitem[\protect\citeauthoryear{Mardia and
  Sutton}{1978}]{mardia1978cylindrical}
\begin{barticle}
\bauthor{\bsnm{Mardia}, \binits{K.V.}},
\bauthor{\bsnm{Sutton}, \binits{T.W.}}:
\batitle{A model for cylindrical variables with applications}.
\bjtitle{Journal of the Royal Statistical Society: Series B (Methodological)}
\bvolume{40}(\bissue{2}),
\bfpage{229}--\blpage{233}
(\byear{1978})
\doiurl{10.1111/j.2517-6161.1978.tb01668.x}
\end{barticle}
\endbibitem

\bibitem[\protect\citeauthoryear{Efron}{1975}]{efron1975curvature}
\begin{barticle}
\bauthor{\bsnm{Efron}, \binits{B.}}:
\batitle{Defining the curvature of a statistical problem (with applications to
  second order efficiency)}.
\bjtitle{The Annals of Statistics}
\bvolume{3}(\bissue{6}),
\bfpage{1189}--\blpage{1242}
(\byear{1975})
\doiurl{10.1214/aos/1176343282}
\end{barticle}
\endbibitem

\bibitem[\protect\citeauthoryear{Amari}{1982}]{amari1982curved}
\begin{barticle}
\bauthor{\bsnm{Amari}, \binits{S.-I.}}:
\batitle{Differential geometry of curved exponential families---curvatures and
  information loss}.
\bjtitle{The Annals of Statistics}
\bvolume{10}(\bissue{2}),
\bfpage{357}--\blpage{385}
(\byear{1982})
\doiurl{10.1214/aos/1176345779}
\end{barticle}
\endbibitem

\bibitem[\protect\citeauthoryear{Kent}{1982}]{kent1982fisherbingham}
\begin{barticle}
\bauthor{\bsnm{Kent}, \binits{J.T.}}:
\batitle{The {F}isher--{B}ingham distribution on the sphere}.
\bjtitle{Journal of the Royal Statistical Society: Series B (Methodological)}
\bvolume{44}(\bissue{1}),
\bfpage{71}--\blpage{80}
(\byear{1982})
\doiurl{10.1111/j.2517-6161.1982.tb01189.x}
\end{barticle}
\endbibitem

\bibitem[\protect\citeauthoryear{Barndorff-Nielsen}{1978}]{barndorffnielsen1978information}
\begin{bbook}
\bauthor{\bsnm{Barndorff-Nielsen}, \binits{O.E.}}:
\bbtitle{Information and Exponential Families in Statistical Theory}.
\bpublisher{Wiley},
\blocation{Chichester}
(\byear{1978}).
\doiurl{10.1002/9781118857281}
\end{bbook}
\endbibitem

\bibitem[\protect\citeauthoryear{Amari}{1985}]{amari1985differential}
\begin{bbook}
\bauthor{\bsnm{Amari}, \binits{S.-i.}}:
\bbtitle{Differential-Geometrical Methods in Statistics}.
\bsertitle{Lecture Notes in Statistics},
vol. \bseriesno{28}.
\bpublisher{Springer},
\blocation{New York}
(\byear{1985}).
\doiurl{10.1007/978-1-4612-5056-2}
\end{bbook}
\endbibitem

\bibitem[\protect\citeauthoryear{Johnson and
  Wehrly}{1977}]{johnsonwehrly1977correlation}
\begin{barticle}
\bauthor{\bsnm{Johnson}, \binits{R.A.}},
\bauthor{\bsnm{Wehrly}, \binits{T.}}:
\batitle{Measures and models for angular correlation and angular--linear
  correlation}.
\bjtitle{Journal of the Royal Statistical Society: Series B (Methodological)}
\bvolume{39}(\bissue{2}),
\bfpage{222}--\blpage{229}
(\byear{1977})
\doiurl{10.1111/j.2517-6161.1977.tb01619.x}
\end{barticle}
\endbibitem

\bibitem[\protect\citeauthoryear{Johnson and
  Wehrly}{1978}]{johnsonwehrly1978angularlinear}
\begin{barticle}
\bauthor{\bsnm{Johnson}, \binits{R.A.}},
\bauthor{\bsnm{Wehrly}, \binits{T.E.}}:
\batitle{Some angular-linear distributions and related regression models}.
\bjtitle{Journal of the American Statistical Association}
\bvolume{73}(\bissue{363}),
\bfpage{602}--\blpage{606}
(\byear{1978})
\doiurl{10.1080/01621459.1978.10480062}
\end{barticle}
\endbibitem

\bibitem[\protect\citeauthoryear{Jupp and
  Mardia}{1980}]{juppmardia1980correlation}
\begin{barticle}
\bauthor{\bsnm{Jupp}, \binits{P.E.}},
\bauthor{\bsnm{Mardia}, \binits{K.V.}}:
\batitle{A general correlation coefficient for directional data and related
  regression problems}.
\bjtitle{Biometrika}
\bvolume{67}(\bissue{1}),
\bfpage{163}--\blpage{173}
(\byear{1980})
\doiurl{10.1093/biomet/67.1.163}
\end{barticle}
\endbibitem

\bibitem[\protect\citeauthoryear{Kato and
  Shimizu}{2008}]{katoshimizu2008dependent}
\begin{barticle}
\bauthor{\bsnm{Kato}, \binits{S.}},
\bauthor{\bsnm{Shimizu}, \binits{K.}}:
\batitle{Dependent models for observations which include angular ones}.
\bjtitle{Journal of Statistical Planning and Inference}
\bvolume{138}(\bissue{11}),
\bfpage{3538}--\blpage{3549}
(\byear{2008})
\doiurl{10.1016/j.jspi.2006.12.009}
\end{barticle}
\endbibitem

\bibitem[\protect\citeauthoryear{Abe and Ley}{2017}]{abe2017tractable}
\begin{barticle}
\bauthor{\bsnm{Abe}, \binits{T.}},
\bauthor{\bsnm{Ley}, \binits{C.}}:
\batitle{A tractable, parsimonious and flexible model for cylindrical data,
  with applications}.
\bjtitle{Econometrics and Statistics}
\bvolume{4},
\bfpage{91}--\blpage{104}
(\byear{2017})
\doiurl{10.1016/j.ecosta.2016.04.001}
\end{barticle}
\endbibitem

\bibitem[\protect\citeauthoryear{Luengo-S{\'a}nchez
  et~al.}{2019}]{luengo2019directional}
\begin{barticle}
\bauthor{\bsnm{Luengo-S{\'a}nchez}, \binits{S.}},
\bauthor{\bsnm{Larra{\~n}aga}, \binits{P.}},
\bauthor{\bsnm{Bielza}, \binits{C.}}:
\batitle{A directional-linear {Bayesian} network and its application for
  clustering and simulation of neural somas}.
\bjtitle{IEEE Access}
\bvolume{7},
\bfpage{69907}--\blpage{69921}
(\byear{2019})
\doiurl{10.1109/ACCESS.2019.2918494}
\end{barticle}
\endbibitem

\bibitem[\protect\citeauthoryear{Kume
  et~al.}{2013}]{kumeprestonwood2013fisherbingham}
\begin{barticle}
\bauthor{\bsnm{Kume}, \binits{A.}},
\bauthor{\bsnm{Preston}, \binits{S.P.}},
\bauthor{\bsnm{Wood}, \binits{A.T.A.}}:
\batitle{Saddlepoint approximations for the normalizing constant of
  {F}isher--{B}ingham distributions on products of spheres and {S}tiefel
  manifolds}.
\bjtitle{Biometrika}
\bvolume{100}(\bissue{4}),
\bfpage{971}--\blpage{984}
(\byear{2013})
\doiurl{10.1093/biomet/ast021}
\end{barticle}
\endbibitem

\bibitem[\protect\citeauthoryear{Mastrantonio
  et~al.}{2015}]{mastrantonio2015hmm}
\begin{barticle}
\bauthor{\bsnm{Mastrantonio}, \binits{G.}},
\bauthor{\bsnm{Maruotti}, \binits{A.}},
\bauthor{\bsnm{Jona-Lasinio}, \binits{G.}}:
\batitle{Bayesian hidden {M}arkov modelling using circular-linear general
  projected normal distribution}.
\bjtitle{Environmetrics}
\bvolume{26}(\bissue{2}),
\bfpage{145}--\blpage{158}
(\byear{2015})
\doiurl{10.1002/env.2326}
\end{barticle}
\endbibitem

\bibitem[\protect\citeauthoryear{Mastrantonio}{2018}]{mastrantonio2018polycylindrical}
\begin{barticle}
\bauthor{\bsnm{Mastrantonio}, \binits{G.}}:
\batitle{The joint projected normal and skew-normal: A distribution for
  poly-cylindrical data}.
\bjtitle{Journal of Multivariate Analysis}
\bvolume{165},
\bfpage{14}--\blpage{26}
(\byear{2018})
\doiurl{10.1016/j.jmva.2017.11.006}
\end{barticle}
\endbibitem

\bibitem[\protect\citeauthoryear{García-Portugués
  et~al.}{2013}]{garciaportugues2013kde}
\begin{barticle}
\bauthor{\bsnm{García-Portugués}, \binits{E.}},
\bauthor{\bsnm{Crujeiras}, \binits{R.M.}},
\bauthor{\bsnm{González-Manteiga}, \binits{W.}}:
\batitle{Kernel density estimation for directional--linear data}.
\bjtitle{Journal of Multivariate Analysis}
\bvolume{121},
\bfpage{152}--\blpage{175}
(\byear{2013})
\doiurl{10.1016/j.jmva.2013.06.009}
\end{barticle}
\endbibitem

\bibitem[\protect\citeauthoryear{Sadeghianpourhamami
  et~al.}{2019}]{sadeghianpourhamami2019abeley}
\begin{barticle}
\bauthor{\bsnm{Sadeghianpourhamami}, \binits{N.}},
\bauthor{\bsnm{Benoit}, \binits{D.F.}},
\bauthor{\bsnm{Deschrijver}, \binits{D.}},
\bauthor{\bsnm{Develder}, \binits{C.}}:
\batitle{Bayesian cylindrical data modeling using {A}be--{L}ey mixtures}.
\bjtitle{Applied Mathematical Modelling}
\bvolume{68},
\bfpage{629}--\blpage{642}
(\byear{2019})
\doiurl{10.1016/j.apm.2018.11.039}
\end{barticle}
\endbibitem

\bibitem[\protect\citeauthoryear{Roy et~al.}{2016}]{roy2016swgmm}
\begin{barticle}
\bauthor{\bsnm{Roy}, \binits{A.}},
\bauthor{\bsnm{Parui}, \binits{S.K.}},
\bauthor{\bsnm{Roy}, \binits{U.}}:
\batitle{{SWGMM}: a semi-wrapped {G}aussian mixture model for clustering of
  circular--linear data}.
\bjtitle{Pattern Analysis and Applications}
\bvolume{19}(\bissue{3}),
\bfpage{631}--\blpage{645}
(\byear{2016})
\doiurl{10.1007/s10044-014-0418-2}
\end{barticle}
\endbibitem

\bibitem[\protect\citeauthoryear{Roy et~al.}{2017}]{roy2017jclmm}
\begin{barticle}
\bauthor{\bsnm{Roy}, \binits{A.}},
\bauthor{\bsnm{Pal}, \binits{A.}},
\bauthor{\bsnm{Garain}, \binits{U.}}:
\batitle{{JCLMM}: A finite mixture model for clustering of circular--linear
  data and its application to psoriatic plaque segmentation}.
\bjtitle{Pattern Recognition}
\bvolume{66},
\bfpage{160}--\blpage{173}
(\byear{2017})
\doiurl{10.1016/j.patcog.2016.12.016}
\end{barticle}
\endbibitem

\bibitem[\protect\citeauthoryear{Hasnat et~al.}{2016}]{hasnat2016jcsd}
\begin{barticle}
\bauthor{\bsnm{Hasnat}, \binits{M.A.}},
\bauthor{\bsnm{Alata}, \binits{O.}},
\bauthor{\bsnm{Tr{\'e}meau}, \binits{A.}}:
\batitle{Joint color-spatial-directional clustering and region merging
  ({JCSD-RM}) for unsupervised {RGB-D} image segmentation}.
\bjtitle{IEEE Transactions on Pattern Analysis and Machine Intelligence}
\bvolume{38}(\bissue{11}),
\bfpage{2255}--\blpage{2268}
(\byear{2016})
\doiurl{10.1109/TPAMI.2015.2513407}
\end{barticle}
\endbibitem

\bibitem[\protect\citeauthoryear{Peel et~al.}{2001}]{peel2001kent}
\begin{barticle}
\bauthor{\bsnm{Peel}, \binits{D.}},
\bauthor{\bsnm{Whiten}, \binits{W.J.}},
\bauthor{\bsnm{McLachlan}, \binits{G.J.}}:
\batitle{Fitting mixtures of {K}ent distributions to aid in joint set
  identification}.
\bjtitle{Journal of the American Statistical Association}
\bvolume{96}(\bissue{453}),
\bfpage{56}--\blpage{63}
(\byear{2001})
\doiurl{10.1198/016214501750332974}
\end{barticle}
\endbibitem

\bibitem[\protect\citeauthoryear{Banerjee
  et~al.}{2005}]{banerjee2005clustering}
\begin{barticle}
\bauthor{\bsnm{Banerjee}, \binits{A.}},
\bauthor{\bsnm{Dhillon}, \binits{I.S.}},
\bauthor{\bsnm{Ghosh}, \binits{J.}},
\bauthor{\bsnm{Sra}, \binits{S.}}:
\batitle{Clustering on the unit hypersphere using von {M}ises--{F}isher
  distributions}.
\bjtitle{Journal of Machine Learning Research}
\bvolume{6}(\bissue{46}),
\bfpage{1345}--\blpage{1382}
(\byear{2005}).
\bcomment{\url{https://www.jmlr.org/papers/v6/banerjee05a.html}}
\end{barticle}
\endbibitem

\bibitem[\protect\citeauthoryear{Sra}{2012}]{sra2012vmf}
\begin{barticle}
\bauthor{\bsnm{Sra}, \binits{S.}}:
\batitle{A short note on parameter approximation for von {M}ises--{F}isher
  distributions: and a fast implementation of {$I_s(x)$}}.
\bjtitle{Computational Statistics}
\bvolume{27}(\bissue{1}),
\bfpage{177}--\blpage{190}
(\byear{2012})
\doiurl{10.1007/s00180-011-0232-x}
\end{barticle}
\endbibitem

\bibitem[\protect\citeauthoryear{Gershenfeld}{1997}]{gershenfeld1997cwm}
\begin{barticle}
\bauthor{\bsnm{Gershenfeld}, \binits{N.}}:
\batitle{Nonlinear inference and cluster-weighted modeling}.
\bjtitle{Annals of the New York Academy of Sciences}
\bvolume{808}(\bissue{1}),
\bfpage{18}--\blpage{24}
(\byear{1997})
\doiurl{10.1111/j.1749-6632.1997.tb51651.x}
\end{barticle}
\endbibitem

\bibitem[\protect\citeauthoryear{Ingrassia et~al.}{2012}]{ingrassia2012cwm}
\begin{barticle}
\bauthor{\bsnm{Ingrassia}, \binits{S.}},
\bauthor{\bsnm{Minotti}, \binits{S.C.}},
\bauthor{\bsnm{Vittadini}, \binits{G.}}:
\batitle{Local statistical modeling via a cluster-weighted approach with
  elliptical distributions}.
\bjtitle{Journal of Classification}
\bvolume{29}(\bissue{3}),
\bfpage{363}--\blpage{401}
(\byear{2012})
\doiurl{10.1007/s00357-012-9114-3}
\end{barticle}
\endbibitem

\bibitem[\protect\citeauthoryear{Bulla et~al.}{2012}]{bulla2012searegimes}
\begin{barticle}
\bauthor{\bsnm{Bulla}, \binits{J.}},
\bauthor{\bsnm{Lagona}, \binits{F.}},
\bauthor{\bsnm{Maruotti}, \binits{A.}},
\bauthor{\bsnm{Picone}, \binits{M.}}:
\batitle{A multivariate hidden {M}arkov model for the identification of sea
  regimes from incomplete skewed and circular time series}.
\bjtitle{Journal of Agricultural, Biological, and Environmental Statistics}
\bvolume{17}(\bissue{4}),
\bfpage{544}--\blpage{567}
(\byear{2012})
\doiurl{10.1007/s13253-012-0110-1}
\end{barticle}
\endbibitem

\bibitem[\protect\citeauthoryear{Lagona
  et~al.}{2015}]{lagona2015hmmcylindrical}
\begin{barticle}
\bauthor{\bsnm{Lagona}, \binits{F.}},
\bauthor{\bsnm{Picone}, \binits{M.}},
\bauthor{\bsnm{Maruotti}, \binits{A.}}:
\batitle{A hidden {M}arkov model for the analysis of cylindrical time series}.
\bjtitle{Environmetrics}
\bvolume{26}(\bissue{8}),
\bfpage{534}--\blpage{544}
(\byear{2015})
\doiurl{10.1002/env.2355}
\end{barticle}
\endbibitem

\bibitem[\protect\citeauthoryear{Wang et~al.}{2015}]{wang2015wavesadriatic}
\begin{barticle}
\bauthor{\bsnm{Wang}, \binits{F.}},
\bauthor{\bsnm{Gelfand}, \binits{A.E.}},
\bauthor{\bsnm{Jona-Lasinio}, \binits{G.}}:
\batitle{Joint spatio-temporal analysis of a linear and a directional variable:
  space-time modeling of wave heights and wave directions in the {A}driatic
  {S}ea}.
\bjtitle{Statistica Sinica}
\bvolume{25}(\bissue{1}),
\bfpage{25}--\blpage{39}
(\byear{2015})
\doiurl{10.5705/ss.2013.204w}
\end{barticle}
\endbibitem

\bibitem[\protect\citeauthoryear{Bingham}{1974}]{bingham1974antipodal}
\begin{barticle}
\bauthor{\bsnm{Bingham}, \binits{C.}}:
\batitle{An antipodally symmetric distribution on the sphere}.
\bjtitle{The Annals of Statistics}
\bvolume{2}(\bissue{6}),
\bfpage{1201}--\blpage{1225}
(\byear{1974})
\doiurl{10.1214/aos/1176342874}
\end{barticle}
\endbibitem

\bibitem[\protect\citeauthoryear{Gatto and
  Jammalamadaka}{2007}]{gattojammalamadaka2007gvm}
\begin{barticle}
\bauthor{\bsnm{Gatto}, \binits{R.}},
\bauthor{\bsnm{Jammalamadaka}, \binits{S.R.}}:
\batitle{The generalized von {M}ises distribution}.
\bjtitle{Statistical Methodology}
\bvolume{4}(\bissue{3}),
\bfpage{341}--\blpage{353}
(\byear{2007})
\doiurl{10.1016/j.stamet.2006.11.003}
\end{barticle}
\endbibitem

\bibitem[\protect\citeauthoryear{Kume and Wood}{2005}]{kumewood2005saddlepoint}
\begin{barticle}
\bauthor{\bsnm{Kume}, \binits{A.}},
\bauthor{\bsnm{Wood}, \binits{A.T.A.}}:
\batitle{Saddlepoint approximations for the {B}ingham and {F}isher--{B}ingham
  normalising constants}.
\bjtitle{Biometrika}
\bvolume{92}(\bissue{2}),
\bfpage{465}--\blpage{476}
(\byear{2005})
\doiurl{10.1093/biomet/92.2.465}
\end{barticle}
\endbibitem

\bibitem[\protect\citeauthoryear{Kume and Sei}{2018}]{kumesei2018holonomic}
\begin{barticle}
\bauthor{\bsnm{Kume}, \binits{A.}},
\bauthor{\bsnm{Sei}, \binits{T.}}:
\batitle{On the exact maximum likelihood inference of {F}isher--{B}ingham
  distributions using an adjusted holonomic gradient method}.
\bjtitle{Statistics and Computing}
\bvolume{28}(\bissue{4}),
\bfpage{835}--\blpage{847}
(\byear{2018})
\doiurl{10.1007/s11222-017-9765-3}
\end{barticle}
\endbibitem

\bibitem[\protect\citeauthoryear{Abbahaddou and
  Aboussalah}{2025}]{abbahaddou2025geometry}
\begin{bchapter}
\bauthor{\bsnm{Abbahaddou}, \binits{Y.}},
\bauthor{\bsnm{Aboussalah}, \binits{A.}}:
\bctitle{A geometry-aware metric for mode collapse in time series generative
  models}.
In: \bbtitle{Advances in Neural Information Processing Systems 38 ({NeurIPS}
  2025)},
pp. \bfpage{168149}--\blpage{168179}.
\bpublisher{Neural Information Processing Systems Foundation},
\blocation{San Diego, CA, USA}
(\byear{2025}).
\doiurl{10.52202/085713-5065}
\end{bchapter}
\endbibitem

\bibitem[\protect\citeauthoryear{Cox}{1975}]{cox1975discussion}
\begin{barticle}
\bauthor{\bsnm{Cox}, \binits{D.R.}}:
\batitle{Contribution to the discussion of ``{S}tatistics of directional data''
  by {K. V. Mardia}}.
\bjtitle{Journal of the Royal Statistical Society: Series B (Methodological)}
\bvolume{37}(\bissue{3}),
\bfpage{380}--\blpage{381}
(\byear{1975})
\doiurl{10.1111/j.2517-6161.1975.tb01551.x}
\end{barticle}
\endbibitem

\bibitem[\protect\citeauthoryear{Pewsey}{2002}]{pewsey2002symmetry}
\begin{barticle}
\bauthor{\bsnm{Pewsey}, \binits{A.}}:
\batitle{Testing circular symmetry}.
\bjtitle{The Canadian Journal of Statistics}
\bvolume{30}(\bissue{4}),
\bfpage{591}--\blpage{600}
(\byear{2002})
\doiurl{10.2307/3316098}
\end{barticle}
\endbibitem

\bibitem[\protect\citeauthoryear{Mardia et~al.}{1984}]{mardiaholmeskent1984gof}
\begin{barticle}
\bauthor{\bsnm{Mardia}, \binits{K.V.}},
\bauthor{\bsnm{Holmes}, \binits{D.}},
\bauthor{\bsnm{Kent}, \binits{J.T.}}:
\batitle{A goodness-of-fit test for the von {M}ises--{F}isher distribution}.
\bjtitle{Journal of the Royal Statistical Society: Series B (Methodological)}
\bvolume{46}(\bissue{1}),
\bfpage{72}--\blpage{78}
(\byear{1984})
\doiurl{10.1111/j.2517-6161.1984.tb01278.x}
\end{barticle}
\endbibitem

\bibitem[\protect\citeauthoryear{Rivest}{1986}]{rivest1986kent}
\begin{barticle}
\bauthor{\bsnm{Rivest}, \binits{L.-P.}}:
\batitle{Modified {K}ent's statistics for testing goodness of fit for the
  {F}isher distribution in small concentrated samples}.
\bjtitle{Statistics \& Probability Letters}
\bvolume{4}(\bissue{1}),
\bfpage{1}--\blpage{4}
(\byear{1986})
\doiurl{10.1016/0167-7152(86)90028-3}
\end{barticle}
\endbibitem

\bibitem[\protect\citeauthoryear{Mardia and
  Jupp}{2000}]{mardiajupp2000directional}
\begin{bbook}
\bauthor{\bsnm{Mardia}, \binits{K.V.}},
\bauthor{\bsnm{Jupp}, \binits{P.E.}}:
\bbtitle{Directional Statistics}.
\bsertitle{Wiley Series in Probability and Statistics}.
\bpublisher{Wiley},
\blocation{Chichester}
(\byear{2000}).
\doiurl{10.1002/9780470316979}
\end{bbook}
\endbibitem

\bibitem[\protect\citeauthoryear{Efron}{1978}]{efron1978geometry}
\begin{barticle}
\bauthor{\bsnm{Efron}, \binits{B.}}:
\batitle{The geometry of exponential families}.
\bjtitle{The Annals of Statistics}
\bvolume{6}(\bissue{2}),
\bfpage{362}--\blpage{376}
(\byear{1978})
\doiurl{10.1214/aos/1176344130}
\end{barticle}
\endbibitem

\bibitem[\protect\citeauthoryear{Amari}{1982}]{amari1982ancillarity}
\begin{barticle}
\bauthor{\bsnm{Amari}, \binits{S.-i.}}:
\batitle{Geometrical theory of asymptotic ancillarity and conditional
  inference}.
\bjtitle{Biometrika}
\bvolume{69}(\bissue{1}),
\bfpage{1}--\blpage{17}
(\byear{1982})
\doiurl{10.1093/biomet/69.1.1}
\end{barticle}
\endbibitem

\bibitem[\protect\citeauthoryear{Efron and
  Hinkley}{1978}]{efronhinkley1978observed}
\begin{barticle}
\bauthor{\bsnm{Efron}, \binits{B.}},
\bauthor{\bsnm{Hinkley}, \binits{D.V.}}:
\batitle{Assessing the accuracy of the maximum likelihood estimator: Observed
  versus expected {F}isher information}.
\bjtitle{Biometrika}
\bvolume{65}(\bissue{3}),
\bfpage{457}--\blpage{483}
(\byear{1978})
\doiurl{10.1093/biomet/65.3.457}
\end{barticle}
\endbibitem

\bibitem[\protect\citeauthoryear{Kumon and Amari}{1983}]{kumonamari1983higher}
\begin{barticle}
\bauthor{\bsnm{Kumon}, \binits{M.}},
\bauthor{\bsnm{Amari}, \binits{S.-i.}}:
\batitle{Geometrical theory of higher-order asymptotics of test, interval
  estimator and conditional inference}.
\bjtitle{Proceedings of the Royal Society of London. Series A}
\bvolume{387}(\bissue{1793}),
\bfpage{429}--\blpage{458}
(\byear{1983})
\doiurl{10.1098/rspa.1983.0068}
\end{barticle}
\endbibitem

\bibitem[\protect\citeauthoryear{Amari and
  Nagaoka}{2000}]{amarinagaoka2000methods}
\begin{bbook}
\bauthor{\bsnm{Amari}, \binits{S.-i.}},
\bauthor{\bsnm{Nagaoka}, \binits{H.}}:
\bbtitle{Methods of Information Geometry}.
\bsertitle{Translations of Mathematical Monographs},
vol. \bseriesno{191}.
\bpublisher{American Mathematical Society},
\blocation{Providence, RI}
(\byear{2000}).
\doiurl{10.1090/mmono/191}
\end{bbook}
\endbibitem

\bibitem[\protect\citeauthoryear{Kass and Vos}{1997}]{kassvos1997geometrical}
\begin{bbook}
\bauthor{\bsnm{Kass}, \binits{R.E.}},
\bauthor{\bsnm{Vos}, \binits{P.W.}}:
\bbtitle{Geometrical Foundations of Asymptotic Inference}.
\bsertitle{Wiley Series in Probability and Statistics}.
\bpublisher{Wiley},
\blocation{New York}
(\byear{1997}).
\doiurl{10.1002/9781118165980}
\end{bbook}
\endbibitem

\bibitem[\protect\citeauthoryear{Amari}{1998}]{amari1998natural}
\begin{barticle}
\bauthor{\bsnm{Amari}, \binits{S.-i.}}:
\batitle{Natural gradient works efficiently in learning}.
\bjtitle{Neural Computation}
\bvolume{10}(\bissue{2}),
\bfpage{251}--\blpage{276}
(\byear{1998})
\doiurl{10.1162/089976698300017746}
\end{barticle}
\endbibitem

\bibitem[\protect\citeauthoryear{Attar et~al.}{2026}]{attar2026quantum}
\begin{botherref}
\oauthor{\bsnm{Attar}, \binits{A.}},
\oauthor{\bsnm{Aboussalah}, \binits{A.M.}},
\oauthor{\bsnm{Hibat-Allah}, \binits{M.}}:
Quantum Geometric Tensor Preconditioning for Stable Training of Recurrent
  Neural Quantum States.
arXiv:2608.18065
(2026)
\end{botherref}
\endbibitem

\bibitem[\protect\citeauthoryear{Ed-dib et~al.}{2025}]{edib2025gelora}
\begin{bchapter}
\bauthor{\bsnm{Ed-dib}, \binits{A.}},
\bauthor{\bsnm{Datbayev}, \binits{Z.}},
\bauthor{\bsnm{Aboussalah}, \binits{A.M.}}:
\bctitle{{G}e{L}o{RA}: Geometric adaptive ranks for efficient {L}o{RA}
  fine-tuning}.
In: \beditor{\bsnm{Christodoulopoulos}, \binits{C.}},
\beditor{\bsnm{Chakraborty}, \binits{T.}},
\beditor{\bsnm{Rose}, \binits{C.}},
\beditor{\bsnm{Peng}, \binits{V.}} (eds.)
\bbtitle{Findings of the {A}ssociation for {C}omputational {L}inguistics:
  {EMNLP} 2025},
pp. \bfpage{25174}--\blpage{25196}.
\bpublisher{Association for Computational Linguistics},
\blocation{Suzhou, China}
(\byear{2025}).
\doiurl{10.18653/v1/2025.findings-emnlp.1372}
\end{bchapter}
\endbibitem

\bibitem[\protect\citeauthoryear{Rao}{1948}]{rao1948score}
\begin{barticle}
\bauthor{\bsnm{Rao}, \binits{C.R.}}:
\batitle{Large sample tests of statistical hypotheses concerning several
  parameters with applications to problems of estimation}.
\bjtitle{Mathematical Proceedings of the Cambridge Philosophical Society}
\bvolume{44}(\bissue{1}),
\bfpage{50}--\blpage{57}
(\byear{1948})
\doiurl{10.1017/S0305004100023987}
\end{barticle}
\endbibitem

\bibitem[\protect\citeauthoryear{Neyman}{1959}]{neyman1959optimal}
\begin{bchapter}
\bauthor{\bsnm{Neyman}, \binits{J.}}:
\bctitle{Optimal asymptotic tests of composite statistical hypotheses}.
In: \beditor{\bsnm{Grenander}, \binits{U.}} (ed.)
\bbtitle{Probability and Statistics: The {H}arald {C}ram\'er Volume},
pp. \bfpage{213}--\blpage{234}.
\bpublisher{Almqvist \& Wiksell},
\blocation{Stockholm}
(\byear{1959})
\end{bchapter}
\endbibitem

\bibitem[\protect\citeauthoryear{Dempster et~al.}{1977}]{dempster1977em}
\begin{barticle}
\bauthor{\bsnm{Dempster}, \binits{A.P.}},
\bauthor{\bsnm{Laird}, \binits{N.M.}},
\bauthor{\bsnm{Rubin}, \binits{D.B.}}:
\batitle{Maximum likelihood from incomplete data via the {EM} algorithm}.
\bjtitle{Journal of the Royal Statistical Society: Series B (Methodological)}
\bvolume{39}(\bissue{1}),
\bfpage{1}--\blpage{22}
(\byear{1977})
\doiurl{10.1111/j.2517-6161.1977.tb01600.x}
\end{barticle}
\endbibitem

\bibitem[\protect\citeauthoryear{Wu}{1983}]{wu1983em}
\begin{barticle}
\bauthor{\bsnm{Wu}, \binits{C.F.J.}}:
\batitle{On the convergence properties of the {EM} algorithm}.
\bjtitle{The Annals of Statistics}
\bvolume{11}(\bissue{1}),
\bfpage{95}--\blpage{103}
(\byear{1983})
\doiurl{10.1214/aos/1176346060}
\end{barticle}
\endbibitem

\bibitem[\protect\citeauthoryear{Csisz{\'a}r and
  Tusn{\'a}dy}{1984}]{csiszartusnady1984alternating}
\begin{barticle}
\bauthor{\bsnm{Csisz{\'a}r}, \binits{I.}},
\bauthor{\bsnm{Tusn{\'a}dy}, \binits{G.}}:
\batitle{Information geometry and alternating minimization procedures}.
\bjtitle{Statistics \& Decisions}
\bvolume{Supplement Issue 1},
\bfpage{205}--\blpage{237}
(\byear{1984})
\end{barticle}
\endbibitem

\bibitem[\protect\citeauthoryear{Amari}{1995}]{amari1995em}
\begin{barticle}
\bauthor{\bsnm{Amari}, \binits{S.-i.}}:
\batitle{Information geometry of the {EM} and em algorithms for neural
  networks}.
\bjtitle{Neural Networks}
\bvolume{8}(\bissue{9}),
\bfpage{1379}--\blpage{1408}
(\byear{1995})
\doiurl{10.1016/0893-6080(95)00003-8}
\end{barticle}
\endbibitem

\bibitem[\protect\citeauthoryear{Neal and Hinton}{1998}]{nealhinton1998em}
\begin{bchapter}
\bauthor{\bsnm{Neal}, \binits{R.M.}},
\bauthor{\bsnm{Hinton}, \binits{G.E.}}:
\bctitle{A view of the {EM} algorithm that justifies incremental, sparse, and
  other variants}.
In: \beditor{\bsnm{Jordan}, \binits{M.I.}} (ed.)
\bbtitle{Learning in Graphical Models},
pp. \bfpage{355}--\blpage{368}.
\bpublisher{Kluwer},
\blocation{Dordrecht}
(\byear{1998}).
\doiurl{10.1007/978-94-011-5014-9_12}
\end{bchapter}
\endbibitem

\bibitem[\protect\citeauthoryear{Yakowitz and
  Spragins}{1968}]{yakowitz1968identifiability}
\begin{barticle}
\bauthor{\bsnm{Yakowitz}, \binits{S.J.}},
\bauthor{\bsnm{Spragins}, \binits{J.D.}}:
\batitle{On the identifiability of finite mixtures}.
\bjtitle{The Annals of Mathematical Statistics}
\bvolume{39}(\bissue{1}),
\bfpage{209}--\blpage{214}
(\byear{1968})
\doiurl{10.1214/aoms/1177698520}
\end{barticle}
\endbibitem

\bibitem[\protect\citeauthoryear{Kent}{1983}]{kent1983identifiability}
\begin{barticle}
\bauthor{\bsnm{Kent}, \binits{J.T.}}:
\batitle{Identifiability of finite mixtures for directional data}.
\bjtitle{The Annals of Statistics}
\bvolume{11}(\bissue{3}),
\bfpage{984}--\blpage{988}
(\byear{1983})
\doiurl{10.1214/aos/1176346264}
\end{barticle}
\endbibitem

\bibitem[\protect\citeauthoryear{Hathaway}{1985}]{hathaway1985constrained}
\begin{barticle}
\bauthor{\bsnm{Hathaway}, \binits{R.J.}}:
\batitle{A constrained formulation of maximum-likelihood estimation for normal
  mixture distributions}.
\bjtitle{The Annals of Statistics}
\bvolume{13}(\bissue{2}),
\bfpage{795}--\blpage{800}
(\byear{1985})
\doiurl{10.1214/aos/1176349557}
\end{barticle}
\endbibitem

\bibitem[\protect\citeauthoryear{Abbahaddou et~al.}{2025}]{abbahaddou2025graph}
\begin{bchapter}
\bauthor{\bsnm{Abbahaddou}, \binits{Y.}},
\bauthor{\bsnm{Malliaros}, \binits{F.D.}},
\bauthor{\bsnm{Lutzeyer}, \binits{J.F.}},
\bauthor{\bsnm{Aboussalah}, \binits{A.M.}},
\bauthor{\bsnm{Vazirgiannis}, \binits{M.}}:
\bctitle{Graph neural network generalization with {G}aussian mixture model
  based augmentation}.
In: \bbtitle{Proceedings of the 42nd International Conference on Machine
  Learning}.
\bsertitle{Proceedings of Machine Learning Research},
vol. \bseriesno{267},
pp. \bfpage{102}--\blpage{125}.
\bpublisher{PMLR},
\blocation{Vancouver, Canada}
(\byear{2025}).
\burl{https://proceedings.mlr.press/v267/abbahaddou25a.html}
\end{bchapter}
\endbibitem

\bibitem[\protect\citeauthoryear{Malsiner-Walli
  et~al.}{2017}]{malsinerwalli2017identifying}
\begin{barticle}
\bauthor{\bsnm{Malsiner-Walli}, \binits{G.}},
\bauthor{\bsnm{Fr{\"u}hwirth-Schnatter}, \binits{S.}},
\bauthor{\bsnm{Gr{\"u}n}, \binits{B.}}:
\batitle{Identifying mixtures of mixtures using {B}ayesian estimation}.
\bjtitle{Journal of Computational and Graphical Statistics}
\bvolume{26}(\bissue{2}),
\bfpage{285}--\blpage{295}
(\byear{2017})
\doiurl{10.1080/10618600.2016.1200472}
\end{barticle}
\endbibitem

\bibitem[\protect\citeauthoryear{Decroos et~al.}{2021}]{decroos2021soccermix}
\begin{bchapter}
\bauthor{\bsnm{Decroos}, \binits{T.}},
\bauthor{\bsnm{{Van Roy}}, \binits{M.}},
\bauthor{\bsnm{Davis}, \binits{J.}}:
\bctitle{{SoccerMix}: Representing soccer actions with mixture models}.
In: \beditor{\bsnm{Dong}, \binits{Y.}},
\beditor{\bsnm{Ifrim}, \binits{G.}},
\beditor{\bsnm{Mladeni{\'c}}, \binits{D.}},
\beditor{\bsnm{Saunders}, \binits{C.}},
\beditor{\bsnm{Van~Hoecke}, \binits{S.}} (eds.)
\bbtitle{Machine Learning and Knowledge Discovery in Databases. Applied Data
  Science and Demo Track. {ECML} {PKDD} 2020}.
\bsertitle{Lecture Notes in Computer Science},
vol. \bseriesno{12461},
pp. \bfpage{459}--\blpage{474}.
\bpublisher{Springer},
\blocation{Cham}
(\byear{2021}).
\doiurl{10.1007/978-3-030-67670-4_28}
\end{bchapter}
\endbibitem

\bibitem[\protect\citeauthoryear{Bolck et~al.}{2004}]{bolck2004threestep}
\begin{barticle}
\bauthor{\bsnm{Bolck}, \binits{A.}},
\bauthor{\bsnm{Croon}, \binits{M.}},
\bauthor{\bsnm{Hagenaars}, \binits{J.}}:
\batitle{Estimating latent structure models with categorical variables:
  One-step versus three-step estimators}.
\bjtitle{Political Analysis}
\bvolume{12}(\bissue{1}),
\bfpage{3}--\blpage{27}
(\byear{2004})
\doiurl{10.1093/pan/mph001}
\end{barticle}
\endbibitem

\bibitem[\protect\citeauthoryear{Vermunt}{2010}]{vermunt2010threestep}
\begin{barticle}
\bauthor{\bsnm{Vermunt}, \binits{J.K.}}:
\batitle{Latent class modeling with covariates: Two improved three-step
  approaches}.
\bjtitle{Political Analysis}
\bvolume{18}(\bissue{4}),
\bfpage{450}--\blpage{469}
(\byear{2010})
\doiurl{10.1093/pan/mpq025}
\end{barticle}
\endbibitem

\bibitem[\protect\citeauthoryear{Bakk and Kuha}{2018}]{bakkkuha2018twostep}
\begin{barticle}
\bauthor{\bsnm{Bakk}, \binits{Z.}},
\bauthor{\bsnm{Kuha}, \binits{J.}}:
\batitle{Two-step estimation of models between latent classes and external
  variables}.
\bjtitle{Psychometrika}
\bvolume{83}(\bissue{4}),
\bfpage{871}--\blpage{892}
(\byear{2018})
\doiurl{10.1007/s11336-017-9592-7}
\end{barticle}
\endbibitem

\bibitem[\protect\citeauthoryear{Garay~Vergara}{2026}]{garayvergara2026thesis}
\begin{botherref}
\oauthor{\bsnm{Garay~Vergara}, \binits{J.E.}}:
Geometry-aware latent variable models for cylindrical data.
Master's thesis,
New York University Tandon School of Engineering,
Brooklyn, NY
(August 2026)
\end{botherref}
\endbibitem

\bibitem[\protect\citeauthoryear{Kiefer and
  Vogelsang}{2005}]{kiefervogelsang2005fixedb}
\begin{barticle}
\bauthor{\bsnm{Kiefer}, \binits{N.M.}},
\bauthor{\bsnm{Vogelsang}, \binits{T.J.}}:
\batitle{A new asymptotic theory for heteroskedasticity-autocorrelation robust
  tests}.
\bjtitle{Econometric Theory}
\bvolume{21}(\bissue{6}),
\bfpage{1130}--\blpage{1164}
(\byear{2005})
\doiurl{10.1017/S0266466605050565}
\end{barticle}
\endbibitem

\bibitem[\protect\citeauthoryear{Anderson}{2003}]{anderson2003multivariate}
\begin{bbook}
\bauthor{\bsnm{Anderson}, \binits{T.W.}}:
\bbtitle{An Introduction to Multivariate Statistical Analysis},
\bedition{3}rd edn.
\bpublisher{John Wiley \& Sons},
\blocation{Hoboken, NJ}
(\byear{2003})
\end{bbook}
\endbibitem

\bibitem[\protect\citeauthoryear{Fraser et~al.}{1981}]{fraser1981vonmises}
\begin{barticle}
\bauthor{\bsnm{Fraser}, \binits{M.D.}},
\bauthor{\bsnm{Hsu}, \binits{Y.-S.}},
\bauthor{\bsnm{Walker}, \binits{J.J.}}:
\batitle{Identifiability of finite mixtures of von {M}ises distributions}.
\bjtitle{The Annals of Statistics}
\bvolume{9}(\bissue{5}),
\bfpage{1130}--\blpage{1131}
(\byear{1981})
\doiurl{10.1214/aos/1176345595}
\end{barticle}
\endbibitem

\bibitem[\protect\citeauthoryear{Hennig}{2000}]{hennig2000identifiability}
\begin{barticle}
\bauthor{\bsnm{Hennig}, \binits{C.}}:
\batitle{Identifiability of models for clusterwise linear regression}.
\bjtitle{Journal of Classification}
\bvolume{17}(\bissue{2}),
\bfpage{273}--\blpage{296}
(\byear{2000})
\doiurl{10.1007/s003570000022}
\end{barticle}
\endbibitem

\bibitem[\protect\citeauthoryear{Ingrassia et~al.}{2015}]{ingrassia2015glmcwm}
\begin{barticle}
\bauthor{\bsnm{Ingrassia}, \binits{S.}},
\bauthor{\bsnm{Punzo}, \binits{A.}},
\bauthor{\bsnm{Vittadini}, \binits{G.}},
\bauthor{\bsnm{Minotti}, \binits{S.C.}}:
\batitle{The generalized linear mixed cluster-weighted model}.
\bjtitle{Journal of Classification}
\bvolume{32}(\bissue{1}),
\bfpage{85}--\blpage{113}
(\byear{2015})
\doiurl{10.1007/s00357-015-9175-1}
\end{barticle}
\endbibitem

\bibitem[\protect\citeauthoryear{Louis}{1982}]{louis1982em}
\begin{barticle}
\bauthor{\bsnm{Louis}, \binits{T.A.}}:
\batitle{Finding the observed information matrix when using the {EM}
  algorithm}.
\bjtitle{Journal of the Royal Statistical Society: Series B (Methodological)}
\bvolume{44}(\bissue{2}),
\bfpage{226}--\blpage{233}
(\byear{1982})
\doiurl{10.1111/j.2517-6161.1982.tb01203.x}
\end{barticle}
\endbibitem

\bibitem[\protect\citeauthoryear{Amari}{2016}]{amari2016information}
\begin{bbook}
\bauthor{\bsnm{Amari}, \binits{S.-i.}}:
\bbtitle{Information Geometry and Its Applications}.
\bsertitle{Applied Mathematical Sciences},
vol. \bseriesno{194}.
\bpublisher{Springer},
\blocation{Tokyo}
(\byear{2016}).
\doiurl{10.1007/978-4-431-55978-8}
\end{bbook}
\endbibitem

\bibitem[\protect\citeauthoryear{Ay et~al.}{2017}]{ay2017information}
\begin{bbook}
\bauthor{\bsnm{Ay}, \binits{N.}},
\bauthor{\bsnm{Jost}, \binits{J.}},
\bauthor{\bsnm{L{\^e}}, \binits{H.V.}},
\bauthor{\bsnm{Schwachh{\"o}fer}, \binits{L.}}:
\bbtitle{Information Geometry}.
\bsertitle{Ergebnisse der Mathematik und ihrer Grenzgebiete}.
\bpublisher{Springer},
\blocation{Cham}
(\byear{2017}).
\doiurl{10.1007/978-3-319-56478-4}
\end{bbook}
\endbibitem

\bibitem[\protect\citeauthoryear{Banerjee et~al.}{2005}]{banerjee2005bregman}
\begin{barticle}
\bauthor{\bsnm{Banerjee}, \binits{A.}},
\bauthor{\bsnm{Merugu}, \binits{S.}},
\bauthor{\bsnm{Dhillon}, \binits{I.S.}},
\bauthor{\bsnm{Ghosh}, \binits{J.}}:
\batitle{Clustering with {B}regman divergences}.
\bjtitle{Journal of Machine Learning Research}
\bvolume{6}(\bissue{58}),
\bfpage{1705}--\blpage{1749}
(\byear{2005}).
\bcomment{\url{https://www.jmlr.org/papers/v6/banerjee05b.html}}
\end{barticle}
\endbibitem

\bibitem[\protect\citeauthoryear{Hino et~al.}{2024}]{hino2022geometryem}
\begin{barticle}
\bauthor{\bsnm{Hino}, \binits{H.}},
\bauthor{\bsnm{Akaho}, \binits{S.}},
\bauthor{\bsnm{Murata}, \binits{N.}}:
\batitle{Geometry of {EM} and related iterative algorithms}.
\bjtitle{Information Geometry}
\bvolume{7}(\bissue{S1}),
\bfpage{39}--\blpage{77}
(\byear{2024})
\doiurl{10.1007/s41884-022-00080-y}
\end{barticle}
\endbibitem

\bibitem[\protect\citeauthoryear{Schou}{1978}]{schou1978concentration}
\begin{barticle}
\bauthor{\bsnm{Schou}, \binits{G.}}:
\batitle{Estimation of the concentration parameter in von {M}ises--{F}isher
  distributions}.
\bjtitle{Biometrika}
\bvolume{65}(\bissue{2}),
\bfpage{369}--\blpage{377}
(\byear{1978})
\doiurl{10.1093/biomet/65.2.369}
\end{barticle}
\endbibitem

\bibitem[\protect\citeauthoryear{Pennec et~al.}{2006}]{pennec2006riemannian}
\begin{barticle}
\bauthor{\bsnm{Pennec}, \binits{X.}},
\bauthor{\bsnm{Fillard}, \binits{P.}},
\bauthor{\bsnm{Ayache}, \binits{N.}}:
\batitle{A {R}iemannian framework for tensor computing}.
\bjtitle{International Journal of Computer Vision}
\bvolume{66}(\bissue{1}),
\bfpage{41}--\blpage{66}
(\byear{2006})
\doiurl{10.1007/s11263-005-3222-z}
\end{barticle}
\endbibitem

\bibitem[\protect\citeauthoryear{Aboussalah and
  Ed-dib}{2025}]{aboussalah2025geohnns}
\begin{botherref}
\oauthor{\bsnm{Aboussalah}, \binits{A.M.}},
\oauthor{\bsnm{Ed-dib}, \binits{A.}}:
{GeoHNN}s: Geometric {H}amiltonian Neural Networks.
arXiv:2507.15678
(2025)
\end{botherref}
\endbibitem

\bibitem[\protect\citeauthoryear{Aboussalah
  et~al.}{2023}]{aboussalah2023recursive}
\begin{bchapter}
\bauthor{\bsnm{Aboussalah}, \binits{A.M.}},
\bauthor{\bsnm{Kwon}, \binits{M.}},
\bauthor{\bsnm{Patel}, \binits{R.}},
\bauthor{\bsnm{Chi}, \binits{C.}},
\bauthor{\bsnm{Lee}, \binits{C.-G.}}:
\bctitle{Recursive time series data augmentation}.
In: \bbtitle{The Eleventh International Conference on Learning Representations
  ({ICLR} 2023)}
(\byear{2023}).
\burl{https://openreview.net/forum?id=5lgD4vU-l24s}
\end{bchapter}
\endbibitem

\bibitem[\protect\citeauthoryear{{van der
  Vaart}}{1998}]{vandervaart1998asymptotic}
\begin{bbook}
\bauthor{\bsnm{{van der Vaart}}, \binits{A.W.}}:
\bbtitle{Asymptotic Statistics}.
\bpublisher{Cambridge University Press},
\blocation{Cambridge}
(\byear{1998}).
\doiurl{10.1017/CBO9780511802256}
\end{bbook}
\endbibitem

\bibitem[\protect\citeauthoryear{Ingrassia
  et~al.}{2014}]{ingrassia2014linearcwm}
\begin{barticle}
\bauthor{\bsnm{Ingrassia}, \binits{S.}},
\bauthor{\bsnm{Minotti}, \binits{S.C.}},
\bauthor{\bsnm{Punzo}, \binits{A.}}:
\batitle{Model-based clustering via linear cluster-weighted models}.
\bjtitle{Computational Statistics \& Data Analysis}
\bvolume{71},
\bfpage{159}--\blpage{182}
(\byear{2014})
\doiurl{10.1016/j.csda.2013.02.012}
\end{barticle}
\endbibitem

\bibitem[\protect\citeauthoryear{Jupp}{2005}]{jupp2005sobolev}
\begin{barticle}
\bauthor{\bsnm{Jupp}, \binits{P.E.}}:
\batitle{Sobolev tests of goodness of fit of distributions on compact
  {R}iemannian manifolds}.
\bjtitle{The Annals of Statistics}
\bvolume{33}(\bissue{6}),
\bfpage{2957}--\blpage{2966}
(\byear{2005})
\doiurl{10.1214/009053605000000697}
\end{barticle}
\endbibitem

\bibitem[\protect\citeauthoryear{Redner and
  Walker}{1984}]{rednerwalker1984mixture}
\begin{barticle}
\bauthor{\bsnm{Redner}, \binits{R.A.}},
\bauthor{\bsnm{Walker}, \binits{H.F.}}:
\batitle{Mixture densities, maximum likelihood and the {EM} algorithm}.
\bjtitle{SIAM Review}
\bvolume{26}(\bissue{2}),
\bfpage{195}--\blpage{239}
(\byear{1984})
\doiurl{10.1137/1026034}
\end{barticle}
\endbibitem

\bibitem[\protect\citeauthoryear{Watanabe}{2013}]{watanabe2013wbic}
\begin{barticle}
\bauthor{\bsnm{Watanabe}, \binits{S.}}:
\batitle{A widely applicable {B}ayesian information criterion}.
\bjtitle{Journal of Machine Learning Research}
\bvolume{14}(\bissue{27}),
\bfpage{867}--\blpage{897}
(\byear{2013}).
\bcomment{\url{https://jmlr.org/papers/v14/watanabe13a.html}}
\end{barticle}
\endbibitem

\bibitem[\protect\citeauthoryear{Newey and West}{1987}]{neweywest1987hac}
\begin{barticle}
\bauthor{\bsnm{Newey}, \binits{W.K.}},
\bauthor{\bsnm{West}, \binits{K.D.}}:
\batitle{A simple, positive semi-definite, heteroskedasticity and
  autocorrelation consistent covariance matrix}.
\bjtitle{Econometrica}
\bvolume{55}(\bissue{3}),
\bfpage{703}--\blpage{708}
(\byear{1987})
\doiurl{10.2307/1913610}
\end{barticle}
\endbibitem

\bibitem[\protect\citeauthoryear{K{\"u}nsch}{1989}]{kunsch1989block}
\begin{barticle}
\bauthor{\bsnm{K{\"u}nsch}, \binits{H.R.}}:
\batitle{The jackknife and the bootstrap for general stationary observations}.
\bjtitle{The Annals of Statistics}
\bvolume{17}(\bissue{3}),
\bfpage{1217}--\blpage{1241}
(\byear{1989})
\doiurl{10.1214/aos/1176347265}
\end{barticle}
\endbibitem

\bibitem[\protect\citeauthoryear{Shao}{2010}]{shao2010selfnormalized}
\begin{barticle}
\bauthor{\bsnm{Shao}, \binits{X.}}:
\batitle{A self-normalized approach to confidence interval construction in time
  series}.
\bjtitle{Journal of the Royal Statistical Society: Series B (Statistical
  Methodology)}
\bvolume{72}(\bissue{3}),
\bfpage{343}--\blpage{366}
(\byear{2010})
\doiurl{10.1111/j.1467-9868.2009.00737.x}
\end{barticle}
\endbibitem

\bibitem[\protect\citeauthoryear{Presnell et~al.}{1998}]{presnell1998projected}
\begin{barticle}
\bauthor{\bsnm{Presnell}, \binits{B.}},
\bauthor{\bsnm{Morrison}, \binits{S.P.}},
\bauthor{\bsnm{Littell}, \binits{R.C.}}:
\batitle{Projected multivariate linear models for directional data}.
\bjtitle{Journal of the American Statistical Association}
\bvolume{93}(\bissue{443}),
\bfpage{1068}--\blpage{1077}
(\byear{1998})
\doiurl{10.1080/01621459.1998.10473768}
\end{barticle}
\endbibitem

\bibitem[\protect\citeauthoryear{Hubert and Arabie}{1985}]{hubertarabie1985ari}
\begin{barticle}
\bauthor{\bsnm{Hubert}, \binits{L.}},
\bauthor{\bsnm{Arabie}, \binits{P.}}:
\batitle{Comparing partitions}.
\bjtitle{Journal of Classification}
\bvolume{2}(\bissue{1}),
\bfpage{193}--\blpage{218}
(\byear{1985})
\doiurl{10.1007/bf01908075}
\end{barticle}
\endbibitem

\bibitem[\protect\citeauthoryear{{National Data Buoy Center}}{2026}]{ndbc46042}
\begin{botherref}
\oauthor{\bsnm{{National Data Buoy Center}}}:
Station 46042: {M}onterey, 27 Nautical Miles West-Northwest of {M}onterey,
  {C}alifornia -- Historical Data.
\url{https://www.ndbc.noaa.gov/station_history.php?station=46042}.
Accessed 3 November 2026
(2026)
\end{botherref}
\endbibitem

\bibitem[\protect\citeauthoryear{Aboussalah and
  Neal}{2016}]{aboussalah2016forecasting}
\begin{botherref}
\oauthor{\bsnm{Aboussalah}, \binits{A.M.}},
\oauthor{\bsnm{Neal}, \binits{C.}}:
Forecasting local warming: Missing data generation and future temperature
  prediction.
Les Cahiers du GERAD G-2016-76,
GERAD
(2016).
\url{https://www.gerad.ca/en/papers/G-2016-76}
\end{botherref}
\endbibitem

\bibitem[\protect\citeauthoryear{Yeo and Johnson}{2000}]{yeojohnson2000power}
\begin{barticle}
\bauthor{\bsnm{Yeo}, \binits{I.-K.}},
\bauthor{\bsnm{Johnson}, \binits{R.A.}}:
\batitle{A new family of power transformations to improve normality or
  symmetry}.
\bjtitle{Biometrika}
\bvolume{87}(\bissue{4}),
\bfpage{954}--\blpage{959}
(\byear{2000})
\doiurl{10.1093/biomet/87.4.954}
\end{barticle}
\endbibitem

\bibitem[\protect\citeauthoryear{MacQueen}{1967}]{macqueen1967methods}
\begin{bchapter}
\bauthor{\bsnm{MacQueen}, \binits{J.}}:
\bctitle{Some methods for classification and analysis of multivariate
  observations}.
In: \beditor{\bsnm{Le~Cam}, \binits{L.M.}},
\beditor{\bsnm{Neyman}, \binits{J.}} (eds.)
\bbtitle{Proceedings of the Fifth Berkeley Symposium on Mathematical Statistics
  and Probability, Volume 1: Statistics},
pp. \bfpage{281}--\blpage{297}.
\bpublisher{University of California Press},
\blocation{Berkeley, CA}
(\byear{1967}).
\burl{https://projecteuclid.org/euclid.bsmsp/1200512992}
\end{bchapter}
\endbibitem

\bibitem[\protect\citeauthoryear{Dhillon and Modha}{2001}]{dhillon2001concept}
\begin{barticle}
\bauthor{\bsnm{Dhillon}, \binits{I.S.}},
\bauthor{\bsnm{Modha}, \binits{D.S.}}:
\batitle{Concept decompositions for large sparse text data using clustering}.
\bjtitle{Machine Learning}
\bvolume{42}(\bissue{1--2}),
\bfpage{143}--\blpage{175}
(\byear{2001})
\doiurl{10.1023/A:1007612920971}
\end{barticle}
\endbibitem

\bibitem[\protect\citeauthoryear{Magnus and
  Neudecker}{2019}]{magnusneudecker2019matrix}
\begin{bbook}
\bauthor{\bsnm{Magnus}, \binits{J.R.}},
\bauthor{\bsnm{Neudecker}, \binits{H.}}:
\bbtitle{Matrix Differential Calculus with Applications in Statistics and
  Econometrics},
\bedition{3}rd edn.
\bsertitle{Wiley Series in Probability and Statistics}.
\bpublisher{John Wiley \& Sons},
\blocation{Hoboken, NJ}
(\byear{2019}).
\doiurl{10.1002/9781119541219}
\end{bbook}
\endbibitem

\bibitem[\protect\citeauthoryear{Amos}{1974}]{amos1974bessel}
\begin{barticle}
\bauthor{\bsnm{Amos}, \binits{D.E.}}:
\batitle{Computation of modified {Bessel} functions and their ratios}.
\bjtitle{Mathematics of Computation}
\bvolume{28}(\bissue{125}),
\bfpage{239}--\blpage{251}
(\byear{1974})
\doiurl{10.1090/s0025-5718-1974-0333287-7}
\end{barticle}
\endbibitem

\bibitem[\protect\citeauthoryear{Wainwright and
  Jordan}{2008}]{wainwright2008graphical}
\begin{barticle}
\bauthor{\bsnm{Wainwright}, \binits{M.J.}},
\bauthor{\bsnm{Jordan}, \binits{M.I.}}:
\batitle{Graphical models, exponential families, and variational inference}.
\bjtitle{Foundations and Trends in Machine Learning}
\bvolume{1}(\bissue{1--2}),
\bfpage{1}--\blpage{305}
(\byear{2008})
\doiurl{10.1561/2200000001}
\end{barticle}
\endbibitem

\bibitem[\protect\citeauthoryear{Celeux and
  Govaert}{1992}]{CeleuxGovaert1992CEM}
\begin{barticle}
\bauthor{\bsnm{Celeux}, \binits{G.}},
\bauthor{\bsnm{Govaert}, \binits{G.}}:
\batitle{A classification {EM} algorithm for clustering and two stochastic
  versions}.
\bjtitle{Computational Statistics \& Data Analysis}
\bvolume{14}(\bissue{3}),
\bfpage{315}--\blpage{332}
(\byear{1992})
\doiurl{10.1016/0167-9473(92)90042-E}
\end{barticle}
\endbibitem

\bibitem[\protect\citeauthoryear{Ingrassia and
  Rocci}{2007}]{ingrassia2007constrained}
\begin{barticle}
\bauthor{\bsnm{Ingrassia}, \binits{S.}},
\bauthor{\bsnm{Rocci}, \binits{R.}}:
\batitle{Constrained monotone {EM} algorithms for finite mixture of
  multivariate {G}aussians}.
\bjtitle{Computational Statistics \& Data Analysis}
\bvolume{51}(\bissue{11}),
\bfpage{5339}--\blpage{5351}
(\byear{2007})
\doiurl{10.1016/j.csda.2006.10.011}
\end{barticle}
\endbibitem

\bibitem[\protect\citeauthoryear{Best and Fisher}{1981}]{best1981bias}
\begin{barticle}
\bauthor{\bsnm{Best}, \binits{D.J.}},
\bauthor{\bsnm{Fisher}, \binits{N.I.}}:
\batitle{The bias of the maximum likelihood estimators of the von
  {M}ises--{F}isher concentration parameters}.
\bjtitle{Communications in Statistics -- Simulation and Computation}
\bvolume{10}(\bissue{5}),
\bfpage{493}--\blpage{502}
(\byear{1981})
\doiurl{10.1080/03610918108812225}
\end{barticle}
\endbibitem

\bibitem[\protect\citeauthoryear{Akaike}{1974}]{akaike1974aic}
\begin{barticle}
\bauthor{\bsnm{Akaike}, \binits{H.}}:
\batitle{A new look at the statistical model identification}.
\bjtitle{IEEE Transactions on Automatic Control}
\bvolume{19}(\bissue{6}),
\bfpage{716}--\blpage{723}
(\byear{1974})
\doiurl{10.1109/tac.1974.1100705}
\end{barticle}
\endbibitem

\bibitem[\protect\citeauthoryear{Schwarz}{1978}]{schwarz1978bic}
\begin{barticle}
\bauthor{\bsnm{Schwarz}, \binits{G.}}:
\batitle{Estimating the dimension of a model}.
\bjtitle{The Annals of Statistics}
\bvolume{6}(\bissue{2}),
\bfpage{461}--\blpage{464}
(\byear{1978})
\doiurl{10.1214/aos/1176344136}
\end{barticle}
\endbibitem

\bibitem[\protect\citeauthoryear{Andrews}{1991}]{andrews1991hac}
\begin{barticle}
\bauthor{\bsnm{Andrews}, \binits{D.W.K.}}:
\batitle{Heteroskedasticity and autocorrelation consistent covariance matrix
  estimation}.
\bjtitle{Econometrica}
\bvolume{59}(\bissue{3}),
\bfpage{817}--\blpage{858}
(\byear{1991})
\doiurl{10.2307/2938229}
\end{barticle}
\endbibitem

\bibitem[\protect\citeauthoryear{Davison and
  Hinkley}{1997}]{davisonhinkley1997bootstrap}
\begin{bbook}
\bauthor{\bsnm{Davison}, \binits{A.C.}},
\bauthor{\bsnm{Hinkley}, \binits{D.V.}}:
\bbtitle{Bootstrap Methods and Their Application}.
\bpublisher{Cambridge University Press},
\blocation{Cambridge}
(\byear{1997}).
\doiurl{10.1017/CBO9780511802843}
\end{bbook}
\endbibitem

\bibitem[\protect\citeauthoryear{Wood}{1994}]{wood1994vmfsim}
\begin{barticle}
\bauthor{\bsnm{Wood}, \binits{A.T.A.}}:
\batitle{Simulation of the von {M}ises {F}isher distribution}.
\bjtitle{Communications in Statistics -- Simulation and Computation}
\bvolume{23}(\bissue{1}),
\bfpage{157}--\blpage{164}
(\byear{1994})
\doiurl{10.1080/03610919408813161}
\end{barticle}
\endbibitem

\end{thebibliography}

\clearpage

\setcounter{section}{0}
\setcounter{subsection}{0}
\setcounter{subsubsection}{0}
\setcounter{paragraph}{0}
\setcounter{subparagraph}{0}
\setcounter{equation}{0}
\setcounter{figure}{0}
\setcounter{table}{0}
\setcounter{theorem}{0}
\setcounter{remark}{0}
\setcounter{example}{0}
\setcounter{algorithm}{0}
\setcounter{footnote}{0}
\setcounter{lstlisting}{0}

\renewcommand{\thesection}{S\arabic{section}}
\renewcommand{\thesubsection}{S\arabic{section}.\arabic{subsection}}
\renewcommand{\thesubsubsection}{\thesubsection.\arabic{subsubsection}}
\renewcommand{\thetable}{S\arabic{table}}
\renewcommand{\thefigure}{S\arabic{figure}}
\renewcommand{\theequation}{S\arabic{equation}}
\renewcommand{\thetheorem}{S\arabic{theorem}}
\renewcommand{\thealgorithm}{S\arabic{algorithm}}

\makeatletter
\def\Hy@chapapp{\Hy@chapterstring}
\makeatother
\def\theHsection{S\arabic{section}}
\def\theHsubsection{S\arabic{section}.\arabic{subsection}}
\def\theHsubsubsection{S\arabic{section}.\arabic{subsection}.\arabic{subsubsection}}
\def\theHparagraph{S\arabic{section}.\arabic{subsection}.\arabic{subsubsection}.\arabic{paragraph}}
\def\theHsubparagraph{S\arabic{section}.\arabic{subsection}.\arabic{subsubsection}.\arabic{paragraph}.\arabic{subparagraph}}
\def\theHtable{S\arabic{table}}
\def\theHfigure{S\arabic{figure}}
\def\theHequation{S\arabic{equation}}
\def\theHtheorem{S\arabic{theorem}}
\def\theHremark{S\arabic{remark}}
\def\theHexample{S\arabic{example}}
\def\theHalgorithm{S\arabic{algorithm}}
\def\theHlstlisting{S\arabic{lstlisting}}

\pdfbookmark[0]{Supplementary Material (Online Resource 1)}{online-resource-1}
\section*{Supplementary Material (Online Resource 1)}

\noindent Sections~\ref{sec:esm-generators}--\ref{sec:esm-ocean} give further experimental results;
Sections~\ref{sec:esm-background}--\ref{sec:esm-hier} and~\ref{sec:esm-geomtest} give background, derivations and
proofs; Sections~\ref{sec:esm-synth-design}--\ref{sec:esm-ocean-preprocessing} give design and preprocessing details.
Section, result, equation, table and figure numbers without the prefix S refer to the article above, and the
references are listed in its References.


\section{Generator parameters of the simulation studies}\label{sec:esm-generators}

This section lists the arrays that define the two simulation generators of Section~7.1 of the main text: Generator~1, the within-component-dependence generator with component label $z$, and Generator~2, the generator with dependence through a latent class. Both generators draw a Gaussian vector $\mathbf u_2$ and project it onto the sphere, $\bx_2=\mathbf u_2/\norm{\mathbf u_2}$. A \emph{pre-projection} directional mean is therefore the mean of $\mathbf u_2$, not the mean direction of $\bx_2$. The separability level changes only the component means. The dependence level changes only the cross-covariance blocks $\bGamma_{12,j}$ in Generator~1, and only the class angles $\boldsymbol\gamma^{(r)}$, which place the directional means, in Generator~2. Every full covariance matrix below ($\bGamma_j$ in Generator~1; $\bPsi_j$ and $\bLambda_k$ in Generator~2) is symmetric and strictly positive definite.

The code encodes exactly the arrays printed below, for both dimensions (module \texttt{cylmix.generators}), and every simulation result was generated from them; a unit test checks the separability, dependence and added-correlation rules stated below and the positive definiteness of every covariance matrix. For $d=3$ these are the arrays of the thesis~\cite{garayvergara2026thesis} (Section~\ref{subsec:esm-study1-3d}). The rules below that use the factors $6$ and $2.25$ (separability), $1/3$ (weak dependence) and $0.1$ (added correlation) hold to the five-decimal storage precision of the displayed arrays. The paired observation noise is described in Section~\ref{sec:esm-synth-design}.

\begingroup\footnotesize\setlength{\abovedisplayskip}{4pt}\setlength{\belowdisplayskip}{4pt}\setlength{\abovedisplayshortskip}{2pt}\setlength{\belowdisplayshortskip}{2pt}
\subsection{Generator 1, \texorpdfstring{$d=2$}{d=2}}\label{subsec:esm-study1-2d}

Given its label $z=j\in\{1,2,3\}$, Generator~1 draws $(\mathbf u_1,\mathbf u_2)\sim\calN_{2d}(\bnu_j,\bGamma_j)$ with $\mathbf u_1,\mathbf u_2\in\R^d$ (so $m=d$) and keeps $\bx_1=\mathbf u_1$. For every dimension, separability level and dependence level, the generating mixture weights are
\begin{equation}
(\pi_1,\pi_2,\pi_3)=(0.36,\,0.34,\,0.30).
\end{equation}
Let $\mathbf V^{(d)}_{1,s}$ and $\mathbf V^{(d)}_{2,s}$ denote the matrices whose $j$th rows are $(\bnu^{(d)}_{1j,s})^\top$ and $(\bnu^{(d)}_{2j,s})^\top$, the Euclidean and the pre-projection directional mean of component $j$, where $d\in\{2,3\}$ and $s\in\{\mathrm{low},\mathrm{high}\}$.

\paragraph{Component means.}
The low-separability Euclidean and pre-projection directional means are
\begin{equation}
\mathbf V^{(2)}_{1,\mathrm{low}}=\begin{pmatrix} -0.00033 & -0.00033\\ 0.19767 & -0.08833\\ -0.18733 & 0.09867 \end{pmatrix},\qquad
\mathbf V^{(2)}_{2,\mathrm{low}}=\begin{pmatrix} 0.73000 & 0.66033\\ 0.65300 & 0.77033\\ 0.80700 & 0.53933 \end{pmatrix}.
\end{equation}
The high-separability means are
\begin{equation}
\mathbf V^{(2)}_{1,\mathrm{high}}=\begin{pmatrix} -0.01866 & -0.01866\\ 1.16934 & -0.54666\\ -1.14066 & 0.57534 \end{pmatrix},\qquad
\mathbf V^{(2)}_{2,\mathrm{high}}=\begin{pmatrix} 0.72618 & 0.66452\\ 0.51226 & 0.87029\\ 0.88934 & 0.38883 \end{pmatrix}.
\end{equation}

\paragraph{Separability transformations.}
For $d\in\{2,3\}$, the unweighted centre of the low-separability Euclidean means is
\begin{equation}
\overline{\bnu}^{(d)}_{1,\mathrm{low}}:=\frac13\sum_{j=1}^{3}\bnu^{(d)}_{1j,\mathrm{low}},
\end{equation}
and the stored high-separability Euclidean means, six times further from this centre, are the five-decimal representations of
\begin{equation}\label{eq:esm-high-linear-means}
\bnu^{(d)}_{1j,\mathrm{high}}=\operatorname{round}_5\Bigl[\overline{\bnu}^{(d)}_{1,\mathrm{low}}+6\bigl(\bnu^{(d)}_{1j,\mathrm{low}}-\overline{\bnu}^{(d)}_{1,\mathrm{low}}\bigr)\Bigr],
\end{equation}
where $\operatorname{round}_5$ denotes rounding to five decimal places. The high-separability pre-projection directional means keep the low-separability norms to storage precision, while every pairwise angle grows by the factor $2.25$, again to storage precision:
\begin{equation}
\angle\bigl(\bnu^{(d)}_{2j,\mathrm{high}},\bnu^{(d)}_{2\ell,\mathrm{high}}\bigr)\approx 2.25\,\angle\bigl(\bnu^{(d)}_{2j,\mathrm{low}},\bnu^{(d)}_{2\ell,\mathrm{low}}\bigr),\qquad 1\le j<\ell\le 3,
\end{equation}
where $\angle(\mathbf a,\mathbf b):=\arccos\bigl(\mathbf a^\top\mathbf b/(\norm{\mathbf a}\,\norm{\mathbf b})\bigr)$.

\paragraph{Strong covariance matrices.}
For $d=2$ the latent coordinate ordering is $(x_{1,1},x_{1,2},u_{2,1},u_{2,2})$, so each $\bGamma_j$ has a Euclidean block $\bGamma_{11,j}$, a pre-projection block $\bGamma_{22,j}$ and a cross-covariance block $\bGamma_{12,j}$. The three strong-dependence covariance matrices are
\begin{equation}
\bGamma^{(2)}_{1,\mathrm{strong}}=\begin{pmatrix} 0.90250 & 0.12112 & 0.36765 & -0.04275\\ 0.12112 & 0.72250 & 0.04590 & 0.32130\\ 0.36765 & 0.04590 & 0.20250 & -0.02025\\ -0.04275 & 0.32130 & -0.02025 & 0.20250 \end{pmatrix},
\end{equation}
\begin{equation}
\bGamma^{(2)}_{2,\mathrm{strong}}=\begin{pmatrix} 0.77440 & 0.12144 & 0.34320 & 0.06653\\ 0.12144 & 0.84640 & -0.07360 & 0.31685\\ 0.34320 & -0.07360 & 0.25000 & -0.02100\\ 0.06653 & 0.31685 & -0.02100 & 0.17640 \end{pmatrix},
\end{equation}
\begin{equation}
\bGamma^{(2)}_{3,\mathrm{strong}}=\begin{pmatrix} 1.04040 & 0.24480 & 0.35965 & -0.10771\\ 0.24480 & 0.64000 & 0.06880 & 0.29184\\ 0.35965 & 0.06880 & 0.18490 & -0.05160\\ -0.10771 & 0.29184 & -0.05160 & 0.23040 \end{pmatrix}.
\end{equation}
Equivalently, their strong cross-covariance blocks are
\begin{equation}
\bGamma^{(2)}_{12,1,\mathrm{strong}}=\begin{pmatrix} 0.36765 & -0.04275\\ 0.04590 & 0.32130 \end{pmatrix},\quad
\bGamma^{(2)}_{12,2,\mathrm{strong}}=\begin{pmatrix} 0.34320 & 0.06653\\ -0.07360 & 0.31685 \end{pmatrix},
\end{equation}
\begin{equation}
\bGamma^{(2)}_{12,3,\mathrm{strong}}=\begin{pmatrix} 0.35965 & -0.10771\\ 0.06880 & 0.29184 \end{pmatrix}.
\end{equation}
Because the projection is nonlinear, the dependence levels describe these latent blocks, not any fitted $\bSigma_{12,j}=\bB_j/\kappa_j$ (Section~3 of the main text).

\paragraph{Dependence transformations.}
For either dimension, the weak and conditionally independent cross-covariance blocks are
\begin{equation}\label{eq:esm-cross-covariance-rules}
\bGamma^{(d)}_{12,j,\mathrm{weak}}=\operatorname{round}_5\Bigl(\tfrac13\,\bGamma^{(d)}_{12,j,\mathrm{strong}}\Bigr),\qquad
\bGamma^{(d)}_{12,j,\mathrm{ind}}=\mathbf 0_{d\times d},\qquad d\in\{2,3\},
\end{equation}
and the within-block covariance matrices $\bGamma^{(d)}_{11,j}$ and $\bGamma^{(d)}_{22,j}$ are shared across the strong, weak and conditionally independent cases. Conditional independence holds only within a component: through the label $z$, the blocks remain associated after mixing. The exact stored two-dimensional weak cross-covariance blocks are
\begin{equation}
\bGamma^{(2)}_{12,1,\mathrm{weak}}=\begin{pmatrix} 0.12255 & -0.01425\\ 0.01530 & 0.10710 \end{pmatrix},\quad
\bGamma^{(2)}_{12,2,\mathrm{weak}}=\begin{pmatrix} 0.11440 & 0.02218\\ -0.02453 & 0.10562 \end{pmatrix},
\end{equation}
\begin{equation}
\bGamma^{(2)}_{12,3,\mathrm{weak}}=\begin{pmatrix} 0.11988 & -0.03590\\ 0.02293 & 0.09728 \end{pmatrix}.
\end{equation}

\subsection{Generator 1, \texorpdfstring{$d=3$}{d=3}}\label{subsec:esm-study1-3d}

\paragraph{Component means.}
The low-separability Euclidean and pre-projection directional means are
\begin{equation}
\mathbf V^{(3)}_{1,\mathrm{low}}=\begin{pmatrix} -0.00033 & -0.00033 & 0.00000\\ 0.19767 & -0.08833 & 0.12400\\ -0.18733 & 0.09867 & -0.12400 \end{pmatrix},\qquad
\mathbf V^{(3)}_{2,\mathrm{low}}=\begin{pmatrix} 0.73000 & 0.66033 & 0.12000\\ 0.65300 & 0.77033 & 0.03000\\ 0.80700 & 0.53933 & 0.21000 \end{pmatrix}.
\end{equation}
The high-separability means are
\begin{equation}
\mathbf V^{(3)}_{1,\mathrm{high}}=\begin{pmatrix} -0.01866 & -0.01866 & 0.00000\\ 1.16934 & -0.54666 & 0.74400\\ -1.14066 & 0.57534 & -0.74400 \end{pmatrix},\qquad
\mathbf V^{(3)}_{2,\mathrm{high}}=\begin{pmatrix} 0.72722 & 0.66356 & 0.11906\\ 0.51254 & 0.86627 & -0.08720\\ 0.86716 & 0.36932 & 0.31284 \end{pmatrix}.
\end{equation}
The first two coordinates of $\mathbf V^{(3)}_{1,\mathrm{low}}$, $\mathbf V^{(3)}_{2,\mathrm{low}}$ and $\mathbf V^{(3)}_{1,\mathrm{high}}$ coincide with their two-dimensional counterparts; $\mathbf V^{(3)}_{2,\mathrm{high}}$, however, is constructed directly in three dimensions and does not preserve $\mathbf V^{(2)}_{2,\mathrm{high}}$ as its first two columns. The separability rule \eqref{eq:esm-high-linear-means} and the angular rule of Section~\ref{subsec:esm-study1-2d} apply unchanged.

\paragraph{Covariance construction.}
For $d=3$ the latent coordinate ordering is $(x_{1,1},x_{1,2},x_{1,3},u_{2,1},u_{2,2},u_{2,3})$. The three-dimensional extension preserves the leading two-dimensional within-block and cross-block submatrices. The third-coordinate variances and the strong cross-covariance between $x_{1,3}$ and $u_{2,3}$ are
\begin{equation}
\begin{pmatrix}a_1\\a_2\\a_3\end{pmatrix}=\begin{pmatrix}0.60840\\0.67240\\0.81000\end{pmatrix},\qquad
\begin{pmatrix}b_1\\b_2\\b_3\end{pmatrix}=\begin{pmatrix}0.18720\\0.23386\\0.22968\end{pmatrix},\qquad
\begin{pmatrix}c_1\\c_2\\c_3\end{pmatrix}=\begin{pmatrix}0.16000\\0.21160\\0.19360\end{pmatrix},
\end{equation}
where $a_j=\operatorname{var}(x_{1,3}\mid z=j)$, $b_j=\operatorname{cov}(x_{1,3},u_{2,3}\mid z=j)$ in the strong case, and $c_j=\operatorname{var}(u_{2,3}\mid z=j)$. Every other new within-block covariance and every other new strong-case cross-block covariance uses the added correlation $\rho_{\mathrm{add}}=0.1$:
\begin{equation}
\gamma^{\mathrm{new}}_{rs}=\operatorname{round}_5\bigl(\rho_{\mathrm{add}}\sqrt{\gamma_{rr}\gamma_{ss}}\bigr),
\end{equation}
where $\gamma_{rr}$ and $\gamma_{ss}$ are the corresponding marginal variances, so that the newly introduced pairs have correlation $0.1$ before five-decimal rounding. The weak and conditionally independent blocks follow \eqref{eq:esm-cross-covariance-rules}, applied to the complete rounded strong cross-covariance block.

\paragraph{Euclidean covariance blocks.}
\begin{equation}
\bGamma^{(3)}_{11,1}=\begin{pmatrix} 0.90250 & 0.12112 & 0.07410\\ 0.12112 & 0.72250 & 0.06630\\ 0.07410 & 0.06630 & 0.60840 \end{pmatrix},\qquad
\bGamma^{(3)}_{11,2}=\begin{pmatrix} 0.77440 & 0.12144 & 0.07216\\ 0.12144 & 0.84640 & 0.07544\\ 0.07216 & 0.07544 & 0.67240 \end{pmatrix},
\end{equation}
\begin{equation}
\bGamma^{(3)}_{11,3}=\begin{pmatrix} 1.04040 & 0.24480 & 0.09180\\ 0.24480 & 0.64000 & 0.07200\\ 0.09180 & 0.07200 & 0.81000 \end{pmatrix}.
\end{equation}

\paragraph{Pre-projection directional covariance blocks.}
\begin{equation}
\bGamma^{(3)}_{22,1}=\begin{pmatrix} 0.20250 & -0.02025 & 0.01800\\ -0.02025 & 0.20250 & 0.01800\\ 0.01800 & 0.01800 & 0.16000 \end{pmatrix},\qquad
\bGamma^{(3)}_{22,2}=\begin{pmatrix} 0.25000 & -0.02100 & 0.02300\\ -0.02100 & 0.17640 & 0.01932\\ 0.02300 & 0.01932 & 0.21160 \end{pmatrix},
\end{equation}
\begin{equation}
\bGamma^{(3)}_{22,3}=\begin{pmatrix} 0.18490 & -0.05160 & 0.01892\\ -0.05160 & 0.23040 & 0.02112\\ 0.01892 & 0.02112 & 0.19360 \end{pmatrix}.
\end{equation}

\paragraph{Strong cross-covariance blocks.}
\begin{equation}
\bGamma^{(3)}_{12,1,\mathrm{strong}}=\begin{pmatrix} 0.36765 & -0.04275 & 0.03800\\ 0.04590 & 0.32130 & 0.03400\\ 0.03510 & 0.03510 & 0.18720 \end{pmatrix},\qquad
\bGamma^{(3)}_{12,2,\mathrm{strong}}=\begin{pmatrix} 0.34320 & 0.06653 & 0.04048\\ -0.07360 & 0.31685 & 0.04232\\ 0.04100 & 0.03444 & 0.23386 \end{pmatrix},
\end{equation}
\begin{equation}
\bGamma^{(3)}_{12,3,\mathrm{strong}}=\begin{pmatrix} 0.35965 & -0.10771 & 0.04488\\ 0.06880 & 0.29184 & 0.03520\\ 0.03870 & 0.04320 & 0.22968 \end{pmatrix}.
\end{equation}

\paragraph{Weak cross-covariance blocks.}
The exact stored weak cross-covariance blocks are
\begin{equation}
\bGamma^{(3)}_{12,1,\mathrm{weak}}=\begin{pmatrix} 0.12255 & -0.01425 & 0.01267\\ 0.01530 & 0.10710 & 0.01133\\ 0.01170 & 0.01170 & 0.06240 \end{pmatrix},\qquad
\bGamma^{(3)}_{12,2,\mathrm{weak}}=\begin{pmatrix} 0.11440 & 0.02218 & 0.01349\\ -0.02453 & 0.10562 & 0.01411\\ 0.01367 & 0.01148 & 0.07795 \end{pmatrix},
\end{equation}
\begin{equation}
\bGamma^{(3)}_{12,3,\mathrm{weak}}=\begin{pmatrix} 0.11988 & -0.03590 & 0.01496\\ 0.02293 & 0.09728 & 0.01173\\ 0.01290 & 0.01440 & 0.07656 \end{pmatrix}.
\end{equation}
For $r\in\{\mathrm{strong},\mathrm{weak},\mathrm{ind}\}$ the complete three-dimensional covariance matrix is
\begin{equation}
\bGamma^{(3)}_{j,r}=\begin{pmatrix}\bGamma^{(3)}_{11,j} & \bGamma^{(3)}_{12,j,r}\\ (\bGamma^{(3)}_{12,j,r})^\top & \bGamma^{(3)}_{22,j}\end{pmatrix}.
\end{equation}

\subsection{Generator 2}\label{subsec:esm-study2}

Generator~2 (Section~7.1 of the main text) has first-layer latent classes $Y_1=j\in\{1,2,3\}$ and local directional subtypes $Y_2=k\in\{1,2\}$. Its blocks are independent given the class, so the class alone links them. Its parameters are the class weights $\bpi$, the first-layer means $\bmu_j$ and covariances $\bPsi_j$, the subtype weights $\varpi_{jk}$, the pre-projection directional means $\mathbf c_{jk}$ and the directional covariances $\bLambda_k$. For every dimension, separability level and class--direction dependence level,
\begin{equation}
(\pi_1,\pi_2,\pi_3)=(0.36,\,0.34,\,0.30),\qquad (\varpi_{j1},\varpi_{j2})=(0.58,\,0.42),\quad j=1,2,3.
\end{equation}

\paragraph{First-layer means.}
Let $\mathbf V^{(q)}_s$ denote the matrix whose $j$th row is $(\bmu^{(q,s)}_j)^\top$, where $q\in\{\mathrm{2d},\mathrm{3d}\}$ and $s\in\{\mathrm{low},\mathrm{high}\}$. The two-dimensional means are
\begin{equation}
\mathbf V^{(\mathrm{2d})}_{\mathrm{low}}=\begin{pmatrix}-0.00033 & -0.00033\\ 0.19767 & -0.08833\\ -0.18733 & 0.09867\end{pmatrix},\qquad
\mathbf V^{(\mathrm{2d})}_{\mathrm{high}}=\begin{pmatrix}-0.01866333 & -0.01866333\\ 1.16933667 & -0.54666333\\ -1.14066333 & 0.57533667\end{pmatrix},
\end{equation}
and the three-dimensional means are
\begin{equation}
\mathbf V^{(\mathrm{3d})}_{\mathrm{low}}=\begin{pmatrix}-0.00033 & -0.00033 & 0.00000\\ 0.19767 & -0.08833 & 0.12400\\ -0.18733 & 0.09867 & -0.12400\end{pmatrix},\qquad
\mathbf V^{(\mathrm{3d})}_{\mathrm{high}}=\begin{pmatrix}-0.01866333 & -0.01866333 & 0.00000\\ 1.16933667 & -0.54666333 & 0.74400\\ -1.14066333 & 0.57533667 & -0.74400\end{pmatrix}.
\end{equation}
If $\overline{\bmu}^{(q,\mathrm{low})}:=3^{-1}\sum_{j=1}^{3}\bmu^{(q,\mathrm{low})}_j$ denotes the unweighted centre of the low-separability means, then
\begin{equation}
\bmu^{(q,\mathrm{high})}_j=\overline{\bmu}^{(q,\mathrm{low})}+6\bigl(\bmu^{(q,\mathrm{low})}_j-\overline{\bmu}^{(q,\mathrm{low})}\bigr).
\end{equation}
The low-separability means are the Euclidean means $\mathbf V^{(d)}_{1,\mathrm{low}}$ of Generator~1; the high-separability means are displayed here to eight decimals, whereas the Generator~1 arrays are stored to five.

\paragraph{First-layer covariance matrices.}
The two-dimensional covariance matrices are
\begin{equation}
\bPsi^{(\mathrm{2d})}_1=\begin{pmatrix}0.90250 & 0.12112\\ 0.12112 & 0.72250\end{pmatrix},\quad
\bPsi^{(\mathrm{2d})}_2=\begin{pmatrix}0.77440 & 0.12144\\ 0.12144 & 0.84640\end{pmatrix},\quad
\bPsi^{(\mathrm{2d})}_3=\begin{pmatrix}1.04040 & 0.24480\\ 0.24480 & 0.64000\end{pmatrix}.
\end{equation}
The three-dimensional matrices keep each two-dimensional principal block and give the added coordinate correlation $0.1$ with each original coordinate:
\begin{equation}
\bPsi^{(\mathrm{3d})}_1=\begin{pmatrix}0.90250 & 0.12112 & 0.07410\\ 0.12112 & 0.72250 & 0.06630\\ 0.07410 & 0.06630 & 0.60840\end{pmatrix},\qquad
\bPsi^{(\mathrm{3d})}_2=\begin{pmatrix}0.77440 & 0.12144 & 0.07216\\ 0.12144 & 0.84640 & 0.07544\\ 0.07216 & 0.07544 & 0.67240\end{pmatrix},
\end{equation}
\begin{equation}
\bPsi^{(\mathrm{3d})}_3=\begin{pmatrix}1.04040 & 0.24480 & 0.09180\\ 0.24480 & 0.64000 & 0.07200\\ 0.09180 & 0.07200 & 0.81000\end{pmatrix}.
\end{equation}
Equivalently, for $r\in\{1,2\}$,
\begin{equation}
\bigl(\bPsi^{(\mathrm{3d})}_j\bigr)_{r3}=\bigl(\bPsi^{(\mathrm{3d})}_j\bigr)_{3r}=\operatorname{round}_5\Bigl(0.1\sqrt{\bigl(\bPsi^{(\mathrm{2d})}_j\bigr)_{rr}\bigl(\bPsi^{(\mathrm{3d})}_j\bigr)_{33}}\Bigr).
\end{equation}
These matrices are not block diagonal, and they coincide with the Euclidean blocks $\bGamma^{(d)}_{11,j}$ of Generator~1.

\paragraph{Directional mean construction.}
The subtype-specific pre-projection mean norms are $(\rho_1,\rho_2)=(1.04,\,0.98)$. For each dimension, the means lie in the plane of an anchor direction $\mathbf q^{(q)}$ and an orthogonal unit tangent $\mathbf t^{(q)}$, so their directions lie on the great circle through these two vectors:
\begin{equation}
\mathbf q^{(\mathrm{2d})}=\begin{pmatrix}0.74160966\\ 0.67083166\end{pmatrix},\qquad
\mathbf t^{(\mathrm{2d})}=\begin{pmatrix}-0.67083166\\ 0.74160966\end{pmatrix},
\end{equation}
\begin{equation}
\mathbf q^{(\mathrm{3d})}=\begin{pmatrix}0.73615956\\ 0.66590170\\ 0.12101253\end{pmatrix},\qquad
\mathbf t^{(\mathrm{3d})}=\begin{pmatrix}0.10749436\\ -0.29156614\\ 0.95049152\end{pmatrix}.
\end{equation}
At low separability, the angle of $\mathbf c_{jk}$ from the anchor is a class angle $\gamma_j$ plus a subtype offset $\delta_k$. The within-class subtype offsets are $(\delta_1,\delta_2)=(-11^\circ,11^\circ)$, and the low-separability class angles are
\begin{equation}
\boldsymbol\gamma^{(\mathrm{strong})}=(-25^\circ,0^\circ,25^\circ),\qquad
\boldsymbol\gamma^{(\mathrm{weak})}=\tfrac13\,\boldsymbol\gamma^{(\mathrm{strong})},\qquad
\boldsymbol\gamma^{(\mathrm{ind})}=(0^\circ,0^\circ,0^\circ).
\end{equation}
For dependence level $r\in\{\mathrm{strong},\mathrm{weak},\mathrm{ind}\}$,
\begin{equation}
\varphi^{(\mathrm{low},r)}_{jk}=\gamma^{(r)}_j+\delta_k,\qquad \varphi^{(\mathrm{high},r)}_{jk}=2.25\,\varphi^{(\mathrm{low},r)}_{jk},
\end{equation}
and the pre-projection mean is
\begin{equation}\label{eq:esm-study2-directional-means}
\mathbf c^{(q,s,r)}_{jk}=\rho_k\Bigl[\cos\bigl(\varphi^{(s,r)}_{jk}\bigr)\,\mathbf q^{(q)}+\sin\bigl(\varphi^{(s,r)}_{jk}\bigr)\,\mathbf t^{(q)}\Bigr],
\end{equation}
where angles are converted to radians when the expression is evaluated. The complete angular offsets $\varphi^{(s,r)}_{jk}$ from the common anchor, in degrees, are
\begin{equation}\label{eq:esm-study2-angle-offsets}
\begin{array}{llcc}
\text{Separability } s & \text{Dependence } r & (\varphi_{11},\varphi_{21},\varphi_{31}) \text{ for } k=1 & (\varphi_{12},\varphi_{22},\varphi_{32}) \text{ for } k=2\\[2pt]
\text{Low}  & \text{Strong}      & (-36,\,-11,\,14)              & (-14,\,11,\,36)\\
\text{Low}  & \text{Weak}        & (-19.3333,\,-11,\,-2.6667)    & (2.6667,\,11,\,19.3333)\\
\text{Low}  & \text{Independent} & (-11,\,-11,\,-11)             & (11,\,11,\,11)\\
\text{High} & \text{Strong}      & (-81,\,-24.75,\,31.5)         & (-31.5,\,24.75,\,81)\\
\text{High} & \text{Weak}        & (-43.5,\,-24.75,\,-6)         & (6,\,24.75,\,43.5)\\
\text{High} & \text{Independent} & (-24.75,\,-24.75,\,-24.75)    & (24.75,\,24.75,\,24.75)
\end{array}
\end{equation}

\paragraph{Directional covariance matrices.}
The pre-projection covariance depends only on the subtype $k$. In two dimensions,
\begin{equation}
\bLambda^{(\mathrm{2d})}_1=\begin{pmatrix}0.140 & -0.025\\ -0.025 & 0.220\end{pmatrix},\qquad
\bLambda^{(\mathrm{2d})}_2=\begin{pmatrix}0.210 & 0.035\\ 0.035 & 0.130\end{pmatrix},
\end{equation}
and in three dimensions,
\begin{equation}
\bLambda^{(\mathrm{3d})}_1=\begin{pmatrix}0.140 & -0.025 & 0.015\\ -0.025 & 0.220 & -0.012\\ 0.015 & -0.012 & 0.170\end{pmatrix},\qquad
\bLambda^{(\mathrm{3d})}_2=\begin{pmatrix}0.210 & 0.035 & -0.018\\ 0.035 & 0.130 & 0.010\\ -0.018 & 0.010 & 0.190\end{pmatrix}.
\end{equation}
No cross-covariance matrix between the Euclidean and the pre-projection directional block is used, so positive definiteness is checked separately for each $\bPsi^{(q)}_j$ and $\bLambda^{(q)}_k$, and no Schur-complement restriction is required.

\paragraph{Independent condition.}
When $r=\mathrm{ind}$ the class angles are zero, and equation \eqref{eq:esm-study2-directional-means} gives
\begin{equation}
\mathbf c^{(q,s,\mathrm{ind})}_{1k}=\mathbf c^{(q,s,\mathrm{ind})}_{2k}=\mathbf c^{(q,s,\mathrm{ind})}_{3k},\qquad k=1,2.
\end{equation}
Because the subtype weights and the directional covariances $\bLambda_k$ are also identical across classes, the law $g$ of $\bx_2$ given $Y_1=j$ is the same for every $j$, so $g=p(\bx_2)$. With $\bx_1\indep\bx_2\mid Y_1$ this gives $p(\bx_1,\bx_2)=\sum_{j}\pi_j\,p(\bx_1\mid Y_1=j)\,g(\bx_2)=p(\bx_1)\,p(\bx_2)$: the blocks are marginally independent, not only conditionally independent as at the corresponding level of Generator~1.

\paragraph{Sampling and projection.}
For every observation, sampling follows
\begin{equation}
Y_1\longrightarrow(\bx_1,Y_2)\mid Y_1\longrightarrow\mathbf u_2\mid Y_1,Y_2\longrightarrow\bx_2=\operatorname{unit}_{10^{-12}}(\mathbf u_2),\qquad
\operatorname{unit}_{10^{-12}}(\mathbf v):=\frac{\mathbf v}{\max\{\norm{\mathbf v},10^{-12}\}},
\end{equation}
where the floor $10^{-12}$ guards against division by zero. The sampler retains $(\bx_1,\mathbf u_2)$ with the label pair $(Y_1,Y_2)$, so that the directional noise of Section~\ref{sec:esm-synth-design} can be added before projection; the projection of Generator~1 uses the same floor.
\endgroup

Sections~\ref{sec:esm-study1}--\ref{sec:esm-recovery} report further results from these generators.

\section{Additional results of the simulation studies}\label{sec:esm-study1}

This section adds to Section~7 of the main text: the direct-dependence contrasts by separability level, the agreement of the ten starts and the fitted boundaries, the hierarchy against flat mixtures with matched component counts, and the test of the von Mises--Fisher component on the simulated fits. Every fit is the best of the ten starts of Algorithm~\ref{alg:em-cyl}, each run by exact EM to the relative tolerance $10^{-8}$, on clean samples (Section~\ref{sec:esm-synth-design}).

\subsection{Direct dependence by separability}\label{subsec:esm-study1-q1}

Table~\ref{tab:esm-q1} splits panel A of Table~3 of the main text by separability. Under conditional independence at low separability, estimating $\bB_j$ costs $0.003$ nats per observation, about twice the asymptotic price $3dm/(2N_{\mathrm{train}})$ ($0.0015$ on average) of its $3dm$ additional parameters; under weak and strong dependence the gain holds at both levels. The recovery gain under strong dependence is large only at high separability ($82.5$ percentage points), where the labels can be recovered at all. The Bayesian information criterion (BIC) prefers the cylindrical mixture in every pair under strong dependence, in $50$\% (low separability) and $100$\% (high) of the pairs under weak dependence, and in none under conditional independence; under Generator~2 it prefers it in at most $10$\% of the pairs of a cell.

\begin{table}[htbp]
\caption{Cylindrical minus independent mixture at $K=3$, by generator, dependence and separability, pooled over dimension ($2$ data sets per seed, $20$ seeds). Held-out log-density $\HLL$ and its parts, in nats per observation; first-layer adjusted Rand index ($\ARI$) in percentage points; CI: two-sided 95\% Student-$t$ interval across seeds; BIC: percentage of pairs in which the cylindrical mixture has the smaller training BIC}\label{tab:esm-q1}
\centering\footnotesize
\setlength{\tabcolsep}{3pt}
\begin{tabular}{@{}lllrcrrrr@{}}
\toprule
Gen. & Dependence & Sep. & $\Delta\HLL$ & 95\% CI & $\Delta\HLL_{1\mid2}$ & $\Delta\HLL_2$ & $\Delta\ARI$ & BIC\\
\midrule
1 & Cond.\ independent & Low & $-0.003$ & $[-0.003,\,-0.002]$ & $-0.002$ & $-0.0002$ & $-0.5$ & $0$\\
1 & Weak & Low & $0.009$ & $[0.007,\,0.010]$ & $0.008$ & $0.0004$ & $0.8$ & $50$\\
1 & Strong & Low & $0.282$ & $[0.273,\,0.292]$ & $0.260$ & $0.023$ & $4.8$ & $100$\\
1 & Cond.\ independent & High & $0.003$ & $[0.001,\,0.004]$ & $0.001$ & $0.002$ & $-5.0$ & $0$\\
1 & Weak & High & $0.023$ & $[0.021,\,0.025]$ & $0.019$ & $0.004$ & $2.7$ & $100$\\
1 & Strong & High & $0.319$ & $[0.310,\,0.327]$ & $0.296$ & $0.023$ & $82.5$ & $100$\\
2 & Marg.\ independent & Low & $-0.002$ & $[-0.003,\,-0.001]$ & $-0.002$ & $-0.0004$ & $0.2$ & $0$\\
2 & Weak & Low & $-0.002$ & $[-0.002,\,-0.001]$ & $-0.002$ & $-0.0002$ & $0.1$ & $0$\\
2 & Strong & Low & $-0.001$ & $[-0.002,\,-0.001]$ & $-0.001$ & $-0.0001$ & $-1.1$ & $0$\\
2 & Marg.\ independent & High & $-0.0003$ & $[-0.002,\,0.001]$ & $0.0001$ & $-0.0003$ & $1.5$ & $0$\\
2 & Weak & High & $0.002$ & $[0.0003,\,0.004]$ & $0.003$ & $-0.001$ & $0.2$ & $10$\\
2 & Strong & High & $0.001$ & $[-0.0002,\,0.002]$ & $-0.002$ & $0.003$ & $-6.8$ & $0$\\
\bottomrule
\end{tabular}
\end{table}

\subsection{Agreement of the starts and fitted boundaries}\label{subsec:esm-starts}

In the direct-dependence study ($960$ fits, $9{,}600$ starts) no start was degenerate and every start converged, after a median of $278$--$496$ iterations and at most $5{,}935$. The ten starts of a fit reached training log-likelihoods more than $10^{-3}$ nats per observation apart in $18$\% (cylindrical) and $5$\% (independent) of the Generator~1 fits and in $15$\% and $8$\% of the Generator~2 fits; their held-out scores differed by $0.002$ nats on average and by at most $0.106$, so a single start would often have stopped at a different stationary point. Without floors, the smallest fitted weight was $0.003$ (about $18$ training observations), the smallest eigenvalue of a fitted $\hat\bS_j$ was $0.0017$ and the largest concentration $37$.

\subsection{The hierarchy against flat mixtures}\label{subsec:esm-gap}

Table~\ref{tab:full-vs-iso} splits panel B of Table~3 of the main text by dependence level (Generator~2, high separability, $20$ seeds, $2$ data sets per seed and row). Joint estimation beats isolated estimation at every level, by more as the class rotates the directions more ($0.007$, $0.016$ and $0.090$ nats), and so does the variational gap of Proposition~19 at the isolated fits, $\lhier-\liso$ per observation: $0.0003$, $0.005$ and $0.033$ nats (intervals $[0.0002,0.0004]$, $[0.005,0.006]$ and $[0.032,0.033]$). The gap is what the direction tells the first stage about the class; it is about a third of the held-out advantage of joint estimation under weak and strong dependence, and the rest comes from re-estimating the parameters. Against the independent mixture with $K=6$, which has the same six von Mises--Fisher components and contains the $(3,2)$ hierarchy as the submodel whose Gaussian components are shared in pairs, the hierarchy's held-out density is higher by at most $0.002$ nats at every level, with $15$ ($d=2$) or $27$ ($d=3$) fewer parameters, and its first-layer recovery is better by $9$--$24$ percentage points. Against the cylindrical mixture with $K=3$ it gains $0.012$--$0.026$ nats, most of it (pooled, $0.010$ of $0.017$) in the directional part. None of the $3{,}600$ starts of the $360$ hierarchy and $K=6$ fits was degenerate and all converged; the isolated estimator needed a median of $1{,}843$ iterations over its stages, the full one $524$.

\begin{table}[htbp]
\caption{Full $(3,2)$ hierarchy minus three alternatives under Generator~2 at high separability, by dependence level, pooled over dimension ($2$ data sets per seed, $20$ seeds). Held-out log-density and its parts in nats per observation, first-layer $\ARI$ in percentage points; CI: two-sided 95\% Student-$t$ interval across seeds; BIC: percentage of pairs in which the hierarchy has the smaller training BIC}\label{tab:full-vs-iso}
\centering\footnotesize
\setlength{\tabcolsep}{2.0pt}
\begin{tabular}{@{}llrcrrrr@{}}
\toprule
Minus & Dependence & $\Delta\HLL$ & 95\% CI & $\Delta\HLL_{1\mid2}$ & $\Delta\HLL_2$ & $\Delta\ARI$ & BIC\\
\midrule
isolated $(3,2)$ hierarchy & Marg.\ independent & $0.007$ & $[0.005,\,0.008]$ & $-0.003$ & $0.010$ & $-3.5$ & $100$\\
 & Weak & $0.016$ & $[0.015,\,0.018]$ & $0.010$ & $0.006$ & $4.4$ & $100$\\
 & Strong & $0.090$ & $[0.083,\,0.096]$ & $0.060$ & $0.030$ & $25.6$ & $100$\\
\addlinespace
independent mixture, $K=6$ & Marg.\ independent & $0.002$ & $[0.001,\,0.003]$ & $0.002$ & $-0.0001$ & $8.9$ & $100$\\
 & Weak & $0.001$ & $[0.0003,\,0.002]$ & $0.003$ & $-0.001$ & $14.2$ & $100$\\
 & Strong & $-0.0001$ & $[-0.001,\,0.001]$ & $0.002$ & $-0.002$ & $23.7$ & $100$\\
\addlinespace
cylindrical mixture, $K=3$ & Marg.\ independent & $0.012$ & $[0.011,\,0.014]$ & $0.004$ & $0.008$ & $7.5$ & $100$\\
 & Weak & $0.012$ & $[0.010,\,0.013]$ & $0.007$ & $0.005$ & $12.5$ & $100$\\
 & Strong & $0.026$ & $[0.024,\,0.028]$ & $0.009$ & $0.017$ & $10.2$ & $100$\\
\bottomrule
\end{tabular}
\end{table}

\subsection{The test of the von Mises--Fisher component on the simulated fits}\label{subsec:esm-geomtest-synthfits}

On the recovery fits of Section~\ref{sec:esm-recovery}, where the hypothesis holds, the mixture statistic of Theorem~17 with the expected information from $2\times10^4$ draws rejected at the 5\% level in $13$ of $240$ fits at $K=3$ ($3$ of $120$ for $d=2$, $10$ of $120$ for $d=3$; two-sided binomial $p=0.77$). Under Generator~1, whose directions are projected-normal, it rejected in $29$ of $30$ fits for $(d,m)=(3,3)$ but in only $9$ of $30$ for $(2,2)$ (seeds $0$--$4$, both separability levels and all three dependence levels), with median KL effect sizes of $0.0036$ and $0.0002$ nats: on the circle these projected-normal components are close to von Mises--Fisher at component sizes of about $2{,}000$.

\section{Leaf recovery under Generator 2}\label{sec:esm-study2}

Under Generator~2 each observation has a latent class $Y_1\in\{1,2,3\}$ and a subtype $Y_2\in\{1,2\}$ local to that class. The first-layer score compares the most probable fitted class with $Y_1$; the leaf score compares the joint maximum a posteriori leaf $(\hat Y_1,\hat Y_2)$ with the six-class partition $(Y_1,Y_2)$. The leaf score does not register how the fitted leaves are grouped under the classes, and a perfect first-layer score does not determine how each class is split, so neither score follows from the other. At high separability ($20$ seeds), the full $(3,2)$ hierarchy recovers the classes and the leaves better than the isolated one under strong dependence (first layer $62$ and $68$ percentage points for $d=2$ and $3$, against $35$ and $44$; leaves $42$ and $50$, against $30$ and $36$), where the direction carries information about the class that the isolated estimator's first stage ignores. Under marginal independence, where $p(\bx_2\mid Y_1)$ does not depend on $Y_1$ and the joint first-layer posterior equals the $\bx_1$-only posterior in the population, the two fitted estimators are within $6.1$ percentage points in either score; the identity holds for the population, not exactly for the fits.

\section{Parameter recovery under exact specification}\label{sec:esm-recovery}

Section~7 of the main text compares misspecified approximations. Here we ask two narrower implementation questions: when the model is exactly right, are its parameters recovered, for one component and for a mixture?

\subsection{One component}\label{subsec:esm-recovery-one}

For one cylindrical component with $d=m\in\{2,3\}$, $\kappa\in\{0.5,2,10\}$, $\balpha$ with equally spaced entries in $[-0.3,0.3]$, $\bB=0.6\bI_d+0.15\mathbf P_d$ ($\mathbf P_d$ the cyclic shift), $\bS=0.45\bI_d+0.08\,\mathbf 1\mathbf 1^{\top}/d$ and $\bmu_2\propto(1,\dots,d)^{\top}$, we drew $200$ samples of each size $N\in\{100,500,2{,}000,8{,}000\}$ and computed the estimator of Theorem~6 and, for comparison, the optional safeguarded update of Section~\ref{sec:esm-mix}, which the reported fits do not use. Under correct specification $2N\KL[f:\hat f]$ is asymptotically $\chi^2_q$, with $q=d+dm+d(d+1)/2+m$ ($11$ and $21$), so its mean should approach $q$. At $N=8{,}000$ the mean of $2N\KL$ over the $200$ samples is $10.9$--$11.3$ for $d=2$ and $20.6$--$20.9$ for $d=3$, within one standard error of $q$ in all six settings; at $N=100$ it is $12.3$--$13.5$ and $23.0$--$24.8$, the excess expected from higher-order terms. In every sample the exact estimator attained a training log-likelihood at least as large as the safeguarded update (the largest advantage was $5.7\times10^{-6}$ nats per observation), and the median divergences of the two differ by less than $1\%$. The closed-form divergence used here, $\KL[f:\hat f]=\KL[\vMF:\widehat{\vMF}]+\E_{\vMF}\KL[\calN(\balpha+\bB\bx_2,\bS):\calN(\hat{\balpha}+\hat{\bB}\bx_2,\hat{\bS})]$, needs only the first two moments of the direction (Lemma~\ref{lem:vmf-moments}) and is checked against Monte Carlo in the unit tests.

\subsection{Mixture: design and generator}\label{subsec:esm-recovery-design}\label{subsec:esm-recovery-generator}

Observations are generated exactly from a three-component cylindrical mixture and fitted by a three-component cylindrical mixture with Algorithm~\ref{alg:em-cyl}, so no model-selection uncertainty enters. The study crosses $(d,m)\in\{(2,2),(3,3)\}$ with the two separability levels and the three dependence levels of Generator~1, giving $12$ configurations, each for seeds $0,\dots,19$. Let $Z\in\{1,2,3\}$ be the mixture label, with weights $\bpi=(0.36,0.34,0.30)$. Given $Z=j$, the generator is the cylindrical distribution of Section~3 of the main text in the coordinates $\bmu_{1,j}=\balpha_j+\bB_j\bmu_{2,j}$ and $\bSigma_{12,j}=\bB_j/\kappa_j$:
\begin{equation}\label{eq:esm-recovery-generator}
\bx_2\mid Z=j\sim\vMF_m(\bmu_{2,j},\kappa_j),\qquad
\bx_1\mid\bx_2,Z=j\sim\calN_d\bigl(\bmu_{1,j}+\kappa_j\bSigma_{12,j}(\bx_2-\bmu_{2,j}),\,\bS_j\bigr).
\end{equation}
The $12$ configurations are translated from the corresponding projected-normal configurations of Section~\ref{sec:esm-generators}: the weights are copied, $\bmu_{1,j}$ is the Euclidean part of the source mean and $\bmu_{2,j}$ its normalized directional part, and with $\bSigma^{\mathrm{PG}}_{11,j}$, $\bSigma^{\mathrm{PG}}_{12,j}$, $\bSigma^{\mathrm{PG}}_{22,j}$ the source covariance blocks,
\begin{equation}\label{eq:esm-recovery-mapping}
\bSigma_{12,j}=\bSigma^{\mathrm{PG}}_{12,j},\qquad
\kappa_j=\lambda_{\max}\bigl(\bSigma^{\mathrm{PG}}_{22,j}\bigr)^{-1},\qquad
\bS_j=\bSigma^{\mathrm{PG}}_{11,j}-\kappa_j\bSigma_{12,j}\bSigma_{12,j}^\top.
\end{equation}
Because $\kappa_j\bI_m\preceq(\bSigma^{\mathrm{PG}}_{22,j})^{-1}$, $\bS_j$ is no smaller than the positive-definite Schur complement of the source covariance, so it is positive definite. The rule is a heuristic, not an information projection: it keeps the Euclidean and cross blocks of the source and replaces its directional block by $\kappa_j^{-1}\bI_m$, and by Proposition~2 the result is the normalized restriction of a Gaussian density to $\R^d\times\sphere^{m-1}$. For a fixed dimension and dependence level, the two separability levels share $\bpi$, $\bSigma_{12,j}$, $\bS_j$ and $\kappa_j$; the weak cross-dependence matrices are one third of the strong ones and the conditionally independent ones are zero.

\subsection{Protocol, alignment and error}\label{subsec:esm-recovery-protocol}

Each data set has $N=8{,}000$ observations, of which the first $6{,}400$ are fitted. Fitted components are aligned to generator components by the one-to-one assignment that minimizes the summed discrepancy $\|\hat{\bS}_i-\bS_j\|_F+\|\hat{\bSigma}_{12,i}-\bSigma_{12,j}\|_F+\|\hat{\bmu}_{1,i}-\bmu_{1,j}\|_2+|\hat\kappa_i-\kappa_j|+\arccos(\hat{\bmu}_{2,i}^\top\bmu_{2,j})$. For each block $A\in\{\bSigma_{12},\bB,\bS,\bmu_1,\bmu_2,\kappa\}$, with the true and aligned fitted component arrays stacked into $A^\star$ and $\hat A^\star$,
\begin{equation}\label{eq:esm-recovery-error}
E_A=\frac{\|\hat A^\star-A^\star\|_F}{\max\{\|A^\star\|_F,\,1\}},
\end{equation}
a relative error when $\|A^\star\|_F\ge1$ and an absolute one otherwise. The floor matters for $\bSigma_{12}$, whose stack norm is below one in every cell (about $0.84$ and $0.93$ under strong and $0.28$ and $0.31$ under weak dependence, for $d=2$ and $3$; zero under conditional independence), and for $\bmu_1$ at low separability (about $0.30$ and $0.35$, against $1.82$ and $2.10$ at high separability).

\subsection{Findings}\label{subsec:esm-recovery-findings}

Table~\ref{tab:esm-recovery} gives the mean normalized errors \eqref{eq:esm-recovery-error} over the $20$ seeds. Recovery is best in the strong-dependence, high-separability cells, with errors of $0.06$ or less for every block in both dimensions. Overlap degrades the location and the cross-dependence most: at low separability without strong dependence, the error of $\bmu_1$ (an absolute error there) is $0.66$--$1.12$ and that of $\bSigma_{12}$ $0.36$--$0.75$. Under conditional independence, where the true $\bSigma_{12}$ is zero, its error measures spurious fitted dependence ($0.21$--$0.75$). In regression coordinates the errors of $\bB=\kappa\bSigma_{12}$ are larger, up to $5.4$, because $\kappa_j\approx4$--$5$ multiplies the error of $\bSigma_{12}$. None of the $2{,}400$ starts was degenerate and all converged; the ten starts of a fit reached training log-likelihoods more than $10^{-3}$ nats apart in $24$\% of the fits. On these fits, where the von Mises--Fisher hypothesis holds, the mixture test of Section~5.3 rejected at the 5\% level in $13$ of $240$ fits (two-sided binomial $p=0.77$).

\begin{table}[htbp]
\caption{Parameter recovery of the $K=3$ cylindrical mixture fitted by Algorithm~\ref{alg:em-cyl} to data from a three-component cylindrical generator: mean normalized error \eqref{eq:esm-recovery-error} over $20$ seeds, by dimension, separability and dependence; lower is better}\label{tab:esm-recovery}
\centering\footnotesize
\begin{tabular}{@{}lll rrrrrr@{}}
\toprule
$d$ & Sep. & Dependence & $\bSigma_{12}$ & $\bB$ & $\bS$ & $\bmu_1$ & $\bmu_2$ & $\kappa$\\
\midrule
2 & Low & Cond.\ ind. & $0.51$ & $5.40$ & $0.30$ & $0.83$ & $0.17$ & $0.82$\\
2 & Low & Weak & $0.36$ & $1.55$ & $0.26$ & $0.66$ & $0.20$ & $0.24$\\
2 & Low & Strong & $0.29$ & $0.32$ & $0.23$ & $0.54$ & $0.18$ & $0.12$\\
2 & High & Cond.\ ind. & $0.24$ & $1.18$ & $0.23$ & $0.28$ & $0.11$ & $0.21$\\
2 & High & Weak & $0.21$ & $0.71$ & $0.20$ & $0.21$ & $0.08$ & $0.08$\\
2 & High & Strong & $0.05$ & $0.06$ & $0.04$ & $0.03$ & $0.01$ & $0.03$\\
3 & Low & Cond.\ ind. & $0.75$ & $2.63$ & $0.40$ & $1.12$ & $0.27$ & $0.16$\\
3 & Low & Weak & $0.60$ & $1.90$ & $0.40$ & $1.04$ & $0.25$ & $0.16$\\
3 & Low & Strong & $0.22$ & $0.23$ & $0.21$ & $0.39$ & $0.14$ & $0.06$\\
3 & High & Cond.\ ind. & $0.21$ & $0.92$ & $0.18$ & $0.22$ & $0.08$ & $0.07$\\
3 & High & Weak & $0.15$ & $0.49$ & $0.11$ & $0.12$ & $0.05$ & $0.05$\\
3 & High & Strong & $0.06$ & $0.06$ & $0.04$ & $0.03$ & $0.01$ & $0.02$\\
\bottomrule
\end{tabular}
\end{table}

\section{Additional material of the buoy study}\label{sec:esm-ocean}

This section adds to Section~8 of the main text: the transformed sample, the fitting protocol and its audit, including the second-best optima, the annual contrasts with their across-window inference and the ambient-component contrasts by $K$, the test of the von Mises--Fisher component in every window, a buoy-like simulation, and the fitting time.

\subsection{Transformed sample}\label{subsec:esm-ocean-sample}

\begin{figure}[htbp]
\centering
\includegraphics[width=0.9\textwidth]{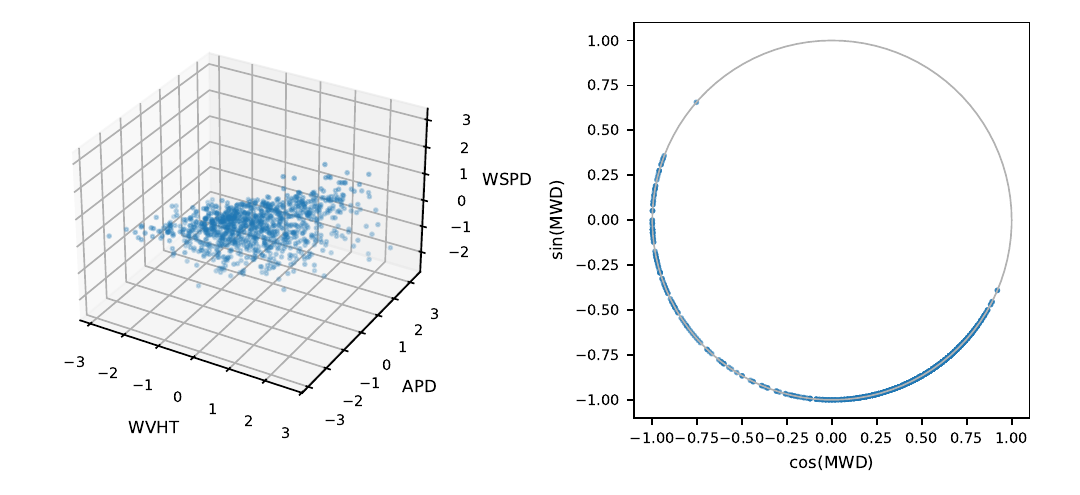}
\caption{$1{,}000$ training observations of the window with test year $2022$ (training years $2018$--$2021$). Left: the three Euclidean coordinates after Yeo--Johnson transformation and standardization (WVHT, significant wave height; APD, average wave period; WSPD, wind speed). Right: the wave direction on the unit circle}
\label{fig:esm-ocean-scatter}
\end{figure}

Figure~\ref{fig:esm-ocean-scatter} shows $1{,}000$ randomly chosen training observations of the window with test year $2022$, after the Yeo--Johnson transformation and standardization fitted on that window's training block (Section~\ref{sec:esm-ocean-preprocessing}).

\subsection{Fitting protocol and audit}\label{subsec:esm-ocean-audit}

\emph{Protocol.} In each window, the three families are fitted at every $K\in\{1,\dots,8\}$ with equal start budgets. The cylindrical and independent mixtures receive $k$-means++ starts with seeds $1,\dots,n_K$ (\texttt{numpy.random.default\_rng}; $n_K=1$ at $K=1$, $20$ at $K=3,\dots,6$ and $10$ at $K\in\{2,7,8\}$; the budget of $20$ was set after a first run with ten starts, at the $K$ where those starts disagreed most), each run by exact EM until the increase of $\bar\ell$ is at most $10^{-8}\max\{1,|\bar\ell|\}$ (Algorithm~\ref{alg:em-cyl}), and at $K\ge2$ the split start of Section~\ref{sec:esm-mix} from the retained $(K-1)$-component fit of the same family, so that the retained fits form a chain along $K$. The generalized von Mises (GvM) mixtures, whose directional factor $\exp\{\boldsymbol\zeta_j^{\top}\mathbf h_{\mathrm G}(\bx_2)\}/c_{\mathrm G}(\boldsymbol\zeta_j)$ has the sufficient statistic $\mathbf h_{\mathrm G}(\bx_2)=(u_1,u_2,u_1^2-u_2^2,2u_1u_2)^{\top}$, $\bx_2=(u_1,u_2)^{\top}$, of the Fisher--Bingham family on the circle, are started from every distinct converged cylindrical optimum at the same $K$, with $\boldsymbol\zeta_j=(\kappa_j\bmu_{2,j},0,0)$, and from the split of their own retained $(K-1)$-component fit; this gives them as many starts as the cylindrical mixture has distinct optima, plus one. Their M-step keeps the least-squares regression block and maximizes the concave directional log-likelihood by damped Newton steps (step halving, at most $100$ iterations, stopped when every coordinate of the gradient is below $10^{-11}$), with $\log c_{\mathrm G}$ and its first two derivatives computed by the $2{,}048$-point periodic trapezoidal rule, which converges geometrically for these smooth periodic integrands; a unit test checks the normalization of the component density by adaptive numerical integration. The nested starts guarantee that the training log-likelihood of the GvM fit is at least that of the cylindrical fit at the same $K$. In every family the run with the largest training log-likelihood is retained, and no complexity is selected; two runs reach distinct optima when their training log-likelihoods differ by more than $10^{-7}\max\{1,|\bar\ell|\}$. Because the GvM starts are nested in the cylindrical optima, the GvM--cylindrical contrast compares a refined fit with its starting point; the cylindrical--independent contrast compares equal start budgets from an initializer that ignores $\bB_j$.

Table~\ref{tab:esm-ocean-audit} summarizes the $5{,}031$ runs; none was degenerate and every run converged. Up to $K=2$ the $k$-means++ starts reach one optimum (the split start a second one in three windows at $K=2$); from $K=4$ on, and in some windows at $K=3$, they reach several, up to $8$ in a configuration. More starts do not reduce the spread: at $K=6$ the held-out scores of the first ten starts span $0.028$ and $0.029$ nats on average and those of all twenty $0.028$ and $0.031$ (cylindrical and independent), because the extra starts find further optima of the same likelihood surface. Twenty starts instead of ten raised the best training log-likelihood in $4$ of the $152$ cylindrical and independent configurations with $K=3,\dots,6$, and changed the held-out score of the best $k$-means++ start by at most $0.024$ nats. Every held-out contrast therefore rests on the retained optimum. With the second-best distinct optimum of each family in place of the best, the direct-dependence contrast becomes ($K=3$: $0.054$ and $0.069$; $K=4$: $0.029$ and $0.045$; $K=5$: $0.022$ and $0.020$; $K=6$: $0.012$ and $0.019$; $K=7$: $0.012$ and $0.009$; $K=8$: $0.006$ and $0.011$, best and second best), and the GvM contrast ($K=3$: $0.057$ and $0.070$; $K=4$: $0.047$ and $0.025$; $K=5$: $0.029$ and $0.029$; $K=6$: $0.021$ and $0.029$; $K=7$: $0.021$ and $0.026$; $K=8$: $0.018$ and $0.017$; at $K=2$: $0.191$ and $0.106$). With the second-best optimum of the cylindrical mixture against the best of the independent one, the direct-dependence contrast is $0.032$, $0.035$, $0.016$, $-0.003$, $0.005$ and $0.005$ at $K=3,\dots,8$, so at $K\ge6$ the contrasts change by up to $0.015$ nats and can change sign at $K=6$; with the second-best GvM optimum against the best cylindrical one, the GvM contrast at $K=2$ is $0.105$ instead of $0.191$. The largest fitted concentration $\hat\kappa_j$ of a retained fit is $89.1$ (cylindrical) and $84.2$ (independent), and $106.3$ over all stored cylindrical starts; the smallest summed responsibility of a retained component is $206$ hours (Section~\ref{sec:esm-ocean-preprocessing}).

\begin{table}[htbp]
\caption{Audit of the buoy fits, over the $19$ windows, for the cylindrical (Cyl.), independent (Ind.) and GvM mixtures. Spread: maximum minus minimum, over the $k$-means++ starts of a configuration, of the final training or held-out average log-likelihood, in nats per observation (mean and maximum over windows; GvM: over the nested starts). Optima: median and maximum number of distinct training optima reached. Iterations: median and maximum over all runs. $\min\hat\pi_j$ and $\min\lambda_{\min}(\hat\bS_j)$: smallest fitted weight and smallest eigenvalue of a fitted conditional covariance, over windows and components}\label{tab:esm-ocean-audit}
\centering\scriptsize
\setlength{\tabcolsep}{1.4pt}
\begin{tabular}{@{}lrccccrr@{}}
\toprule
Family & $K$ & Spread, train & Spread, held out & Optima & Iterations & $\min\hat\pi_j$ & $\min\lambda_{\min}(\hat\bS_j)$\\
\midrule
Cyl. & 1 & -- / -- & -- / -- & $1$ / $1$ & $2$ / $2$ & $1.000$ & $0.147$\\
Cyl. & 2 & $3{\times}10^{-11}$ / $5{\times}10^{-10}$ & $1{\times}10^{-8}$ / $2{\times}10^{-7}$ & $1$ / $2$ & $46$ / $251$ & $0.126$ & $0.112$\\
Cyl. & 3 & $0.015$ / $0.152$ & $0.015$ / $0.201$ & $1$ / $2$ & $81$ / $347$ & $0.108$ & $0.022$\\
Cyl. & 4 & $6{\times}10^{-3}$ / $0.019$ & $0.013$ / $0.053$ & $2$ / $3$ & $117$ / $400$ & $0.054$ & $0.021$\\
Cyl. & 5 & $0.010$ / $0.023$ & $0.016$ / $0.072$ & $3$ / $3$ & $242$ / $792$ & $0.034$ & $0.016$\\
Cyl. & 6 & $0.024$ / $0.037$ & $0.028$ / $0.065$ & $4$ / $7$ & $311$ / $1{,}494$ & $0.035$ & $0.015$\\
Cyl. & 7 & $0.022$ / $0.034$ & $0.026$ / $0.088$ & $4$ / $7$ & $409$ / $1{,}424$ & $0.007$ & $0.008$\\
Cyl. & 8 & $0.020$ / $0.043$ & $0.025$ / $0.062$ & $5$ / $8$ & $525$ / $1{,}442$ & $0.009$ & $0.009$\\
Ind. & 1 & -- / -- & -- / -- & $1$ / $1$ & $2$ / $2$ & $1.000$ & $0.157$\\
Ind. & 2 & $2{\times}10^{-11}$ / $3{\times}10^{-10}$ & $9{\times}10^{-9}$ / $2{\times}10^{-7}$ & $1$ / $1$ & $27$ / $80$ & $0.124$ & $0.124$\\
Ind. & 3 & $0.038$ / $0.197$ & $0.037$ / $0.228$ & $1$ / $2$ & $70$ / $189$ & $0.111$ & $0.021$\\
Ind. & 4 & $0.025$ / $0.050$ & $0.029$ / $0.086$ & $2$ / $3$ & $95$ / $411$ & $0.112$ & $0.021$\\
Ind. & 5 & $0.015$ / $0.031$ & $0.021$ / $0.065$ & $2$ / $3$ & $122$ / $855$ & $0.045$ & $0.016$\\
Ind. & 6 & $0.024$ / $0.036$ & $0.031$ / $0.069$ & $3$ / $5$ & $204$ / $1{,}123$ & $0.035$ & $0.015$\\
Ind. & 7 & $0.023$ / $0.041$ & $0.030$ / $0.072$ & $4$ / $7$ & $284$ / $973$ & $0.007$ & $0.008$\\
Ind. & 8 & $0.022$ / $0.043$ & $0.031$ / $0.100$ & $5$ / $7$ & $343$ / $1{,}159$ & $0.006$ & $0.008$\\
GvM & 1 & -- / -- & -- / -- & $1$ / $1$ & $3$ / $3$ & $1.000$ & $0.147$\\
GvM & 2 & -- / -- & -- / -- & $2$ / $2$ & $85$ / $285$ & $0.304$ & $0.025$\\
GvM & 3 & -- / -- & -- / -- & $2$ / $3$ & $156$ / $511$ & $0.187$ & $0.016$\\
GvM & 4 & -- / -- & -- / -- & $3$ / $3$ & $226$ / $1{,}108$ & $0.059$ & $0.015$\\
GvM & 5 & -- / -- & -- / -- & $4$ / $4$ & $289$ / $1{,}070$ & $0.011$ & $0.014$\\
GvM & 6 & -- / -- & -- / -- & $4$ / $7$ & $351$ / $1{,}125$ & $0.009$ & $0.010$\\
GvM & 7 & -- / -- & -- / -- & $4$ / $8$ & $424$ / $1{,}807$ & $0.010$ & $0.008$\\
GvM & 8 & -- / -- & -- / -- & $6$ / $8$ & $495$ / $2{,}368$ & $0.008$ & $0.007$\\
\bottomrule
\end{tabular}
\end{table}

\subsection{Annual contrasts and across-window inference}\label{subsec:esm-ocean-q1}

\begin{figure}[htbp]
\centering
\includegraphics[width=0.55\textwidth]{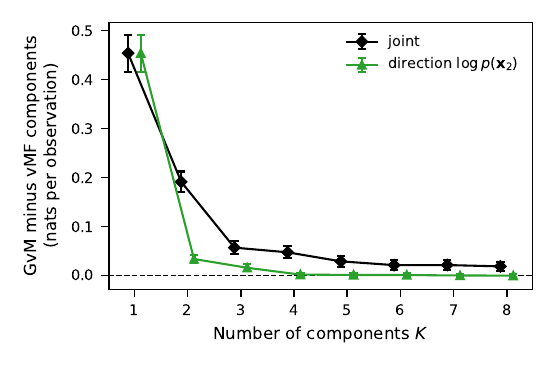}
\caption{Ambient directional components on the buoy records: macro held-out contrast of the GvM over the cylindrical mixture at fixed $K$ (black) and its directional part, the contrast of $\log p(\bx_2)$ (green), with 95\% Student-$t$ intervals over the $19$ test years, in nats per observation}
\label{fig:esm-ocean-gvm}
\end{figure}

Table~\ref{tab:esm-ocean-annual} compares five ways of putting an interval on a macro contrast; adjacent windows share training years, so none is exact. The four every-fourth-year sets have disjoint training blocks but are not independent, since each set's windows train on test years of the other sets, and the direct-dependence contrasts have negative lag-one autocorrelation at six of the eight $K$; the four-set intervals, with $3$ degrees of freedom, are therefore unstable (at $K=7$ the set means nearly coincide and the interval is five times narrower than the iid-$t$ one). For direct dependence every procedure separates the contrast from zero at $K\le5$; at $K=6$ the sign test does not ($p=0.064$), and at $K=8$ two of the four intervals include zero. For the GvM contrast every procedure excludes zero at every $K$. The main text therefore states clear direct-dependence gains for $K\le5$ only. The moving-block bootstrap uses $2{,}000$ resamples of blocks of four consecutive test years (seed $20261104$). Figure~\ref{fig:esm-ocean-gvm} shows the ambient-component contrasts of Table~5 of the main text with their directional part.

\begin{table}[htbp]
\caption{Sensitivity of the across-window inference for the fixed-$K$ buoy contrasts, in nats per observation. Macro: average over the $19$ test years. 95\% intervals: Student-$t$ over the $19$ annual contrasts (iid-$t$, used in the main text); Student-$t$ with $3$ degrees of freedom over the means of the four sets of every fourth test year (four sets); Student-$t$ with a Newey--West variance of lag $3$ (NW); and a moving-block bootstrap of the annual contrasts with blocks of four years (MBB). Sign: two-sided sign-test $p$-value over the $19$ years. $\rho_1$: lag-one autocorrelation of the annual contrasts. Dep.: cylindrical minus independent; GvM: GvM minus cylindrical}\label{tab:esm-ocean-annual}
\centering\scriptsize
\setlength{\tabcolsep}{1.1pt}
\begin{tabular}{@{}lrrccccrr@{}}
\toprule
Contrast & $K$ & Macro & iid-$t$ & Four sets & NW & MBB & Sign & $\rho_1$\\
\midrule
Dep. & 1 & $0.121$ & $[0.097,\,0.145]$ & $[0.100,\,0.143]$ & $[0.107,\,0.136]$ & $[0.104,\,0.133]$ & $8\times10^{-5}$ & $-0.49$\\
Dep. & 2 & $0.080$ & $[0.068,\,0.092]$ & $[0.066,\,0.094]$ & $[0.068,\,0.092]$ & $[0.070,\,0.094]$ & $4\times10^{-6}$ & $0.05$\\
Dep. & 3 & $0.054$ & $[0.045,\,0.064]$ & $[0.043,\,0.065]$ & $[0.043,\,0.065]$ & $[0.044,\,0.062]$ & $4\times10^{-6}$ & $0.11$\\
Dep. & 4 & $0.029$ & $[0.018,\,0.040]$ & $[0.017,\,0.041]$ & $[0.020,\,0.037]$ & $[0.022,\,0.036]$ & $7\times10^{-4}$ & $-0.39$\\
Dep. & 5 & $0.022$ & $[0.014,\,0.030]$ & $[0.014,\,0.029]$ & $[0.015,\,0.029]$ & $[0.015,\,0.027]$ & $8\times10^{-5}$ & $-0.25$\\
Dep. & 6 & $0.012$ & $[0.005,\,0.019]$ & $[0.007,\,0.016]$ & $[0.006,\,0.017]$ & $[0.007,\,0.016]$ & $0.064$ & $-0.29$\\
Dep. & 7 & $0.012$ & $[0.005,\,0.019]$ & $[0.010,\,0.013]$ & $[0.007,\,0.016]$ & $[0.008,\,0.017]$ & $7\times10^{-4}$ & $-0.33$\\
Dep. & 8 & $0.006$ & $[-0.001,\,0.013]$ & $[-0.002,\,0.013]$ & $[0.002,\,0.010]$ & $[0.003,\,0.009]$ & $0.019$ & $-0.38$\\
GvM & 1 & $0.454$ & $[0.416,\,0.492]$ & $[0.441,\,0.468]$ & $[0.401,\,0.506]$ & $[0.400,\,0.504]$ & $4\times10^{-6}$ & $0.38$\\
GvM & 2 & $0.191$ & $[0.170,\,0.212]$ & $[0.164,\,0.219]$ & $[0.175,\,0.207]$ & $[0.180,\,0.209]$ & $4\times10^{-6}$ & $-0.02$\\
GvM & 3 & $0.057$ & $[0.043,\,0.070]$ & $[0.047,\,0.067]$ & $[0.040,\,0.073]$ & $[0.050,\,0.076]$ & $8\times10^{-5}$ & $0.24$\\
GvM & 4 & $0.047$ & $[0.035,\,0.059]$ & $[0.037,\,0.058]$ & $[0.038,\,0.056]$ & $[0.044,\,0.058]$ & $4\times10^{-6}$ & $-0.42$\\
GvM & 5 & $0.029$ & $[0.017,\,0.040]$ & $[0.015,\,0.042]$ & $[0.018,\,0.039]$ & $[0.022,\,0.039]$ & $7\times10^{-4}$ & $0.10$\\
GvM & 6 & $0.021$ & $[0.010,\,0.032]$ & $[0.008,\,0.035]$ & $[0.011,\,0.032]$ & $[0.012,\,0.034]$ & $7\times10^{-4}$ & $0.17$\\
GvM & 7 & $0.021$ & $[0.011,\,0.031]$ & $[0.012,\,0.031]$ & $[0.011,\,0.032]$ & $[0.012,\,0.034]$ & $7\times10^{-4}$ & $0.02$\\
GvM & 8 & $0.018$ & $[0.010,\,0.027]$ & $[0.009,\,0.029]$ & $[0.010,\,0.027]$ & $[0.011,\,0.028]$ & $8\times10^{-5}$ & $0.08$\\
\bottomrule
\end{tabular}
\end{table}

\subsection{The test of the von Mises--Fisher component on the buoy fits}\label{subsec:esm-ocean-geomtest}

Table~\ref{tab:esm-ocean-geomtest} reports the test in every window. At $K=1$ it is the closed-form statistic of Theorem~13, and its effect size $\KL[\FB^*:\vMF]$ equals the training log-likelihood gain of the GvM fit at $K=1$ to within $10^{-13}$ nats, as the identity $\LRstat_N/(2N)=\KL[\FB^*:\vMF]$ requires (at $m=2$ the generalized von Mises family is the Fisher--Bingham family). For $K>1$ it is the mixture statistic of Theorem~17 at the retained fit, with the expected information estimated from $2\times10^4$ draws from the fit; its null hypothesis includes the conditional Gaussian part, so only the per-component directional effect sizes are read as directional. At $K>1$ the fixed-$b$ $p$-values (below $0.05$ in every window at $K\le7$ and in $18$ of $19$ at $K=8$) rest on a heuristic calibration (Section~5.3 of the main text) and refer to the whole mixture, Gaussian conditionals included; they are not read as evidence about the directions. At $K=1$ the long-run (Newey--West, lag $168$) variance of the efficient scores is $32$--$46$ times their sample variance, and $113$--$233$ times with the Bartlett bandwidth $0.1N$ of the fixed-$b$ statistic, whose weights reach further; the lag-$168$ ratio lies between the values $10.3$ and $48.3$ of the null simulation of Section~\ref{sec:esm-geomtest} at lag-one correlation $0.95$ and $0.99$, at which the fixed-$b$ calibration rejected $5.7$ and $4.7\%$ of null series. The ratio of the same long-run variance to the model's expected efficient information, $50$--$98$, also contains a misfit factor $1.1$--$2.2$ (sample over model variance of the scores). Under dependence the null scale of the effect size is about $\lambda q_m/(2N)$, with $\lambda$ the variance ratio; with $\lambda\le233$ it is at most $0.007$ nats at $K=1$, against observed values $0.344$--$0.548$. The fixed-$b$ $p$-values use $20{,}000$ simulated Gaussian series of length $1{,}000$ (seed $12345$), so the smallest attainable value is $1/20{,}001\approx5\times10^{-5}$. The Newey--West calibration with lag $168$, which over-rejects at this persistence (Section~5.3), gives statistics $216$--$634$ at $K=1$ and is not used. The per-component effect sizes have a median of $0.005$ nats at $K=6$ but reach $0.111$ for some components. The worked example of Section~8 (the $2022$ window at $K=1$) is reproduced by \texttt{experiments/worked\_example\_2022.py} of the code, which computes each entry a second way: the closed-form fit against the stored EM fit, $T_N$ and its parts from $\bDelta$ and the closed forms (25) of the main text against the test code, $\iota_\parallel$ and $\iota_\perp$ against their implementation, the observed-information intervals against least squares and the gains against the stored held-out scores; all agree to $2\times10^{-9}$ or better.

\begin{table}[htbp]
\caption{The test of Section~5 on the fitted cylindrical mixtures of the $19$ windows (training blocks). $T$: statistic with independent-sampling calibration; $T^{\mathrm{fb}}$: fixed-$b$ statistic (Bartlett kernel, bandwidth $0.1N$; $2K$ restrictions) with the Kiefer--Vogelsang null of Section~5.3, descriptive for $K>1$; $p<0.05$: windows with fixed-$b$ $p$-value below $0.05$, and the largest fixed-$b$ $p$-value ($5\times10^{-5}=1/20{,}001$ is the smallest attainable value, so there $p\le5\times10^{-5}$); inflation: trace ratio of the Newey--West (lag $168$) long-run covariance to the sample covariance of the efficient scores, which reflects serial dependence only; KL: $\sum_j\hat\pi_j\KL[\FB^*_j:\vMF_j]$ in nats (median and range over windows), equal to $\LRstat_N/(2N)$ at $K=1$; component KL: median over windows and components of the per-component directional effect size}\label{tab:esm-ocean-geomtest}
\centering\scriptsize
\setlength{\tabcolsep}{1.0pt}
\begin{tabular}{@{}rrccrrccc@{}}
\toprule
$K$ & $2K$ & $T$ & $T^{\mathrm{fb}}$ & $p<0.05$ & max $p$ & Inflation & KL & Component KL\\
\midrule
1 & 2 & $17{,}321$--$31{,}374$ & $117$--$494$ & $19$ & $5\times10^{-5}$ & $32$--$46$ & $0.449$ $[0.344,\,0.548]$ & $0.449$\\
2 & 4 & $1{,}729$--$3{,}803$ & $81$--$273$ & $19$ & $0.0001$ & $8$--$21$ & $0.050$ $[0.033,\,0.070]$ & $0.044$\\
3 & 6 & $1{,}453$--$3{,}467$ & $109$--$560$ & $19$ & $5\times10^{-5}$ & $5$--$15$ & $0.027$ $[0.011,\,0.047]$ & $0.026$\\
4 & 8 & $899$--$2{,}883$ & $72$--$348$ & $19$ & $0.003$ & $3$--$14$ & $0.010$ $[0.006,\,0.039]$ & $0.011$\\
5 & 10 & $569$--$1{,}915$ & $90$--$615$ & $19$ & $0.004$ & $3$--$12$ & $0.010$ $[0.004,\,0.019]$ & $0.007$\\
6 & 12 & $502$--$1{,}412$ & $97$--$417$ & $19$ & $0.01$ & $2$--$7$ & $0.006$ $[0.003,\,0.010]$ & $0.005$\\
7 & 14 & $387$--$1{,}435$ & $124$--$582$ & $19$ & $0.01$ & $2$--$8$ & $0.004$ $[0.003,\,0.010]$ & $0.004$\\
8 & 16 & $472$--$1{,}379$ & $85$--$548$ & $18$ & $0.2$ & $2$--$8$ & $0.005$ $[0.002,\,0.008]$ & $0.004$\\
\bottomrule
\end{tabular}
\end{table}

\subsection{A buoy-like simulation}\label{subsec:esm-ocean-buoylike}

Two designs (Section~\ref{sec:esm-synth-design}) mimic the buoy records, with $34{,}000$ training and $8{,}700$ test hours. \emph{Design A (null).} In $200$ series whose direction is exactly $\vMF_2(\kappa=3)$ marginally, made serially dependent by a Gaussian AR(1) copula with lag-one correlation $0.95$, the independent-sampling calibration of the test rejected at the 5\% level in $75.0\%$ of the series (at 1\%: $66.0\%$), the Newey--West calibration with lag $168$ in $5.0\%$ (1\%: $2.5\%$; 10\%: $14.0\%$) and a moving-block bootstrap with $168$-hour blocks in $4.5\%$ (1\%: $2.0\%$; 10\%: $13.5\%$). This is an independent run of the setting of Section~\ref{sec:esm-geomtest} at lag-one correlation $0.95$, with a regression block and $200$ series (Monte Carlo standard error $1.5$ points), consistent with the $7.0\%$ found there for Newey--West. The median ratio of long-run (lag $168$) to independent-sampling score variance was $9.5$, against $32$--$46$ for the ratio to the sample variance on the buoy records (Section~\ref{subsec:esm-ocean-geomtest}), so this design is milder than the data; in the null simulation of Section~\ref{sec:esm-geomtest}, whose ratios $48.3$ and $87.6$ at lag-one correlation $0.99$ and $0.995$ reach and exceed the data's, Newey--West with lag $168$ rejects in $8.0$ and $18.7\%$ and the fixed-$b$ calibration used in Section~8 in $4.7$ and $6.3\%$. The median KL effect size in design A, $0.0002$ nats, gives the scale of a null effect at this $N$. \emph{Design B} (Section~\ref{sec:esm-synth-design}). In five series with three latent classes that follow a Markov chain (stay probability $0.99$ per hour) and skewed generalized von Mises directions, the cylindrical mixture beats its independent restriction by $0.315$ nats at $K=1$ (range over series $0.300$--$0.334$) and $0.057$ at $K=3$ ($0.053$--$0.063$); the Newey--West statistic is $130$--$177$ at $K=1$ and $108$--$413$ at $K=3$; and GvM components gain $0.007$ ($0.002$--$0.011$) and $0.017$ ($0.008$--$0.022$) nats out of sample, against training effect sizes of $0.009$ and $0.011$. These five series are consistent with the three findings of Section~8 under buoy-like dependence, at a smaller directional effect; with five replicates they illustrate rather than test them.

\subsection{Fitting time}\label{subsec:esm-ocean-time}

Process CPU times, under the conditions of Section~\ref{sec:esm-mix}: $7.3$ hours for the $19$ windows with all three families, $K=1,\dots,8$ and all starts; about $25$ minutes for the test in every window and at every $K$ (the sum of the recorded times of the $152$ window-and-$K$ runs); and a few minutes for each buoy-like design.

\section{Background details for Section~2 of the main text}\label{sec:esm-background}

This section fills in the steps that Section~2 of the main text skipped. The notation is that of the main text: $\bt$, $\psi$, $\calM$, $\phi$, and responsibilities $r_{nj}$ with masses $N_j=\sum_nr_{nj}$.

\paragraph{Free energy and the conditions for monotone EM.}
Fix one observation $\bx$ and any probability mass function $q$ on the labels $\{1,\dots,K\}$ whose support lies inside that of the posterior $p(y\mid\bx;\bTheta)$, so that both terms below are finite. Then
\begin{gather*}
\log p(\bx;\bTheta)=F(q,\bTheta)+\KL[q:p(\cdot\mid\bx;\bTheta)],\\ F(q,\bTheta):=\E_{Y\sim q}[\log p(\bx,Y;\bTheta)]+H[q],
\end{gather*}
where $H[q]$ is the entropy of $q$. Since the divergence is nonnegative and vanishes only at the posterior, the free energy $F$ of Neal and Hinton~\cite{nealhinton1998em} is a lower bound on the log-density, attained exactly when $q$ is the posterior. The E-step closes this gap: it sets $q$ to the posterior at the current iterate $\bTheta^{\mathrm{old}}$, which gives
\begin{equation*}
r_{nj}=p(Y_n=j\mid\bx_n;\bTheta^{\mathrm{old}})=\frac{\pi_j^{\mathrm{old}}f(\bx_n;\btheta_j^{\mathrm{old}})}{\sum_{j'=1}^K\pi_{j'}^{\mathrm{old}}f(\bx_n;\btheta_{j'}^{\mathrm{old}})}.
\end{equation*}
The M-step then maximizes the summed bound over $\bTheta$, with $q$ fixed. Since $H[q]$ does not depend on $\bTheta$, it equivalently maximizes the expected complete-data log-likelihood
\begin{equation*}
Q(\bTheta^{\mathrm{old}},\bTheta)=\sum_n\sum_jr_{nj}\log\bigl(\pi_jf(\bx_n;\btheta_j)\bigr)=\sum_jN_j\log\pi_j+\sum_j\sum_nr_{nj}\log f(\bx_n;\btheta_j).
\end{equation*}
The weight part is maximized over the simplex by $\hat\pi_j=N_j/N$, with the convention $0\log0=0$; this maximizer is unique, and interior when every $N_j>0$. The rest splits into one weighted likelihood problem per component. In the geometric reading of Section~2 of the main text~\cite{amari1995em,csiszartusnady1984alternating,hino2022geometryem}, the E-step is an e-projection onto the complete-data laws (of observation and label) whose observed marginal is the empirical distribution, and the M-step an m-projection back onto the model. Both hold up to an infinite term common to all candidates, because the empirical marginal has no density; the remainder of each divergence is minus the summed bound divided by $N$.

Monotonicity, $\ell(\bTheta^{\mathrm{new}})\ge\ell(\bTheta^{\mathrm{old}})$, holds when the observed density is finite and positive at the iterates, the E-step is exact, and the M-step returns a global maximizer of the same $Q(\bTheta^{\mathrm{old}},\cdot)$ over a fixed parameter space. The identity above gives the reason: the summed bound equals $\ell(\bTheta^{\mathrm{old}})$ after the E-step, the M-step does not lower it, and it never exceeds $\ell(\bTheta^{\mathrm{new}})$. A generalized EM step, which need not maximize $Q(\bTheta^{\mathrm{old}},\cdot)$ but does not decrease it, therefore keeps monotonicity. The conclusion that limit points of the iterates are stationary points of $\ell$ needs further assumptions: a finite interior accumulation point, differentiability, and a continuous or closed update map~\cite{dempster1977em,wu1983em}. Neither conclusion transfers to an update that clips, floors, regularizes or reinitializes a component and is not shown to be a generalized EM step for the same $Q(\bTheta^{\mathrm{old}},\cdot)$: the second step of the chain is then unproved, so the guarantees no longer apply. This is not a claim that the likelihood decreases.

\paragraph{The dual-coordinate M-step.} Section~2 of the main text uses the following classical lemma.

\begin{lemma}[Dual-coordinate M-step for exponential-family components \cite{banerjee2005bregman}]
\label{lem:dual-mstep}
Consider a finite mixture whose component densities share the same minimal regular exponential-family representation (2). Fix responsibilities $r_{nj}$ and define $N_j:=\sum_{n=1}^Nr_{nj}$ and $\bar{\bt}_j:=N_j^{-1}\sum_{n=1}^Nr_{nj}\,\bt(\bx_n)$. If $N_j>0$ and $\bar{\bt}_j\in\calM$, then the unique finite componentwise M-step satisfies $\bfeta_j=\nabla\psi(\btheta_j)=\bar{\bt}_j$ and $\btheta_j=\nabla\phi(\bar{\bt}_j)$.
\end{lemma}

\begin{proof}
For fixed $j$, the parameter-dependent part of the weighted log-likelihood is $\sum_nr_{nj}\log f(\bx_n;\btheta)=N_j(\langle\bar{\bt}_j,\btheta\rangle-\psi(\btheta))$. It is strictly concave on the natural parameter space because $N_j>0$ and $\psi$ is strictly convex, so a stationary point is the unique maximizer. Its gradient vanishes exactly where $\nabla\psi(\btheta)=\bar{\bt}_j$, that is, where the expected statistic equals the weighted average. Since $\bar{\bt}_j\in\calM$, this equation has the unique solution $\btheta_j=\nabla\phi(\bar{\bt}_j)$ by Legendre duality.
\end{proof}

\paragraph{Boundary and curved components.}
This argument can fail in two ways. If $\bar{\bt}_j$ lies on the boundary of $\calM$, the weighted likelihood may have only a boundary supremum, approached but not attained, so a finite natural-parameter update need not exist. If the component family is curved or otherwise constrained to a subset $\mathcal E$ of the natural parameter space, the unrestricted mean matching of Lemma~\ref{lem:dual-mstep} is not a valid M-step either, because $\nabla\phi(\bar{\bt}_j)$ need not lie on $\mathcal E$. The M-step is then the maximization of $N_j(\langle\bar{\bt}_j,\btheta\rangle-\psi(\btheta))$ over $\btheta\in\mathcal E$: the constrained information projection of the weighted empirical distribution $\hat P_j$ onto $\mathcal E$, not mean matching. The cylindrical family of Section~3 of the main text is such a curved family.

Section~2.1 of the main text gives the interior conditions for the Gaussian and von Mises--Fisher families. For the latter, the two extreme resultant lengths behave differently. A weighted resultant of norm zero is interior to $\calM$, so Lemma~\ref{lem:dual-mstep} gives $\btheta_j=\nabla\phi(\mathbf 0)=\mathbf 0$: concentration zero, the uniform law, with undetermined mean direction. A norm of one lies on the boundary of $\calM$ and admits no finite concentration, because $A_m^{-1}(r)\to\infty$ as $r\uparrow1$.

\paragraph{Further von Mises--Fisher facts.}
With $\psi(\btheta)=-\log C_m(\norm{\btheta})$ and $\btheta=\kappa\bmu$, inverting the strictly increasing map $A_m$~\cite{schou1978concentration,mardiajupp2000directional} is the radial part of the expectation-to-natural Legendre map $\nabla\phi(\bfeta)=A_m^{-1}(\norm{\bfeta})\,\bfeta/\norm{\bfeta}$ ($\bfeta\neq\mathbf 0$) of Section~2.1 of the main text: it keeps the direction of $\bfeta$ and turns its length $A_m(\kappa)$ into $\kappa$, while $\nabla\phi(\mathbf 0)=\mathbf 0$ is the natural parameter of the uniform law. For $m=2$ the family is the von Mises distribution on the circle.

\paragraph{The conditional Gaussian regression.}
Section~3 of the main text uses the conditional Gaussian regression $\calN(\bmu_c(\bx_2),\bS)$ of $\bx_1\in\R^d$ given $\bx_2\in\sphere^{m-1}$, with $\bmu_c(\bx_2):=\balpha+\bB\bx_2$. Its statistic is $(\bx_1,\,\bx_1\bx_1^\top,\,\bx_1\bx_2^\top)$ and its natural parameter is $(\bS^{-1}\balpha,\,-\tfrac12\bS^{-1},\,\bS^{-1}\bB)$. For each fixed $\bx_2$ it is an exponential family with this natural parameter, but its log-partition function
\begin{equation*}
\psi_{\bx_2}=\tfrac12\bigl\{d\log2\pi+\log|\bS|+\bmu_c(\bx_2)^\top\bS^{-1}\bmu_c(\bx_2)\bigr\}
\end{equation*}
depends on $\bx_2$. Given the observed directions $\bx_{21},\dots,\bx_{2N}$, the conditional model of $(\bx_{1n})_n$ is an exponential family with log-partition function $\sum_n\psi_{\bx_{2n}}$. Its dual potential and mean space depend on these directions, and its moment matching is the set of conditional moment identities \eqref{eq:cond-moments}.

\paragraph{Starting values.}
$k$-means~\cite{macqueen1967methods} alternates nearest-centroid assignment in squared Euclidean distance with the centroid update. Spherical $k$-means~\cite{dhillon2001concept} uses the cosine dissimilarity $1-\bx^\top\bmu=\tfrac12\norm{\bx-\bmu}^2$ and the normalized resultant. Heuristically, away from ties, the responsibilities of an isotropic Gaussian mixture with common variance $\sigma^2$ concentrate on the nearest mean as $\sigma^2\downarrow0$, and those of a von Mises--Fisher mixture with common concentration $\kappa$ on the largest inner product as $\kappa\to\infty$~\cite{banerjee2005clustering}. These limits motivate $k$-means-type starting values and say nothing about the fitted mixtures. If $\bx$ is split into blocks $\bx^{(i)}$ and $\phi(\bx)=\sum_i\alpha_i\phi_i(\bx^{(i)})$ with $\alpha_i\ge0$, the Bregman divergence separates over the blocks~\cite{banerjee2005bregman}:
\begin{equation*}
D_\phi[\bx:\mathbf y]=\sum_i\alpha_iD_{\phi_i}[\bx^{(i)}:\mathbf y^{(i)}].
\end{equation*}
A Euclidean and a directional block may therefore carry different divergences with a relative weight, as in the initializer of Section~4 of the main text, whose directional centroid is constrained to the sphere. \paragraph{Moments of the von Mises--Fisher law.} Section~2.1 of the main text summarizes the following classical facts, which Section~\ref{sec:esm-proofs} proves.

\begin{lemma}[Second moment of the von Mises--Fisher law \cite{mardiajupp2000directional,schou1978concentration}]
\label{lem:vmf-moments}
Let $\bx\sim\vMF_m(\bmu,\kappa)$ on $\sphere^{m-1}$ with $\kappa>0$. Then
\begin{align}
\E[\bx]&=A_m(\kappa)\,\bmu,\label{eq:vmf-m1}\\
\E[\bx\bx^\top]&=\frac{A_m(\kappa)}\kappa\,\bI_m+\Bigl(1-\frac{m\,A_m(\kappa)}\kappa\Bigr)\bmu\bmu^\top,\label{eq:vmf-m2}\\
\cov(\bx)&=\frac{A_m(\kappa)}\kappa\bigl(\bI_m-\bmu\bmu^\top\bigr)+A_m'(\kappa)\,\bmu\bmu^\top.\label{eq:vmf-cov}
\end{align}
The eigenvalues of $\E[\bx\bx^\top]$ are $A_m(\kappa)/\kappa$, with multiplicity $m-1$ on the tangent space $\bmu^{\perp}$, and $1-(m-1)A_m(\kappa)/\kappa$, on $\bmu$; as $\kappa\downarrow0$, $\E[\bx]\to\mathbf 0$ and $\E[\bx\bx^\top]\to\bI_m/m$.
\end{lemma}

Section~\ref{sec:esm-cyl} turns from these general tools to one cylindrical component.

\section{The cylindrical family: moments, special cases and computations}\label{sec:esm-cyl}

This section gives the moments of one component, the case $d=1$, $m=2$, and the computations of Section~3 of the main text behind Proposition~3(a) and Theorem~6, the hypotheses of that theorem and the reporting of dependence, with the notation and hypotheses (A0w)--(A3w) of Assumption~5 of the main text; $\psi(\bvartheta):=-\log\Cm(\norm{\bvartheta})$ is the von Mises--Fisher log-partition function on $\R^m$.

\paragraph{The cylindrical density.}\ The density of Definition~1 of the main text is normalized:

\begin{proposition}[Cylindrical density]\label{prop:cyl-density}
For every $(\balpha,\bB,\bS,\bmu_2,\kappa)\in\calP$,
\begin{equation*}
\fcyl(\bx_1,\bx_2)=\frac{\Cm(\kappa)}{\sqrt{(2\pi)^d|\bS|}}\exp\Bigl(-\tfrac12\bigl(\bx_1-\bmu_c(\bx_2)\bigr)^{\top}\bS^{-1}\bigl(\bx_1-\bmu_c(\bx_2)\bigr)+\kappa\bmu_2^{\top}\bx_2\Bigr)
\end{equation*}
is a probability density on $\R^d\times\sphere^{m-1}$ with respect to $\nu$, and so is its extension to $\Pext$.
\end{proposition}

\begin{proof}
By Tonelli's theorem, a nonnegative function can be integrated one variable at a time. For fixed $\bx_2$ the Gaussian factor integrates to one over $\R^d$, and the von Mises--Fisher factor then integrates to one over $\sphere^{m-1}$.
\end{proof}

\paragraph{The Fisher metric.} Proposition~9 of the main text is part (i) and the first half of (iii) of the following proposition, which Section~\ref{sec:esm-proofs} proves.

\begin{proposition}[Fisher metric, complete form]\label{prop:esm-cyl-fisher}
At $(\balpha,\bB,\bS,\bvartheta_2)\in\Pext$, let $\bfeta_2:=\E[\bx_2]$ and $\mathbf V:=\cov(\bx_2)$ (Lemma~\ref{lem:vmf-moments}; $\mathbf 0$ and $\bI_m/m$ at $\bvartheta_2=\mathbf 0$), $\bz:=(1,\bx_2^{\top})^{\top}$, $\mathbf M_z:=\E[\bz\bz^{\top}]$ and $\bbeta:=[\balpha\ \ \bB]$; let $\operatorname{vec}$ stack the columns of a matrix, $\operatorname{vech}$ the distinct entries of a symmetric matrix, $\otimes$ be the Kronecker product, and $\mathbf D_d$ the duplication matrix, $\operatorname{vec}(\bS)=\mathbf D_d\operatorname{vech}(\bS)$ \cite{magnusneudecker2019matrix}.
\begin{enumerate}
\item[(i)] In the coordinates $(\operatorname{vec}\bbeta,\operatorname{vech}\bS,\bvartheta_2)$ the Fisher information is
\begin{equation}\label{eq:esm-fisher-info}
\calI=\blockdiag\Bigl(\mathbf M_z\otimes\bS^{-1},\ \ \tfrac12\mathbf D_d^{\top}(\bS^{-1}\otimes\bS^{-1})\mathbf D_d,\ \ \mathbf V\Bigr),
\end{equation}
positive definite on all of $\Pext$: a tangent vector $(\dot{\balpha},\dot{\bB},\dot{\bS},\dot{\bvartheta}_2)$ has squared length $\dot{\bvartheta}_2^{\top}\mathbf V\dot{\bvartheta}_2+(\dot{\balpha}+\dot{\bB}\bfeta_2)^{\top}\bS^{-1}(\dot{\balpha}+\dot{\bB}\bfeta_2)+\tr(\bS^{-1}\dot{\bB}\mathbf V\dot{\bB}^{\top})+\tfrac12\tr(\bS^{-1}\dot{\bS}\bS^{-1}\dot{\bS})$.
\item[(ii)] On $\calP$, in the polar coordinates $(\bmu_2,\kappa)$ with $\bmu_2$ varying in $T_{\bmu_2}\sphere^{m-1}$, the directional block is $\blockdiag(\kappa\Am(\kappa)\,\Pperp|_{T_{\bmu_2}\sphere^{m-1}},\ \Am'(\kappa))$, $\Pperp:=\bI_m-\bmu_2\bmu_2^{\top}$.
\item[(iii)] With $\partial\btheta_{\cyl}$ the Jacobian of the map $\btheta_{\cyl}$ given by (9) on $\Pext$, $\calI=(\partial\btheta_{\cyl})^{\top}\nabla^2\psi_{\mathrm{amb}}\,\partial\btheta_{\cyl}$, whereas $\nabla^2(\psi_{\mathrm{amb}}\circ\btheta_{\cyl})=\calI+\langle\bfeta,\nabla^2\btheta_{\cyl}\rangle$, with $\bfeta$ the expectation coordinates (12).
\end{enumerate}
\end{proposition}

\paragraph{Moments of one component.}
Let $a:=\Am(\kappa)$. By Lemma~\ref{lem:vmf-moments}, the direction has mean $\bfeta_2:=\E[\bx_2]=a\bmu_2$ and covariance $\cov(\bx_2)=\tfrac a\kappa(\bI_m-\bmu_2\bmu_2^{\top})+\Am'(\kappa)\bmu_2\bmu_2^{\top}$. The regression form $\bx_1=\balpha+\bB\bx_2+\beps$, with $\beps\sim\calN(\mathbf 0,\bS)$ independent of $\bx_2$, then gives
\begin{equation*}
\E[\bx_1]=\balpha+\bB\bfeta_2,\qquad \cov(\bx_1,\bx_2)=\bB\cov(\bx_2),\qquad \cov(\bx_1)=\bS+\bB\cov(\bx_2)\bB^{\top}.
\end{equation*}
In the original coordinates $\bSigma_{12}=\bB/\kappa$ and $\bmu_1=\balpha+\bB\bmu_2$, these read $\E[\bx_1]=\bmu_1+(\Am(\kappa)-1)\bB\bmu_2$ and $\cov(\bx_1,\bx_2)=\kappa\bSigma_{12}\cov(\bx_2)$. Since $\E[\bx_2]=a\bmu_2\neq\bmu_2$, $\bmu_1$ is the mean of $\bx_1$ if and only if $\bB\bmu_2=\mathbf 0$. For the other two parameters, $\cov(\bx_1,\bx_2)-\bSigma_{12}=\bSigma_{12}(\kappa\cov(\bx_2)-\bI_m)$, and $\kappa\cov(\bx_2)-\bI_m$ has the eigenvalues $a-1$ on $\bmu_2^{\perp}$ and $\kappa\Am'(\kappa)-1$ along $\bmu_2$. Both are negative: $a<1$, and $\kappa\Am'(\kappa)<a$, because by the Riccati identity (eq.~(4) of the main text) this strict inequality is equivalent to $a>\kappa/\{m/2+(m^2/4+\kappa^2)^{1/2}\}$, a lower bound of Amos~\cite{amos1974bessel} for $I_{m/2}(\kappa)/I_{m/2-1}(\kappa)$. Hence $\cov(\bx_2)\prec\kappa^{-1}\bI_m$, so $\kappa\cov(\bx_2)-\bI_m$ is invertible, and $\bSigma_{12}$ is the cross-covariance if and only if $\bB=\mathbf 0$. For the same reason, the auxiliary block $\bSigma_{11}=\bS+\kappa\bSigma_{12}\bSigma_{12}^{\top}=\bS+\kappa^{-1}\bB\bB^{\top}$ of the Gaussian kernel whose restriction is the component (Proposition~2 of the main text) exceeds $\cov(\bx_1)=\bS+\bB\cov(\bx_2)\bB^{\top}$ by $\bB(\kappa^{-1}\bI_m-\cov(\bx_2))\bB^{\top}\succeq\mathbf 0$, which is zero if and only if $\bB=\mathbf 0$. Thus $\bmu_1$ is the conditional mean of $\bx_1$ at the mean direction, and $\bSigma_{12}$ and $\bSigma_{11}$ are parameters of that kernel.

\paragraph{Reporting dependence through $\bB$.}\label{rem:sigma12-reporting}
Because $\bSigma_{12}=\bB/\kappa$, a fitted concentration near zero or very large (or near the bounds $\kappa_{\min}$, $\kappa_{\max}$ of the optional safeguards of Section~\ref{sec:esm-mix}, not used in the reported fits) turns a moderate $\bB$ into an extreme or vanishing $\bSigma_{12}$ without changing the fitted conditional law. The paper therefore reports dependence through $\bB$; only the exact-specification recovery study of Section~\ref{sec:esm-recovery}, whose generator is specified in original coordinates, scores $\bSigma_{12}$.

\paragraph{The case \texorpdfstring{$d=1$, $m=2$}{d=1, m=2}: Mardia--Sutton and Kato--Shimizu.}
Write $\bx_2=(\cos\theta,\sin\theta)^{\top}$ and $\bmu_2=(\cos\mu,\sin\mu)^{\top}$. The directional factor is then a von Mises law for the angle $\theta$, with mean direction $\mu$ and concentration $\kappa$, and, given $\theta$, the Euclidean variable $\bx_1$ is normal with a mean affine in $(\cos\theta,\sin\theta)$: the cylindrical model of Mardia and Sutton~\cite{mardia1978cylindrical}, with $\bB=\mathbf 0$ as their independence hypothesis.

In the same case the ambient family of the main text has $d+m+\tfrac{d(d+1)}2+dm+\tfrac{m(m+1)}2-1=8$ natural parameters: one in $\btheta_1$, two in $\btheta_2$, one in $\btheta_{11}$, two in $\btheta_{12}$ and two in the traceless block $\btheta_{22}$. On the circle, a traceless symmetric $\bOmega=\bigl(\begin{smallmatrix}u&v\\ v&-u\end{smallmatrix}\bigr)$ gives $\bx_2^{\top}\bOmega\bx_2=u\cos2\theta+v\sin2\theta$. The circular marginal of an ambient member, its Fisher--Bingham directional factor, therefore carries two harmonics of the angle: the first harmonic $\bvartheta_2(\btheta)^{\top}\bx_2$, with concentration $\kappa_1=\norm{\bvartheta_2(\btheta)}$, and a second harmonic with coefficients $(u,v)$ read from $\bOmega(\btheta)$ and concentration $\kappa_2=(u^2+v^2)^{1/2}$. The cylindrical family is the subfamily $\bOmega(\btheta)=\mathbf 0$, $\bvartheta_2(\btheta)\neq\mathbf 0$, that is, $\kappa_2=0$ and $\kappa_1>0$. Its codimension $\tfrac{m(m+1)}2-1$ at $m=2$ counts the two coefficients $u$ and $v$ that this constraint sets to zero. Abe and Ley~\cite{abe2017tractable} write the conditional mean of $\bx_1$ given $\theta$ as $\mu'+\lambda\cos(\theta-\nu)$, with intercept $\mu'$, phase $\nu$ (not the base measure) and amplitude $\lambda$; on the cylindrical family, $\lambda=\norm{\bB}$. Up to reparametrization, the ambient family is the eight-parameter family that they attribute to Kato and Shimizu~\cite{katoshimizu2008dependent}, together with the boundary values $\kappa_1=0$, $\kappa_2=0$ and $\lambda=0$ that their parametrization excludes; their family is the open dense subset on which $\kappa_1$, $\kappa_2$ and $\lambda$ are all positive. As Abe and Ley note, the Mardia--Sutton model is the case $\kappa_2=0$ of the Kato--Shimizu density. The cylindrical family is the part $\kappa_2=0$, $\kappa_1>0$ of this boundary, with $\lambda\ge0$ free, and $\lambda=0$ is the independence hypothesis $\bB=\mathbf 0$; its extension $\Pext$ adds $\kappa_1=0$.

\paragraph{Proof of Proposition~3(a) of the main text (curved family).}
Expanding the quadratic form in $\log\fcyl$ (Proposition~\ref{prop:cyl-density}) and reading off the coefficients of $\bx_1\bx_1^{\top}$, $\bx_1\bx_2^{\top}$, $\bx_1$ and $\bx_2$ gives $\btheta_{11}=-\tfrac12\bS^{-1}$, $\btheta_{12}=\bS^{-1}\bB$, $\btheta_1=\bS^{-1}\balpha$ and $\btheta_2=\kappa\bmu_2-\bB^{\top}\bS^{-1}\balpha$. The coefficient of $\bx_2\bx_2^{\top}$ is $-\tfrac12\bB^{\top}\bS^{-1}\bB$. Because $\bx_2^{\top}\bx_2=1$, we have $\bx_2^{\top}\mathbf M\bx_2=\bx_2^{\top}\mathbf M^{\circ}\bx_2+\tfrac1m\tr\mathbf M$ on the sphere for every symmetric $\mathbf M$. Hence only the traceless part $\btheta_{22}=(-\tfrac12\bB^{\top}\bS^{-1}\bB)^{\circ}$ pairs with the statistic, and the constant $-\tfrac1{2m}\tr(\bB^{\top}\bS^{-1}\bB)$ joins the other constants, whose sum is $-\psi_{\cyl}$. Since $\btheta_{11}^{-1}=-2\bS$, we have $\tfrac14\btheta_{12}^{\top}\btheta_{11}^{-1}\btheta_{12}=-\tfrac12\bB^{\top}\bS^{-1}\bB$, so $\bOmega(\btheta)=\mathbf 0$. Likewise $\tfrac12\btheta_{12}^{\top}\btheta_{11}^{-1}\btheta_1=-\bB^{\top}\bS^{-1}\balpha$, so $\bvartheta_2(\btheta)=\kappa\bmu_2\neq\mathbf 0$. So integrating $\bx_1$ out cancels the quadratic term in $\bx_2$ of the joint log-density exactly, and the directional marginal is $\vMF_m(\bmu_2,\kappa)$.

\paragraph{Proof of Theorem~6 of the main text (maximum likelihood) in full.}
Observations with $w_n=0$ do not enter $\ell_w$. Write $\bz_n:=(1,\bx_{2n}^{\top})^{\top}$, $\bbeta:=[\balpha\ \ \bB]$, $\mathbf M^{w}:=\sum_nw_n\bz_n\bz_n^{\top}$ and $E_w(\bbeta):=\sum_nw_n(\bx_{1n}-\bbeta\bz_n)(\bx_{1n}-\bbeta\bz_n)^{\top}$, so that Lemma~\ref{lem:ols-loewner} for the pairs $(\sqrt{w_n}\bz_n,\sqrt{w_n}\bx_{1n})$ has design moment $\mathbf M^{w}$ and residual scatter $E^{w}$.

\emph{(i) Splitting.} By (5), $\ell_w=\ell_{\CG}^{w}(\balpha,\bB,\bS)+\ell_{\vMF}^{w}(\bvartheta_2)$ with $\ell_{\CG}^{w}=-\tfrac{Wd}2\log(2\pi)-\tfrac W2\log|\bS|-\tfrac12\tr(\bS^{-1}E_w(\bbeta))$ and $\ell_{\vMF}^{w}=\sum_nw_n\{\bvartheta_2^{\top}\bx_{2n}-\psi(\bvartheta_2)\}=W\{\langle\bvartheta_2,\bar{\br}\rangle-\psi(\bvartheta_2)\}$, $\bar{\br}:=\br^{w}/W$. The two blocks range independently over $\Pext=(\R^d\times\R^{d\times m}\times\Sympp{d})\times\R^m$, so $\sup\ell_w=\sup\ell_{\CG}^{w}+\sup\ell_{\vMF}^{w}$, a maximizer exists if and only if each part has one, and the maximizers are the pairs of maximizers.

\emph{(ii) Regression coefficients.} By Lemma~\ref{lem:ols-loewner}, for every fixed $\bS\succ\mathbf 0$, $\ell_{\CG}^{w}(\cdot,\bS)$ is maximized exactly at the solutions $\hat{\bbeta}$ of the normal equations $\bbeta\mathbf M^{w}=\sum_nw_n\bx_{1n}\bz_n^{\top}$, the weighted least-squares coefficients, with maximum $-\tfrac{Wd}2\log(2\pi)+G(\bS)$, where $G$ is the function of Lemma~\ref{lem:cov-profile} with $(N,\mathbf C)=(W,E^{w})$. The set of maximizers does not depend on $\bS$, which makes the profile argument (first $\bbeta$, then $\bS$) legitimate. The matrix $\mathbf M^{w}$ has leading entry $W>0$ and Schur complement $\sum_nw_n\bx_{2n}\bx_{2n}^{\top}-W\bar{\bx}_2^{w}(\bar{\bx}_2^{w})^{\top}=\sct{22}^{w}$, so the solution is unique if and only if $\sct{22}^{w}\succ\mathbf 0$, that is, (A1w). The normal equations then read $\bar{\bx}_1^{w}=\hat{\balpha}+\hat{\bB}\bar{\bx}_2^{w}$ and $\sum_nw_n\bx_{1n}\bx_{2n}^{\top}=W\hat{\balpha}(\bar{\bx}_2^{w})^{\top}+\hat{\bB}\sum_nw_n\bx_{2n}\bx_{2n}^{\top}$; eliminating $\hat{\balpha}$ gives $\sct{12}^{w}=\hat{\bB}\sct{22}^{w}$, hence $\hat{\bB}$ and $\hat{\balpha}$ of (13). For a unit $\mathbf a\in\R^d$, Lemma~\ref{lem:ols-loewner} applied to the scalar responses $\mathbf a^{\top}\bx_{1n}$ gives $\mathbf a^{\top}E^{w}\mathbf a=\min_{\boldsymbol\gamma}\sum_nw_n(\mathbf a^{\top}\bx_{1n}-\boldsymbol\gamma^{\top}\bz_n)^2$, which is zero exactly when $\mathbf a^{\top}\bx_1$ is an affine function of $\bx_2$ on the observations with positive weight: this is the verbal form of (A2w).

\emph{(iii) Conditional covariance.} If $E^{w}\succ\mathbf 0$, Lemma~\ref{lem:cov-profile} gives the unique maximizer $\hat{\bS}=E^{w}/W$. Under (A1w), the residuals are $\hat{\beps}_n:=\bx_{1n}-\hat{\balpha}-\hat{\bB}\bx_{2n}=(\bx_{1n}-\bar{\bx}_1^{w})-\hat{\bB}(\bx_{2n}-\bar{\bx}_2^{w})$, and expanding with $\hat{\bB}\sct{22}^{w}=\sct{12}^{w}$ gives $E^{w}=\sum_nw_n\hat{\beps}_n\hat{\beps}_n^{\top}=\sct{11}^{w}-\sct{12}^{w}(\sct{22}^{w})^{-1}(\sct{12}^{w})^{\top}$, which is $\hat{\bS}$ of (13). If $E^{w}$ is singular, take a unit $\mathbf v$ with $E^{w}\mathbf v=\mathbf 0$ and $\bS_t:=t\mathbf v\mathbf v^{\top}+(\bI_d-\mathbf v\mathbf v^{\top})$, $t>0$. Then $\bS_t^{-1}=t^{-1}\mathbf v\mathbf v^{\top}+(\bI_d-\mathbf v\mathbf v^{\top})$ and $\mathbf v^{\top}E^{w}\mathbf v=0$, so $G(\bS_t)=-\tfrac W2\log t-\tfrac12\tr\{(\bI_d-\mathbf v\mathbf v^{\top})E^{w}\}\to+\infty$ as $t\downarrow0$: $\sup\ell_{\CG}^{w}=+\infty$, and $\ell_w$ has no maximizer.

\emph{(iv) Direction and concentration.} The function $\ell_{\vMF}^{w}$ is smooth and strictly concave on $\R^m$, because $\psi$ is strictly convex (Section~2 of the main text), with gradient $W(\bar{\br}-\nabla\psi(\bvartheta_2))$. By Lemma~\ref{lem:dual-mstep} for the von Mises--Fisher family, whose mean space is the open unit ball, it has the unique maximizer $\hat{\bvartheta}_2=\nabla\phi(\bar{\br})$ whenever $\norm{\bar{\br}}<1$; by Section~2.1 of the main text, $\nabla\phi(\bar{\br})=\Am^{-1}(\norm{\bar{\br}})\,\bar{\br}/\norm{\bar{\br}}$ if $\bar{\br}\neq\mathbf 0$ and $\nabla\phi(\mathbf 0)=\mathbf 0$. In polar terms: for fixed $\kappa>0$, the Cauchy--Schwarz inequality gives $\bmu_2^{\top}\br^{w}\le\norm{\br^{w}}$, with equality if and only if $\bmu_2=\br^{w}/\norm{\br^{w}}$ when $\br^{w}\neq\mathbf 0$, and $L(\kappa):=W\log\Cm(\kappa)+\kappa\norm{\br^{w}}$ has $L'(\kappa)=W(\norm{\bar{\br}}-\Am(\kappa))$ and $L''(\kappa)=-W\Am'(\kappa)<0$ by (3), because $\Am'(\kappa)$ is the variance of $\bmu_2^{\top}\bx_2$ under $\vMF_m(\bmu_2,\kappa)$. So $\hat\kappa$ is the unique root of $\Am(\kappa)=\norm{\bar{\br}}$ in $(0,\infty)$, the range of $\Am$ there being $(0,1)$. By the triangle inequality $\norm{\br^{w}}\le\sum_nw_n\norm{\bx_{2n}}=W$, with equality if and only if all $\bx_{2n}$ with $w_n>0$ coincide; then $\sct{22}^{w}=\mathbf 0$, so (A1w) implies (A0w). If they all equal $\mathbf v$, then $\ell_{\vMF}^{w}(t\mathbf v)=W\{t-\psi(t\mathbf v)\}=-W\log\int_{\sphere^{m-1}}\exp\{t(\mathbf v^{\top}\mathbf u-1)\}\,d\sigma(\mathbf u)$; the integrand is at most one and tends to zero for every $\mathbf u\neq\mathbf v$, so by dominated convergence the integral tends to zero and $\ell_{\vMF}^{w}(t\mathbf v)\to+\infty$ as $t\to\infty$.

\emph{(v) Conclusion on $\Pext$.} By (i)--(iv), a maximizer exists if and only if (A0w) and (A2w) hold, and otherwise $\sup\ell_w=+\infty$. The covariance and directional maximizers are unique, so the maximizer is unique if and only if the least-squares coefficient is, that is, under (A1w), where (13) holds. Under (A1w) the directions with positive weight span $\R^m$ affinely, so (A0w) holds automatically.

\emph{(vi) The polar chart.} $\calP$ is identified with the open set $\{\bvartheta_2\neq\mathbf 0\}$ of $\Pext$. If a maximizer over $\Pext$ has $\hat{\bvartheta}_2\neq\mathbf 0$, that is, under (A3w), it lies in $\calP$, and the maximizers over $\calP$ are exactly those over $\Pext$. If (A0w) or (A2w) fails, the unbounded sequences of (iii) and (iv) can be taken in $\calP$ (fix $\bvartheta_2\neq\mathbf 0$ in (iii)), so $\sup_{\calP}\ell_w=+\infty$. If (A2w) holds and $\br^{w}=\mathbf 0$, the unique directional maximizer is $\mathbf 0\notin\calP$, so every point of $\calP$ has $\ell_w<\max_{\Pext}\ell_w$, while $\ell_w$ at $(\hat{\balpha},\hat{\bB},\hat{\bS},\kappa\bmu_2)$ tends to $\max_{\Pext}\ell_w$ as $\kappa\downarrow0$, by continuity of $\psi$ at $\mathbf 0$: the supremum over $\calP$ is finite and not attained. The original coordinates follow from the bijection (8) for fixed $(\bmu_2,\kappa)$.

\paragraph{The hypotheses of Theorem~6, boundary cases and nonuniqueness.}\label{rem:cyl-conditions}
(A1) says that the affine hull of $\{\bx_{2n}\}$ is $\R^m$, the empirical analogue of the fact used in Proposition~7; it needs $N\ge m+1$, and with (A2) $N\ge m+d+1$, sizes that are necessary, not sufficient. If (A1) fails, the directions lie in a hyperplane $\{\mathbf u:\mathbf v^{\top}\mathbf u=c\}$, and under (A0) and (A2) every $(\hat{\balpha}-c\mathbf b,\hat{\bB}+\mathbf b\mathbf v^{\top})$, $\mathbf b\in\R^d$, is a maximizing regression block, as the data do not say how $\bx_1$ varies across the hyperplane: for instance, two distinct angles on the circle, or directions on one circle of $\sphere^2$. (A1) implies (A0) but not (A3): the directions $\pm\mathbf e_1,\pm\mathbf e_2$ span the plane and cancel, so $\hat{\bvartheta}_2=\mathbf 0$, whereas $\hat{\bB}/\kappa$ diverges as $\kappa\downarrow0$ unless $\hat{\bB}=\mathbf 0$. The reported fits use exact EM without bounds on the concentrations (Section~4.2 of the main text); the optional safeguards of Section~\ref{sec:esm-mix}, not used in the reported fits, would instead keep concentrations in $[\kappa_{\min},\kappa_{\max}]$. The following arguments support these statements.
The centred directions $\bx_{2n}-\bar{\bx}_2$ sum to zero, so they span at most $N-1$ dimensions, and (A1) requires $N\ge m+1$. Under (A1) the design matrix $\mathbf Z:=[\bz_1\ \cdots\ \bz_N]^{\top}$ has rank $m+1$, and by the normal equations each of the $d$ columns of the residual matrix $\hat{\mathbf E}:=[\hat{\beps}_1\ \cdots\ \hat{\beps}_N]^{\top}$ lies in the orthogonal complement in $\R^N$ of the columns of $\mathbf Z$, of dimension $N-m-1$; hence $E=\hat{\mathbf E}^{\top}\hat{\mathbf E}$ has rank at most $N-m-1$, and (A2) requires $N\ge m+d+1$. If (A1) fails, the directions lie in an affine hyperplane $\{\mathbf u:\mathbf v^{\top}\mathbf u=c\}$ with $\norm{\mathbf v}=1$. Then $\mathbf M(-c,\mathbf v^{\top})^{\top}=\sum_n\bz_n(\mathbf v^{\top}\bx_{2n}-c)=\mathbf 0$, so for every $\mathbf b\in\R^d$ the matrix $\hat{\bbeta}+\mathbf b(-c,\mathbf v^{\top})$ also solves the normal equations; that is, $(\hat{\balpha}-c\mathbf b,\hat{\bB}+\mathbf b\mathbf v^{\top})$ is again a least-squares coefficient, and under (A0) and (A2) a maximizing regression block, by part~(i) of the theorem. The set of maximizing regression blocks is an affine space of dimension $d\,(m+1-\rank\mathbf M)\ge d$. On the circle, an affine line meets $\sphere^1$ in at most two points, so (A1) fails exactly when at most two distinct angles are observed; on $\sphere^2$, a plane meets the sphere in a circle. (A1) does not imply (A3): for $m=2$ and $N=4$, the directions $\pm\mathbf e_1,\pm\mathbf e_2$ span the plane affinely but cancel in pairs, giving $\sct{22}=2\bI_2$ and $\br=\mathbf 0$. Then $\hat{\bvartheta}_2=\mathbf 0$, and, by part~(iii) of the theorem, the likelihood over $\calP$ approaches its finite supremum only as $\kappa\downarrow0$; the regression block stays finite, while $\bSigma_{12}=\hat{\bB}/\kappa$ has no finite limit unless $\hat{\bB}=\mathbf 0$. If $E$ is singular, a zero resultant does not help: the likelihood is already unbounded through the covariance. Section~\ref{sec:esm-proofs} gives the remaining proofs of lemmas, propositions and corollaries of Sections~2--4 of the main text.

\section{Proofs for Sections 2--4 of the main text}\label{sec:esm-proofs}

This section has two parts. The first part writes out the longer computations of five proofs, in the order in which Section~\ref{subsec:esm-moved-proofs} cites them: Lemma~\ref{lem:vmf-moments}, Proposition~2, Proposition~3(b)--(c) with the extension to $\Pext$, and Proposition~8. Section~\ref{subsec:esm-moved-proofs} then gives the formal proofs of the statements of Sections~2--4 that are proved neither in Appendix~A of the main text (Theorems~6 and~12) nor in Sections~\ref{sec:esm-background} (Lemma~\ref{lem:dual-mstep}), \ref{sec:esm-cyl} (Proposition~\ref{prop:cyl-density}, Proposition~3(a), the full computations for Theorem~6 and its hypotheses) and~\ref{sec:esm-mix} (the details of the mixture procedure of Section~4: the conditions of Lemma~\ref{lem:cyl-mle-weighted}, the optional safeguards and the initializer as classification EM). In this section, numbers without the prefix S refer to the main text, whose notation and hypotheses (A0w)--(A3w) are used; $\langle\cdot,\cdot\rangle$ is the Euclidean inner product (Frobenius on matrix blocks), and $\mathbf A^{\circ}:=\mathbf A-\tfrac1m\tr(\mathbf A)\bI_m$ is the traceless part of a symmetric $m\times m$ matrix.

\paragraph{Second moment of the von Mises--Fisher law (Lemma~\ref{lem:vmf-moments}).}
The lemma is classical \cite{mardiajupp2000directional,schou1978concentration}. The idea: the log-partition function depends on the natural parameter only through its length, so its gradient and Hessian (the mean and the covariance of $\bx$) follow from one scalar function. The natural parameter of $\vMF_m(\bmu,\kappa)$ is $\btheta:=\kappa\bmu$, and its log-partition function is the radial function $\psi(\btheta)=\psi_{\mathrm r}(\norm{\btheta})$ with $\psi_{\mathrm r}(\kappa):=-\log\Cm(\kappa)$ and, by (3), $\psi_{\mathrm r}'=\Am$. At a nonzero argument, $\nabla\psi=\psi_{\mathrm r}'(\kappa)\bmu$ and, since $\nabla(\btheta/\kappa)=(\bI_m-\bmu\bmu^{\top})/\kappa$,
\begin{equation*}
\nabla^2\psi(\btheta)=\psi_{\mathrm r}''(\kappa)\,\bmu\bmu^{\top}+\frac{\psi_{\mathrm r}'(\kappa)}{\kappa}\bigl(\bI_m-\bmu\bmu^{\top}\bigr),
\end{equation*}
so in the $m-1$ tangent directions the eigenvalue is the slope of $\psi_{\mathrm r}$ divided by the radius, $\psi_{\mathrm r}'(\kappa)/\kappa=\Am(\kappa)/\kappa$. The family is regular (the sphere is compact), so $\nabla\psi=\E[\bx]$ and $\nabla^2\psi=\cov(\bx)$. With $a=\Am(\kappa)$ and $\psi_{\mathrm r}''=\Am'$, this gives $\E[\bx]=a\bmu$ and the covariance formula of the lemma. The second moment needs the Riccati identity (4). With the Bessel order $\nu:=m/2-1$ (not the base measure of Section~3.1) and $A:=\Am=I_{\nu+1}/I_\nu$, the Bessel recurrences $I_\nu'=I_{\nu+1}+\tfrac\nu\kappa I_\nu$ and $I_{\nu+1}'=I_\nu-\tfrac{\nu+1}\kappa I_{\nu+1}$ give $A'=I_{\nu+1}'/I_\nu-A\,I_\nu'/I_\nu=(1-\tfrac{\nu+1}\kappa A)-A(A+\tfrac\nu\kappa)=1-A^2-\tfrac{m-1}\kappa A$, which is (4). Adding $\E[\bx]\E[\bx]^{\top}=a^2\bmu\bmu^{\top}$ and inserting (4),
\begin{equation*}
\E[\bx\bx^{\top}]=\frac a\kappa\bI_m+\Bigl(\Am'(\kappa)-\frac a\kappa+a^2\Bigr)\bmu\bmu^{\top}
=\frac a\kappa\bI_m+\Bigl(1-\frac{m\,a}{\kappa}\Bigr)\bmu\bmu^{\top}.
\end{equation*}
Its trace is $1$, the trace of the covariance is $(m-1)a/\kappa+\Am'=1-a^2$, and $\bmu^{\top}\E[\bx\bx^{\top}]\bmu=1-(m-1)a/\kappa$. The eigenvalues are read off the display.

\paragraph{Restriction of a Gaussian kernel to the cylinder (Proposition~2, Section~3.1).}
Completing the square in $\bx_1$ splits the kernel into a Gaussian factor in $\bx_1$, whose mean depends on $\bx_2$, and a Fisher--Bingham factor in $\bx_2$ (step (i), for a general kernel); step (ii) reads the parameters off the block inverse when $\bSigma_{22}=\kappa^{-1}\bI_m$.

\emph{(i) General kernel.} Let $\mathbf c=(\mathbf c_1,\mathbf c_2)\in\R^d\times\R^m$, let $\bLambda$ be symmetric of order $d+m$ with blocks $\bLambda_{11}\in\Sympp{d}$, $\bLambda_{12}=\bLambda_{21}^{\top}$ and symmetric $\bLambda_{22}$, let $K(\bx):=\exp\{-\tfrac12(\bx-\mathbf c)^{\top}\bLambda(\bx-\mathbf c)\}$, and $\mathbf Q:=\bLambda_{22}-\bLambda_{21}\bLambda_{11}^{-1}\bLambda_{12}$. Put $\mathbf u_i:=\bx_i-\mathbf c_i$. Expanding the right-hand side of
\begin{equation*}
(\bx-\mathbf c)^{\top}\bLambda(\bx-\mathbf c)
=\bigl(\mathbf u_1+\bLambda_{11}^{-1}\bLambda_{12}\mathbf u_2\bigr)^{\top}\bLambda_{11}\bigl(\mathbf u_1+\bLambda_{11}^{-1}\bLambda_{12}\mathbf u_2\bigr)+\mathbf u_2^{\top}\mathbf Q\mathbf u_2
\end{equation*}
gives $\mathbf u_1^{\top}\bLambda_{11}\mathbf u_1+2\mathbf u_1^{\top}\bLambda_{12}\mathbf u_2+\mathbf u_2^{\top}(\bLambda_{21}\bLambda_{11}^{-1}\bLambda_{12}+\mathbf Q)\mathbf u_2$, which is the left-hand side by the definition of the Schur complement $\mathbf Q$. Hence $K(\bx)=\exp\{-\tfrac12(\bx_1-\mathbf g(\bx_2))^{\top}\bLambda_{11}(\bx_1-\mathbf g(\bx_2))\}\exp\{-\tfrac12\mathbf u_2^{\top}\mathbf Q\mathbf u_2\}$ with $\mathbf g(\bx_2):=\mathbf c_1-\bLambda_{11}^{-1}\bLambda_{12}(\bx_2-\mathbf c_2)$. For fixed $\bx_2$ the first factor integrates over $\R^d$ to $(2\pi)^{d/2}|\bLambda_{11}|^{-1/2}<\infty$, because $\bLambda_{11}\succ\mathbf 0$. The second factor is continuous on the compact sphere. By Tonelli's theorem (a nonnegative function can be integrated one block of variables at a time), $\int K\,d\nu$ is therefore finite and positive. After division, the $\bx_1$-factor is $\fG(\bx_1;\mathbf g(\bx_2),\bLambda_{11}^{-1})$, and the $\bx_2$-factor is the normalization on the sphere of $\exp\{-\tfrac12\bx_2^{\top}\mathbf Q\bx_2+(\mathbf Q\mathbf c_2)^{\top}\bx_2-\tfrac12\mathbf c_2^{\top}\mathbf Q\mathbf c_2\}$. The last term is a constant, so this factor is $\FB_m(\mathbf Q\mathbf c_2,-\tfrac12\mathbf Q)$.

\emph{(ii) The case $\bLambda=\bSigma^{-1}$, $\bSigma_{22}=\kappa^{-1}\bI_m$, $\norm{\mathbf c_2}=1$.} Since $\bSigma\succ\mathbf 0$, also $\bLambda\succ\mathbf 0$ and $\bLambda_{11}\succ\mathbf 0$. Write $\bSigma\bLambda=\bI_{d+m}$ blockwise \cite{magnusneudecker2019matrix}. The $(2,1)$ block gives $\bLambda_{21}=-\bSigma_{22}^{-1}\bSigma_{12}^{\top}\bLambda_{11}$, hence $-\bLambda_{11}^{-1}\bLambda_{12}=\bSigma_{12}\bSigma_{22}^{-1}$. Substituting $\bLambda_{21}$ into the $(1,1)$ block gives $\bLambda_{11}^{-1}=\bSigma_{11}-\bSigma_{12}\bSigma_{22}^{-1}\bSigma_{12}^{\top}$. The $(2,2)$ block gives $\bLambda_{22}=\bSigma_{22}^{-1}+\bSigma_{22}^{-1}\bSigma_{12}^{\top}\bLambda_{11}\bSigma_{12}\bSigma_{22}^{-1}$, whose second term is $\bLambda_{21}\bLambda_{11}^{-1}\bLambda_{12}$, so $\mathbf Q=\bSigma_{22}^{-1}$. With $\bSigma_{22}=\kappa^{-1}\bI_m$ these are the three identities claimed. Inserted in (i), they make the conditional factor $\fG(\bx_1;\mathbf c_1+\kappa\bSigma_{12}(\bx_2-\mathbf c_2),\bLambda_{11}^{-1})$. Its mean is the conditional mean of the original parametrization (8), with $\bmu_1=\mathbf c_1$, $\bmu_2=\mathbf c_2$ and $\bS=\bLambda_{11}^{-1}\succ\mathbf 0$. The directional factor is $\FB_m(\kappa\mathbf c_2,-\tfrac\kappa2\bI_m)=\FB_m(\kappa\mathbf c_2,\mathbf 0)=\vMF_m(\mathbf c_2,\kappa)$, because $(-\tfrac\kappa2\bI_m)^{\circ}=\mathbf 0$ and $\norm{\mathbf c_2}=1$. The product is therefore $\fcyl$ with $\bB=\kappa\bSigma_{12}$ and $\balpha=\mathbf c_1-\bB\mathbf c_2$.

\paragraph{Minimality of the ambient family and non-e-flatness of \texorpdfstring{$\Ecyl$}{E\_cyl} (Proposition~3, the curved family, Section~3.2, parts (b)--(c)).}
Here $U:=\R^d\times\R^m\times\Symmm{d}\times\R^{d\times m}$ holds the first four blocks of $\btheta$, so that $\Hnat_{\mathrm{amb}}=U\times\Symzero{m}$, and $\bOmega$, $\bvartheta_2$ are the maps of the constraint (11).

\emph{Minimality.} The idea: a linear combination $\langle\mathbf a,\bt\rangle$ of the statistic is a polynomial of degree at most two in $(\bx_1,\bx_2)$ that is constant on the cylinder only if $\mathbf a=\mathbf 0$. Let $\langle\mathbf a,\bt(\bx)\rangle=b$ $\nu$-a.e., with $\mathbf a=(\mathbf b_1,\mathbf b_2,\mathbf M_{11},\mathbf M_{12},\mathbf M_{22})$ and $\mathbf M_{22}$ traceless, like the block of $\bt$ it multiplies; by continuity the identity holds on all of $\R^d\times\sphere^{m-1}$. A polynomial in $\bx_1$ that is constant on $\R^d$ has no quadratic or linear part, so $\mathbf M_{11}=\mathbf 0$ and $\mathbf b_1+\mathbf M_{12}\bx_2=\mathbf 0$ for all unit $\bx_2$; the choices $\bx_2=\pm\mathbf e_i$ give $\mathbf b_1=\mathbf 0$ and $\mathbf M_{12}=\mathbf 0$. Comparing $\bx_2$ with $-\bx_2$ gives $\mathbf b_2=\mathbf 0$. Then $\bx_2^{\top}\mathbf M_{22}\bx_2=b$ at every unit eigenvector forces $\mathbf M_{22}=b\bI_m$, which is traceless only if $b=0$. The same argument on the $\bx_2$-terms shows that the Fisher--Bingham family, with statistic $\bt_2(\bx_2):=(\bx_2,\bx_2\bx_2^{\top}-\tfrac1m\bI_m)$, is minimal; it is regular because the sphere is compact.

\emph{Not e-flat.} The idea: along the curves used below, $\btheta_{12}$ moves linearly in a parameter $s$ while the constraint makes $\btheta_{22}$ quadratic in $s$, so an affine equation that holds on $\Ecyl$ cannot involve $\btheta_{22}$; the other blocks fill an open set, so it cannot involve them either. We show that no nonzero affine functional $L(\btheta)=\ell_1(\btheta_1)+\ell_2(\btheta_2)+\ell_{11}(\btheta_{11})+\ell_{12}(\btheta_{12})+\ell_{22}(\btheta_{22})+c$ vanishes on $\Ecyl$. Then $\Ecyl$ lies in no proper affine subspace. Because $\Ecyl$ has positive codimension (Proposition~3(c)), an affine subspace of its dimension is proper, so $\Ecyl$ is not an open subset of one, that is, not e-flat. For a unit $\mathbf e\in\R^m$ and $\mathbf E\in\R^{d\times m}$, the curve $\btheta(s):=(\mathbf 0,\mathbf e,-\tfrac12\bI_d,s\mathbf E,-\tfrac12s^2(\mathbf E^{\top}\mathbf E)^{\circ})$, $s\in\R$, has $\bOmega=\mathbf 0$ and $\bvartheta_2=\mathbf e$, so it lies in $\Ecyl$. If $L$ vanishes on $\Ecyl$, the coefficient of $s^2$ in $L(\btheta(s))$ gives $\ell_{22}((\mathbf E^{\top}\mathbf E)^{\circ})=0$. For $\mathbf E$ with a single nonzero row $\mathbf v^{\top}$, $\mathbf E^{\top}\mathbf E=\mathbf v\mathbf v^{\top}$; these matrices span the symmetric matrices, and the traceless-part map is onto $\Symzero{m}$, so $\ell_{22}=0$. The rest of $L$ vanishes on the first four blocks of the points of $\Ecyl$, which form the nonempty open set $U\cap\{\bvartheta_2\ne\mathbf 0\}$, so it is zero.

\paragraph{\texorpdfstring{$\Ecyl$}{E\_cyl} is not m-flat (Proposition~3, Section~3.2, part (c)).}
The idea: varying only $\kappa$ moves the expectation coordinates along the linearly independent functions $1$, $\Am(\kappa)$ and $\Am(\kappa)/\kappa$. On $\Ecyl$ the expectation coordinates $\bfeta=\nabla\psi_{\mathrm{amb}}(\btheta)=\E_{\btheta}[\bt]$ are given by (12), with last block $\bfeta_{22}-\tfrac1m\bI_m$. Suppose the affine functional $L=\ell_1(\bfeta_1)+\ell_2(\bfeta_2)+\ell_{11}(\bfeta_{11})+\ell_{12}(\bfeta_{12})+\ell_{22}(\bfeta_{22}-\tfrac1m\bI_m)+c$ vanishes on the $\bfeta$-image of $\Ecyl$.

\emph{Step 1.} Fix $\balpha=\mathbf 0$, $\bB=\mathbf 0$, $\bS$ and a unit $\bmu_2=\mathbf e$, and let $\kappa>0$ vary. Then $\bfeta_1=\mathbf 0$, $\bfeta_{12}=\mathbf 0$, $\bfeta_{11}=\bS$, $\bfeta_2=\Am(\kappa)\mathbf e$ and, by Lemma~\ref{lem:vmf-moments}, $\bfeta_{22}-\tfrac1m\bI_m=(1-\tfrac{m\Am(\kappa)}{\kappa})(\mathbf e\mathbf e^{\top}-\tfrac1m\bI_m)$, so
\begin{equation*}
\ell_2(\mathbf e)\,\Am(\kappa)+\ell_{22}(\mathbf e\mathbf e^{\top}-\tfrac1m\bI_m)\bigl(1-m\Am(\kappa)/\kappa\bigr)+\ell_{11}(\bS)+c=0\qquad(\kappa>0).
\end{equation*}
To prove that $1$, $\Am(\kappa)$ and $\Am(\kappa)/\kappa$ are linearly independent on $(0,\infty)$, let $c_0+c_1\Am+c_2\Am/\kappa\equiv0$. The limit $\kappa\to\infty$, where $\Am\to1$, gives $c_0+c_1=0$. The limit $\kappa\downarrow0$, where $\Am(\kappa)/\kappa\to1/m$ by $I_\nu(\kappa)\sim(\kappa/2)^\nu/\Gamma(\nu+1)$ at the Bessel orders $\nu=m/2-1$ and $\nu=m/2$ \cite{mardiajupp2000directional}, gives $c_0+c_2/m=0$. Substituting $c_1=-c_0$ and $c_2=-mc_0$ and multiplying by $-\kappa$ gives $c_0[(\kappa+m)\Am(\kappa)-\kappa]\equiv0$. If $c_0\ne0$, then $\Am(\kappa)=\kappa/(\kappa+m)$ for all $\kappa>0$, which is impossible: its derivative $m/(\kappa+m)^2$ differs from the value $[(m+1)\kappa+m]/(\kappa+m)^2$ that the Riccati identity assigns to it. So $c_0=c_1=c_2=0$. The display is such a combination, with $c_1=\ell_2(\mathbf e)$, $c_2=-m\,\ell_{22}(\mathbf e\mathbf e^{\top}-\tfrac1m\bI_m)$ and $c_0=\ell_{22}(\mathbf e\mathbf e^{\top}-\tfrac1m\bI_m)+\ell_{11}(\bS)+c$. Hence $\ell_2(\mathbf e)=0$, $\ell_{22}(\mathbf e\mathbf e^{\top}-\tfrac1m\bI_m)=0$ and $\ell_{11}(\bS)+c=0$ for every unit $\mathbf e$. The unit vectors span $\R^m$, the matrices $\mathbf e\mathbf e^{\top}$ span the symmetric matrices, and the traceless-part map is onto $\Symzero{m}$, so $\ell_2=0$ and $\ell_{22}=0$.

\emph{Step 2.} Varying $\bS$ over the open cone $\Sympp{d}$ at $\balpha=\mathbf 0$, $\bB=\mathbf 0$ gives $\ell_{11}=0$ and $c=0$.

\emph{Step 3.} At $\bB=\mathbf 0$ with $\balpha$ varying, $\bfeta_1=\balpha$ and $\bfeta_{12}=a\,\balpha\mathbf e^{\top}$, with $a=\Am(\kappa)$, so $\ell_1(\balpha)+a\,\ell_{12}(\balpha\mathbf e^{\top})=0$. The same identity at a second concentration, with $a'\ne a$ because $\Am$ is strictly increasing, gives $\ell_{12}(\balpha\mathbf e^{\top})=0=\ell_1(\balpha)$. The rank-one matrices $\balpha\mathbf e^{\top}$ span $\R^{d\times m}$, so $\ell_{12}=0$ and $\ell_1=0$.

Thus $L=0$: the $\bfeta$-image of $\Ecyl$ lies in no proper affine subspace of the expectation space. That space has dimension larger than $\dim\Ecyl$, so $\Ecyl$ is not m-flat.

\paragraph{Extension to \texorpdfstring{$\Pext$}{the extended parameter space} (Section~3.2).}
With $\kappa\bmu_2$ replaced by $\bvartheta_2$, the map $\btheta_{\cyl}:\Pext\to\Hnat_{\mathrm{amb}}$ of (9) is smooth. The proof of Proposition~3(c) below, with the condition $\bvartheta_2\neq\mathbf 0$ dropped, shows that $\btheta_{\cyl}$ is a bijection of $\Pext$ onto $\{\bOmega=\mathbf 0\}$, with the smooth inverse $\bS=-\tfrac12\btheta_{11}^{-1}$, $\bB=\bS\btheta_{12}$, $\balpha=\bS\btheta_1$, $\bvartheta_2=\bvartheta_2(\btheta)$. The set $\{\bOmega=\mathbf 0\}$ is closed in $\Hnat_{\mathrm{amb}}$, as the zero set of a continuous map, and is the graph of a smooth map over the open set $U$, hence an embedded submanifold; so $\btheta_{\cyl}$ is a diffeomorphism onto it, and $\Ecyl=\btheta_{\cyl}(\calP)$ is its open dense subset $\{\bvartheta_2\neq\mathbf 0\}$. By Proposition~3(b), $p_{\btheta_{\cyl}(\cdot)}=\fcyl(\cdot;\balpha,\bB,\bS,\bvartheta_2)$ and $\psi_{\mathrm{amb}}\circ\btheta_{\cyl}$ is (10) with $-\log\Cm(\kappa)$ replaced by $\psi(\bvartheta_2)$, which is smooth on $\R^m$ because the von Mises--Fisher family is regular with natural parameter space $\R^m$. The expectation coordinates are (12) with $\bfeta_2=\E[\bx_2]$ and $\bfeta_{22}=\E[\bx_2\bx_2^{\top}]$ of Lemma~\ref{lem:vmf-moments}; at $\bvartheta_2=\mathbf 0$ the direction is uniform, so $\bfeta_2=\mathbf 0$ and, by rotation invariance and $\tr\E[\bx_2\bx_2^{\top}]=1$, $\bfeta_{22}=\bI_m/m$, the limit of Lemma~\ref{lem:vmf-moments} as $\kappa\downarrow0$, because $\Am(\kappa)/\kappa\to1/m$.

\paragraph{Existence of \texorpdfstring{$p_{\bar{\bt}}$}{p\_t-bar} (Proposition~8, information projection, Section~3.3, part (i)).}
Part (i) needs an ambient member $p_{\bar{\bt}}$ whose expected statistic is the sample average $\bar{\bt}$. The idea: if the directional part of $\bar{\bt}$ were a boundary point of the Fisher--Bingham mean space, every observed direction would maximize one function $\mathbf a_2^{\top}\bx_2+\bx_2^{\top}\mathbf A_{22}\bx_2$ that is not constant on the sphere, and so lie in a proper affine subspace, which (A1) excludes.

\emph{Interiority of the directional statistic.}
The Fisher--Bingham family on $\sphere^{m-1}$, with statistic $\bt_2$, is regular and minimal (paragraph on minimality above). Let $\bar{\bt}_2:=N^{-1}\sum_n\bt_2(\bx_{2n})=(\bar{\bx}_2,\bar{\mathbf M}_{22}-\tfrac1m\bI_m)$. The mean space $\calM_{\FB}$ is the interior of the convex hull $C$ of $\bt_2(\sphere^{m-1})$, which equals the interior of $\bar C$ \cite{barndorffnielsen1978information}. We show $\bar{\bt}_2\in\calM_{\FB}$.

Suppose not. Being an average of points of $\bt_2(\sphere^{m-1})$, $\bar{\bt}_2$ lies in $C$, hence on the boundary of $\bar C$, and a supporting hyperplane exists \cite{barndorffnielsen1978information}: some $\boldsymbol\lambda=(\mathbf a_2,\mathbf A_{22})\ne\mathbf 0$ with $\mathbf A_{22}\in\Symzero{m}$ has $\langle\boldsymbol\lambda,\bt_2(\bx_2)\rangle\le\langle\boldsymbol\lambda,\bar{\bt}_2\rangle$ on the sphere. Since $\mathbf A_{22}$ is traceless, $\langle\boldsymbol\lambda,\bt_2(\bx_2)\rangle=\tilde q(\bx_2):=\mathbf a_2^{\top}\bx_2+\bx_2^{\top}\mathbf A_{22}\bx_2$ and $\langle\boldsymbol\lambda,\bar{\bt}_2\rangle=N^{-1}\sum_n\tilde q(\bx_{2n})$. The $N$ nonnegative differences between this average and the $\tilde q(\bx_{2n})$ therefore sum to zero, so every $\bx_{2n}$ maximizes $\tilde q$ on the sphere.

Let $\bx^{*}$ be one maximizer. The Lagrange condition on the sphere gives $\mathbf a_2=2(c^{*}\bI_m-\mathbf A_{22})\bx^{*}$ for some $c^{*}\in\R$. Substituting this into $\tilde q(\bx^{*})-\tilde q(\mathbf y)$ and using $\norm{\bx^{*}-\mathbf y}^2=2-2\bx^{*\top}\mathbf y$ for unit $\mathbf y$,
\begin{equation*}
\tilde q(\bx^{*})-\tilde q(\mathbf y)=(\bx^{*}-\mathbf y)^{\top}(c^{*}\bI_m-\mathbf A_{22})(\bx^{*}-\mathbf y)\ \ge0\qquad(\norm{\mathbf y}=1),
\end{equation*}
where the sign holds because $\bx^{*}$ is a maximizer. For a unit $\mathbf v$ with $\mathbf v^{\top}\bx^{*}>0$, the reflection $\mathbf y:=\bx^{*}-2(\mathbf v^{\top}\bx^{*})\mathbf v$ is a unit vector with $\bx^{*}-\mathbf y\propto\mathbf v$, so $\mathbf v^{\top}(c^{*}\bI_m-\mathbf A_{22})\mathbf v\ge0$. By symmetry in $\pm\mathbf v$ and continuity, $c^{*}\bI_m-\mathbf A_{22}\succeq\mathbf 0$. At every other maximizer $\mathbf y=\bx_{2n}$ the left side of the display is zero, and a positive semidefinite form vanishes only on the kernel of its matrix. So all $\bx_{2n}$ lie in the affine subspace $\bx^{*}+\ker(c^{*}\bI_m-\mathbf A_{22})$. By (A1) their affine hull is $\R^m$ (Section~\ref{rem:cyl-conditions}), so the kernel is $\R^m$ and $\mathbf A_{22}=c^{*}\bI_m$. This matrix is traceless only for $c^{*}=0$; then $\mathbf A_{22}=\mathbf 0$ and $\mathbf a_2=\mathbf 0$, so $\boldsymbol\lambda=(\mathbf a_2,\mathbf A_{22})=\mathbf 0$, a contradiction.

Hence $\bar{\bt}_2\in\calM_{\FB}$: a unique $\FB_m(\bvartheta^{*},\bOmega^{*})$ has $\E[\bx_2]=\bar{\bx}_2$ and $\E[\bx_2\bx_2^{\top}]=\bar{\mathbf M}_{22}$.

\emph{Construction of $p_{\bar{\bt}}$.} Since $\hat{\bS}\succ\mathbf 0$ by (A2), the factorization of Proposition~3(b) and the constraint (11) show that $q:=\fCG(\cdot\mid\bx_2;\hat{\balpha},\hat{\bB},\hat{\bS})\,\FB_m(\bvartheta^{*},\bOmega^{*})$ is the member $p_{\btheta^{*}}$ of $\Eamb$ with $\btheta^{*}_{11}=-\tfrac12\hat{\bS}^{-1}$, $\btheta^{*}_{12}=\hat{\bS}^{-1}\hat{\bB}$, $\btheta^{*}_1=\hat{\bS}^{-1}\hat{\balpha}$, $\btheta^{*}_2=\bvartheta^{*}-\hat{\bB}^{\top}\hat{\bS}^{-1}\hat{\balpha}$ and $\btheta^{*}_{22}=\bOmega^{*}-\tfrac12(\hat{\bB}^{\top}\hat{\bS}^{-1}\hat{\bB})^{\circ}$. We check that $q$ matches every block of $\bar{\bt}$, with $\bz_n$, $\hat{\bbeta}$, $\mathbf M$ and $\hat{\beps}_n$ as in the proof of Theorem~6. Under $q$ the directional moments match, so $\E_q[\bz\bz^{\top}]=\mathbf M/N=:\bar{\mathbf M}_z$, with blocks $1$, $\bar{\bx}_2^{\top}$, $\bar{\bx}_2$, $\bar{\mathbf M}_{22}$. Conditioning on $\bx_2$, $\E_q[\bx_1\bz^{\top}]=\hat{\bbeta}\bar{\mathbf M}_z=N^{-1}\sum_n\bx_{1n}\bz_n^{\top}$ by the normal equations, which matches $(\bar{\bx}_1,\bar{\bt}_{12})$. Likewise $\E_q[\bx_1\bx_1^{\top}]=\hat{\bS}+\hat{\bbeta}\bar{\mathbf M}_z\hat{\bbeta}^{\top}=\bar{\bt}_{11}$ by the residual identity $\sum_n\hat{\beps}_n\hat{\beps}_n^{\top}=\sum_n\bx_{1n}\bx_{1n}^{\top}-\hat{\bbeta}\mathbf M\hat{\bbeta}^{\top}$. Thus $\bar{\bt}=\nabla\psi_{\mathrm{amb}}(\btheta^{*})\in\calM_{\mathrm{amb}}$, and $p_{\bar{\bt}}=q$ because $\nabla\psi_{\mathrm{amb}}$ is a bijection (Section~2).

\emph{Converse.} Conversely, $\bar{\bt}\in\calM_{\mathrm{amb}}$ only if (A1) and (A2) hold, so $\bar{\bt}\in\calM_{\mathrm{amb}}$ if and only if (A1)--(A2), as Proposition~8(i) of the main text states; the argument is the same for a weighted average over the observations with positive weight, with (A1w)--(A2w). An interior point of the convex hull of $\bt(\R^d\times\sphere^{m-1})$ lies on no supporting hyperplane, so it suffices to find, when (A1) or (A2) fails, a nonzero $\boldsymbol\lambda$ with $\langle\boldsymbol\lambda,\bt(\bx)\rangle\le\langle\boldsymbol\lambda,\bar{\bt}\rangle$ for every $\bx$. On the sphere, $\bx_2^{\top}\mathbf A\bx_2=\langle\mathbf A^{\circ},\bx_2\bx_2^{\top}-\tfrac1m\bI_m\rangle+\tfrac1m\tr\mathbf A$ for symmetric $\mathbf A$, so every quadratic polynomial in $(\bx_1,\bx_2)$ is $\langle\boldsymbol\lambda,\bt(\bx)\rangle$ plus a constant. If (A1) fails, all $\bx_{2n}$ satisfy $\mathbf v^{\top}\bx_{2n}=c$ for a unit $\mathbf v$; take $q_1(\bx):=-(\mathbf v^{\top}\bx_2-c)^2=\langle\boldsymbol\lambda,\bt(\bx)\rangle+\mathrm{const}$, with $\boldsymbol\lambda=(\mathbf 0,2c\mathbf v,\mathbf 0,\mathbf 0,-(\mathbf v\mathbf v^{\top})^{\circ})\neq\mathbf 0$, because $(\mathbf v\mathbf v^{\top})^{\circ}\neq\mathbf 0$ for $m\ge2$. If (A2) fails, a unit $\mathbf a\in\R^d$ and $\boldsymbol\gamma=(\gamma_0,\boldsymbol\gamma_2)\in\R\times\R^m$ satisfy $\mathbf a^{\top}\bx_{1n}=\gamma_0+\boldsymbol\gamma_2^{\top}\bx_{2n}$ for all $n$ (step~(ii) of the proof of Theorem~6, Section~\ref{sec:esm-cyl}); take $q_2(\bx):=-(\mathbf a^{\top}\bx_1-\gamma_0-\boldsymbol\gamma_2^{\top}\bx_2)^2$, whose $\bx_1\bx_1^{\top}$-coefficient $-\mathbf a\mathbf a^{\top}$ makes $\boldsymbol\lambda\neq\mathbf 0$. In both cases $q_i\le0$ everywhere and $q_i(\bx_n)=0$ for every $n$, so $\langle\boldsymbol\lambda,\bt(\bx)\rangle\le\langle\boldsymbol\lambda,\bt(\bx_n)\rangle=\langle\boldsymbol\lambda,\bar{\bt}\rangle$ for all $\bx$, and $\bar{\bt}\notin\calM_{\mathrm{amb}}$.

Section~\ref{subsec:esm-moved-proofs} now proves the results, citing these steps.

\subsection{Proofs of results of Sections~2--4 of the main text}\label{subsec:esm-moved-proofs}

The main text uses two classical lemmas, whose proofs follow below~\cite[Chapters~3 and~8]{anderson2003multivariate}.

\begin{lemma}[Loewner-order optimality of multivariate least squares]\label{lem:ols-loewner}
For $\bz_n\in\R^p$, $\mathbf y_n\in\R^d$, $\mathbf M:=\sum_n\bz_n\bz_n^{\top}$ and $E(\bbeta):=\sum_n(\mathbf y_n-\bbeta\bz_n)(\mathbf y_n-\bbeta\bz_n)^{\top}$, the normal equations $\bbeta\mathbf M=\sum_n\mathbf y_n\bz_n^{\top}$ have a solution $\hat{\bbeta}$, and $E(\bbeta)=E(\hat{\bbeta})+(\bbeta-\hat{\bbeta})\mathbf M(\bbeta-\hat{\bbeta})^{\top}$. Hence $E(\bbeta)\succeq E(\hat{\bbeta})$ and $\tr(\bS^{-1}E(\bbeta))\ge\tr(\bS^{-1}E(\hat{\bbeta}))$ for $\bS\succ\mathbf 0$, with equality if and only if $\bbeta$ solves the normal equations; $E(\hat{\bbeta})$ does not depend on the solution, which is unique, $(\sum_n\mathbf y_n\bz_n^{\top})\mathbf M^{-1}$, if $\mathbf M\succ\mathbf 0$.
\end{lemma}

\begin{lemma}[Maximizer of the Gaussian covariance profile]\label{lem:cov-profile}
Let $\mathbf C\in\Sympp{d}$ and $N>0$. The function $G(\bS):=-\tfrac N2\log|\bS|-\tfrac12\tr(\bS^{-1}\mathbf C)$ on $\Sympp{d}$ has the unique maximizer $\bS=\mathbf C/N$, with $G(\mathbf C/N)=-\tfrac N2\log|\mathbf C/N|-\tfrac{Nd}2$ and $G(\bS)-G(\mathbf C/N)=-N\,\KL[\calN(\mathbf 0,\mathbf C/N):\calN(\mathbf 0,\bS)]$.
\end{lemma}

This subsection proves Lemma~\ref{lem:vmf-moments} (von Mises--Fisher moments, Section~\ref{sec:esm-background}) and, of the main text, Proposition~9 (Fisher metric), Proposition~2 (Gaussian kernel restricted to the cylinder), Proposition~3(b)--(d) (the curved family), Lemma~\ref{lem:ols-loewner} (least squares), the divergence identity of Lemma~\ref{lem:cov-profile} (covariance profile), both stated below, Proposition~10 (divergence between components), Corollary~11 (likelihood gain of $\bB$), Propositions~7 (identifiability of one component) and~8 (m-projection) and the lemmas used in the proof of Theorem~12 (identifiability of finite mixtures). Theorems~6 and~12 are proved in Appendix~A of the main text, and Lemma~\ref{lem:cyl-mle-weighted} in Section~\ref{sec:esm-mix}. The opening of Section~\ref{sec:esm-proofs} says where the other results of Sections~2--4 are proved.

\begin{proof}[Proof of Proposition~7 of the main text]
The von Mises--Fisher family is regular and minimal, with natural parameter space $\R^m$ and mean space the open unit ball, so $\nabla\psi$ is a smooth bijection between them with inverse $\nabla\phi$; its Jacobian $\nabla^2\psi=\mathbf V$ is positive definite, as the sphere lies in no affine hyperplane, so $\nabla\phi$ is smooth, also at $\mathbf 0$, where $\mathbf V=\bI_m/m$. By (6), $\E[\bx_1]=\balpha+\bB\bfeta_2$, $\mathbf K_{12}=\bB\mathbf V$ and $\mathbf K_{11}=\bS+\bB\mathbf V\bB^{\top}$; equal densities have equal moments.
\end{proof}

\begin{proof}[Proof of Proposition~8 of the main text]
\emph{(i), only if.} An interior point of the convex hull of $\bt(\R^d\times\sphere^{m-1})$ lies on no supporting hyperplane, and on the sphere $\bx_2^{\top}\mathbf A\bx_2=\langle\mathbf A^{\circ},\bx_2\bx_2^{\top}-\tfrac1m\bI_m\rangle+\tfrac1m\tr\mathbf A$, so every quadratic polynomial in $\bx$ is $\langle\boldsymbol\lambda,\bt(\bx)\rangle$ plus a constant. If (A1) fails, $\mathbf v^{\top}\bx_{2n}=c$ for all $n$ and a unit $\mathbf v$, and $q(\bx):=-(\mathbf v^{\top}\bx_2-c)^2$ has $\boldsymbol\lambda=(\mathbf 0,2c\mathbf v,\mathbf 0,\mathbf 0,-(\mathbf v\mathbf v^{\top})^{\circ})\neq\mathbf 0$ ($m\ge2$). If (A2) fails, $\mathbf a^{\top}\bx_{1n}=\gamma_0+\boldsymbol\gamma_2^{\top}\bx_{2n}$ for all $n$, a unit $\mathbf a$ and some $(\gamma_0,\boldsymbol\gamma_2)$, and $q(\bx):=-(\mathbf a^{\top}\bx_1-\gamma_0-\boldsymbol\gamma_2^{\top}\bx_2)^2$ has $\bx_1\bx_1^{\top}$-block $-\mathbf a\mathbf a^{\top}\neq\mathbf 0$. In both cases $q\le0=q(\bx_n)$, so $\langle\boldsymbol\lambda,\bt(\bx)\rangle\le\langle\boldsymbol\lambda,\bar{\bt}\rangle$ for every $\bx$, and $\bar{\bt}\notin\calM_{\mathrm{amb}}$.

\emph{(i), if.} Let (A1)--(A2) hold. The Fisher--Bingham family with statistic $(\bx_2,\bx_2\bx_2^{\top}-\tfrac1m\bI_m)$ is regular and minimal. If its sample average $(\bar{\bx}_2,\bar{\mathbf M}_{22}-\tfrac1m\bI_m)$ were not in its mean space, a nonzero $(\mathbf a_2,\mathbf A_{22})$, $\mathbf A_{22}$ traceless, would make every $\bx_{2n}$ a maximizer of $\tilde q(\mathbf u):=\mathbf a_2^{\top}\mathbf u+\mathbf u^{\top}\mathbf A_{22}\mathbf u$ on the sphere. At the maximizer $\bx^{*}:=\bx_{21}$, $\mathbf a_2=2(c^{*}\bI_m-\mathbf A_{22})\bx^{*}$ for a Lagrange multiplier $c^{*}$, so $\tilde q(\bx^{*})-\tilde q(\mathbf u)=(\bx^{*}-\mathbf u)^{\top}(c^{*}\bI_m-\mathbf A_{22})(\bx^{*}-\mathbf u)\ge0$ for unit $\mathbf u$; the reflections $\mathbf u=\bx^{*}-2(\mathbf e^{\top}\bx^{*})\mathbf e$ of $\bx^{*}$, $\norm{\mathbf e}=1$, give $c^{*}\bI_m-\mathbf A_{22}\succeq\mathbf 0$, and every $\bx_{2n}-\bx^{*}$ lies in its kernel. By (A1) these vectors span $\R^m$, so the kernel is $\R^m$, so $\mathbf A_{22}=c^{*}\bI_m$ is traceless only if $c^{*}=0$, and then $(\mathbf a_2,\mathbf A_{22})=\mathbf 0$, a contradiction. So a unique $\FB_m(\bvartheta^{*},\bOmega^{*})$ has $\E[\bx_2]=\bar{\bx}_2$ and $\E[\bx_2\bx_2^{\top}]=\bar{\mathbf M}_{22}$. By (A2), $\hat{\bS}\succ\mathbf 0$, and by Proposition~3(b), $q:=\fCG(\cdot\mid\bx_2;\hat{\balpha},\hat{\bB},\hat{\bS})\,\FB_m(\bvartheta^{*},\bOmega^{*})$ is a member of $\Eamb$. Under $q$, $\E[\bz\bz^{\top}]=\bar{\mathbf M}_z:=N^{-1}\sum_n\bz_n\bz_n^{\top}$, so $\E_q[\bx_1\bz^{\top}]=\hat{\bbeta}\bar{\mathbf M}_z=N^{-1}\sum_n\bx_{1n}\bz_n^{\top}$ by the normal equations, and $\E_q[\bx_1\bx_1^{\top}]=\hat{\bS}+\hat{\bbeta}\bar{\mathbf M}_z\hat{\bbeta}^{\top}=\bar{\bt}_{11}$ by the residual identity. So $\E_q[\bt]=\bar{\bt}\in\calM_{\mathrm{amb}}$ and $p_{\bar{\bt}}=q$, as $\nabla\psi_{\mathrm{amb}}$ is injective.

\emph{(i), projection.} $\E_{p_{\bar{\bt}}}[\log p_{\btheta}]=\langle\btheta,\bar{\bt}\rangle-\psi_{\mathrm{amb}}(\btheta)$, and the Fenchel--Young equality gives $\E_{p_{\bar{\bt}}}[\log p_{\bar{\bt}}]=\phi_{\mathrm{amb}}(\bar{\bt})$; subtracting gives the divergence. On $\Ecyl$, $\psi_{\mathrm{amb}}=\psi_{\cyl}$, so averaging $\log p_{\btheta}(\bx_n)=\langle\bt(\bx_n),\btheta\rangle-\psi_{\mathrm{amb}}(\btheta)$ gives $\ell(\btheta)/N$, and $\KL[p_{\bar{\bt}}:p_{\btheta}]=\phi_{\mathrm{amb}}(\bar{\bt})-\ell(\btheta)/N$ there. Minimizing over $\Ecyl$ is maximizing $\ell$ over $\calP$, whose unique solution under (A1)--(A3) is $\hat{\btheta}$ (Theorem~6). Both $p_{\bar{\bt}}$ and $p_{\hat{\btheta}}$ have the fitted conditional block, and their directional marginals share the mean $\bar{\bx}_2$.

\emph{(ii).} By (12) and $\hat{\balpha}=\bar{\bx}_1-\hat{\bB}\bar{\bx}_2$, $\hat{\bfeta}_2=\bar{\bx}_2$ and $\hat{\bfeta}_1=\bar{\bx}_1$. The normal equations and the residual identity give $\bar{\bt}_{12}=\hat{\balpha}\bar{\bx}_2^{\top}+\hat{\bB}\bar{\mathbf M}_{22}$ and $\bar{\bt}_{11}=\hat{\bS}+\hat{\bbeta}\bar{\mathbf M}_z\hat{\bbeta}^{\top}$, and $\hat{\bfeta}_{12}$, $\hat{\bfeta}_{11}$ of (12) are the same expressions with $\bar{\mathbf M}_{22}$ replaced by $\hat{\bfeta}_{22}$, which gives (14); $\bDelta$ is traceless as both matrices have unit trace. The ambient score is $N(\bar{\bt}-\nabla\psi_{\mathrm{amb}}(\hat{\btheta}))=N(\bar{\bt}-\hat{\bfeta})$, with the stated blocks. On $\Ecyl$ the ambient log-likelihood is $\ell$ composed with the inverse of the smooth bijection (9), which $\hat{\btheta}$ maximizes, so the score annihilates every tangent vector of $\Ecyl$ at $\hat{\btheta}$.
\end{proof}

\begin{proof}[Proof of Lemma~\ref{lem:vmf-moments}]
The von Mises--Fisher family is a regular exponential family with the radial log-partition function $\psi(\btheta)=-\log\Cm(\norm{\btheta})$, so its mean $\E[\bx]=\nabla\psi$ and covariance $\cov(\bx)=\nabla^2\psi$ follow by differentiating at $\btheta=\kappa\bmu$ with (3) and (4). The lemma is classical \cite[Chapter~9]{mardiajupp2000directional}, and (4) follows from the Bessel recurrences \cite{schou1978concentration}. The paragraph on this lemma earlier in Section~\ref{sec:esm-proofs} writes out the computation.
\end{proof}

\begin{proof}[Proof of Proposition~2]
The paragraph on this proposition earlier in Section~\ref{sec:esm-proofs} writes out both computations. In its step (i), completing the square in $\bx_1$ writes $K$ as a Gaussian kernel in $\bx_1$, whose integral over $\R^d$ does not depend on $\bx_2$, times a function of $\bx_2$ that is continuous on the compact sphere, so $K$ is $\nu$-integrable by Tonelli's theorem, and the normalized restriction is a conditional Gaussian times a Fisher--Bingham law. In its step (ii), the block-inverse formulas for $\bLambda=\bSigma^{-1}$ \cite{magnusneudecker2019matrix} give $\mathbf Q=\bSigma_{22}^{-1}=\kappa\bI_m$ and the other two identities; as $(-\tfrac\kappa2\bI_m)^{\circ}=\mathbf 0$ and $\norm{\mathbf c_2}=1$, the directional factor is $\vMF_m(\mathbf c_2,\kappa)$, and (8) identifies the product with $\fcyl$. Conversely, given $\fcyl$ with parameter in $\calP$, put $\mathbf c:=(\bmu_1,\bmu_2)$, $\bSigma_{12}:=\bB/\kappa$, $\bSigma_{11}:=\bS+\kappa\bSigma_{12}\bSigma_{12}^{\top}$ and $\bSigma_{22}:=\kappa^{-1}\bI_m$. Then $\bSigma_{22}\succ\mathbf 0$ and the Schur complement $\bSigma_{11}-\bSigma_{12}\bSigma_{22}^{-1}\bSigma_{12}^{\top}=\bS$ is positive definite, so $\bSigma\succ\mathbf 0$, and the identities above return $\fcyl$.
\end{proof}

\begin{proof}[Proof of Proposition~3(b)--(d)]
\emph{(b)} For fixed $\bx_2$, the exponent $\langle\bt,\btheta\rangle$ is quadratic in $\bx_1$, so its exponential is integrable in $\bx_1$ exactly when $\btheta_{11}\prec\mathbf 0$; the compact sphere adds no condition, and completing the square gives the factorization. If $\btheta_{11}$ has an eigenvalue $\lambda\ge0$ with eigenvector $\mathbf v$, the exponent along each line $\bx_1^{0}+s\mathbf v$ is $\lambda s^2+bs+c$, whose exponential is not integrable in $s$; by Tonelli, $\psi_{\mathrm{amb}}(\btheta)=\infty$. If $\btheta_{11}\prec\mathbf 0$, put $\mathbf c(\bx_2):=-\tfrac12\btheta_{11}^{-1}(\btheta_1+\btheta_{12}\bx_2)$; completing the square writes the $\bx_1$-part of the exponent as
\begin{equation}\label{eq:esm-x1-square}
(\bx_1-\mathbf c(\bx_2))^{\top}\btheta_{11}(\bx_1-\mathbf c(\bx_2))-\tfrac14(\btheta_1+\btheta_{12}\bx_2)^{\top}\btheta_{11}^{-1}(\btheta_1+\btheta_{12}\bx_2).
\end{equation}
The first term integrates over $\R^d$ to $\pi^{d/2}|{-\btheta_{11}}|^{-1/2}$. The second depends on $\bx_2$ only. Expanding it and adding $\btheta_2^{\top}\bx_2+\bx_2^{\top}\btheta_{22}\bx_2$ leaves $-\tfrac14\btheta_1^{\top}\btheta_{11}^{-1}\btheta_1+\bvartheta_2^{\top}\bx_2+\bx_2^{\top}(\btheta_{22}-\tfrac14\btheta_{12}^{\top}\btheta_{11}^{-1}\btheta_{12})\bx_2$, with $\bvartheta_2$ as in (11). On the sphere $\bx_2^{\top}\mathbf A\bx_2=\bx_2^{\top}\mathbf A^{\circ}\bx_2+\tfrac1m\tr\mathbf A$ and $\tr\btheta_{22}=0$, which give $\bOmega$ and the trace term below. Because the sphere is compact,
\begin{equation}\label{eq:esm-psi-amb}
\begin{split}
\psi_{\mathrm{amb}}(\btheta)&=\tfrac d2\log\pi-\tfrac12\log|{-\btheta_{11}}|-\tfrac14\btheta_1^{\top}\btheta_{11}^{-1}\btheta_1-\tfrac1{4m}\tr\bigl(\btheta_{12}^{\top}\btheta_{11}^{-1}\btheta_{12}\bigr)\\
&\quad+\log\int_{\sphere^{m-1}}\exp\bigl(\bvartheta_2^{\top}\bx_2+\bx_2^{\top}\bOmega\bx_2\bigr)\,d\sigma(\bx_2)
\end{split}
\end{equation}
is finite exactly when $\btheta_{11}\prec\mathbf 0$. Hence $\Hnat_{\mathrm{amb}}$ is the stated open set and $\Eamb$ is regular. The last term is the Fisher--Bingham normalizer, and the two integrals normalize the two factors of $\exp\langle\bt,\btheta\rangle$ separately. When $\bOmega=\mathbf 0$ the integral is $1/\Cm(\norm{\bvartheta_2})$, and substituting (9) ($\btheta_{11}^{-1}=-2\bS$, $|{-\btheta_{11}}|=2^{-d}|\bS|^{-1}$) turns \eqref{eq:esm-psi-amb} into (10).

\emph{Minimality.} The paragraph on minimality earlier in Section~\ref{sec:esm-proofs} shows that $\langle\mathbf a,\bt\rangle$ is $\nu$-a.e.\ constant only for $\mathbf a=\mathbf 0$, and that the Fisher--Bingham family, with the statistic $\bt_2$ defined there, is minimal and regular.

\emph{(c)} Part (a) maps $\calP$ into $\{\bOmega=\mathbf 0,\bvartheta_2\neq\mathbf 0\}$. Conversely, take such a $\btheta$ and put $\bS:=-\tfrac12\btheta_{11}^{-1}\succ\mathbf 0$, $\bB:=\bS\btheta_{12}$, $\balpha:=\bS\btheta_1$, $\kappa:=\norm{\bvartheta_2}$, $\bmu_2:=\bvartheta_2/\kappa$. This point of $\calP$ is mapped to $\btheta$ by (9): $\btheta_2=\kappa\bmu_2-\bB^{\top}\bS^{-1}\balpha$ by the definition of $\bvartheta_2$, and $\btheta_{22}=(-\tfrac12\bB^{\top}\bS^{-1}\bB)^{\circ}$ by $\bOmega=\mathbf 0$ and $\tr\btheta_{22}=0$. So (9) is a bijection of $\calP$ onto $\Ecyl$, $p_{\btheta}=\fcyl$ by (b), and $\bB=\mathbf 0$ if and only if $\btheta_{12}=\mathbf 0$. Counting the entries of $(\balpha,\bB,\bS,\bmu_2,\kappa)$ and of $\btheta_{22}$ gives the three dimensions. For the submanifold structure, the constraint $\bOmega=\mathbf 0$ fixes $\btheta_{22}$ as a smooth function of the other four blocks: $\{\bOmega=\mathbf 0\}$ is the graph of the map $(\tfrac14\btheta_{12}^{\top}\btheta_{11}^{-1}\btheta_{12})^{\circ}$ over the open set $U:=\R^d\times\R^m\times\Symmm{d}\times\R^{d\times m}$. Such a graph is an embedded submanifold of $\Hnat_{\mathrm{amb}}=U\times\Symzero{m}$, of codimension $\dim\Symzero{m}=\tfrac{m(m+1)}2-1$, and so is its intersection $\Ecyl$ with the open set $\{\bvartheta_2\ne\mathbf 0\}$.

\emph{Flatness.} With $\btheta_{12}=\mathbf 0$, (11) gives $\bOmega=\btheta_{22}$ and $\bvartheta_2=\btheta_2$. So $\Eprod$ is the intersection of $\Hnat_{\mathrm{amb}}$ with the linear subspace $\{\btheta_{12}=\mathbf 0,\btheta_{22}=\mathbf 0\}$, minus only the uniform directional law $\btheta_2=\mathbf 0$ that Definition~1 excludes through $\kappa>0$. Removing this closed set leaves an open subset of the intersection, so $\Eprod$ is e-flat. An e-flat (or m-flat) $\Ecyl$ would lie in a proper affine subspace of the natural (or expectation) coordinates, because both coordinate spaces have dimension larger than $\dim\Ecyl$. The paragraphs on non-flatness earlier in Section~\ref{sec:esm-proofs} show that no nonzero affine functional vanishes on $\Ecyl$, or on its image (12) in the expectation coordinates. So $\Ecyl$ is neither e-flat nor m-flat.

\emph{Constraint.} By (b) the directional marginal of $p_{\btheta}$ is $\FB_m(\bvartheta_2,\bOmega)$. The von Mises--Fisher laws are its members with $\bOmega=\mathbf 0$, and by minimality no member with $\bOmega\ne\mathbf 0$ is one of them.

\emph{(d)} By (6), $\bx_1=\balpha+\bB\bx_2+\beps$ with $\beps\sim\calN(\mathbf 0,\bS)$ independent of $\bx_2$. Lemma~\ref{lem:vmf-moments} gives $\bfeta_2$ and $\bfeta_{22}$, and $\E[\beps]=\mathbf 0$, $\E[\beps\beps^{\top}]=\bS$ give $\bfeta_1$, $\bfeta_{12}$ and $\bfeta_{11}$. They are expectation coordinates because $\nabla\psi_{\mathrm{amb}}=\E_{\btheta}[\bt]$ and $p_{\btheta}=\fcyl$ on $\Ecyl$.
\end{proof}

\begin{proof}[Proof of Lemma~\ref{lem:ols-loewner}]
Let $\mathbf Y$ and $\mathbf Z$ be the matrices with rows $\mathbf y_n^{\top}$ and $\bz_n^{\top}$, so that $\mathbf M=\mathbf Z^{\top}\mathbf Z$ and the normal equations read $\mathbf Z^{\top}\mathbf Z\bbeta^{\top}=\mathbf Z^{\top}\mathbf Y$. The columns of $\mathbf Z^{\top}\mathbf Y$ lie in the range of $\mathbf Z^{\top}$, which is the range of $\mathbf Z^{\top}\mathbf Z$, so a solution $\hat{\bbeta}$ exists. Expand $\mathbf y_n-\bbeta\bz_n=(\mathbf y_n-\hat{\bbeta}\bz_n)-(\bbeta-\hat{\bbeta})\bz_n$; the cross terms vanish because $\sum_n(\mathbf y_n-\hat{\bbeta}\bz_n)\bz_n^{\top}=\sum_n\mathbf y_n\bz_n^{\top}-\hat{\bbeta}\mathbf M=\mathbf 0$. So $E(\bbeta)=E(\hat{\bbeta})+\mathbf D$ with $\mathbf D:=(\bbeta-\hat{\bbeta})\mathbf M(\bbeta-\hat{\bbeta})^{\top}\succeq\mathbf 0$. Since $\mathbf M\succeq\mathbf 0$, $\mathbf D=\mathbf 0$ if and only if $(\bbeta-\hat{\bbeta})\mathbf M^{1/2}=\mathbf 0$, that is, $(\bbeta-\hat{\bbeta})\mathbf M=\mathbf 0$, which says that $\bbeta$ solves the normal equations; and $\tr(\bS^{-1}\mathbf D)=\tr(\bS^{-1/2}\mathbf D\bS^{-1/2})\ge0$, with equality only if $\mathbf D=\mathbf 0$. Two solutions therefore give the same $E(\hat{\bbeta})$. If $\mathbf M\succ\mathbf 0$ the solution is $(\sum_n\mathbf y_n\bz_n^{\top})\mathbf M^{-1}$; if $\mathbf M\mathbf c=\mathbf 0$ for some $\mathbf c\neq\mathbf 0$, every $\hat{\bbeta}+\mathbf b\mathbf c^{\top}$ is another one.
\end{proof}

\begin{proof}[Proof of the divergence identity of Lemma~\ref{lem:cov-profile}]
The idea is that the gap $G(\bS)-G(\mathbf C/N)$ is $-N$ times a Kullback--Leibler divergence between two centred Gaussians. Taking the expectation of $\log\fG(\bx;\mathbf 0,\bSigma_0)-\log\fG(\bx;\mathbf 0,\bS)$ under $\calN(\mathbf 0,\bSigma_0)$, with $\E[\bx^{\top}\mathbf A\bx]=\tr(\mathbf A\bSigma_0)$, gives
\begin{equation}\label{eq:esm-gauss-kl}
\KL[\calN(\mathbf 0,\bSigma_0):\calN(\mathbf 0,\bS)]=\tfrac12\bigl[\tr(\bS^{-1}\bSigma_0)-d+\log|\bS|-\log|\bSigma_0|\bigr].
\end{equation}
With $\bSigma_0=\mathbf C/N$, $-N$ times \eqref{eq:esm-gauss-kl} is $G(\bS)-G(\mathbf C/N)$. The divergence is nonnegative and vanishes only when the two Gaussians coincide, so $\mathbf C/N$ is the unique maximizer. With means $\mathbf m_0$ and $\mathbf m_1$, the same computation gives
\begin{equation}\label{eq:esm-gauss-kl-mean}
\begin{split}
\KL[\calN(\mathbf m_0,\bSigma_0):\calN(\mathbf m_1,\bSigma_1)]={}&\tfrac12\bigl[\tr(\bSigma_1^{-1}\bSigma_0)-d+\log|\bSigma_1|-\log|\bSigma_0|\\
&+(\mathbf m_0-\mathbf m_1)^{\top}\bSigma_1^{-1}(\mathbf m_0-\mathbf m_1)\bigr],
\end{split}
\end{equation}
used in the proofs of Proposition~10 and Corollary~11.
\end{proof}

\paragraph{Cross-entropy form of the estimator.}\label{subsec:esm-cross-entropy}
Under (A1)--(A3) of Section~3.3 of the main text, the estimator of Theorem~6 uniquely minimizes over $\calP$ the empirical cross-entropy $H_N:=-\ell/N$, which splits by the chain rule into a directional cross-entropy, minimized at the von Mises--Fisher member with expectation coordinate $\hat{\bfeta}_2=\bar{\bx}_2$ (the information projection of Section~2 of the empirical directional distribution), and a conditional one, minimized at the regression satisfying the conditional moment identities
\begin{equation}\label{eq:cond-moments}
\begin{gathered}
\sum_n\bx_{1n}=\sum_n\hat{\bmu}_c(\bx_{2n}),\qquad
\sum_n\bx_{1n}\bx_{2n}^{\top}=\sum_n\hat{\bmu}_c(\bx_{2n})\bx_{2n}^{\top},\\
\sum_n\bx_{1n}\bx_{1n}^{\top}=\sum_n\bigl(\hat{\bS}+\hat{\bmu}_c(\bx_{2n})\hat{\bmu}_c(\bx_{2n})^{\top}\bigr),
\end{gathered}
\end{equation}
$\hat{\bmu}_c(\bx_2)=\hat{\balpha}+\hat{\bB}\bx_2$, the normal equations of Lemma~\ref{lem:ols-loewner} and the residual identity. Here $H_N$ and the information projection in the cross-entropy sense are as in Section~2 of the main text: the minimizer of $\psi(\btheta)-\langle\E_P[\bt],\btheta\rangle$ for the empirical measure $P$.

\begin{proof}
Since $H_N=-\ell/N$, the unique maximizer of Theorem~6 is the unique minimizer of $H_N$. The split $\ell=\ell_{\CG}+\ell_{\vMF}$ over variation-independent blocks separates $H_N$ into the conditional and the directional cross-entropy, minimized separately. The directional term is $\psi(\btheta)-\langle\bar{\bx}_2,\btheta\rangle$ in $\btheta=\kappa\bmu_2$, with $\bar{\bx}_2=\br/N\in\{\norm{\bfeta}<1\}$ (step (iv) of the theorem); by Section~2 its unique minimizer has expectation coordinate $\bar{\bx}_2$.

For the conditional term, fix $\bx_2$. By \eqref{eq:esm-x1-square}, $\fCG(\cdot\mid\bx_2)$ is a regular exponential family in $\boldsymbol\omega:=(\btheta_1,\btheta_{11},\btheta_{12})$ on the convex open set $\{\btheta_{11}\prec\mathbf 0\}$, with statistic $\bt_{\CG}(\bx):=(\bx_1,\bx_1\bx_1^{\top},\bx_1\bx_2^{\top})$ and log-partition function $\psi_{\bx_2}(\boldsymbol\omega)=\tfrac d2\log\pi-\tfrac12\log|{-\btheta_{11}}|-\tfrac14(\btheta_1+\btheta_{12}\bx_2)^{\top}\btheta_{11}^{-1}(\btheta_1+\btheta_{12}\bx_2)$. Hence $\psi_{\bx_2}$ is convex with gradient $\E_{\boldsymbol\omega}[\bt_{\CG}\mid\bx_2]$ \cite{barndorffnielsen1978information,wainwright2008graphical}, and the conditional cross-entropy $N^{-1}\sum_n[\psi_{\bx_{2n}}(\boldsymbol\omega)-\langle\bt_{\CG}(\bx_n),\boldsymbol\omega\rangle]$ is convex and differentiable. For such a function every stationary point is a minimizer. Its unique minimizer, the theorem's estimator carried to $\boldsymbol\omega$ through (9), is therefore its unique stationary point. Stationarity, $\sum_n\E_{\boldsymbol\omega}[\bt_{\CG}\mid\bx_{2n}]=\sum_n\bt_{\CG}(\bx_n)$, is \eqref{eq:cond-moments}, because $\E[\bx_1\mid\bx_2]=\bmu_c(\bx_2)$ and $\E[\bx_1\bx_1^{\top}\mid\bx_2]=\bS+\bmu_c(\bx_2)\bmu_c(\bx_2)^{\top}$. Its first two identities are the normal equations $\sum_n\bx_{1n}\bz_n^{\top}=\hat{\bbeta}\mathbf M$ of step (ii), written blockwise. The third is step (iii), because $\sum_n\hat{\beps}_n\hat{\beps}_n^{\top}=\sum_n\bx_{1n}\bx_{1n}^{\top}-\hat{\bbeta}\mathbf M\hat{\bbeta}^{\top}$ by the normal equations and $\hat{\bbeta}\mathbf M\hat{\bbeta}^{\top}=\sum_n\hat{\bmu}_c(\bx_{2n})\hat{\bmu}_c(\bx_{2n})^{\top}$.
\end{proof}

\paragraph{Proposition~8.} Its proof is given above; the paragraphs on the existence of $p_{\bar{\bt}}$ and on its converse earlier in this section give the supporting-hyperplane argument in more detail.

\begin{proof}[Proof of Proposition~\ref{prop:esm-cyl-fisher}, hence of Proposition~9 of the main text]
The score splits into a conditional and a directional part, which are uncorrelated, so each block is computed separately. Write $\log\fcyl=\log\fCG(\bx_1\mid\bx_2;\bbeta,\bS)+\bvartheta_2^{\top}\bx_2-\psi(\bvartheta_2)$, with scores $\mathbf s_{\CG}$ and $\mathbf s_{\vMF}=\bx_2-\bfeta_2$ in their own parameters, so $\calI=\E[\mathbf s\mathbf s^{\top}]$ for $\mathbf s=(\mathbf s_{\CG},\mathbf s_{\vMF})$. By (6), $\beps:=\bx_1-\bbeta\bz\sim\calN(\mathbf 0,\bS)$ is independent of $\bz$. The family is a smooth curved exponential family (see (iii) below), so the scores have mean zero and finite variance, and the information identity holds.

\emph{Cross block.} $\mathbf s_{\vMF}$ depends on $\bx_2$ only, and $\E[\mathbf s_{\CG}\mid\bx_2]=\mathbf 0$ (score of a density in $\bx_1$), so conditioning on $\bx_2$ first gives $\E[\mathbf s_{\CG}\mathbf s_{\vMF}^{\top}]=\mathbf 0$.

\emph{Regression block.} The $\bbeta$-score is $\bS^{-1}\beps\bz^{\top}$, with $\operatorname{vec}(\bS^{-1}\beps\bz^{\top})=(\bz\otimes\bS^{-1})\beps$ \cite{magnusneudecker2019matrix}. By the mixed-product rule and independence, its second moment is $\E[\bz\bz^{\top}]\otimes\bS^{-1}\bS\bS^{-1}=\mathbf M_z\otimes\bS^{-1}$. As a quadratic form, $\E[(\dot{\bbeta}\bz)^{\top}\bS^{-1}(\dot{\bbeta}\bz)]$ with $\dot{\bbeta}\bz=(\dot{\balpha}+\dot{\bB}\bfeta_2)+\dot{\bB}(\bx_2-\bfeta_2)$ is $(\dot{\balpha}+\dot{\bB}\bfeta_2)^{\top}\bS^{-1}(\dot{\balpha}+\dot{\bB}\bfeta_2)+\tr(\bS^{-1}\dot{\bB}\mathbf V\dot{\bB}^{\top})$, the cross term vanishing because $\E[\bx_2-\bfeta_2]=\mathbf 0$.

\emph{Covariance block.} By $d\log|\bS|=\tr(\bS^{-1}d\bS)$ and $d\bS^{-1}=-\bS^{-1}d\bS\,\bS^{-1}$ \cite{magnusneudecker2019matrix}, the $\operatorname{vech}(\bS)$-score is $\mathbf D_d^{\top}\operatorname{vec}(\tfrac12\bS^{-1}\beps\beps^{\top}\bS^{-1}-\tfrac12\bS^{-1})$. Its product with the $\bbeta$-score involves only first and third moments of the centred Gaussian $\beps$, which vanish, so the $(\bbeta,\bS)$ block is zero. By the information identity, the $\bS$ block is also minus the expected second differential of $\log\fCG$ in $\bS$, which by $\E[\beps\beps^{\top}]=\bS$ is $\tfrac12\tr(\bS^{-1}d\bS\bS^{-1}d\bS)-\tr(\bS^{-1}d\bS\bS^{-1}d\bS)=-\tfrac12\operatorname{vec}(d\bS)^{\top}(\bS^{-1}\otimes\bS^{-1})\operatorname{vec}(d\bS)$, and $\operatorname{vec}(d\bS)=\mathbf D_d\operatorname{vech}(d\bS)$ gives the information $\tfrac12\mathbf D_d^{\top}(\bS^{-1}\otimes\bS^{-1})\mathbf D_d$, that is, the quadratic form $\tfrac12\tr(\bS^{-1}\dot{\bS}\bS^{-1}\dot{\bS})$.

\emph{Directional block, part (i).} In $\bvartheta_2$ the information is $\E[\mathbf s_{\vMF}\mathbf s_{\vMF}^{\top}]=\cov(\bx_2)=\mathbf V=\nabla^2\psi(\bvartheta_2)$ (Section~2). For $\bvartheta_2\neq\mathbf 0$, $\bfeta_2$ and $\mathbf V$ are given by Lemma~\ref{lem:vmf-moments}; at $\bvartheta_2=\mathbf 0$ the direction is uniform, so $\bfeta_2=\mathbf 0$ and $\mathbf V=\E[\bx_2\bx_2^{\top}]=\bI_m/m$, by rotation invariance and unit trace.

\emph{Directional block, part (ii).} A tangent vector $(d\bmu_2,d\kappa)$, $\bmu_2^{\top}d\bmu_2=0$, is pushed to $d\bvartheta_2=d\kappa\,\bmu_2+\kappa\,d\bmu_2$, and with $\mathbf V=\tfrac a\kappa\Pperp+\Am'(\kappa)\bmu_2\bmu_2^{\top}$ (Lemma~\ref{lem:vmf-moments}) we get $d\bvartheta_2^{\top}\mathbf V\,d\bvartheta_2=\Am'(\kappa)\,d\kappa^2+\kappa\Am(\kappa)\norm{d\bmu_2}^2$, with no cross term because $d\bmu_2\perp\bmu_2$.

\emph{Positive definiteness.} $\mathbf V\succ\mathbf 0$ at every $\bvartheta_2\in\R^m$, because a nonzero linear function of $\bx_2$ is not almost surely constant under a law with positive density on the sphere, which lies in no affine hyperplane. Hence $\mathbf M_z\succ\mathbf 0$ (its Schur complement is $\mathbf V$), $\mathbf M_z\otimes\bS^{-1}\succ\mathbf 0$, and, $\mathbf D_d$ having full column rank, the covariance block is positive definite. In the polar chart also $\kappa\Am(\kappa)>0$ and $\Am'(\kappa)>0$.

\emph{Part (iii): pullback and Hessian.} Write $\btheta_{\cyl}$ in linear coordinates of $\Hnat_{\mathrm{amb}}$ in which the pairing $\langle\bt,\btheta\rangle$ is Euclidean. By Proposition~3(a)--(b) and its extension to $\Pext$ (Section~\ref{sec:esm-proofs}), $\log\fcyl(\bx;\boldsymbol\zeta)=\langle\bt(\bx),\btheta_{\cyl}(\boldsymbol\zeta)\rangle-\psi_{\mathrm{amb}}(\btheta_{\cyl}(\boldsymbol\zeta))$ for $\boldsymbol\zeta\in\Pext$. Differentiating, the score along the $i$th coordinate of $\boldsymbol\zeta$ is $\langle\bt-\bfeta,\partial_i\btheta_{\cyl}\rangle$, with $\bfeta=\nabla\psi_{\mathrm{amb}}(\btheta_{\cyl}(\boldsymbol\zeta))=\E[\bt]$. Hence $\calI_{ij}=\partial_i\btheta_{\cyl}^{\top}\cov(\bt)\,\partial_j\btheta_{\cyl}=(\mathbf J^{\top}\nabla^2\psi_{\mathrm{amb}}\mathbf J)_{ij}$, because $\cov(\bt)=\nabla^2\psi_{\mathrm{amb}}$ for a regular family (Section~2). The chain rule gives $\partial_{ij}(\psi_{\mathrm{amb}}\circ\btheta_{\cyl})=\sum_{a,b}\partial_{ab}\psi_{\mathrm{amb}}\,\partial_i\iota_a\,\partial_j\iota_b+\sum_a\partial_a\psi_{\mathrm{amb}}\,\partial_{ij}\iota_a=(\mathbf J^{\top}\nabla^2\psi_{\mathrm{amb}}\mathbf J)_{ij}+\langle\bfeta,\partial_{ij}\btheta_{\cyl}\rangle$. Along $\dot{\bB}$, (10) gives the Hessian $\tfrac1m\tr(\bS^{-1}\dot{\bB}\dot{\bB}^{\top})$ and part (i) the metric $\tr(\bS^{-1}\dot{\bB}(\mathbf V+\bfeta_2\bfeta_2^{\top})\dot{\bB}^{\top})=\tr(\bS^{-1}\dot{\bB}\,\E[\bx_2\bx_2^{\top}]\dot{\bB}^{\top})$; they agree at $\bvartheta_2=\mathbf 0$, where $\E[\bx_2\bx_2^{\top}]=\bI_m/m$. They differ, for example, at $d=1$, $m=2$, $\bS=1$, $\bvartheta_2=(2,0)^{\top}$, $\dot{\bB}=(1,0)$, where the Hessian is $1/2$ and the metric $\E[x_{21}^2]=1-A_2(2)/2\approx0.6511$ by (\ref{eq:vmf-m2}).

\emph{Original coordinates.} Here $(\bmu_1,\bSigma_{12})$ and $(\bmu_2,\kappa)$ can be coupled, because (8) gives $\bB=\kappa\bSigma_{12}$ and $\balpha=\bmu_1-\kappa\bSigma_{12}\bmu_2$, so $d\bB=d\kappa\,\bSigma_{12}+\kappa\,d\bSigma_{12}$ and $d\balpha=d\bmu_1-d\kappa\,\bSigma_{12}\bmu_2-\kappa\,d\bSigma_{12}\bmu_2-\kappa\bSigma_{12}d\bmu_2$. On the regression block the pulled-back information is $\E[(d\bbeta\bz)^{\top}\bS^{-1}(d\bbeta'\bz)]$. For the tangent vectors $u$ ($d\bSigma_{12}=\bSigma_{12}$) and $v$ ($d\kappa=1$), all else fixed, $d\bbeta_u\bz=\kappa\bSigma_{12}(\bx_2-\bmu_2)$ and $d\bbeta_v\bz=\bSigma_{12}(\bx_2-\bmu_2)$, and $u$ has no $\bS$, $\bmu_2$ or $\kappa$ component, so only the regression block contributes. Hence $\calI(u,v)=\kappa\,\tr\bigl(\bS^{-1}\bSigma_{12}\E[(\bx_2-\bmu_2)(\bx_2-\bmu_2)^{\top}]\bSigma_{12}^{\top}\bigr)$, which is positive for $\bSigma_{12}\ne\mathbf 0$ because $\E[(\bx_2-\bmu_2)(\bx_2-\bmu_2)^{\top}]\succeq\cov(\bx_2)\succ\mathbf 0$. If $\bSigma_{12}=\mathbf 0$, then $d\bB=\kappa\,d\bSigma_{12}$ and $d\balpha=d\bmu_1-\kappa\,d\bSigma_{12}\bmu_2$ do not mix the groups, and block-diagonality is preserved.
\end{proof}

\begin{proof}[Proof of Proposition~10]
Write $f=\fcyl(\cdot;\btheta)$ and $f'=\fcyl(\cdot;\btheta')$. Both are positive, and $\log(f/f')$ is a quadratic polynomial in $(\bx_1,\bx_2)$, integrable under $f$ because $\bx_2$ is bounded and $\bx_1$ has Gaussian tails. By (5), $\log(f/f')=\log(\fvMF/\fvMF')(\bx_2)+\log(\fCG/\fCG')(\bx_1\mid\bx_2)$, and taking expectations, first over $\bx_1$ given $\bx_2$, gives the chain rule $\KL[f:f']=\KL[\fvMF:\fvMF']+\E_{\btheta}\bigl[\KL[\fCG(\cdot\mid\bx_2):\fCG'(\cdot\mid\bx_2)]\bigr]$. The first term is $\E_{\btheta}[\log\Cm(\norm{\bvartheta_2})-\log\Cm(\norm{\bvartheta_2'})+(\bvartheta_2-\bvartheta_2')^{\top}\bx_2]$, the first two terms of (16). Given $\bx_2$, the conditional laws are $\calN(\balpha+\bB\bx_2,\bS)$ and $\calN(\balpha'+\bB'\bx_2,\bS')$, whose mean difference is $\boldsymbol\delta+(\bB-\bB')(\bx_2-\bfeta_2)$ with $\boldsymbol\delta:=\balpha-\balpha'+(\bB-\bB')\bfeta_2$; by \eqref{eq:esm-gauss-kl-mean} their divergence is $\tfrac12[\tr(\bS'^{-1}\bS)-d+\log(|\bS'|/|\bS|)]$ plus half the quadratic form of this difference in $\bS'^{-1}$. Since $\E_{\btheta}[\bx_2-\bfeta_2]=\mathbf 0$ and $\cov(\bx_2)=\mathbf V$, the quadratic form has expectation $\boldsymbol\delta^{\top}\bS'^{-1}\boldsymbol\delta+\tr(\bS'^{-1}(\bB-\bB')\mathbf V(\bB-\bB')^{\top})$, which gives the remaining terms. The first two terms are $\psi(\bvartheta_2')-\psi(\bvartheta_2)-\langle\bvartheta_2'-\bvartheta_2,\nabla\psi(\bvartheta_2)\rangle$, the Bregman divergence of $\psi$, because $\nabla\psi(\bvartheta_2)=\bfeta_2$. Finally, $\KL[f_{\btheta}:f_{\btheta'}]=\E_{\btheta}[\log f_{\btheta}]-\E_{\btheta}[\log f_{\btheta'}]$ is smooth in $\btheta'$, with zero gradient and Hessian $-\E_{\btheta}[\nabla^2\log f_{\btheta}]=\calI(\btheta)$ at $\btheta'=\btheta$, by the information identity.
\end{proof}

\begin{proof}[Proof of Corollary~11]
Because the directional factor is the same with and without $\bB$, the whole gain comes from the conditional Gaussian factor.

\emph{Independent restriction.} With $\bB=\mathbf 0$, Lemma~\ref{lem:ols-loewner} with $\bz_n=1$ gives $\hat{\balpha}_{\mathrm{ind}}=\bar{\bx}_1$, and Lemma~\ref{lem:cov-profile} gives $\hat{\bS}_{\mathrm{ind}}=\sct{11}/N$, since $\sct{11}\succeq\sct{11}-\sct{12}\sct{22}^{-1}\sct{12}^{\top}\succ\mathbf 0$ by (A2). The directional part is unaffected: by Theorem~6(i), both fits have the same $\hat{\bvartheta}_2$, which exists because (A1) implies (A0), and $\ell_{\vMF}$ cancels.

\emph{Determinant form.} By Lemma~\ref{lem:cov-profile} the maximized conditional log-likelihoods are $-\tfrac{Nd}2\log(2\pi)-\tfrac N2\log|\hat{\bS}|-\tfrac{Nd}2$ and the same with $\hat{\bS}_{\mathrm{ind}}$, which gives the first expression of (17).

\emph{Canonical correlations.} The squared sample canonical correlations $\hat\rho_i^2$ are the eigenvalues of $\sct{11}^{-1}\sct{12}\sct{22}^{-1}\sct{12}^{\top}$. This matrix is similar to a positive semidefinite matrix of rank at most $\min(d,m)$ \cite{anderson2003multivariate}. By (A2), $\bI_d$ minus this matrix is similar to $\sct{11}^{-1/2}(\sct{11}-\sct{12}\sct{22}^{-1}\sct{12}^{\top})\sct{11}^{-1/2}\succ\mathbf 0$, so each $\hat\rho_i^2<1$. Taking determinants, $|\sct{11}-\sct{12}\sct{22}^{-1}\sct{12}^{\top}|=|\sct{11}|\prod_{i\le\min(d,m)}(1-\hat\rho_i^2)$, which gives the second expression.

\emph{Divergence form.} By \eqref{eq:esm-gauss-kl-mean}, and since $\hat{\bmu}_c(\bx_{2n})-\bar{\bx}_1=\hat{\bB}(\bx_{2n}-\bar{\bx}_2)$, summing over $n$ gives
\begin{equation*}
\sum_{n}\KL\bigl[\calN(\hat{\bmu}_c(\bx_{2n}),\hat{\bS}):\calN(\bar{\bx}_1,\hat{\bS}_{\mathrm{ind}})\bigr]
=\frac N2\Bigl[\tr\bigl(\hat{\bS}_{\mathrm{ind}}^{-1}(\hat{\bS}+\hat{\bB}\sct{22}\hat{\bB}^{\top}/N)\bigr)-d+\log\frac{|\hat{\bS}_{\mathrm{ind}}|}{|\hat{\bS}|}\Bigr].
\end{equation*}
Here $\hat{\bS}+\hat{\bB}\sct{22}\hat{\bB}^{\top}/N=N^{-1}(\sct{11}-\sct{12}\sct{22}^{-1}\sct{12}^{\top}+\sct{12}\sct{22}^{-1}\sct{12}^{\top})=\hat{\bS}_{\mathrm{ind}}$, so the trace is $d$ and the bracket reduces to the determinant form. This identity is the projection reading stated after the corollary: the average $N^{-1}\sum_n\calN(\hat{\bmu}_c(\bx_{2n}),\hat{\bS})$ of the fitted conditional laws has mean $\bar{\bx}_1$ and, by the same identity, covariance $\hat{\bS}_{\mathrm{ind}}$. So $\calN(\bar{\bx}_1,\hat{\bS}_{\mathrm{ind}})$ is its moment match, that is, its m-projection onto the Gaussian family (Section~2).
\end{proof}

\paragraph{Exact null law of the likelihood gain.}
Condition on the directions. Under $\bB=\mathbf 0$ the $\bx_{1n}$ are then independent $\calN(\balpha,\bS)$ observations in a multivariate regression on the fixed design $\bz_n=(1,\bx_{2n}^{\top})^{\top}$, of rank $m+1$ by (A1). Its residual scatter $\sct{11}-\sct{12}\sct{22}^{-1}\sct{12}^{\top}$ has $N-m-1$ degrees of freedom and the scatter $\sct{12}\sct{22}^{-1}\sct{12}^{\top}$ of the hypothesis $\bB=\mathbf 0$ has $m$. Hence $\Lambda:=|\sct{11}-\sct{12}\sct{22}^{-1}\sct{12}^{\top}|/|\sct{11}|=\prod_i(1-\hat\rho_i^2)$ is the likelihood-ratio criterion of this linear hypothesis, and for $N\ge d+m+1$ it has Wilks' distribution $\Lambda(d,m,N-m-1)$, the law of a product of $d$ independent variables with the Beta distributions $\mathrm{Beta}(\tfrac{N-m-i}2,\tfrac m2)$, $i=1,\dots,d$~\cite[Ch.~8]{anderson2003multivariate}. This law does not depend on the directions, so it is also the unconditional null law, for every $\kappa\ge0$. By (17), $2(\hat\ell_{\cyl}-\hat\ell_{\mathrm{ind}})=-N\log\Lambda$, and $-(N-1-\tfrac{d+m+1}2)\log\Lambda$ is Bartlett's corrected statistic~\cite[Ch.~8]{anderson2003multivariate}. Computed from the exact law by numerical integration, the test that compares $-N\log\Lambda$ with the $95\%$ quantile of $\chi^2_{dm}$ has size $22.9\%$ at $d=m=3$, $N=15$, $11.0\%$ at $N=30$ and $6.4\%$ at $N=100$; with Bartlett's correction the sizes are $5.35\%$, $5.06\%$ and $5.00\%$. A simulation with von Mises--Fisher directions ($\kappa=0$, $2$, $20$; $40{,}000$ null samples each, $d=m=3$, $N=15$) agrees: $22.8$--$23.1\%$ for $\chi^2_9$ and $4.8$--$5.1\%$ for the exact law.

\paragraph{Identifiability of finite cylindrical mixtures (Theorem~12 of the main text).}\label{subsec:esm-mix-identifiable}
Appendix~A of the main text proves the theorem; this paragraph adds the two lemmas it uses and the null-set step in detail.

\emph{Gaussian densities.} Finitely many Gaussian densities $\fG(\cdot;\mathbf m_g,\bSigma_g)$ on $\R^d$ with distinct pairs $(\mathbf m_g,\bSigma_g)$ are linearly independent~\cite{yakowitz1968identifiability}. For completeness: if $\sum_gc_g\fG(\cdot;\mathbf m_g,\bSigma_g)=0$, the characteristic functions give $\sum_gc_g\exp(\mathrm i\mathbf s^{\top}\mathbf m_g-\tfrac12\mathbf s^{\top}\bSigma_g\mathbf s)=0$ for all $\mathbf s\in\R^d$. Choose a unit $\mathbf v$ for which the pairs $(\mathbf v^{\top}\bSigma_g\mathbf v,\mathbf v^{\top}\mathbf m_g)$ are distinct; this excludes the zero sets of finitely many nonzero quadratic or linear forms, a Lebesgue-null set. With $\mathbf s=s\mathbf v$, $s\in\R$, $\sigma_g^2:=\mathbf v^{\top}\bSigma_g\mathbf v$ and $\mu_g:=\mathbf v^{\top}\mathbf m_g$, $\sum_gc_g\exp(\mathrm is\mu_g-\tfrac12s^2\sigma_g^2)=0$. Let $\sigma_*^2$ be the smallest $\sigma_g^2$ and $G_*$ the indices attaining it, whose $\mu_g$ are distinct. Multiplying by $\exp(\tfrac12s^2\sigma_*^2)$ and letting $s\to\infty$, the trigonometric polynomial $P(s):=\sum_{g\in G_*}c_ge^{\mathrm is\mu_g}$ tends to zero; but $T^{-1}\int_T^{2T}|P(s)|^2\,ds\to\sum_{g\in G_*}|c_g|^2$, so $c_g=0$ on $G_*$, and induction on the number of groups gives $c=\mathbf 0$.

\emph{Null sets.} For components $j\not\sim k$ with $\bS_j=\bS_k$ and $\bB_j\neq\bB_k$, the directions at which the conditional means agree solve $(\bB_j-\bB_k)\mathbf u=\balpha_k-\balpha_j$, an affine subspace of dimension at most $m-1$. A proper affine subspace meets $\sphere^{m-1}$ in a sphere of lower dimension, a point or the empty set, all $\sigma$-null, because the sphere lies in no affine hyperplane.

\emph{Exponentials on the sphere.} Distinct $\bvartheta_1,\dots,\bvartheta_L\in\R^m$ give linearly independent functions $\mathbf u\mapsto\exp(\bvartheta_l^{\top}\mathbf u)$ on $\sphere^{m-1}$; this is the identifiability of finite von Mises--Fisher mixtures~\cite{kent1983identifiability}, including the uniform law $\bvartheta_l=\mathbf 0$. Appendix~A proves it by analytic continuation of the restriction to a great circle $\mathbf u(t)=\cos t\,\mathbf v+\sin t\,\mathbf w$ along which the $\mathbf v^{\top}\bvartheta_l$ are distinct: at $t=-\mathrm i\tau$ the term with the largest $\mathbf v^{\top}\bvartheta_l$ dominates as $\tau\to\infty$. The script \texttt{check\_mixture\_identifiability.py} in the folder \texttt{theorychecks/} of the code confirms the theorem numerically: for eight distinct components, including pairs that share the regression part, the direction, or conditional means on a great circle, and a uniform direction, the matrix of their densities at 4{,}000 random points has full column rank, with smallest singular value $0.04$--$0.11$ of the largest, while duplicating one component makes it singular to machine precision.

Section~\ref{sec:esm-mix} turns from proofs to the details of the mixture procedure of Section~4 of the main text.

\section{Details of the cylindrical mixture procedure}\label{sec:esm-mix}

This section gives the M-step and its conditions, the algorithm of Section~4.2 of the main text (Algorithm~\ref{alg:em-cyl}), the implementation and controls, the $k$-means initializer, the optional safeguards and the parameter counts. The notation is that of Section~4.2 of the main text: with responsibilities $r_{nj}$, $N_j=\sum_n r_{nj}$ is the component mass, $\bar{\bx}_{i,j}$ and $\mathbf C_{ik,j}$ are the weighted means and the weighted centred scatters divided by $N_j$, and $\bs_{2,j}=\sum_nr_{nj}\bx_{2n}$ is the weighted resultant, with mean length $\bar R_j=\norm{\bs_{2,j}}/N_j$.

\paragraph{The M-step for one component.} Section~4.2 of the main text uses the following consequence of Theorem~6.

\begin{lemma}[The M-step for one component]
\label{lem:cyl-mle-weighted}
Let $r_{nj}>0$ for every $n$, as after every E-step. The weighted problem $\max_{\btheta_j}\sum_nr_{nj}\log\fcyl(\bx_{1n},\bx_{2n};\btheta_j)$ has a maximizer over $\Pext$ if and only if the sample satisfies (A0) and (A2), and the maximizer is unique if and only if the sample also satisfies (A1). It is then
\begin{gather*}
\hat{\bB}_j=\mathbf C_{12,j}\mathbf C_{22,j}^{-1},\quad
\hat{\balpha}_j=\bar{\bx}_{1,j}-\hat{\bB}_j\bar{\bx}_{2,j},\\
\hat{\bS}_j=\mathbf C_{11,j}-\mathbf C_{12,j}\mathbf C_{22,j}^{-1}\mathbf C_{12,j}^{\top},\quad
\hat{\bvartheta}_{2,j}=\Am^{-1}(\bar R_j)\,\frac{\bs_{2,j}}{\norm{\bs_{2,j}}},
\end{gather*}
with $\hat{\bvartheta}_{2,j}=\mathbf 0$ if $\bs_{2,j}=\mathbf 0$; over $\calP$ a maximizer exists if and only if, in addition, $\bs_{2,j}\neq\mathbf 0$. All blocks except $\hat\kappa_j=\Am^{-1}(\bar R_j)$ are in closed form, and $\hat\kappa_j$ is the root of the strictly increasing scalar equation $\Am(\kappa)=\bar R_j$.
\end{lemma}

\begin{proof}
With positive weights, (A0w), (A1w) and (A2w) are equivalent to (A0), (A1) and (A2): a weighted resultant has norm $W$ only if all directions coincide, the rank of a weighted centred scatter is the dimension of the affine hull of the points, and $E^{w}\succ\mathbf 0$ says that no nonzero linear combination of $\bx_1$ is an affine function of $\bx_2$ on the sample (see below). Theorem~6 then gives the claims.
\end{proof}

\begin{algorithm}[htb]
\caption{Exact EM for cylindrical mixtures, with multiple starts}
\label{alg:em-cyl}
\footnotesize
\begin{algorithmic}[1]
\REQUIRE Data $\{(\bx_{1n},\bx_{2n})\}_{n=1}^N$ satisfying (A0)--(A2); $K$; $n_{\mathrm{start}}=10$ starts; $\mathrm{tol}=10^{-8}$; $T_{\max}=10^{4}$; $N_{\min}=d+m+1$.
\FOR{$b=1$ \TO $n_{\mathrm{start}}$}
\STATE Set $r^{(0)}_{nj}=\mathbf 1\{\zeta^{(b)}_n=j\}$ from the labels $\zeta^{(b)}_n$ of the $b$th $k$-means run of Section~4.3 of the main text; one M-step (lines~\ref{alg:em-cyl-mstart}--\ref{alg:em-cyl-mend}) gives $\bTheta^{(0)}$.
\FOR{$t=0$ \TO $T_{\max}-1$}
\STATE E-step: compute $r_{nj}$ by (19) at $\bTheta^{(t)}$, by log-sum-exp, and $\bar\ell^{(t)}$ and $N_{j,\mathrm{eff}}$ by (20)--(21).
\STATE \label{alg:em-cyl-ess} If $N_{j,\mathrm{eff}}<N_{\min}$ for some $j$, discard start $b$ as degenerate.
\STATE If $t\ge1$ and $\bar\ell^{(t)}-\bar\ell^{(t-1)}\le\mathrm{tol}\max\{1,|\bar\ell^{(t)}|\}$, stop: start $b$ has converged at $\bTheta^{(t)}$.
\STATE \label{alg:em-cyl-mstart} M-step, weights: $N_j\leftarrow\sum_n r_{nj}$, $\pi_j\leftarrow N_j/N$.
\STATE \label{alg:em-cyl-mend} M-step, components: for each $j$, the exact maximizer of Lemma~\ref{lem:cyl-mle-weighted} with weights $r_{nj}$, with $\hat\kappa_j$ the root of $\Am(\kappa)=\bar R_j$; this gives $\bTheta^{(t+1)}$.
\ENDFOR
\ENDFOR
\RETURN among the non-degenerate starts, the fit with the largest final $\bar\ell$, together with every start's final $\bar\ell$ and number of iterations.
\end{algorithmic}
\end{algorithm}

\paragraph{Implementation of Algorithm~\ref{alg:em-cyl}.} The root $\hat\kappa_j$ is computed by a safeguarded Newton iteration, with $\Am$ evaluated through exponentially scaled Bessel functions, and $\hat\kappa_j=0$ when $\bar R_j<10^{-12}$; it agrees with Brent's method to $10^{-11}$, which a unit test checks. Every iteration of Algorithm~\ref{alg:em-cyl} is an exact EM step, so $\bar\ell^{(t)}$ is nondecreasing~\cite{dempster1977em}, which the implementation checks to $10^{-12}\max\{1,|\bar\ell^{(t)}|\}$ at every iteration; and if the iterates stay in a compact subset of the parameter space, with weights bounded away from zero, covariances away from singularity and concentrations bounded, every limit point is a stationary point of $\ell$ and $\ell(\bTheta^{(t)})$ converges to its value at a stationary point~\cite{wu1983em}. Nothing more is claimed: the limit need not be a global or a local maximum, and different starts may reach different stationary points, which is why the algorithm uses ten. A start that reaches $T_{\max}$ is reported as not converged. Optionally, when a fit with $K-1$ components is available, one further start splits its component $j^{*}$ of largest mass into two with weights $\pi_{j^{*}}/2$, the same $\bB_{j^{*}}$, $\bS_{j^{*}}$ and $\bvartheta_{2,j^{*}}$, and intercepts $\balpha_{j^{*}}\pm\lambda^{1/2}\mathbf v$, where $(\lambda,\mathbf v)$ is the leading eigenpair of $\bS_{j^{*}}$; it enters Algorithm~\ref{alg:em-cyl} at the E-step with this $\bTheta^{(0)}$. The generalized von Mises (GvM) components of Section~8 of the main text start from the converged von Mises--Fisher fit at the same $K$ and from the split start, and use the same degeneracy rule, with $N_{\min}=d+m+1$ (Section~\ref{subsec:esm-ocean-audit}).

\paragraph{Weighted rank conditions.}
Lemma~\ref{lem:cyl-mle-weighted} (the M-step for one component) assumes that every weight is positive. Then (A0w), (A1w) and (A2w) of Theorem~6 are equivalent to their unweighted forms (A0), (A1) and (A2). For (A0w), equality in the triangle inequality $\norm{\br^{w}}\le W$ holds if and only if all directions coincide, whatever the positive weights. For (A1w), the range of $\sct{22}^{w}=\sum_nw_n(\bx_{2n}-\bar{\bx}_2^{w})(\bx_{2n}-\bar{\bx}_2^{w})^{\top}$ is the span of the vectors $\bx_{2n}-\bar{\bx}_2^{w}$, which, because $\bar{\bx}_2^{w}$ lies in the affine hull of the $\bx_{2n}$, is the direction space of that affine hull; its dimension does not depend on the weights. For (A2w), by step~(ii) of the proof of Theorem~6 (Section~\ref{sec:esm-cyl}), $\mathbf a^{\top}E^{w}\mathbf a=0$ for a unit $\mathbf a$ exactly when $\mathbf a^{\top}\bx_{1n}$ is an affine function of $\bx_{2n}$ on the sample, again a condition free of the weights. Only (A3w) depends on the positive weights: on the circle ($m=2$), for the directions $\pm\mathbf e_1,\pm\mathbf e_2$, equal weights give $\br^{w}=\mathbf 0$, while unequal positive weights can give $\br^{w}\neq\mathbf 0$; on $\Pext$ a zero resultant is admissible and gives $\hat{\bvartheta}_2=\mathbf 0$. In Algorithm~\ref{alg:em-cyl}, the E-step responsibilities $r_{nj}$ are positive, because the mixture weights and the component densities are, so in every M-step after an E-step the conditions hold for every component once the full sample satisfies (A0)--(A2). The first M-step is different: its weights are the hard $k$-means labels $r^{(0)}_{nj}=\mathbf 1\{\zeta_n=j\}$ of Algorithm~\ref{alg:em-cyl}, which are zero outside cluster $j$, so there the conditions are conditions on each $k$-means cluster and can fail even when the full sample satisfies them; Algorithm~\ref{alg:em-cyl} then discards the start.

\paragraph{Implementation and controls.}\label{app:details}
The code (Python, using NumPy, SciPy and scikit-learn) reproduces every table and figure from fixed seeds and contains unit tests of the densities (normalization by numerical integration), of the exact M-step and of the monotonicity of exact EM. Table~\ref{tab:controls} lists the controls of Algorithm~\ref{alg:em-cyl} and of its hierarchical counterpart (Section~\ref{subsec:esm-hier-stabilized}). Every density, held-out score and information criterion uses the exact, normalized mixture density, evaluated by log-sum-exp. Times are process CPU times under Python~3.14 and NumPy~2.5 on a shared Apple M2 computer (eight cores, 24~GB of RAM) under heavy external load.

\begin{table}[htbp]
\caption{Numerical controls of the fitting procedures}
\label{tab:controls}
\footnotesize
\setlength{\tabcolsep}{3pt}
\begin{tabular}{@{}p{3.6cm}p{9.25cm}@{}}
\toprule
Control & Value \\
\midrule
Algorithm & exact EM (Algorithm~\ref{alg:em-cyl}); no floors, clips, ridges or approximate inverses \\
Stopping & $\bar\ell^{(t)}-\bar\ell^{(t-1)}\le10^{-8}\max\{1,|\bar\ell^{(t)}|\}$, at most $T_{\max}=10^{4}$ iterations; a start reaching $T_{\max}$ is reported as not converged \\
Monotonicity check & $\bar\ell^{(t)}\ge\bar\ell^{(t-1)}-10^{-12}\max\{1,|\bar\ell^{(t)}|\}$ at every iteration \\
Concentrations & exact root of $\Am(\kappa)=\bar R$ by safeguarded Newton iteration, agreeing with Brent's method to $10^{-11}$; $\kappa=0$ if $\bar R<10^{-12}$; $\Am$ from exponentially scaled Bessel functions \\
Degeneracy & a start is discarded if some $N_{j,\mathrm{eff}}<N_{\min}$, with $N_{\min}=d+m+1$ (cylindrical or GvM component), $d+1$ (Gaussian component of a hierarchy or of the first stage of the isolated estimator) and $2$ (von Mises--Fisher component), or if a $k$-means cluster violates (A0) or (A2) \\
Starts & $R=10$ per fitted model, from $k$-means with $k$-means++ seeds $1,\dots,10$, shared across models; optionally one split start from the $(K-1)$ fit (Section~4.2); the non-degenerate start with the largest training $\bar\ell$ is retained, and every start's final training and held-out $\bar\ell$ and iteration count are recorded \\
Initialization & $k$-means under \eqref{eq:kmeans-distance} with $c_{\mathrm{dir}}=1$, $\epsilon_R=10^{-6}$, at most $100$ iterations \\
Grids & Sections~7 and~8 of the main text and Section~\ref{sec:esm-synth-design} \\
GvM M-step & damped Newton iteration on the directional natural parameter, stopped when every coordinate of the moment residual is below $10^{-11}$, at most $100$ iterations \\
Distinct optima & two starts reach distinct optima if their final training $\bar\ell$ differ by more than $10^{-7}\max\{1,|\bar\ell|\}$ \\
Software & Python~3.14.7, NumPy~2.5.4, SciPy~1.18.1, scikit-learn~1.9.1, pandas~3.1.0 (the versions pinned in the code); random numbers from NumPy's \texttt{default\_rng} (PCG64) \\
\bottomrule
\end{tabular}
\end{table}

\paragraph{Initialization.}
Algorithm~\ref{alg:em-cyl} starts from $k$-means on the product space with the dissimilarity
\begin{equation}
\label{eq:kmeans-distance}
\Dcyl[\bx:\mathbf{y}]:=\tfrac1{2d}(\bx_1-\mathbf{y}_1)^\top\bS_{\mathrm{global}}^{-1}(\bx_1-\mathbf{y}_1)+c_{\mathrm{dir}}\,(1-\bx_2^\top\mathbf{y}_2)/\max\{1-\hat R^2,\epsilon_R\},
\end{equation}
where $c_{\mathrm{dir}}>0$ weights the directional block and $\epsilon_R>0$ guards against degeneracy. Let $\mathbf V:=N^{-1}\sum_n(\bx_{1n}-\bar{\bx}_1)(\bx_{1n}-\bar{\bx}_1)^\top$ be the full-sample Euclidean covariance, $\bS_{\mathrm{global}}:=\Floor_{\epsilon_S}(\sym(\mathbf V))$ its floored version, and $\hat R:=N^{-1}\norm{\sum_n\bx_{2n}}$ the full-sample mean resultant length. Both normalizers of $\Dcyl$ are computed once from the full sample. Draw two observations $I$ and $J$ independently and uniformly from the sample. Then $\E[(\bx_{1I}-\bx_{1J})(\bx_{1I}-\bx_{1J})^\top]=2\mathbf V$ and $\E[\bx_{2I}^\top\bx_{2J}]=\hat R^2$, so, when the floors are inactive, the two terms of $\Dcyl$ have expectations $1$ and $c_{\mathrm{dir}}$.

Each iteration assigns every observation to the nearest centroid under $\Dcyl$ and resets each centroid to the cluster mean and the normalized cluster resultant. These are the exact minimizers of the within-cluster sum of $\Dcyl$: the mean minimizes the Mahalanobis term, and the normalized resultant minimizes the directional term on the sphere. An empty or directionally cancelled cluster (zero resultant) is reseeded deterministically from the observation worst represented by the valid centroids. The reported fits use $c_{\mathrm{dir}}=1$ and $\epsilon_R=10^{-6}$; the $r$th start of Algorithm~\ref{alg:em-cyl} runs this $k$-means from a $k$-means++ seeding with seed $r$, for at most $100$ iterations or until no label changes, and receives the final labels $\zeta^{(r)}_n$ as $r^{(0)}_{nj}=\mathbf 1\{\zeta^{(r)}_n=j\}$. Here $\Floor_{\epsilon_S}$ raises every eigenvalue below $\epsilon_S=10^{-6}$ to $\epsilon_S$ and $\sym(\mathbf A):=\tfrac12(\mathbf A+\mathbf A^{\top})$; this floor acts only on the fixed metric of the dissimilarity, not on the fitted components.

\paragraph{The initializer as classification EM.}
This paragraph reads $\Dcyl$ as a negative log-density. For the independent component ($\bB_j=\mathbf 0$) with covariance $\bS_{\mathrm{global}}$ and concentration $dc_{\mathrm{dir}}/\max\{1-\hat R^2,\epsilon_R\}$, $-d\,\Dcyl[\bx:(\balpha_j,\bmu_{2,j})]$ equals $\log\fcyl(\bx;\btheta_j)$ up to a constant common to all $j$, because all components share this covariance and concentration. Hence the initializer is classification EM~\cite{CeleuxGovaert1992CEM} for the equal-weight mixture of these components, and it ignores $\bB_j$. Two components with the same location in $\R^d\times\sphere^{m-1}$ that differ only in $\bB_j$ are identifiable (Theorem~12 of the main text) but not separated by this initializer, and the split start of the implementation paragraph above keeps $\bB_{j^{*}}$; no start targets such components. Its spherical $k$-means part~\cite{dhillon2001concept} has the normalized resultant as centroid, not the arithmetic mean of Bregman hard clustering~\cite{banerjee2005bregman}.

\paragraph{Optional safeguards.}
For data on which the exact M-step is ill-conditioned, for instance nearly coplanar directions on $\sphere^2$ or tiny components, the code offers safeguards that are not used in the reported fits. With $\sym(\mathbf A):=\tfrac12(\mathbf A+\mathbf A^\top)$, $\Floor_{\epsilon_S}$ raising every eigenvalue below $\epsilon_S$ to $\epsilon_S$, and $\tilde\pi_j:=N_j/N$, they replace the exact updates by
\begin{equation}
\label{eq:mix-safeguarded}
\begin{gathered}
\hat{\bmu}_{2,j}=\bs_{2,j}/\norm{\bs_{2,j}},\qquad
\hat\kappa_j=\clip\bigl(\widehat{\Am}^{-1}(\min\{\bar R_j,\Am(\kappa_{\max})\}),\kappa_{\min},\kappa_{\max}\bigr),\\
\hat{\bB}_j=\mathbf C_{12,j}(\mathbf C_{22,j}+\epsilon_{22}\bI_m)^{-1},\qquad
\hat{\balpha}_j=\bar{\bx}_{1,j}-\hat{\bB}_j\bar{\bx}_{2,j},\\
\hat{\bS}_j=\Floor_{\epsilon_S}\bigl(\sym(\mathbf C_{11,j}-\mathbf C_{12,j}\hat{\bB}_j^{\top})\bigr),
\end{gathered}
\end{equation}
with the previous direction kept and $\hat\kappa_j=\kappa_{\min}$ if $\bs_{2,j}=\mathbf 0$, and the weights
\begin{equation}
\label{eq:mix-weight-floor}
\pi_j\leftarrow(\tilde\pi_j+\epsilon_\pi)/(1+K\epsilon_\pi)\ \text{ if }\min_{j'}\tilde\pi_{j'}\le\epsilon_\pi,\qquad \pi_j\leftarrow\tilde\pi_j\ \text{ otherwise},
\end{equation}
with default constants $\epsilon_{22}=\epsilon_S=\kappa_{\min}=10^{-6}$, $\kappa_{\max}=10^{3}$ and $\epsilon_\pi=10^{-9}$. Under $\bS_j\succeq\epsilon_S\bI_d$ and $\kappa_j\le\kappa_{\max}$ the likelihood is bounded:

\begin{lemma}[Bounded constrained likelihood]
\label{lem:mix-bounded}
If $\bS_j\succeq\epsilon_S\bI_d$ and $\kappa_j\le\kappa_{\max}$ for every $j$, then for every sample of size $N$
\begin{equation*}
\ell(\bTheta)\le N\Bigl[-\tfrac d2\log(2\pi\epsilon_S)+\log\Cm(\kappa_{\max})+\kappa_{\max}\Bigr].
\end{equation*}
\end{lemma}

\begin{proof}
$\fCG\le(2\pi)^{-d/2}|\bS_j|^{-1/2}\le(2\pi\epsilon_S)^{-d/2}$, and $\fvMF\le\Cm(\kappa_j)e^{\kappa_j}\le\Cm(\kappa_{\max})e^{\kappa_{\max}}$ because $\tfrac{d}{d\kappa}[\log\Cm(\kappa)+\kappa]=1-\Am(\kappa)>0$ by (3). A convex combination of densities is at most their largest value.
\end{proof}

 Table~\ref{tab:safeguards} sets each safeguarded update beside the exact one, with its status: \emph{exact} means the exact maximizer of the stated, possibly constrained, problem, given the statistics used; \emph{approximate}, an approximation to one; \emph{heuristic}, no optimality claimed. Safeguarded updates are in general not generalized EM steps, so they void the monotonicity of exact EM, except in the monotone version below.

\begin{table}[htbp]
\caption{Optional safeguards: the exact M-step and a safeguarded update, block by block, with the status of each safeguarded update relative to the statistics it uses. The reported fits use the exact M-step}
\label{tab:safeguards}
\footnotesize
\setlength{\tabcolsep}{3pt}
\renewcommand{\arraystretch}{1.05}
\begin{tabular}{@{}p{3.35cm}p{4.25cm}p{1.65cm}p{3.15cm}@{}}
\toprule
\raggedright Exact maximizer (Lemma~\ref{lem:cyl-mle-weighted}) & \raggedright Safeguarded update & \raggedright Constant & \raggedright Status \tabularnewline
\midrule
\multicolumn{4}{@{}l}{\textit{(a) Regression} $(\balpha_j,\bB_j)$} \tabularnewline
\raggedright $\hat{\bB}_j=\mathbf C_{12,j}\mathbf C_{22,j}^{-1}$,\newline $\hat{\balpha}_j=\bar{\bx}_{1,j}-\hat{\bB}_j\bar{\bx}_{2,j}$ &
\raggedright $\hat{\bB}_j=\mathbf C_{12,j}(\mathbf C_{22,j}+\epsilon_{22}\bI_m)^{-1}$, a ridge on the per-unit-mass centred directional scatter;\newline $\hat{\balpha}_j=\bar{\bx}_{1,j}-\hat{\bB}_j\bar{\bx}_{2,j}$ &
\raggedright $\epsilon_{22}=10^{-6}$ &
\raggedright Heuristic: not the maximum-likelihood coefficient; the intercept is exact given $\hat{\bB}_j$ \tabularnewline
\addlinespace[2pt]
\multicolumn{4}{@{}l}{\textit{(b) Direction and concentration} $(\bmu_{2,j},\kappa_j)$} \tabularnewline
\raggedright $\hat{\bmu}_{2,j}=\bs_{2,j}/\norm{\bs_{2,j}}$,\newline $\hat\kappa_j=\Am^{-1}(\bar R_j)$ &
\raggedright $\hat{\bmu}_{2,j}=\bs_{2,j}/\norm{\bs_{2,j}}$; if $\bs_{2,j}=\mathbf 0$, the previous direction is kept and $\hat\kappa_j=\kappa_{\min}$;\newline \mbox{$\hat\kappa_j=\clip\bigl(\widehat{\Am}^{-1}(\bar R_j'),\kappa_{\min},\kappa_{\max}\bigr)$},\newline $\bar R_j':=\min\{\bar R_j,\Am(\kappa_{\max})\}$, with $\widehat{\Am}^{-1}$ of \eqref{eq:best-fisher}--\eqref{eq:banerjee} &
\raggedright \mbox{$\kappa_{\min}=10^{-6}$,}\newline \mbox{$\kappa_{\max}=10^{3}$} &
\raggedright Direction exact; concentration approximate for every $m$ (the exact constrained maximizer is the clipped root of $\Am(\kappa)=\bar R_j$) \tabularnewline
\addlinespace[2pt]
\multicolumn{4}{@{}l}{\textit{(c) Conditional covariance} $\bS_j$} \tabularnewline
\raggedright $\hat{\bS}_j=\mathbf C_{11,j}-\mathbf C_{12,j}\mathbf C_{22,j}^{-1}\mathbf C_{12,j}^\top$ (weighted residual covariance) &
\raggedright $\hat{\bS}_j=\Floor_{\epsilon_S}(\sym(\cdot))$ of the weighted conditional covariance implied by the fitted regression; with the ridge active it differs from the fitted-residual covariance by a positive semidefinite matrix of norm at most $\epsilon_{22}\norm{\hat{\bB}_j}^2$ &
\raggedright $\epsilon_S=10^{-6}$ &
\raggedright Exact maximizer of the covariance profile over $\{\bS\succeq\epsilon_S\bI_d\}$ relative to the scatter actually used~\cite{ingrassia2007constrained} \tabularnewline
\addlinespace[2pt]
\multicolumn{4}{@{}l}{\textit{(d) Weights} $\bpi$} \tabularnewline
\raggedright $\hat\pi_j=N_j/N$ &
\raggedright additive floor \eqref{eq:mix-weight-floor}, applied only when $\min_j N_j/N\le\epsilon_\pi$ &
\raggedright $\epsilon_\pi=10^{-9}$ &
\raggedright Heuristic: not in general the constrained maximizer over the floored simplex \tabularnewline
\bottomrule
\end{tabular}
\end{table}

\paragraph{Approximate inverse of $\Am$.}
Row (b) of Table~\ref{tab:safeguards} uses a clipped approximate inverse $\widehat{\Am}^{-1}$ of $\Am$ instead of the exact inverse, which is the radial part of $\nabla\phi$ (Section~2.1 of the main text): for $m=2$ the form \eqref{eq:best-fisher} of Best and Fisher~\cite{best1981bias}, and for $m>2$ a numerical approximation of $\Am^{-1}$ that starts from the value $\kappa_{\mathrm B}(\bar R)$ of Banerjee et al.~\cite{banerjee2005clustering,sra2012vmf}:
\begin{align}
\widehat{A}_2^{-1}(\bar R)&:=
\begin{cases}
\tfrac56\bar R^{5}+\bar R^{3}+2\bar R, & \bar R<0.53,\\
1.39\bar R-0.4+0.43/(1-\bar R), & 0.53\le\bar R<0.85,\\
1/(\bar R^{3}-4\bar R^{2}+3\bar R), & \bar R\ge0.85,
\end{cases} \label{eq:best-fisher}\\
\kappa_{\mathrm B}(\bar R)&:=\bar R\,(m-\bar R^{2})/(1-\bar R^{2}),\qquad 0<\bar R<1,\ m>2 . \label{eq:banerjee}
\end{align} Two Newton steps on $\Am(\kappa)=\bar R$ from $\kappa_{\mathrm B}(\bar R)$ give the inverse for $m>2$.

\paragraph{Status of the safeguarded updates.}
The status entries of rows (b)--(d) of Table~\ref{tab:safeguards} rest on three facts.

\emph{Concentration, row (b).} With the mean direction at $\bs_{2,j}/\norm{\bs_{2,j}}$, the weighted log-likelihood depends on $\kappa$ through $\kappa\mapsto N_j\log\Cm(\kappa)+\kappa\norm{\bs_{2,j}}$. This is the function $L$ of step~(iv) of the proof of Theorem~6 (Section~\ref{sec:esm-cyl}) with $(N,\norm{\br})$ replaced by $(N_j,\norm{\bs_{2,j}})$. It is therefore strictly concave, with derivative $N_j(\bar R_j-\Am(\kappa))$, and its exact maximizer over $[\kappa_{\min},\kappa_{\max}]$ is the exact root of $\Am(\kappa)=\bar R_j$, clipped to this interval. The clipped approximate inverse $\widehat{\Am}^{-1}$ only approximates this root, for every $m$, so the concentration update is an approximate, bounded moment match. The theory of Section~3 uses the exact $\Am^{-1}$ and is unaffected by the approximation.

\emph{Covariance, row (c).} For a given weighted scatter, the covariance profile is the function $G$ of Lemma~\ref{lem:cov-profile}. With the residual scatter of a fixed regression, $G$ is, up to a constant, the weighted Gaussian log-likelihood as a function of $\bS_j$. Over $\{\bS\succeq\epsilon_S\bI_d\}$, the exact maximizer of $G$ is the eigenvalue floor of the scatter divided by $N_j$~\cite{ingrassia2007constrained}. The floor receives the weighted conditional covariance implied by the fitted regression. With the ridge active, this matrix differs from the fitted-residual covariance by a positive semidefinite matrix of norm at most $\epsilon_{22}\norm{\hat{\bB}_j}^2$. The update is therefore exact only relative to the scatter actually used.

\emph{Weights, row (d).} The additive floor keeps every weight at least $\epsilon_\pi/(1+K\epsilon_\pi)$ and the weights summing to one. It is in general not the constrained maximizer of $\sum_jN_j\log\pi_j$ over the floored simplex $\{\bpi:\pi_j\ge\epsilon_\pi,\ \sum_j\pi_j=1\}$: that maximizer keeps the ratios $N_j/N_{j'}$ of the weights above the floor, which the additive floor changes unless these $N_j$ are equal.

\paragraph{Monotone safeguards.}\label{rem:monotone-possible}
Three changes make the safeguarded procedure monotone: (i) the clipped exact root of $\Am(\kappa)=\bar R_j$, for every $m$; (ii) the ridge only as a rank-deficiency fallback, in an iteration in which (A1w) fails for some component; (iii) the exact constrained maximizer of $\sum_jN_j\log\pi_j$ over the floored simplex for the weights. Step (iii) is a Karush--Kuhn--Tucker step: it solves the first-order optimality conditions under the constraints, which identify the exact maximizer because the objective is concave and the constraints are linear. Every iteration without the fallback (ii) is then a generalized EM step and does not decrease the log-likelihood $\ell$.

Consider an iteration in which the fallback (ii) is not invoked. Its M-step maximizes the expected complete-data log-likelihood
\begin{equation*}
Q(\bTheta\mid\bTheta^{(t)}):=\sum_{n=1}^N\sum_{j=1}^K r^{(t)}_{nj}\log\bigl[\pi_j\fcyl(\bx_{1n},\bx_{2n};\btheta_j)\bigr]
\end{equation*}
over the admissible set (the floored simplex, $\bS_j\succeq\epsilon_S\bI_d$ and $\kappa_{\min}\le\kappa_j\le\kappa_{\max}$). Indeed, $Q$ separates over the weights and, within each component, over the Gaussian block $(\balpha_j,\bB_j,\bS_j)$ and the directional block $(\bmu_{2,j},\kappa_j)$. Step (iii) maximizes the weight block. The normalized resultant and, by the strict concavity shown above, the clipped root (i) maximize the directional block. Within the Gaussian block, $\bS_j$ couples the regression and the covariance, but the least-squares coefficient maximizes $Q$ for every $\bS_j$ (Lemma~\ref{lem:ols-loewner}). Flooring the weighted residual covariance of that fit therefore maximizes $Q$ jointly over $(\balpha_j,\bB_j,\bS_j)$. Hence $\bTheta^{(t+1)}$ maximizes $Q(\cdot\mid\bTheta^{(t)})$ over the admissible set, and in particular $Q(\bTheta^{(t+1)}\mid\bTheta^{(t)})\ge Q(\bTheta^{(t)}\mid\bTheta^{(t)})$.

Because the E-step is exact and the admissible set is fixed, the free-energy argument of Section~\ref{sec:esm-background} turns this inequality into $\ell(\bTheta^{(t+1)})\ge\ell(\bTheta^{(t)})$~\cite{dempster1977em,wu1983em}. An unconditional ridge, the approximate inverse or the additive weight floor breaks this argument.

\paragraph{Original coordinates.}
After the update, the original coordinates are $\bSigma_{12,j}=\bB_j/\kappa_j$ and $\bmu_{1,j}=\balpha_j+\bB_j\bmu_{2,j}$. Because a fitted $\kappa_j$ can be near zero or very large, fitted dependence is reported through $\bB_j$, as in Section~3 of the main text (Section~\ref{rem:sigma12-reporting}).

\paragraph{Parameter counts.}
Evaluated at the dimensions of the main text, the counts of Section~4.3 give the following values. The $K=3$ candidates of Section~7 have $q_{\cyl}=35$, $q_{\mathrm{ind}}=23$ and $q_{\mathrm{hier}}(3,2)=32$ for $(d,m)=(2,2)$, and $65$, $38$ and $50$ for $(d,m)=(3,3)$. In the buoy study of Section~8, where $(d,m)=(3,2)$, the counts become $q_{\cyl}=18K-1$, $q_{\mathrm{ind}}=12K-1$, $q_{\mathrm{hier}}=(K-1)+K(8+3L)$ for the Euclidean-first hierarchy and $q^{\mathrm{dir}}_{\mathrm{hier}}=(K-1)+K(1+10L)$ for the directional-first one. At $K=6$ and $L=4$, for example, these are $107$, $71$, $125$ and $251$.

\paragraph{Information criteria.}
With training log-likelihood $\ell_{\mathrm{train}}$ and size $N_{\mathrm{train}}$, the Akaike and Bayesian criteria~\cite{akaike1974aic,schwarz1978bic} are
\begin{equation}
\label{eq:aic-bic}
\mathrm{AIC}=-2\ell_{\mathrm{train}}+2q,\qquad \mathrm{BIC}=-2\ell_{\mathrm{train}}+q\log N_{\mathrm{train}},
\end{equation}
with the counts $q$ of Section~4.3 of the main text; note $q_{\mathrm{hier}}(K,1,d,m)=q_{\mathrm{ind}}(K,d,m)$ and $q_{\mathrm{hier}}(K,L,d,m)-q_{\cyl}(K,d,m)=K[(L-1)(m+1)-dm]$. They are descriptive penalties, not as asymptotic guarantees: finite-mixture likelihoods are singular and may contain boundary or nonidentifiable configurations~\cite{watanabe2013wbic}, so the regular likelihood asymptotics behind the criteria do not hold literally. Because the full and isolated hierarchies at a common $(K,L)$ share one density family and one count, a paired AIC or BIC contrast between them reflects only the training likelihoods attained. Section~\ref{sec:esm-hier} gives the details of these hierarchies.

\section{Details of the two-layer hierarchy}\label{sec:esm-hier}

Section~6 of the main text defines the two-layer feature-block hierarchy (Definition~18) and its two estimators: the full estimator, which targets the joint likelihood $\lhier$, and the isolated two-stage estimator. This section gives the details behind them, the identifiability of the hierarchy and the fitting procedures.

\paragraph{Orderings, counts and identifiability.} Sections~7--8 of the main text take $f=\fG(\cdot;\bmu_j,\bSigma_j)$ and $h=\fvMF(\cdot;\bmu_{jk},\kappa_{jk})$ (the Euclidean-first ordering) or the blocks swapped (directional-first), with counts (22)--(23). With leaf weights $\pi_j\pi_{jk}$, a $(K,L)$ hierarchy is a parameter-sharing submodel of the $KL$-component independent, hence cylindrical, mixture, so comparing it with a $K$-component mixture compares numbers of components as well as structures. Its leaves are cylindrical components with $\bB=\mathbf 0$, so Theorem~12 of the main text identifies it up to relabelling when the $\btheta_j$ and, within each class, the $\btheta_{jk}$ are distinct.

\subsection{Flattened form and parameter sharing of the two-layer hierarchy}\label{subsec:esm-hier-flat}

With the flattened weights $\omega_{jk}:=\pi_j\pi_{jk}$, the hierarchy is a flat mixture over its $\sum_jL_j$ leaves (Section~6), with leaf densities $f(\bx_1;\btheta_j)\,h(\bx_2;\btheta_{jk})$. All leaves below class $j$ share $\btheta_j$, whereas an unrestricted flat mixture would give every leaf its own first-block parameter, so flattening does not remove the sharing constraint.

In the Euclidean-first (Gaussian/von Mises--Fisher) case with a common local size $L$, the count $q_{\mathrm{hier}}(K,L,d,m)$ of Section~4.3 is smaller than the count $q_{\mathrm{ind}}(KL,d,m)$ of the unrestricted $KL$-leaf mixture (the independent cylindrical mixture with $KL$ components) by $(KL-K)[d+d(d+1)/2]$. Both have $KL-1$ free weights and $KL$ von Mises--Fisher components, so the difference consists of the Gaussian parameters removed by the sharing; at $L=1$ it vanishes.

\emph{Identifiability.} The factorization $\omega_{jk}=\pi_j\pi_{jk}$ alone does not identify the hierarchy, since any positive leaf weights factor in this way for a proposed grouping of the leaves ($\pi_j$ is the total weight of group $j$); the grouping is identified through the shared first-block parameters. In the Gaussian/von Mises--Fisher case, in either ordering, each leaf density $f(\bx_1;\btheta_j)h(\bx_2;\btheta_{jk})$ is a cylindrical component with $\bB=\mathbf 0$ (with the roles of the blocks exchanged in the directional-first ordering, where $f$ is the von Mises--Fisher and $h$ the Gaussian density). Suppose the class parameters $\btheta_1,\dots,\btheta_K$ are distinct and, for each $j$, the local parameters $\btheta_{j1},\dots,\btheta_{jL_j}$ are distinct, with positive weights. Then the leaves are distinct cylindrical components, and if two such hierarchies have the same density, Theorem~12 of the main text gives a weight-preserving bijection between their leaves. Leaves with the same first-block parameter belong to the same class, because the class parameters are distinct, so the bijection maps classes to classes; summing leaf weights gives $\pi_j$, and $\pi_{jk}=\omega_{jk}/\pi_j$. The hierarchy is therefore identified up to permutations of the classes and, within each class, of its local components. If two classes have the same first-block parameter they merge into one, and the hierarchy is not identified.

If $f$ and $h$ are exponential families and the labels are treated as observed, the sharing constraints are linear in the natural coordinates of the complete-data exponential family (the joint law of the labels and the observation): they equate the first-block natural parameters of the leaves below each class. The complete-data model of the hierarchy is therefore an e-flat submanifold (Section~2) of the complete-data model of the flat mixture. Nothing is claimed about the geometry of the observed mixture family, in which the labels are summed out.

\subsection{Gaussian/von Mises--Fisher closed forms of the hierarchical M-step}\label{subsec:esm-hier-mstep}

With the leaf responsibilities $\tau_{njk}=r^{(1)}_{nj}r^{(2)}_{njk}$ of Section~6 of the main text at the current iterate $\bTheta^{\mathrm{old}}$, the expected complete-data log-likelihood $Q=Q(\bTheta\mid\bTheta^{\mathrm{old}})$ separates by layer:
\begin{equation}
\label{eq:hier-Q}
Q
=\sum_{n,j} r^{(1)}_{nj}\bigl[\log\pi_j+\log f(\bx_{1n};\btheta_j)\bigr]
+\sum_{n,j,k}\tau_{njk}\bigl[\log\pi_{jk}+\log h(\bx_{2n};\btheta_{jk})\bigr].
\end{equation}
The M-step maximizes \eqref{eq:hier-Q} term by term. In the Euclidean-first specialization, both component problems of the M-step of Section~6 are the moment matches of Lemma~\ref{lem:dual-mstep}. Under the conditions stated there ($\hat{\bSigma}_j\succ\mathbf 0$ and $\norm{\bs_{jk}}<N_{jk}$, the mean-space conditions), the updates $\hat{\bmu}_j$ and $\hat{\bSigma}_j$ are the $r^{(1)}_{nj}$-weighted mean and covariance of $\bx_1$, and, with $\bs_{jk}$ and $N_{jk}$ as in Section~6, the von Mises--Fisher update is the natural parameter $\nabla\phi(\bs_{jk}/N_{jk})$, that is, $\hat{\bmu}_{jk}=\bs_{jk}/\norm{\bs_{jk}}$ and $\hat\kappa_{jk}=\Am^{-1}(\norm{\bs_{jk}}/N_{jk})$, the root of a strictly increasing scalar equation, if $\bs_{jk}\neq\mathbf 0$, and the uniform law if $\bs_{jk}=\mathbf 0$. If $\norm{\bs_{jk}}=N_{jk}$, all directions with positive weight coincide and the weighted likelihood is unbounded in the concentration (Section~\ref{sec:esm-cyl}, proof of Theorem~6, step (iv)). A clipped, floored or otherwise safeguarded version need not be this maximizer.

\subsection{Proof of the evidence lower bound}\label{app:proofs}

\begin{proof}[Proof of Proposition~19 of the main text]
Put $Z_n:=\sum_j\pi_jf(\bx_{1n};\btheta_j)h_j(\bx_{2n})$, which is positive because $\lhier(\bTheta)$ is finite, so that $\lhier=\sum_n\log Z_n$ and $r^{(1)}_{nj}=\pi_jf(\bx_{1n};\btheta_j)h_j(\bx_{2n})/Z_n$. If $a_{nj}>0$ where $r^{(1)}_{nj}=0$, then $-F(a,\bTheta)$ and the divergence are both $+\infty$ and (33) holds in the extended sense; so assume that $a_{nj}>0$ only where $r^{(1)}_{nj}>0$, with $0\log0=0$.

\emph{(i)} Inserting $r^{(1)}_{nj}$ into $\KL[a_{n\cdot}:r^{(1)}_{n\cdot}(\bTheta)]=\sum_ja_{nj}\log(a_{nj}/r^{(1)}_{nj})$ and using $\sum_ja_{nj}=1$ gives $\log Z_n$ minus the $n$th term of (32). Summing over $n$ gives $\lhier(\bTheta)-F(a,\bTheta)$, and Gibbs' inequality gives the sign and the equality case.

\emph{(ii)} Let $p(\bx_{1n}):=\sum_{j'}\pi_{j'}f(\bx_{1n};\btheta_{j'})>0$. For the $\bx_1$-only posterior, $\pi_jf(\bx_{1n};\btheta_j)=a_{nj}p(\bx_{1n})$, so $Z_n=p(\bx_{1n})\sum_ja_{nj}h_j(\bx_{2n})$, which gives the reweighting formula. Where $a_{nj}>0$, the bracket in (32) is therefore $\log p(\bx_{1n})+\log h_j(\bx_{2n})$; summing with weights $a_{nj}$ gives $F(a,\bTheta)=\ell_1(\bTheta^{(1)})+\sum_j\ell_{2,j}(\bTheta^{(2)}_j\mid\mathbf a_j)=\liso(\bTheta)$, and with (i) the first equality in (34). Inserting the reweighting formula into the divergence gives the second equality: the gap in Jensen's inequality for the strictly concave logarithm and the variable $g_J(\bx_{2n})$, $J\sim a_{n\cdot}$, which is positive where $a_{nj}>0$. The gap vanishes if and only if this variable is almost surely constant, that is, if and only if $h_j(\bx_{2n})$ takes one value for all $j$ with $a_{nj}>0$.

\emph{(iii)} At fixed $a$ and $\bTheta^{(1)}$, the terms of $F(a,\bTheta)$ other than $\sum_{n,j}a_{nj}\log h_j(\bx_{2n})=\sum_j\ell_{2,j}(\bTheta^{(2)}_j\mid\mathbf a_j)$ are constant, and the $j$th term of this sum depends on $\bTheta^{(2)}_j$ only.
\end{proof}

\subsection{The isolated estimator as a stationary point of the free energy}\label{subsec:esm-hier-stationary}

By the proof of Proposition~19(ii) of the main text (Section~\ref{app:proofs}), when $a$ is the $\bx_1$-only posterior of Section~6, the first-layer terms of the free energy $F(a,\cdot)$ reduce exactly to the log-likelihood $\ell_1$ of $\bx_1$ alone. This gives a property of the isolated (two-stage) estimator $\hat{\bTheta}_{\mathrm{iso}}$. If both stages reach stationary points of their own criteria, $\hat{\bTheta}_{\mathrm{iso}}$ is a stationary point of the free energy $F(a,\cdot)$ at fixed $a$, the first-stage $\bx_1$-only posterior. The reason is that, at this $a$, the gradient of $\ell_1$ in $\bTheta^{(1)}$ equals the $a$-weighted gradient of the first-layer terms of $F$, while, by part (iii) of Proposition~19, $F(a,\cdot)$ depends on $\bTheta^{(2)}$ only through the second-stage criteria $\ell_{2,j}$.

\subsection{Fitting procedures for the hierarchy}\label{subsec:esm-hier-stabilized}

The full estimator runs exact EM for $\lhier$ (Section~6 of the main text), with the exact M-steps of Section~\ref{subsec:esm-hier-mstep}; each stage of the isolated estimator runs exact EM on its own criterion, $\ell_1$ for the first stage and, for each class, the weighted criterion $\ell_{2,j}$ of the second stage, whose M-steps are the same moment matches with weights $a_{nj}$ and $a_{nj}r^{(2)}_{njk}$. Both orderings use the controls of Table~\ref{tab:controls}: ten starts, the stopping rule and monotonicity check of Algorithm~\ref{alg:em-cyl}, exact concentration roots, and the degeneracy check with $N_{\min}=d+1$ for a Gaussian and $2$ for a von Mises--Fisher component. Initialization is of $k$-means type, layer by layer: Euclidean $k$-means for Gaussian components and spherical $k$-means for von Mises--Fisher components, each on its own block, from $k$-means++ seeds $1,\dots,10$. Each EM run is monotone in its own criterion; the isolated estimator is monitored on $\ell_1$ and each $\ell_{2,j}$, not on $\lhier$, and Section~\ref{subsec:esm-hier-counterexample} shows that its second stage can lower $\lhier$. In the directional-first ordering, $\lhier$ is unbounded if some $L_j\ge 2$, because a local Gaussian component can collapse on one $\bx_{1n}$, and if $K\ge 2$, because a first-layer von Mises--Fisher component can concentrate on one direction; the degeneracy check discards such starts. The optional safeguards of Section~\ref{sec:esm-mix} apply to the hierarchical components in the same way and are not used in the reported fits.

Every reported hierarchical log-density, pseudo-BIC and AIC uses the normalized hierarchical density of Definition~18 of the main text, $p(\bx_1,\bx_2)=\sum_j\pi_jf(\bx_1;\btheta_j)h_j(\bx_2)$ in the Euclidean-first ordering and the same expression with the blocks exchanged in the directional-first ordering, evaluated by log-sum-exp over the leaves (Section~\ref{app:details}).

\subsection{Ascent of the isolated criterion can lower the joint likelihood}\label{subsec:esm-hier-counterexample}\label{rem:hier-ascent}

Take $K=2$, $L_1=L_2=1$, $m=2$, $\pi_1=\pi_2=\tfrac12$, $d=1$, Gaussian components $\calN(\delta,1)$ and $\calN(-\delta,1)$ with $\delta>0$, and $x_{11}=x_{12}=0$, so that $a_{nj}=\tfrac12$ although the classes are distinct, directions $\mathbf e_1$ and $-\mathbf e_1$, and initial local components $\vMF_2(\pm\mathbf e_1,\kappa)$, $\kappa>0$. Each class's weighted resultant is zero, so the exact second-stage maximizer is the uniform law (Lemma~\ref{lem:dual-mstep}). This step raises $\liso$ by $2\log I_0(\kappa)>0$ but changes $\lhier$ by $2\log\{I_0(\kappa)/\cosh\kappa\}<0$, as $1<I_0(\kappa)<\cosh\kappa$: the gap (34) falls from $2\log\cosh\kappa$ to zero. Both signs persist under a lower bound $\kappa_{\min}<\kappa$ on concentrations, as with the optional safeguards (below).

In detail, the example has $K=2$, $L_1=L_2=1$, $m=2$, $\pi_1=\pi_2=\tfrac12$, $d=1$, Gaussian components $\calN(\pm\delta,1)$ and $x_{11}=x_{12}=0$, hence $a_{nj}=\tfrac12$, directions $\bx_{21}=\mathbf e_1$ and $\bx_{22}=-\mathbf e_1$, and local components $\vMF_2(\pm\mathbf e_1,\kappa)$ with normalizer $C_2(\kappa)=1/\{2\pi I_0(\kappa)\}$. The first-stage terms are common to both parameters compared, and $f(0;\delta,1)=f(0;-\delta,1)$, so only $\sum_j\ell_{2,j}$ and the directional factor $\sum_j\tfrac12h_j(\bx_{2n})$ of the joint density matter. Each class's second-stage criterion is $\tfrac12\log h(\mathbf e_1)+\tfrac12\log h(-\mathbf e_1)=\log C_2(\kappa')+\tfrac{\kappa'}2\bmu^{\top}(\mathbf e_1-\mathbf e_1)=\log C_2(\kappa')$ for $h=\vMF_2(\bmu,\kappa')$. Since $\log C_2$ is strictly decreasing, by (3), its maximizer over $\kappa'\ge\kappa_0$, for any $\kappa_0\in[0,\kappa)$, is $\kappa'=\kappa_0$ with any $\bmu$; keeping $\bmu=\pm\mathbf e_1$ gives the starting configuration with $\kappa$ replaced by $\kappa_0$, and $\kappa_0=0$ is the uniform law of the main text. For the configuration with concentration $x$, $\sum_j\ell_{2,j}=2\log C_2(x)$, and the directional mixture density at either observation is $\tfrac12C_2(x)(e^{x}+e^{-x})=\cosh x/\{2\pi I_0(x)\}$. The step from $\kappa$ to $\kappa_0$ therefore changes $\liso$ by $2\log\{I_0(\kappa)/I_0(\kappa_0)\}>0$, as $I_0'=I_1>0$ on $(0,\infty)$, and $\lhier$ by $2\{r(\kappa)-r(\kappa_0)\}$ with $r(x):=\log\{I_0(x)/\cosh x\}$. Now $r'(x)=A_2(x)-\tanh x$, and $A_2(x)<\tanh x$ for $x>0$: the difference $D:=\tanh-A_2$ has $D(0)=0$ and $D'(0)=1-A_2'(0)=\tfrac12$, and if $x_0>0$ were its first zero, then $D'(x_0)\le0$, whereas the Riccati identity (4) at $m=2$ and $\tanh'=1-\tanh^2$ give $D'(x_0)=A_2(x_0)^2-\tanh^2x_0+A_2(x_0)/x_0=A_2(x_0)/x_0>0$. So $r$ is strictly decreasing, and $\lhier$ decreases. At $\kappa_0=0$ the changes are $2\log I_0(\kappa)$ and $2\log\{I_0(\kappa)/\cosh\kappa\}$, and the gap (34) of Proposition~19, $2\log\cosh\kappa$ before the step, is zero after it. The script \texttt{check\_hierarchy.py} in the folder \texttt{theorychecks/} of the code confirms these values and the inequality $A_2<\tanh$ numerically.

Section~\ref{sec:esm-synth-design} gives the design details of the simulation studies.

\section{Design details of the simulation studies}\label{sec:esm-synth-design}

This section completes the design of the simulation studies of Section~7 of the main text; Section~\ref{sec:esm-generators} lists the numerical arrays of the two generators. Every item below is implemented in the code (module \texttt{cylmix.generators}, script \texttt{experiments/synthetic\_converged.py}), which reruns every reported number from fixed seeds. The computations used Python~3.14.7 with NumPy~2.5.4, SciPy~1.18.1, scikit-learn~1.9.1, pandas~3.1.0 and matplotlib~3.11.2; every random stream is a \texttt{numpy.random.default\_rng} generator (PCG64) with the seeds stated below. The Yeo--Johnson exponents of Section~\ref{sec:esm-ocean-preprocessing} are found by a numerical optimizer, so another scikit-learn version may change them in the last digits.

\paragraph{Separability rule of Generator~1.}
The high separability level multiplies by $6$ the displacement of each Euclidean mean from the unweighted centroid of the low-level means, and by $2.25$ every pairwise angle between the pre-projection directional means, keeping their norms and all covariances (Section~\ref{subsec:esm-study1-2d}).

\paragraph{Angular construction of Generator~2.}
Generator~2 draws
\begin{equation}\label{eq:gen-part2}
\begin{aligned}
Y_{1n}&\sim\operatorname{Categorical}(\bpi),\qquad \bx_{1n}\mid Y_{1n}=j\sim\calN_d(\bmu_j,\bPsi_j),\\
Y_{2n}\mid Y_{1n}=j&\sim\operatorname{Categorical}(0.58,\,0.42),\\
\mathbf u_{2n}\mid(Y_{1n},Y_{2n})=(j,k)&\sim\calN_m(\mathbf c_{jk},\bLambda_k),\qquad
\bx_{2n}=\mathbf u_{2n}/\norm{\mathbf u_{2n}},
\end{aligned}
\end{equation}
with $\bmu_j=\bnu_{1j}$ and $\bPsi_j=\bGamma_{11,j}$ of Generator~1 and $m=d$. The latent class $Y_1$ acts on the directional block only through the angles $\varphi_{jk}$ of the six subtype means $\mathbf c_{jk}$, $j\in\{1,2,3\}$, $k\in\{1,2\}$. With norms $(\rho_1,\rho_2)=(1.04,0.98)$, subtype offsets $(\delta_1,\delta_2)=(-11^\circ,11^\circ)$, class angles $\gamma^{(r)}_j$ and fixed orthonormal vectors $\mathbf q^{(d)},\mathbf t^{(d)}$,
\begin{equation*}
\begin{aligned}[b]
\mathbf c_{jk}&=\rho_k\bigl[\cos\varphi_{jk}\,\mathbf q^{(d)}+\sin\varphi_{jk}\,\mathbf t^{(d)}\bigr],\\
\varphi^{(\mathrm{low},r)}_{jk}&=\gamma^{(r)}_j+\delta_k,\qquad \varphi^{(\mathrm{high},r)}_{jk}=2.25\,\varphi^{(\mathrm{low},r)}_{jk},\\
\boldsymbol\gamma^{(\mathrm{strong})}&=(-25^\circ,0^\circ,25^\circ),\qquad \boldsymbol\gamma^{(\mathrm{weak})}=\tfrac13\,\boldsymbol\gamma^{(\mathrm{strong})},\qquad \boldsymbol\gamma^{(\mathrm{ind})}=(0^\circ,0^\circ,0^\circ),
\end{aligned}
\end{equation*}
where $\varphi_{jk}$ stands for $\varphi^{(s,r)}_{jk}$ at separability $s$ and dependence $r$. In the strong and weak conditions the class angles differ, so the conditional directional mixture varies with the latent class $Y_1$. In the independent condition all class angles are zero, so $p(\bx_2\mid Y_1=j)=p(\bx_2)$ and the blocks are marginally independent (Section~\ref{subsec:esm-study2}). Separability is bundled: the Euclidean rule of Generator~1 and the angular factor $2.25$ are applied together.

\paragraph{Samples, replicates and random numbers.}
Every data set has $N=8{,}000$ observations, the first $6{,}400$ for training and the last $1{,}600$ for testing, without noise; every study uses the seeds $0,\dots,19$. For replicate (seed) $s$ the labels are drawn from one stream shared by all twelve configurations of a study and the latent Gaussian draws from a second stream shared by the configurations of the same dimension, so configurations with the same seed share their random numbers: the labels come from \texttt{default\_rng([g, s, 0])}, the latent Gaussian draws from \texttt{default\_rng([g, s, 1, d])} and the (unused) noise from \texttt{default\_rng([g, s, 2, d])}, with $g=1$ for Generator~1, $g=2$ for Generator~2 and $g=3$ for the exact cylindrical generator of Section~\ref{sec:esm-recovery}. The $k$-means++ start $r$ of every fit uses \texttt{default\_rng(r)}, $r=1,\dots,10$. The seed is therefore the independent replicate: every interval averages a paired difference within seed over the factors that a row pools and then takes the two-sided Student-$t$ interval across seeds. A noisy replicate of each data set, perturbed by Gaussian noise of $0.15$ times the largest coordinate standard deviation of each block (the directional perturbation added before projection), is implemented in the code but is not used in the reported fits.

\paragraph{Fitting.}
Every fitted model is estimated by Algorithm~\ref{alg:em-cyl}, with the controls of Table~\ref{tab:controls}: ten starts from $k$-means with $k$-means++ seeds $1,\dots,10$, exact EM steps, the relative stopping rule with $\tau=10^{-8}$ and $T_{\max}=10^4$, no floors, clips or ridges, and the non-degenerate start with the largest training log-likelihood retained. Hierarchies use exact EM with the same controls, their starts from layer-by-layer $k$-means (Section~\ref{sec:esm-hier}); the isolated estimator runs each of its stages in this way. The root of $\Am(\kappa)=\bar R$ is computed by safeguarded Newton iterations from the closed-form approximations, stopped at a relative step of $10^{-14}$; it agrees with the root found by Brent's method on $\log\kappa$ at relative accuracy $10^{-12}$ to within a relative $10^{-11}$ for $\bar R\le0.999$, which a unit test checks.

\paragraph{Fitted models.}
\emph{Direct dependence:} the cylindrical mixture and its independent restriction at $K=3$, the number of generating first-layer classes, under both generators, both dimensions, both separability levels and all dependence levels. \emph{Hierarchy:} under Generator~2 at high separability, the full and the isolated $(3,2)$ hierarchies, which have $3$ Gaussian and $6$ von Mises--Fisher components, and the independent mixture with $K=6$, which has $6$ of each and contains the $(3,2)$ hierarchy as the submodel whose Gaussian components are shared in pairs. \emph{Recovery:} the cylindrical mixture at $K=3$ on the exact cylindrical generator of Section~\ref{sec:esm-recovery}.

\paragraph{Metrics.}
For a fitted parameter $\hat{\bTheta}$ the held-out log-density is
\begin{equation*}
\begin{aligned}
\HLL&=\frac{1}{N_{\mathrm{test}}}\sum_{n=1}^{N_{\mathrm{test}}}\log p\bigl(\bx_n^{\mathrm{test}};\hat{\bTheta}\bigr)=\HLL_2+\HLL_{1\mid2},\\
\HLL_2&=\frac{1}{N_{\mathrm{test}}}\sum_n\log p\bigl(\bx_{2n}^{\mathrm{test}};\hat{\bTheta}\bigr),\qquad
\HLL_{1\mid2}=\frac{1}{N_{\mathrm{test}}}\sum_n\log p\bigl(\bx_{1n}^{\mathrm{test}}\mid\bx_{2n}^{\mathrm{test}};\hat{\bTheta}\bigr),
\end{aligned}
\end{equation*}
with $\HLL_2$ the directional marginal and $\HLL_{1\mid2}$ the conditional part, and $p$ the exact mixture density, evaluated by log-sum-exp. The directional marginal of a cylindrical or independent mixture is the von Mises--Fisher mixture $\sum_j\pi_j\fvMF(\bx_2;\bmu_{2,j},\kappa_j)$ (the Gaussian factors integrate to one), and that of a Euclidean-first hierarchy is $\sum_j\pi_jh_j(\bx_2)$; both are in closed form, so the decomposition is exact. A difference of the conditional part between the cylindrical and the independent mixture isolates what $\bB_j$ adds to the prediction of $\bx_1$ given $\bx_2$; a difference of the marginal part, what the fits do to the law of the direction. The leaf-partition adjusted Rand index (Section~\ref{sec:esm-metrics}) compares the joint maximum a posteriori leaf $(\hat Y_{1n},\hat Y_{2n})$ of a fitted hierarchy with the six-class partition $(Y_{1n},Y_{2n})$.

\paragraph{Summary of the designs.}
Table~\ref{tab:esm-design} summarizes the two designs.

\begin{table}[t]
\caption{Design of the simulation studies. Both generators are projected-normal and lie outside every fitted family; they differ in the mechanism of cross-block dependence}\label{tab:esm-design}
\centering\footnotesize
\begin{tabular}{@{}p{2.1cm}p{4.9cm}p{4.9cm}@{}}
\toprule
 & Generator 1 & Generator 2\\
\midrule
Construction & Gaussian mixture on $\R^{2d}$, second block projected & Gaussian latent classes on $\R^d$ with projected-normal directional subtypes\\
Latent structure & One label $z_n\in\{1,2,3\}$ & $Y_{1n}\in\{1,2,3\}$, local $Y_{2n}\in\{1,2\}$; six leaves\\
Dependence factor & $\bGamma_{12,j}$: strong, weak ($\tfrac13$ strong), cond.\ independent ($\mathbf 0$) & class angles: strong $(-25^\circ,0^\circ,25^\circ)$, weak ($\tfrac13$ strong), marginally independent ($\mathbf 0$)\\
Separability & Low, high: Euclidean displacements $\times6$, angles $\times2.25$ & Same rules, applied together (bundled)\\
Dimensions & $d=m\in\{2,3\}$ & $d=m\in\{2,3\}$\\
Fitted models & cylindrical and independent, $K=3$ & cylindrical and independent, $K=3$; full and isolated $(3,2)$ hierarchies and independent, $K=6$ (high separability)\\
\bottomrule
\end{tabular}
\end{table}

\paragraph{A buoy-like generator.}
Two further designs imitate the buoy records ($d=3$, $m=2$, hourly rows, $N=34{,}000$ training and $8{,}700$ test rows). \emph{A (null):} the direction is exactly von Mises--Fisher with $\kappa=3$ and mean angle $0.6$ radians marginally, made serially dependent through a Gaussian AR(1) copula with lag-one correlation $0.95$, $\omega_t=F^{-1}_{\vMF}(\Phi(z_t))$ ($F_{\vMF}$ inverted numerically on $8{,}192$ grid points); $\bx_{1t}=\bB\bx_{2t}+0.5\,\mathbf L\boldsymbol\varepsilon_t$, with $\bB$ the $3\times2$ matrix with rows $(0.8,0.3)$, $(-0.4,0.5)$, $(0.2,-0.6)$, $\mathbf L$ the Cholesky factor of the correlation matrix with off-diagonal entries $0.3$, $0.2$, $0.1$, and $\boldsymbol\varepsilon_t$ three independent stationary Gaussian AR(1) series with coefficient $0.9$ and unit variance; $200$ series, series $r$ from \texttt{default\_rng([7, r])}. \emph{B (realistic):} three latent classes follow a Markov chain that stays in its class with probability $0.99$ per hour; within class $j$ the direction has the generalized von Mises density proportional to $\exp\{3\cos(\omega-\omega_j)+\cos2(\omega-\omega_j-0.5)\}$, skewed and with first-harmonic concentration $3$, with $(\omega_1,\omega_2,\omega_3)=(290^\circ,315^\circ,250^\circ)$; the chain moves to each other class with probability $0.005$. Given class $j$, $\bx_{1t}=\balpha_j+\bB_j\bx_{2t}+0.6\,\boldsymbol\varepsilon_t$, with $\bB_j=c_j\bB$ ($\bB$ of design A, $c=(1,0.5,-0.7)$), $\balpha_1=\mathbf 0$, $\balpha_2=(1,-0.5,0.5)^\top$, $\balpha_3=(-1,0.8,-0.3)^\top$ and $\boldsymbol\varepsilon_t$ as in design A without the correlation factor. Each of five series has $42{,}700$ hours, series $r$ from \texttt{default\_rng([8, r])}; the first $34{,}000$ are standardized and fitted, the last $8{,}700$ scored. The three families are fitted at $K\in\{1,3\}$ with one start at $K=1$ and ten $k$-means++ starts at $K=3$, the GvM mixture from the converged cylindrical fit. Design A checks the size of the test of the von Mises--Fisher component under serial dependence (Section~\ref{sec:esm-ocean}); design B checks the direct-dependence contrast and the ambient repair under buoy-like conditions.

\section{Adjusted Rand index and iteration counts}\label{sec:esm-metrics}

\subsection{Adjusted Rand index}

The simulation studies measure recovery by the adjusted Rand index~\cite{hubertarabie1985ari}. Let $n_{ij}$ be the number of the $n$ test observations in reference class $i$ and fitted class $j$, with row totals $a_i=\sum_j n_{ij}$ and column totals $b_j=\sum_i n_{ij}$. The sum $\sum_{i,j}\binom{n_{ij}}{2}$ counts the pairs placed together by both partitions, and its expected value over random partitions with the same class sizes is $E=\binom{n}{2}^{-1}\bigl[\sum_i\binom{a_i}{2}\bigr]\bigl[\sum_j\binom{b_j}{2}\bigr]$. The adjusted Rand index is
\begin{equation*}
\ARI=\frac{\sum_{i,j}\binom{n_{ij}}{2}-E}{\tfrac12\bigl[\sum_i\binom{a_i}{2}+\sum_j\binom{b_j}{2}\bigr]-E}.
\end{equation*}
It equals $1$ for partitions that coincide up to relabelling, is about $0$ for chance agreement, and remains defined when the two partitions have different numbers of classes.

\subsection{Iteration counts of the isolated estimator}

For the isolated estimator, the iteration cap applies separately to the first-stage Gaussian mixture and to each of the $K$ local directional mixtures, and its iteration count is the sum $I_{\mathrm{iso}}=I_{\mathrm{layer\,1}}+\sum_{j=1}^{K}I_{\mathrm{layer\,2},j}$. Iteration counts are therefore not comparable across estimators.

\section{Preprocessing and transformation of the buoy records}\label{sec:esm-ocean-preprocessing}

The buoy study uses the hourly standard-meteorological records of National Data Buoy Center (NDBC) station 46042 from 1 January 2000 to 31 December 2022~\cite{ndbc46042}: the 23 yearly files \texttt{46042h2000.txt.gz}, \dots, \texttt{46042h2022.txt.gz} of the NDBC historical archive (\texttt{https://www.ndbc.noaa.gov/data/historical/stdmet/}), retrieved on 3 November 2026; the code lists their SHA-256 checksums. The files of 2000--2004 have no minute column, those of 2005 and 2006 add one, and those from 2007 onwards use a commented two-line header; the loader reads all three layouts. The four variables used are significant wave height (WVHT), average wave period (APD), wind speed (WSPD) and MWD, which the NDBC defines as the direction from which the waves at the dominant period (DPD) are coming. In the $335{,}789$ archive rows, nonnumeric entries and the missing-value codes ($99$ for WVHT, APD and WSPD, $999$ for MWD) were treated as missing. From 10 November 2019 the station reports every ten minutes, and the minute-40 record was taken as the hourly observation; this rule removes $666$ rows with all four variables present ($665$ at minute $10$ and one at minute $50$). Three records report WVHT, APD and MWD all equal to zero (2000-03-03 03:00, 2002-02-11 02:00 and 2002-09-30 20:00), which is physically impossible, and were removed. This leaves $187{,}238$ complete hourly records; no further quality filtering was applied. The records are stamped at minute $00$ until 31 January 2008 ($69{,}007$ complete records), at minute $50$ from then until 9 November 2019 ($91{,}514$) and at minute $40$ afterwards ($26{,}717$); each record is assigned to the calendar year of its time stamp, as given in the archive, and the windows of Section~8 are calendar years. For observation $i$ the raw Euclidean block and the directional embedding are
\begin{gather*}
\bx_{1i}^{\mathrm{raw}}=(\mathrm{WVHT}_i,\mathrm{APD}_i,\mathrm{WSPD}_i)^{\top}\in\R^3,\\
\bx_{2i}=(\cos\omega_i,\sin\omega_i)^{\top}\in\sphere^1,\qquad \omega_i=\tfrac{\pi}{180}\,\mathrm{MWD}_i .
\end{gather*}

\paragraph{Missingness.}
Most years have $8{,}200$--$8{,}750$ complete hourly records out of $8{,}760$ or $8{,}784$ hours. Four years have fewer: 2008 ($6{,}357$), 2014 ($5{,}957$) and 2015 ($5{,}732$), in which $2{,}259$, $2{,}423$ and $2{,}440$ hourly records lack all three wave variables (WVHT, APD and MWD missing together, a wave-sensor gap), and 2017 ($6{,}278$), in which $2{,}377$ records lack wind speed. The analysis uses complete cases, so these years contribute fewer and seasonally unbalanced records to the training blocks that contain them and to their own test year; the annual scores of the test years 2008, 2014, 2015 and 2017 rest on $5{,}700$--$6{,}400$ observations.

\paragraph{Whole-degree directions.}
MWD is recorded in whole degrees. The $190{,}358$ non-missing MWD values of the $23$ files take $272$ distinct values ($310^\circ$ occurs $4{,}746$ times); the $187{,}238$ complete records take $271$ ($310^\circ$: $4{,}686$ times). Along such ties the mixture likelihood is unbounded: a component can put its mass on the records that share one direction and let $\kappa_j\to\infty$, while its conditional Gaussian factor stays regular and its effective size stays large, so the size check of Algorithm~\ref{alg:em-cyl} does not detect it. We therefore checked every retained fit of the buoy study ($19$ windows, three families, $K=1,\dots,8$; $2{,}052$ components) by recomputing its responsibilities on the training block, which reproduces the stored training log-likelihoods to $2\times10^{-15}$. The largest fitted concentration is $89.1$ (cylindrical) and $84.2$ (independent), and $106.3$ over all stored starts of the cylindrical mixture; at $\kappa=89$ a $1^\circ$ bin around the mean holds $6.6\%$ of the directional probability, against $62\%$ at $\kappa=10^4$. For the GvM components the largest probability of any $1^\circ$ bin is $5.9\%$. The smallest summed responsibility of a retained component is $206$ hours (effective size $223$), and no retained component has more than $8.6\%$ of its responsibility mass on a single recorded direction. No reported fit is therefore of the degenerate kind. The rounding also leaves the held-out scores unchanged: replacing each test density by its average over the $1^\circ$ bin of its recorded direction, which is the density of the data dequantized by uniform noise, changes every annual held-out log-density by at most $2.5\times10^{-4}$ nats and every macro contrast of Table~5 of the main text by at most $2\times10^{-5}$ nats.

\paragraph{Period and direction.}
DPD is the period of the most energetic spectral bin, so it is strongly quantized. APD is derived from spectral moments and is smoother, which suits a continuous density. The analysis therefore models the empirical dependence between APD and the dominant-period direction MWD, without treating APD as the period that defines MWD.

\paragraph{Transformation.}
Within each rolling window, each of the three Euclidean coordinates is transformed by the Yeo--Johnson power transformation~\cite{yeojohnson2000power},
\begin{equation*}
u\mapsto\begin{cases}\{(u+1)^{\lambda_{\mathrm{YJ}}}-1\}/\lambda_{\mathrm{YJ}}, & u\ge 0,\\[2pt] -\{(1-u)^{2-\lambda_{\mathrm{YJ}}}-1\}/(2-\lambda_{\mathrm{YJ}}), & u<0,\end{cases}
\end{equation*}
with the logarithmic limits $\log(u+1)$ at $\lambda_{\mathrm{YJ}}=0$ (for $u\ge0$) and $-\log(1-u)$ at $\lambda_{\mathrm{YJ}}=2$ (for $u<0$). One exponent per coordinate is estimated by maximum likelihood on the training block, followed by standardization to mean zero and unit variance on the same block (scikit-learn's \texttt{PowerTransformer}). The fitted map is applied unchanged to the test year. The direction is not transformed.

\section{The geometric test of the component assumption: computations, scores and simulations}\label{sec:esm-geomtest}

This section supports Section~5 of the main text. It gives the proof of Theorem~13 and the complements stated in Section~5.3 (Section~\ref{subsec:esm-geomtest-proofs}), the parts of $T_N$ and the special cases of Proposition~14 (Section~\ref{subsec:esm-geomtest-moments}), the per-observation scores and the computation of the mixture statistic (Section~\ref{subsec:esm-geomtest-scores}), the simulation designs and results (Section~\ref{subsec:esm-geomtest-simstudy}) and the numerical checks of every identity (Section~\ref{subsec:esm-geomtest-checks}). The notation is that of Section~5 of the main text; the code is in the directory \texttt{geomtest}.

\subsection{Complements to Section~5 of the main text}\label{subsec:esm-geomtest-proofs}

Appendix~B of the main text proves Propositions~14 and~15 and Theorems~16 and~17, including the moments of $t$, the limits at $\kappa=0$ and $\kappa\to\infty$ and the second derivative in Proposition~15(ii). This subsection proves Theorem~13 and the complements stated in Section~5.3 of the main text: Propositions~\ref{prop:esm-geomtest-separated} and~\ref{prop:esm-geomtest-hac} and the M-step identity, Corollary~\ref{cor:esm-geomtest-mstep}. The notation is that of Section~5 and Appendix~B; $\mathbf W$ is the Fisher information of $\FB_m$.

\begin{proposition}[Separated components and the plug-in statistic]\label{prop:esm-geomtest-separated}
Under the assumptions of Theorem~17, let $\mathcal J^{\mathrm c}$ be the complete-data information (labels observed) at $(\bTheta_0,\mathbf 0)$ and $\mathcal J^{\mathrm c}_{\mathrm{eff}}$ its efficient $\bOmega$-block. Then (i) $\mathcal J^{\mathrm c}_{\mathrm{eff}}=\blockdiag_j(\pi_j\Ieff(\bvartheta_j))$ and $\mathcal J_{\mathrm{eff}}\preceq\mathcal J^{\mathrm c}_{\mathrm{eff}}$; (ii) at a stationary point the plug-in statistic $P_N:=\sum_jN_j\langle\bDelta_j,\Ieff(\hat{\bvartheta}_j)^{-1}\bDelta_j\rangle$ equals $N^{-1}\mathbf s^{\natural\top}(\hat{\mathcal J}^{\mathrm c}_{\mathrm{eff}})^{-1}\mathbf s^{\natural}$ and converges to $\sum_i\omega_i\chi^2_{1,i}$, with independent $\chi^2_{1,i}$ and weights $\omega_i\in(0,1]$, the eigenvalues of $(\mathcal J^{\mathrm c}_{\mathrm{eff}})^{-1/2}\mathcal J_{\mathrm{eff}}(\mathcal J^{\mathrm c}_{\mathrm{eff}})^{-1/2}$, so it is stochastically smaller than $\chi^2_{Kq_m}$; (iii) if the missing information $\mathcal J^{\mathrm c}-\mathcal J$ tends to zero along a sequence of parameters (separating components), all $\omega_i\to1$ and $P_N$ and $\Tmix_N$ have the same limit.
\end{proposition}

\begin{proof}
The complete-data log-density is $\sum_jz_j\{\log\pi_j+\log f_{\mathrm{CG},j}+\log f_{\mathrm{FB},j}\}$ with label indicators $z_j$. Its scores for different components are orthogonal ($z_jz_k=0$), the weight scores $z_j/\pi_j-z_K/\pi_K$ are orthogonal to the component scores because these have conditional mean zero given the label, and within component $j$ the information is $\pi_j$ times the one-component information, block-diagonal in cut coordinates. The efficient $\bOmega_j$-block is therefore $\pi_j\Ieff(\bvartheta_j)$, and different $j$ do not interact. The missing-information principle~\cite{dempster1977em,louis1982em}, $\cov(\mathbf s^{\mathrm c})=\cov(\E[\mathbf s^{\mathrm c}\mid\bx])+\E\cov(\mathbf s^{\mathrm c}\mid\bx)$ with $\E[\mathbf s^{\mathrm c}\mid\bx]$ the observed score, gives $\mathcal J\preceq\mathcal J^{\mathrm c}$. Efficient blocks are monotone: $\mathcal J_{\mathrm{eff}}=([\mathcal J^{-1}]_{\Omega\Omega})^{-1}$ and $\mathcal J^{-1}\succeq(\mathcal J^{\mathrm c})^{-1}$. This proves (i). At a stationary point $\mathbf s^{\natural}=(N_j\bDelta_j)_j$ and $\hat\pi_j=N_j/N$, so $N^{-1}\sum_j(N_j\bDelta_j)^{\top}(\hat\pi_j\Ieff)^{-1}(N_j\bDelta_j)=P_N$; with $N^{-1/2}\mathbf s^{\natural}\to\calN(\mathbf 0,\mathcal J_{\mathrm{eff}})$ (proof of Theorem~17), $P_N\to\mathbf Y^{\top}(\mathcal J^{\mathrm c}_{\mathrm{eff}})^{-1}\mathbf Y$, $\mathbf Y\sim\calN(\mathbf 0,\mathcal J_{\mathrm{eff}})$, which is the stated weighted sum; $\omega_i\le1$ by (i). Part (iii) follows by continuity of efficient blocks in the information.
\end{proof}

\begin{corollary}[Exact increase in the M-step]\label{cor:esm-geomtest-mstep}
Let weights $r_n\ge0$ with $W:=\sum_nr_n>0$ satisfy (A1w)--(A2w), let $\hat{\btheta}$ maximize $Q(\btheta):=\sum_nr_n\log p_{\btheta}(\bx_n)$ over $\{\bOmega=\mathbf 0\}$ (Theorem~6 with weights), and let $\bDelta_r$ be the weighted unmatched moment. For every $\btheta$ with $\bOmega(\btheta)=\mathbf 0$, regression matrix $\bB$ and covariance $\bS$,
\[
Q(\hat{\btheta})-Q(\btheta)=W\bigl\{\KL[p_{\hat{\btheta}}:p_{\btheta}]+\tfrac12\tr\bigl((\hat{\bB}-\bB)^{\top}\bS^{-1}(\hat{\bB}-\bB)\bDelta_r\bigr)\bigr\}.
\]
\end{corollary}

\begin{proof}
With $\bar{\bt}_r:=W^{-1}\sum_nr_n\bt(\bx_n)$, $Q(\btheta)/W=\langle\btheta,\bar{\bt}_r\rangle-\psi_{\mathrm{amb}}(\btheta)=\phi_{\mathrm{amb}}(\bar{\bt}_r)-\KL[p_{\bar{\bt}_r}:p_{\btheta}]$, the weighted form of Proposition~8. So $\{Q(\hat{\btheta})-Q(\btheta)\}/W=\KL[p_{\bar{\bt}_r}:p_{\btheta}]-\KL[p_{\bar{\bt}_r}:p_{\hat{\btheta}}]$, and Proposition~15(iii), whose proof uses only the moments of $\bar{\bt}_r$, gives the result.
\end{proof}

Applied to component $j$ of an EM iteration, with $r_n=r_{nj}$, $\btheta$ the current and $\hat{\btheta}$ the updated parameter, the corollary is the statement in Section~5.3 of the main text: the M-step gain is a divergence plus the curvature defect, which can be negative, and then does not exceed the divergence in absolute value, since the left side is nonnegative.

\begin{proposition}[Calibration under serial dependence]\label{prop:esm-geomtest-hac}
Let $(\bx_n)_{n\ge1}$ be strictly stationary and ergodic, with marginal law the cylindrical mixture (18) of the main text at $\bTheta_0$, under (M1)--(M3) of Theorem~17 with $\mathcal J$ the marginal information. Let $\tilde{\bTheta}$ satisfy (M2), let $\mathbf u_n:=\mathbf s_n^{\Omega}-\hat{\mathcal J}_{\Omega\varsigma}\hat{\mathcal J}_{\varsigma\varsigma}^{-1}\mathbf s_n^{\varsigma}$ be the effective scores at $\tilde{\bTheta}$ with $\hat{\mathcal J}\to\mathcal J$, and assume (D1) $N^{-1/2}\sum_n\mathbf u_n(\bTheta_0)\to\calN(\mathbf 0,\boldsymbol\Sigma_{\mathrm{LR}})$ with $\boldsymbol\Sigma_{\mathrm{LR}}:=\sum_{k\in\Z}\cov(\mathbf u_0,\mathbf u_k)\succ\mathbf 0$; (D2) the Newey--West estimator $\hat{\boldsymbol\Sigma}_{\mathrm{LR}}$ of the series $\mathbf u_n$, with Bartlett weights and lag $L_N\to\infty$, is consistent for $\boldsymbol\Sigma_{\mathrm{LR}}$. Then the heteroskedasticity- and autocorrelation-consistent (HAC) statistic $T^{\mathrm{HAC}}_N:=N^{-1}\mathbf s^{\natural\top}\hat{\boldsymbol\Sigma}_{\mathrm{LR}}^{-1}\mathbf s^{\natural}\to\chi^2_{Kq_m}$. For $K=1$ the same holds with $\mathbf u_n$ the efficient influence $\boldsymbol\psi_n:=\mathbf Q(\bx_{2n})-\E\mathbf Q-\mathbf W_{21}\mathbf W_{11}^{-1}(\bx_{2n}-\E\bx_2)$ of $\bDelta$, $\mathbf Q(\bx_2):=\bx_2\bx_2^{\top}$, and no condition on $\bx_1$.
\end{proposition}

\begin{proof}
The expansion in the proof of Theorem~17 used only the uniform convergence of $N^{-1}\nabla_\varsigma\mathbf s$, which holds under stationarity and ergodicity by the ergodic theorem with the same integrable envelope, and (M2); it gives $N^{-1/2}\mathbf s^{\natural}=N^{-1/2}\sum_n\mathbf u_n(\bTheta_0)+o_p(1)$. By (D1) the leading term tends to $\calN(\mathbf 0,\boldsymbol\Sigma_{\mathrm{LR}})$ and by (D2) $\hat{\boldsymbol\Sigma}_{\mathrm{LR}}\to\boldsymbol\Sigma_{\mathrm{LR}}$. For $K=1$, $\sqrt N\bDelta=N^{-1/2}\sum_n\boldsymbol\psi_n+o_p(1)$ by the delta method of Theorem~13(iv), whose derivative $D\mathcal T$ is exactly the map defining $\boldsymbol\psi_n$.
\end{proof}

Conditions (D1) and (D2) hold, for instance, for strongly mixing sequences with summable mixing coefficients raised to a suitable power, the scores having all moments, with $L_N\to\infty$ and $L_N=o(N^{1/2})$~\cite{neweywest1987hac,andrews1991hac}. The moving-block bootstrap~\cite{kunsch1989block} of the effective-score series, with block length $\ell_N\to\infty$, $\ell_N/N\to0$, gives an alternative calibration. The code uses the default lag $\lfloor4(N/100)^{2/9}\rfloor$ unless a lag is given, and returns as an inflation factor the ratio of the traces of $\hat{\boldsymbol\Sigma}_{\mathrm{LR}}$ and of the efficient information of the fitted null model, the covariance of the effective scores under independent sampling from that model; on data that depart from the null, this ratio also contains the misfit.

\begin{proof}[Proof of Theorem~13 of the main text]
\emph{(i)} $\ell_{\mathrm{amb}}=\ell_{\CG}(\balpha,\bB,\bS)+N\{\langle(\bvartheta,\bOmega),\bar{\boldsymbol\tau}\rangle-\psi_{\FB}\}$. At $\hat{\bxi}=(\hat{\balpha},\hat{\bB},\hat{\bS},\hat{\bvartheta},\mathbf 0)$ the regression block maximizes $\ell_{\CG}$ and $\nabla\psi_{\vMF}(\hat{\bvartheta})=\bar{\bx}_2$ (Theorem~6, also when $\hat{\bvartheta}=\mathbf 0$), so the cut-coordinate score is $(\mathbf 0,\mathbf 0,N\bDelta)$. The Rao statistic $N^{-1}\mathbf s^{\top}g^{-1}\mathbf s$ is coordinate-free; in natural coordinates $\mathbf s=N(\bar{\bt}-\hat{\bfeta})$, giving $T_N$, and in cut coordinates it is $N\bDelta^{\top}[\mathbf W^{-1}]_{\Omega\Omega}\bDelta=N\langle\bDelta,\Ieff^{-1}\bDelta\rangle$. In expectation coordinates $\dot\gamma(0)=\bar{\bt}-\hat{\bfeta}$, whose natural expression is $(\nabla^2\psi_{\mathrm{amb}})^{-1}(\bar{\bt}-\hat{\bfeta})$, so $T_N=Ng(\dot\gamma(0),\dot\gamma(0))$ and $g(\dot\gamma(0),v)=N^{-1}d\ell_{\mathrm{amb}}(v)$, which vanishes on tangent vectors of $\{\bOmega=\mathbf 0\}$, the image of the open set $\Pext$ on which $\hat{\btheta}$ maximizes $\ell$. In cut coordinates $\dot\gamma(0)=g^{-1}(\mathbf 0,\mathbf 0,\bDelta)$ has no regression component.

\emph{(ii)} By Proposition~8, $\bar{\bt}\in\calM_{\mathrm{amb}}$, $\sup\ell_{\mathrm{amb}}=N\phi_{\mathrm{amb}}(\bar{\bt})$ and $\KL[p_{\bar{\bt}}:p_{\btheta}]=\phi_{\mathrm{amb}}(\bar{\bt})-\ell(\btheta)/N$ on $\{\bOmega=\mathbf 0\}$, so $\LRstat_N=2N\KL[p_{\bar{\bt}}:p_{\hat{\btheta}}]=2ND_{\phi_{\mathrm{amb}}}[\bar{\bt}:\hat{\bfeta}]$. Taylor's formula with integral remainder along the segment gives $D_{\phi_{\mathrm{amb}}}=\int_0^1(1-s)(\bar{\bt}-\hat{\bfeta})^{\top}\nabla^2\phi_{\mathrm{amb}}(\gamma(s))(\bar{\bt}-\hat{\bfeta})\,ds$, whose integrand is $g(\dot\gamma(s),\dot\gamma(s))$. The two laws share their conditional factor, so the chain rule leaves $\KL[\FB^{*}:\vMF_m(\hat{\bvartheta})]$. Since $T_N=2N\int_0^1(1-s)g(\dot\gamma(0),\dot\gamma(0))\,ds$, the bound follows.

\emph{(iii)} By (i)--(ii), $T_N$ and $\LRstat_N$ are functions of $\bx_{21},\dots,\bx_{2N}$, which under $\fCG\,\FB_m(\bvartheta,\bOmega)$ are independent $\FB_m(\bvartheta,\bOmega)$ draws whatever $(\balpha,\bB,\bS)$.

\emph{(iv)} For an absolutely continuous law, $N\ge m+1$ directions lie in an affine hyperplane with probability zero, and given (A1) the residual scatter of the Gaussian regression is positive definite almost surely when $N-m-1\ge d$. Write $\bDelta=\mathcal T(\bar{\boldsymbol\tau})$ with $\mathcal T(\mathbf y,\mathbf Y):=\mathbf Y+\bI_m/m-\bfeta_{22}(h(\mathbf y))$; then $\mathcal T(\bfeta_0)=\mathbf 0$ and $D\mathcal T(\bfeta_0)[(\mathbf y,\mathbf Y)]=\mathbf Y-\mathbf W_{21}\mathbf W_{11}^{-1}\mathbf y$. Under $\FB_m(\bvartheta,N^{-1/2}\boldsymbol\Upsilon)$, $\bfeta_N:=\E\boldsymbol\tau=\bfeta_0+N^{-1/2}\mathbf W(\mathbf 0,\boldsymbol\Upsilon)+O(N^{-1})$, and since $\boldsymbol\tau$ is bounded the Lindeberg--Feller theorem gives $\sqrt N(\bar{\boldsymbol\tau}-\bfeta_N)\to\calN(\mathbf 0,\mathbf W)$. The delta method gives $\sqrt N\bDelta\to\calN(D\mathcal T\,\mathbf W(\mathbf 0,\boldsymbol\Upsilon),D\mathcal T\,\mathbf W D\mathcal T^{\top})=\calN(\Ieff\boldsymbol\Upsilon,\Ieff)$; with $\Ieff(h(\bar{\bx}_2))\to\Ieff(\bvartheta)$, $T_N\to\chi^2_{q_m}(\langle\boldsymbol\Upsilon,\Ieff\boldsymbol\Upsilon\rangle)$, and $\boldsymbol\Upsilon=\mathbf 0$ is the null. By (ii) and the cut, $\LRstat_N=2ND_{\phi_{\FB}}[\bar{\boldsymbol\tau}:\hat{\boldsymbol\eta}]$ with $\hat{\boldsymbol\eta}=(\bar{\bx}_2,\bfeta_{22}(\hat{\bvartheta})-\bI_m/m)$ and residual $(\mathbf 0,\bDelta)$, and $|\LRstat_N-T_N|\le N\norm{\bDelta}^2\sup_s\norm{\nabla^2\phi_{\FB}(\hat{\boldsymbol\eta}+s(\mathbf 0,\bDelta))-\nabla^2\phi_{\FB}(\hat{\boldsymbol\eta})}\to0$ in probability. For a cylindrical $p$ with regression block $\boldsymbol\gamma$ and parameter $\bvartheta'$, the chain rule gives $\KL[p_N:p]\ge\KL[\FB_m(\bvartheta,N^{-1/2}\boldsymbol\Upsilon):\vMF_m(\bvartheta')]$, with equality when $\boldsymbol\gamma$ is the true block; since $\log\fvMF(\cdot;h(\E_N\bx_2))-\log\fvMF(\cdot;\bvartheta')$ is affine in $\bx_2$, the right side is minimized at $\bvartheta'=h(\E_N\bx_2)$, where it equals $D_{\phi_{\FB}}$ between means with residual $(\mathbf 0,\mathcal T(\bfeta_N))$, $\sqrt N\mathcal T(\bfeta_N)\to\Ieff\boldsymbol\Upsilon$; Taylor's formula gives $2N\inf_p\KL\to\langle\Ieff\boldsymbol\Upsilon,\Ieff^{-1}\Ieff\boldsymbol\Upsilon\rangle$.
\end{proof}

\subsection{Moments of the axial cosine and the special cases}\label{subsec:esm-geomtest-moments}

\emph{The parts of $T_N$.} With $t_n:=\hat{\bmu}_2^{\top}\bx_{2n}$, $\hat{\mathbf P}:=\bI_m-\hat{\bmu}_2\hat{\bmu}_2^{\top}$, $\delta_0:=N^{-1}\sum_nt_n^2-\E_{\hat\kappa}t^2$, $\boldsymbol\delta_{\mathrm c}:=N^{-1}\sum_nt_n\hat{\mathbf P}\bx_{2n}$, $\bar{\mathbf T}:=N^{-1}\sum_n\hat{\mathbf P}\bx_{2n}\bx_{2n}^{\top}\hat{\mathbf P}$ and (25) of the main text at $\hat\kappa$,
\begin{equation}\label{eq:esm-geomtest-parts}
T_N=\underbrace{\frac{mN\delta_0^2}{(m-1)\nu_{\mathrm{rad}}}}_{T_{\mathrm{rad}}}+\underbrace{\frac{2N\norm{\boldsymbol\delta_{\mathrm c}}^2}{\nu_{\mathrm{cpl}}}}_{T_{\mathrm{cpl}}}+\underbrace{\frac{N}{\nu_{\mathrm{tan}}}\Bigl\lVert\bar{\mathbf T}-\frac{\tr\bar{\mathbf T}}{m-1}\hat{\mathbf P}\Bigr\rVert_F^2}_{T_{\mathrm{tan}}},
\end{equation}
three terms that are asymptotically independent $\chi^2_1$, $\chi^2_{m-1}$, $\chi^2_{m(m-1)/2-1}$ under $\bOmega=\mathbf 0$ (proof of Proposition~14 above).

The moments $\E t^k$ of Proposition~14 and their limits at $\kappa=0$ are derived at the end of Appendix~B.

The uniform law on the unit sphere of the $(m-1)$-dimensional space $\bmu_2^{\perp}$ has $\E w_iw_jw_kw_l=(\delta_{ij}\delta_{kl}+\delta_{ik}\delta_{jl}+\delta_{il}\delta_{jk})/((m-1)(m+1))$ in an orthonormal basis of $\bmu_2^{\perp}$, so $\E(\mathbf w^{\top}\mathbf C\mathbf w)^2=\{2\tr(\mathbf C^2)+(\tr\mathbf C)^2\}/((m-1)(m+1))$, which is the value used for traceless $\mathbf C$.

\emph{The circle, $m=2$.} With $\bx_{2n}=(\cos\phi_n,\sin\phi_n)^{\top}$ and $\hat\phi$ the angle of $\hat{\bmu}_2$, $t_n=\cos(\phi_n-\hat\phi)$ and $\hat{\mathbf P}\bx_{2n}=\sin(\phi_n-\hat\phi)\,\mathbf e^{\perp}$, so $t_n^2=\{1+\cos2(\phi_n-\hat\phi)\}/2$ and $t_n\sin(\phi_n-\hat\phi)=\tfrac12\sin2(\phi_n-\hat\phi)$. Since $\E t^2=(1+\varrho_2)/2$ when $\nu_0=0$, $\delta_0=(\bar C_2-\varrho_2)/2$ and $\norm{\boldsymbol\delta_{\mathrm c}}=|\bar S_2|/2$, and the tangential subspace is $\{\mathbf 0\}$. With $m/(m-1)=2$, \eqref{eq:esm-geomtest-parts} gives its two parts, $T_{\mathrm{rad}}=N(\bar C_2-\varrho_2)^2/(2\nu_{\mathrm{rad}})$ and $T_{\mathrm{cpl}}=N\bar S_2^2/(2\nu_{\mathrm{cpl}})$, with $\bar C_2+i\bar S_2:=N^{-1}\sum_ne^{2i(\phi_n-\hat\phi)}$. For $\bOmega=\begin{psmallmatrix}\omega_1&\omega_2\\\omega_2&-\omega_1\end{psmallmatrix}$, $\bx_2^{\top}\bOmega\bx_2=\omega_1\cos2\phi+\omega_2\sin2\phi$, so $\FB_2$ is the generalized von Mises family~\cite{gattojammalamadaka2007gvm}, and $T_N$ is a function of the second trigonometric moments about the mean direction, as in Cox's test of the von Mises law~\cite{cox1975discussion}.

\emph{The sphere, $m=3$.} $\bar{\mathbf T}$ has eigenvalue $0$ on $\hat{\bmu}_2$ and eigenvalues $l_1\ge l_2$ on $\hat{\bmu}_2^{\perp}$, so $\norm{\bar{\mathbf T}-\tfrac12\tr(\bar{\mathbf T})\hat{\mathbf P}}_F^2=(l_1-l_2)^2/2$, and $\nu_{\mathrm{tan}}=2\varrho_2/\kappa^2$ with $\varrho_2=I_{5/2}/I_{1/2}$ gives $T_{\mathrm{tan}}=N\kappa^2(l_1-l_2)^2/(4\varrho_2)$, Kent's statistic.

\emph{Vanishing concentration.} At $\hat\kappa=0$, $\hat{\bfeta}_{22}=\bI_m/m$ and $\Ieff=\tfrac2{m(m+2)}\mathrm{Id}$, so $T_N=\tfrac{m(m+2)}2N\norm{\bar{\mathbf M}_{22}-\bI_m/m}_F^2=\tfrac{m(m+2)}2N\{\tr(\bar{\mathbf M}_{22}^2)-1/m\}$, Bingham's statistic~\cite{bingham1974antipodal}. As $\kappa\downarrow0$, $\bfeta_{22}=(\Am/\kappa)\bI_m+(1-m\Am/\kappa)\bmu_2\bmu_2^{\top}\to\bI_m/m$ uniformly in $\bmu_2$, because $\Am(\kappa)/\kappa\to1/m$, and the eigenvalues of $\Ieff$ do not depend on $\bmu_2$, which gives the limit.

\subsection{Scores of the ambient mixture and computation of the statistic}\label{subsec:esm-geomtest-scores}

At a null parameter, with responsibilities $r_{nj}$ computed at that parameter, residuals $\beps_{nj}:=\bx_{1n}-\balpha_j-\bB_j\bx_{2n}$, $\mathbf L_j:=\bS_j^{-1}$ and $\bfeta_{2,j}:=\Am(\kappa_j)\bmu_{2,j}$, the per-observation scores of $\log p(\bx_n;\bTheta,\bomega)$ at $\bomega=\mathbf 0$ are
\begin{itemize}
\item weights, $j<K$ (with $\pi_K=1-\sum_{j<K}\pi_j$): $r_{nj}/\pi_j-r_{nK}/\pi_K$;
\item intercept and regression matrix: $r_{nj}\mathbf L_j\beps_{nj}$ and $r_{nj}\mathbf L_j\beps_{nj}\bx_{2n}^{\top}$;
\item conditional covariance, entry $(a,b)$ with $a\le b$: $(2-\delta_{ab})\,r_{nj}\,\tfrac12\bigl[\mathbf L_j\beps_{nj}\beps_{nj}^{\top}\mathbf L_j-\mathbf L_j\bigr]_{ab}$;
\item directional natural parameter $\bvartheta_j=\kappa_j\bmu_{2,j}$: $r_{nj}(\bx_{2n}-\bfeta_{2,j})$;
\item $\bOmega_j$, in a Frobenius-orthonormal basis $\{\mathbf E_i\}$ of $\Symzero{m}$ adapted to $\bmu_{2,j}$ (radial, coupling, tangential): $r_{nj}\langle\mathbf E_i,\bx_{2n}\bx_{2n}^{\top}-\bfeta_{22,j}\rangle$.
\end{itemize}
The test does not depend on the coordinates chosen for the null parameters, because changing them multiplies the null block of the scores by an invertible matrix, which leaves $\hat{\mathbf J}_{\mathrm{eff}}$ and $\mathbf s^{\natural}$ unchanged.

\emph{Outer-product form.} Let $\mathbf U=[\mathbf U_\varsigma\ \mathbf U_\Omega]$ be the $N\times p$ matrix with rows $\mathbf s_n^{\top}$, $\mathbf 1$ the vector of ones, and $\hat{\mathbf J}=\mathbf U^{\top}\mathbf U/N$. Then $\mathbf s^{\natural}=\mathbf U_\Omega^{\top}(\bI-\mathbf P_\varsigma)\mathbf 1$ and $N\hat{\mathbf J}_{\mathrm{eff}}=\mathbf U_\Omega^{\top}(\bI-\mathbf P_\varsigma)\mathbf U_\Omega$, with $\mathbf P_\varsigma$ the orthogonal projector onto the columns of $\mathbf U_\varsigma$, so $\Tmix_N=\mathbf 1^{\top}\mathbf P_{(\bI-\mathbf P_\varsigma)\mathbf U_\Omega}\mathbf 1=\mathbf 1^{\top}(\mathbf P_{\mathbf U}-\mathbf P_\varsigma)\mathbf 1$, the increase of the explained sum of squares in the regression of $\mathbf 1$ on the scores. The component statistic $\Tmix_{N,j}$ is the same with $\mathbf U_\Omega$ replaced by the columns of $\bOmega_j$, the other $\bOmega_k$ being fixed at $\mathbf 0$, not profiled.

\emph{Expected information.} The recommended estimate is the expected information of the ambient mixture at the fit, $\mathcal J(\tilde{\bTheta})=\E_{\tilde{\bTheta}}[\mathbf s\mathbf s^{\top}]$, which has no closed form for $K\ge2$; the code averages $\mathbf s\mathbf s^{\top}$ over a fixed number $n_{\mathrm{mc}}=2\times10^4$ of draws from the fitted null mixture. With $n_{\mathrm{mc}}$ fixed this approximates $\mathcal J(\tilde{\bTheta})$, with entrywise Monte Carlo error of relative order $n_{\mathrm{mc}}^{-1/2}\approx10^{-2}$, amplified in the statistic by the conditioning of $\hat{\mathcal J}_{\mathrm{eff}}$; Theorem~17 covers it only if $n_{\mathrm{mc}}=n_{\mathrm{mc},N}\to\infty$, and its finite-sample behaviour, including the Monte Carlo variability of the statistic, is reported in Section~\ref{subsec:esm-geomtest-simstudy}.

\emph{Refinement.} Before the scores are taken, the fit is moved to a stationary point of the null likelihood by exact EM steps started at it: the M-step of Lemma~\ref{lem:cyl-mle-weighted} with the exact inverse of $\Am$ and without ridge, floors or clipping, iterated until the relative change of the log-likelihood is below $10^{-10}$. At a stationary point $\sum_n\mathbf s_n^{\varsigma}=\mathbf 0$ and $\mathbf s^{\natural}=(N_j\bDelta_j)_j$. Exact EM does not guarantee that the stationary point reached is the consistent root required by condition (M2) of Theorem~17; this is assumed.

\emph{Parametric bootstrap.} For $b=1,\dots,B$: draw $N$ observations from the fitted null mixture; refit by exact EM started at the fitted parameters; compute $\Tmix_{N,b}$. The bootstrap $p$-value is $(1+\#\{b:\Tmix_{N,b}\ge\Tmix_N\})/(B+1)$~\cite{davisonhinkley1997bootstrap}. For one component the refit is the exact von Mises--Fisher fit of the simulated directions, drawn by Wood's algorithm~\cite{wood1994vmfsim}.

\emph{The plug-in diagnostic.} Applying the one-component statistic with weights $r_{nj}$, $\sum_jN_j\langle\bDelta_j,\Ieff(\hat\kappa_j\hat{\bmu}_{2,j})^{-1}\bDelta_j\rangle$ ignores the uncertainty of the memberships and the estimation of the other components. By Proposition~\ref{prop:esm-geomtest-separated} it is the C$(\alpha)$ statistic with the complete-data information, asymptotically a weighted sum of $\chi^2_1$ variables with weights in $(0,1]$: not a valid test, and in the simulations below it rejected in 0.5--1.5\% of the samples at the 5\% level. The code returns it only when $\bx_1$ is unavailable.

\subsection{Simulation designs and results}\label{subsec:esm-geomtest-simstudy}

All runs use \texttt{numpy} seed sequences spawned from the integer 20261103 and a fixed tag per cell (\texttt{sim\_size\_power.py}); every sample is fitted by exact EM started at the generating parameters, which isolates the behaviour of the test from that of the initialization. The practical procedure, ten $k$-means++ starts and a split start followed by exact EM, was checked only in design M1 at $N=300$ and $1{,}000$, with 200 samples each (\texttt{sim\_multistart.py}; standard error about 1.6 points): the selected fit coincided with the one started at the truth in 199 and 200 samples, and $\Tmix_N$ rejected in 6.5\% of the samples at the 5\% level in both cases (2.0\% and 1.5\% at the 1\% level), against 3.5\% and 3.9\% with starts at the truth (Table~\ref{tab:esm-geomtest-mix}). Designs M2 and M3, in which local optima and label switching matter more, were not checked in this way.

\emph{One component.} Directions were drawn from $\vMF_m(\mathbf e_1,\kappa)$, 4{,}000 samples per cell (Monte Carlo standard error of a 5\% rejection rate: 0.34 percentage points). Table~\ref{tab:esm-geomtest-single} gives the rejection rates of $T_N$ and Table~\ref{tab:esm-geomtest-lr} compares $T_N$ with the likelihood ratio, computed for $m\in\{2,3\}$ by fitting $\FB_m$ to the directions with a product quadrature rule (2{,}000 samples per cell).

At the 5\% level the rates in Table~\ref{tab:esm-geomtest-single} lie between 3.2\% and 6.2\%. The far tail is less accurate at high concentration, where $T_N$ has a heavier tail than $\chi^2_{q_m}$, with an excess that decreases slowly with $N$. At $\kappa=16$, $m=2,3$ and $N\le300$ the rate at the 1\% level is 1.5--2.0\% (standard error 0.16 points), and at the 0.1\% level, computed from the same stored samples, 0.35--0.85\% at $N\le300$ and 0.18--0.30\% at $N=1{,}000$ (standard error 0.05 points); at $m=2$, $\kappa=4$ the excess is largest at $N\le100$ (1.7--1.8\% at the 1\% level). The variance of $T_N$ exceeded $2q_m$ by up to 81\% ($m=2$, $\kappa=16$, $N=100$). The limit in Theorem~13(iv) is pointwise in $\kappa$: at high concentration $\nu_{\mathrm{rad}}$ is of order $\kappa^{-4}$ and the radial summand $t^2$ is close to an affine function of a $\mathrm{Gamma}((m-1)/2)$ variable (Appendix~B), which is strongly skewed. A $\chi^2$ $p$-value far below 1\% is therefore not accurate at high concentration; the parametric bootstrap of Section~\ref{subsec:esm-geomtest-scores} gives an alternative.

\begin{table}[ht]
\caption{Rejection rates (\%) of $T_N$ under $H_0$ at the 5\% and 1\% levels; 4{,}000 samples per cell}
\label{tab:esm-geomtest-single}
\footnotesize
\begin{tabular}{@{}rrrrrrrrrr@{}}
\toprule
& & \multicolumn{2}{c}{$N=30$} & \multicolumn{2}{c}{$N=100$} & \multicolumn{2}{c}{$N=300$} & \multicolumn{2}{c}{$N=1000$}\\
$m$ & $\kappa$ & 5\% & 1\% & 5\% & 1\% & 5\% & 1\% & 5\% & 1\%\\
\midrule
2 & 0 & 5.0 & 1.2 & 5.2 & 0.9 & 5.2 & 0.9 & 5.0 & 1.1 \\
2 & 1 & 4.8 & 0.8 & 5.1 & 0.9 & 5.6 & 0.9 & 4.8 & 1.1 \\
2 & 4 & 4.1 & 1.8 & 4.7 & 1.7 & 4.5 & 1.1 & 5.3 & 1.4 \\
2 & 16 & 3.2 & 1.6 & 4.6 & 2.0 & 5.4 & 1.6 & 5.2 & 1.5 \\
3 & 0 & 5.9 & 1.3 & 5.1 & 1.0 & 4.4 & 0.7 & 4.5 & 1.0 \\
3 & 1 & 5.5 & 1.1 & 5.5 & 0.9 & 5.1 & 1.0 & 4.8 & 0.9 \\
3 & 4 & 4.5 & 1.5 & 4.6 & 1.1 & 4.7 & 0.9 & 5.3 & 1.0 \\
3 & 16 & 4.2 & 2.0 & 5.3 & 1.7 & 4.5 & 1.5 & 4.5 & 1.1 \\
5 & 0 & 6.2 & 1.6 & 5.1 & 1.1 & 4.8 & 1.1 & 4.5 & 0.8 \\
5 & 1 & 5.8 & 1.2 & 5.0 & 1.0 & 5.3 & 1.0 & 4.9 & 0.9 \\
5 & 4 & 6.0 & 1.5 & 4.9 & 1.0 & 5.0 & 0.9 & 4.9 & 0.8 \\
5 & 16 & 5.6 & 1.7 & 5.4 & 1.6 & 5.3 & 1.0 & 5.5 & 1.3 \\
\bottomrule
\end{tabular}
\end{table}

\begin{table}[ht]
\caption{Rejection rates (\%) at the 5\% level of the score statistic $T_N$ and of the likelihood ratio $\LRstat_N$ under $H_0$; 2{,}000 samples per cell}
\label{tab:esm-geomtest-lr}
\footnotesize
\begin{tabular}{@{}rrrrrrrr@{}}
\toprule
& & \multicolumn{2}{c}{$N=30$} & \multicolumn{2}{c}{$N=100$} & \multicolumn{2}{c}{$N=300$}\\
$m$ & $\kappa$ & $T_N$ & $\LRstat_N$ & $T_N$ & $\LRstat_N$ & $T_N$ & $\LRstat_N$\\
\midrule
2 & 1 & 5.5 & 5.8 & 4.5 & 4.9 & 5.4 & 5.5 \\
2 & 4 & 4.3 & 8.6 & 4.2 & 5.7 & 3.9 & 4.8 \\
3 & 1 & 5.4 & 6.9 & 6.2 & 6.2 & 5.2 & 5.3 \\
3 & 4 & 5.1 & 8.2 & 3.8 & 5.0 & 4.3 & 5.0 \\
\bottomrule
\end{tabular}
\end{table}

\emph{Mixtures.} Three null designs, with directions $\vMF_m$ in every component: M1, $d=1$, $m=2$, $K=2$, $\bpi=(0.5,0.5)$, $\balpha=(0,2)$, $\bB_1=(1,0.5)$, $\bB_2=(-0.5,0.8)$, $\bS=(0.5,0.4)$, mean angles $0$ and $2$ radians, $\kappa=(4,2)$; M2, $d=3$, $m=2$, $K=3$, $\bpi=(0.4,0.35,0.25)$, mean angles $0$, $\pi/2$ and $5\pi/4$, $\kappa=(3,5,2)$, $\bS_j\in\{0.5\bI_3,0.4\bI_3,0.6\bI_3\}$; M3, $d=2$, $m=3$, $K=2$, $\bpi=(0.6,0.4)$, mean directions $\mathbf e_3$ and $\mathbf e_1$, $\kappa=(4,6)$. The remaining arrays are in \texttt{sim\_size\_power.py}. Table~\ref{tab:esm-geomtest-mix} compares three estimates of the information (1{,}000 samples per row; standard error 0.7 points).

\begin{table}[ht]
\caption{Rejection rates (\%) at the 5\% level of $\Tmix_N$ (overall) and of $\Tmix_{N,j}$ (averaged over components) under $H_0$, with the expected information at the fit (expected), the outer-product estimate (OPG) and the plug-in diagnostic; 1{,}000 samples per row}
\label{tab:esm-geomtest-mix}
\footnotesize
\begin{tabular}{@{}lrrrrrrrrrrr@{}}
\toprule
& & & & & & \multicolumn{2}{c}{expected} & \multicolumn{2}{c}{OPG} & \multicolumn{2}{c}{plug-in}\\
design & $d$ & $m$ & $K$ & $N$ & df & overall & comp. & overall & comp. & overall & comp.\\
\midrule
M1 & 1 & 2 & 2 & 300 & 4 & 3.5 & 3.7 & 23.6 & 16.8 & 0.7 & 1.2 \\
M1 & 1 & 2 & 2 & 1000 & 4 & 3.9 & 4.2 & 11.5 & 9.0 & 0.7 & 1.2 \\
M1 & 1 & 2 & 2 & 3000 & 4 & 5.9 & 5.3 & 6.7 & 6.5 & 0.8 & 1.8 \\
M2 & 3 & 2 & 3 & 500 & 6 & 5.4 & 5.2 & 30.3 & 18.1 & 1.1 & 1.7 \\
M2 & 3 & 2 & 3 & 2000 & 6 & 6.1 & 5.1 & 10.8 & 7.6 & 0.8 & 1.8 \\
M3 & 2 & 3 & 2 & 500 & 10 & 5.8 & 5.3 & 24.6 & 17.4 & 1.5 & 1.9 \\
M3 & 2 & 3 & 2 & 2000 & 10 & 4.8 & 4.5 & 9.9 & 8.2 & 0.5 & 1.4 \\
\bottomrule
\end{tabular}
\end{table}

\emph{Bootstrap.} With $B=199$, the one-component test at $m=3$, $\kappa=4$, $N=30$ rejected in 4.5\% of 1{,}000 samples (asymptotic: 5.4\%). With $B=99$, the bootstrap of the outer-product statistic rejected in 3.3\% of 400 samples of M1 at $N=300$ and in 1.5\% of 200 samples of M2 at $N=500$ (asymptotic: 25.0\% and 25.5\% in the same samples): it repairs the size of the outer-product form, somewhat conservatively.

\emph{Power, one component} (2{,}000 samples per cell; Table~\ref{tab:esm-geomtest-power}). On the circle, generalized von Mises (GvM) laws with $\kappa=2$ and second-harmonic coefficient $\omega$ aligned with the mean (radial) or at $45^\circ$ to it (coupling); on $\sphere^2$, with $\kappa=4$, the five-parameter Fisher--Bingham (FB5) law with $\bOmega=\beta\,\diag(1,-1,0)$ about $\bmu_2=\mathbf e_3$ and the rotationally symmetric law with $\bOmega=\omega\mathbf R$; and, on the circle, projected normal directions $\mathbf z/\norm{\mathbf z}$, $\mathbf z\sim\calN((1.5,0)^{\top},\diag(1,s^2))$, as in the generators of Section~7 of the main text.

\begin{table}[ht]
\caption{Power (\%) at the 5\% level of $T_N$ and of its parts against Fisher--Bingham and projected-normal alternatives; ``Kent'' is the tangential part for $m=3$; 2{,}000 samples per cell}
\label{tab:esm-geomtest-power}
\footnotesize
\begin{tabular}{@{}llrrrrr@{}}
\toprule
alternative & parameter & $N$ & $T_N$ & Kent & $T_{\mathrm{rad}}$ & $T_{\mathrm{cpl}}$ \\
\midrule
GvM, radial, $m=2$ & $\omega=0.3$ & 100 & 22.1 & -- & 26.1 & 5.6 \\
 & & 200 & 37.8 & -- & 45.4 & 6.0 \\
 & & 500 & 74.0 & -- & 81.1 & 5.2 \\
\addlinespace[2pt]
GvM, coupling, $m=2$ & $\omega=0.3$ & 100 & 23.1 & -- & 8.0 & 28.8 \\
 & & 200 & 44.0 & -- & 10.1 & 50.8 \\
 & & 500 & 81.0 & -- & 15.7 & 85.2 \\
\addlinespace[2pt]
FB5, $m=3$ & $\beta=0.5$ & 100 & 18.9 & 30.8 & 3.5 & 4.5 \\
 & & 200 & 40.0 & 54.8 & 4.3 & 5.3 \\
 & & 500 & 82.5 & 92.3 & 4.5 & 5.3 \\
\addlinespace[2pt]
FB5, $m=3$ & $\beta=1$ & 100 & 69.2 & 84.6 & 3.2 & 5.8 \\
 & & 200 & 97.0 & 99.2 & 4.0 & 6.8 \\
 & & 500 & 100 & 100 & 5.3 & 5.5 \\
\addlinespace[2pt]
radial, $m=3$ & $\omega=1$ & 100 & 24.9 & 4.0 & 27.6 & 12.2 \\
 & & 200 & 37.2 & 5.5 & 42.6 & 11.3 \\
 & & 500 & 62.7 & 5.1 & 74.7 & 12.5 \\
\addlinespace[2pt]
projected normal & $s=1$ & 100 & 19.4 & -- & 23.4 & 5.9 \\
 & & 200 & 35.0 & -- & 40.8 & 7.3 \\
 & & 1000 & 93.1 & -- & 96.2 & 5.9 \\
\addlinespace[2pt]
projected normal & $s=0.8$ & 50 & 41.5 & -- & 46.5 & 9.0 \\
 & & 100 & 69.8 & -- & 76.0 & 10.4 \\
 & & 200 & 94.5 & -- & 96.8 & 9.1 \\
\addlinespace[2pt]
projected normal & $s=0.6$ & 50 & 86.0 & -- & 87.4 & 16.8 \\
 & & 100 & 98.8 & -- & 99.4 & 16.5 \\
 & & 200 & 100 & -- & 100 & 16.2 \\
\bottomrule
\end{tabular}
\end{table}

The coupling part rejects the projected normal law at a rate that does not grow with $N$ (16--17\% for $s=0.6$): the law is symmetric about its mean direction, so the coupling part of $\bDelta$ has mean zero, but its variance is not the von Mises--Fisher one. The parts locate a departure; they are not separate tests under an alternative.

\emph{Power, mixture.} Design M1 with the direction of the second component replaced by $\FB_2(\kappa_2\bmu_{2,2},\bOmega)$, $\bOmega$ of coupling type relative to $\bmu_{2,2}$ with coefficient $\omega$, 500 samples per cell. With the expected information, at $N=1{,}000$ and $\omega=0,0.25,0.5,0.75,1$, $\Tmix_N$ rejected in 3.2, 12.8, 35.8, 54.2 and 66.6\% of the samples, $\Tmix_{N,2}$ in 4.4, 16.4, 48.2, 64.4 and 75.8\%, and $\Tmix_{N,1}$, the unaltered component, in 4.6, 3.6, 4.2, 4.6 and 4.6\%; at $N=300$ the overall rates were 6.2, 4.6, 10.0, 14.4 and 20.8\%.

Figure~2 of the main text shows the size of the mixture statistic, the power of $T_N$ and of Kent's block, and the coverage of Wald intervals for $\bB$.

\begin{table}[ht]
\caption{Coverage (\%) of nominal 95\% Wald intervals for the entries of $\bB$ with the expected information (E) and the observed information (O). One component: $d=1$, $\kappa=4$, 10{,}000 samples per $N$; ``all'' averages over the samples, ``low'' is the fifth with the smallest ancillary $\Delta_{jj}/V_{jj}$ in the frame of $\hat{\bmu}_2$. Mixture: two components on $\R\times\sphere^1$ ($N=2N_j$, 600 samples), E and O the expected and observed information of the mixture; ``weighted design'' uses the responsibility-weighted design of each component, which ignores the uncertainty of the memberships}
\label{tab:esm-geomtest-coverage}
\footnotesize
\begin{tabular}{@{}lccccccccc@{}}
\toprule
& & \multicolumn{2}{c}{$N_j=30$} & \multicolumn{2}{c}{$N_j=50$} & \multicolumn{2}{c}{$N_j=100$} & \multicolumn{2}{c}{$N_j=200$}\\
& & E & O & E & O & E & O & E & O\\
\midrule
one, $m=2$ & radial, all & 88.1 & 92.4 & 91.1 & 93.8 & 93.1 & 94.4 & 94.1 & 94.6 \\
 & radial, low & 80.8 & 93.2 & 85.2 & 93.7 & 90.2 & 94.6 & 91.2 & 94.3 \\
one, $m=2$ & tangential, all & 91.7 & 92.7 & 93.2 & 93.8 & 94.5 & 94.6 & 94.7 & 94.8 \\
 & tangential, low & 90.6 & 93.5 & 92.0 & 94.2 & 94.0 & 94.9 & 93.8 & 94.5 \\
one, $m=3$ & radial, all & 88.5 & 92.2 & 91.8 & 93.5 & 93.2 & 94.1 & 94.7 & 94.8 \\
 & radial, low & 79.5 & 91.5 & 87.5 & 93.5 & 89.2 & 93.8 & 93.4 & 95.0 \\
one, $m=3$ & tangential, all & 90.8 & 92.3 & 92.7 & 93.3 & 93.8 & 94.3 & 94.3 & 94.5 \\
 & tangential, low & 86.7 & 92.0 & 90.2 & 93.6 & 91.2 & 94.0 & 93.2 & 94.5 \\
mixture, $m=2$ & all entries & 81.2 & 84.1 & 85.4 & 86.7 & 90.2 & 91.4 & 92.1 & 92.6 \\
 & weighted design & 74.1 & 76.8 & 76.6 & 77.3 & 80.8 & 81.3 & 83.0 & 83.4 \\
\bottomrule
\end{tabular}
\end{table}

\emph{Coverage of Wald intervals for $\bB$.} One component: $d=1$, $\balpha=0.2$, $\bS=0.5$, $\kappa=4$, $\bmu_2=\mathbf e_1$, $\bB=(1,0.5)$ for $m=2$ and $(1,0.5,-0.5)$ for $m=3$; 10{,}000 samples per $N\in\{30,50,100,200\}$; intervals $\hat B_{1j}\pm1.96\,\mathrm{se}$ with $\mathrm{se}^2=\hat S[\mathbf V(\hat{\bvartheta})^{-1}]_{jj}/N$ (expected) or $\hat S[\bar{\mathbf V}^{-1}]_{jj}/N$ (observed), $\hat S$ the maximum-likelihood residual variance, which explains the undercoverage of both at $N=30$. The ancillary of entry $j$ is $\Delta_{jj}/V_{jj}$ in the frame $(\hat{\bmu}_2,\hat{\bmu}_2^{\perp})$. Mixture: design M1 below with $N=2N_j$, 600 samples per $N$, fitted by exact EM; ``expected'' and ``observed'' are the inverse of $N\mathcal J(\tilde{\bTheta})$ (Monte Carlo, $2\times10^4$ draws) and of the numerical Hessian of the mixture log-likelihood; ``weighted design'' uses $N_j\mathbf V(\hat{\bvartheta}_j)$ and $N_j\bar{\mathbf V}_j$, $\bar{\mathbf V}_j$ the responsibility-weighted covariance of the directions. The information loss $\iota$ of Theorem~16 is only part of the second-order variance of $\hat{\bB}$: the coverage of the expected-information intervals is governed by the full second-order ratio $N(\E[\bar{\mathbf V}^{-1}]_{\parallel}/\E[\mathbf V(\hat{\bvartheta})^{-1}]_{\parallel}-1)$, which expands as $\iota_\parallel-N\E[\Delta_{\parallel}]/\sigma_\parallel^2+o(1)$, $\Delta_\parallel:=\hat{\bmu}_2^{\top}\bDelta\hat{\bmu}_2$ having a mean of order $N^{-1}$; at $m=2$, $\kappa=4$ the two terms are $7.84$ and about 2--3, depending on $N$, and the ratio is about $10$ for $N$ from 30 to 800.

\emph{Non-Gaussian conditionals, conditioning and Monte Carlo error.} Designs M1 ($N=1{,}000$) and M2 ($N=2{,}000$), 400 samples each, with Gaussian or $t_5$ errors of the same covariance in $\bx_1\mid\bx_2$; $\Tmix_N$ with the expected information. Rejection rates at 5\%: M1 gauss 3.2\%, M1 t5 12.0\%, M2 gauss 6.5\%, M2 t5 27.8\%. The condition number of $\hat{\mathcal J}_{\mathrm{eff}}$ had median 4.7, 3.8, 23.0, 24.7 and maximum 513.5; the statistic recomputed with an independent set of $2\times10^4$ Monte Carlo draws differed by a median relative amount of 4.5\%, 4.1\%, 5.4\%, 4.8\%.

\emph{Serial dependence.} Directions $\phi_n$ with von Mises$(0,2)$ marginal obtained from a Gaussian AR(1) series $z_n$, $\phi_n=F^{-1}(\Phi(z_n))$, lag-one correlation $\phi\in\{0,0.5,0.9\}$, $N=2{,}000$, 800 samples each; rejection rates at 5\% of the independent-sampling calibration, of the Newey--West calibration with the default lag ($\lfloor4(N/100)^{2/9}\rfloor=7$) and with lag 50, and of the moving-block bootstrap (199 resamples, block length 8): $\phi=0$: 5.2, 5.2, 6.5, 4.8\%; $\phi=0.5$: 7.2, 5.8, 6.8, 5.2\%; $\phi=0.9$: 54.1, 13.0, 9.0, 12.2\%. The median variance inflation was 1.00, 1.12, 3.17.

\begin{proposition}[Fixed-$b$ and self-normalized calibration, one component]\label{prop:esm-geomtest-fixedb}
Let $(\bx_{2n})_{n\ge1}$ be strictly stationary and ergodic with marginal $\vMF_m(\bvartheta_0)$, let $\boldsymbol\psi_n(\bvartheta)$ be the efficient influence of $\bDelta$ (Proposition~\ref{prop:esm-geomtest-hac}), and assume $\sqrt N(\bar{\bx}_2-\E\bx_2)=O_p(1)$ and the functional central limit theorem $N^{-1/2}\sum_{n\le\lfloor rN\rfloor}\boldsymbol\psi_n(\bvartheta_0)\Rightarrow\boldsymbol\Sigma_{\mathrm{LR}}^{1/2}\mathbf W_q(r)$ in $D[0,1]$, with $\mathbf W_q$ a standard $q$-dimensional Brownian motion, $q=q_m$, and $\boldsymbol\Sigma_{\mathrm{LR}}\succ\mathbf 0$. Let $T^{\mathrm{fb}}_N:=N\bar{\boldsymbol\psi}^{\top}\hat{\boldsymbol\Omega}_b^{-1}\bar{\boldsymbol\psi}$, with $\bar{\boldsymbol\psi}$ the mean of $\boldsymbol\psi_n(\hat{\bvartheta})$ (the coordinates of $\bDelta$) and $\hat{\boldsymbol\Omega}_b:=\sum_{|l|<M_N}(1-|l|/M_N)\hat{\boldsymbol\Gamma}(l)$ the Bartlett estimator, $\hat{\boldsymbol\Gamma}(l)$ the lag-$l$ sample autocovariance (divisor $N$) of the centred series $\boldsymbol\psi_n(\hat{\bvartheta})-\bar{\boldsymbol\psi}$ and $M_N/N\to b\in(0,1]$, $b$ fixed, and $T^{\mathrm{sn}}_N:=N\bar{\boldsymbol\psi}^{\top}\hat{\mathbf C}^{-1}\bar{\boldsymbol\psi}$, $\hat{\mathbf C}:=N^{-2}\sum_t\mathbf S_t\mathbf S_t^{\top}$ with $\mathbf S_t$ the partial sums of the centred $\boldsymbol\psi_n(\hat{\bvartheta})$. Then $T^{\mathrm{fb}}_N$ and $T^{\mathrm{sn}}_N$ converge to the pivotal fixed-$b$ limit of Kiefer and Vogelsang~\cite{kiefervogelsang2005fixedb} for the Bartlett kernel and to the self-normalized limit $\mathbf W_q(1)^{\top}\{\int_0^1\mathbf B_q(r)\mathbf B_q(r)^{\top}dr\}^{-1}\mathbf W_q(1)$, $\mathbf B_q(r):=\mathbf W_q(r)-r\mathbf W_q(1)$~\cite{shao2010selfnormalized}, which depend only on $(q,b)$.
\end{proposition}

\begin{proof}[Proof sketch]
Since $\hat{\bvartheta}=h(\bar{\bx}_2)$ with $h$ smooth, the assumption $\sqrt N(\bar{\bx}_2-\E\bx_2)=O_p(1)$ gives $\sqrt N(\hat{\bvartheta}-\bvartheta_0)=O_p(1)$; the functional central limit theorem for $\boldsymbol\psi_n$ does not imply it, because $\boldsymbol\psi_n$ is uncorrelated with $\bx_{2n}$ by construction. The derivative $\partial_{\bvartheta}\boldsymbol\psi_n$ is bounded and has mean $-\mathbf W_{21}+\mathbf W_{21}\mathbf W_{11}^{-1}\mathbf W_{11}=\mathbf 0$, so by the ergodic theorem $N^{-1}\sum_{n\le\lfloor rN\rfloor}\partial_{\bvartheta}\boldsymbol\psi_n\to\mathbf 0$ uniformly in $r$ (pointwise on a finite grid of $r$, and between grid points by monotonicity of the partial sums of the positive and negative parts of each entry), and a mean-value expansion gives $\sup_r\norm{N^{-1/2}\sum_{n\le\lfloor rN\rfloor}\{\boldsymbol\psi_n(\hat{\bvartheta})-\boldsymbol\psi_n(\bvartheta_0)\}}=o_p(1)$. Both statistics are continuous functionals of the partial-sum process of $\boldsymbol\psi_n(\hat{\bvartheta})$, the Bartlett estimator of the centred series with $M_N/N\to b$ being one by~\cite[Theorem~1]{kiefervogelsang2005fixedb} and $\hat{\mathbf C}$ by construction; the continuous mapping theorem and the invariance of both statistics to the scale $\boldsymbol\Sigma_{\mathrm{LR}}^{1/2}$ give the limits.
\end{proof}

For $K=1$ the score depends on $\bx_2$ alone, so the proposition covers the one-component test applied to directional records whatever the law of $\bx_1\mid\bx_2$. For $K>1$ the fixed-$b$ calibration of the mixture statistic, built on the effective scores, is heuristic: it would need a functional central limit theorem for the effective scores, which involve $\bx_1$ through the responsibilities, and the $\sqrt N$-consistency (M2) of a multi-start fit under dependence, neither of which we prove. The code simulates the quantiles of both limits from 20{,}000 Gaussian series of length 1{,}000. It takes $M_N=\operatorname{round}(bN)+1$, that is, weights $1-l/(\operatorname{round}(bN)+1)$ for $0\le l\le\operatorname{round}(bN)$, and simulates the null law with the same estimator.

\begin{figure}[ht]
\centering
\includegraphics[width=0.62\textwidth]{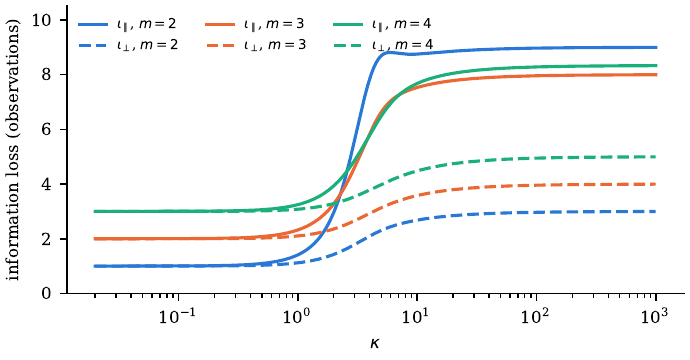}
\caption{Information-loss eigenvalues $\iota_\parallel(\kappa)$ (solid) and $\iota_\perp(\kappa)$ (dashed) of Theorem~16 for $m=2,3,4$. Both equal $m-1$ at $\kappa=0$ and tend to $(m+1)^2/(m-1)$ and $m+1$ as $\kappa\to\infty$ (proved). In between (numerically, $m=2,\dots,8$, $\kappa\in[0,10^3]$) they stay below these limits, $\iota_\perp$ increases, and so does $\iota_\parallel$ for $m\ge3$. At $m=2$, $\iota_\parallel$ is not monotone: $7.84$ at $\kappa=4$, a local maximum $8.81$ at $\kappa\approx5.9$, a local minimum $8.74$ at $\kappa\approx8.7$, then increasing to $9$; $\iota_\perp=2.10$ at $\kappa=4$. At $m=3$, $\kappa=4$, $(\iota_\parallel,\iota_\perp)=(5.84,2.93)$}
\label{fig:esm-geomtest-iota}
\end{figure}

\emph{Serial dependence at the persistence of the buoy records} (\texttt{sim\_hac2.py}, seed 20261105). Directions with a von Mises$(0,3)$ marginal through a Gaussian AR(1) copula, $N=34{,}000$, lag-one correlation $\rho\in\{0.95,0.99,0.995\}$ (median variance inflation 10.3, 48.3, 87.6), 300 series each (Monte Carlo standard error 1.3 points at 5\%). Rejection rates at 5\% (1\%) for $\rho=0.95$, $0.99$, $0.995$: independent sampling 78.3, 94.0, 98.7\%; Newey--West lag 168: 7.0, 8.0, 18.7\% (2.0, 2.7, 5.0\%); lag 1000: 8.0, 7.0, 10.3\%; fixed-$b$ Bartlett with bandwidth $bN$ and the simulated fixed-$b$ null~\cite{kiefervogelsang2005fixedb}, $b=0.05$: 5.7, 5.0, 6.3\%; $b=0.1$: 5.7, 4.7, 6.3\% (2.0, 1.7, 1.3\%); $b=0.2$: 5.0, 6.3, 7.0\%; self-normalized~\cite{shao2010selfnormalized}: 5.3, 6.7, 6.7\% (1.3, 1.0, 0.7\%); moving-block bootstrap variance with blocks of 185, 1000 and 3000: 7.0--16.7, 6.7--9.7 and 11.3--14.3\%. The fixed-$b$ calibration with $b=0.1$ holds its size within Monte Carlo error up to $\rho=0.99$ and is slightly oversized (6.3\%) at $\rho=0.995$. Under the expansion of Proposition~\ref{prop:esm-geomtest-hac} and a functional central limit theorem for the partial sums of the effective scores, the fixed-$b$ and self-normalized statistics have pivotal limits~\cite{kiefervogelsang2005fixedb,shao2010selfnormalized}, whose quantiles the code simulates (20{,}000 draws).

\subsection{Numerical checks}\label{subsec:esm-geomtest-checks}

The script \texttt{checks\_theory.py} (seed 20261103, about one minute) verifies every identity of Section~5 of the main text and of Section~\ref{subsec:esm-geomtest-proofs} on random instances; all 85 checks pass. Largest errors: closed-form $\Ieff$ against the Schur complement of the Fisher--Bingham information computed by quadrature, $m\in\{2,3\}$, $\kappa\in\{0,0.7,3,12\}$, $6\times10^{-16}$; against $2\times10^6$ Monte Carlo draws at $m=5$, relative $1.8\times10^{-3}$; Bessel forms (25) of the main text against quadrature for $m\in\{2,3,4,7\}$, $\kappa\le30$, relative $8\times10^{-10}$; the limit $2/(m(m+2))$, $6\times10^{-16}$. Ambient score statistic $N(\bar{\bt}-\hat{\bfeta})^{\top}\nabla^2\psi_{\mathrm{amb}}^{-1}(\bar{\bt}-\hat{\bfeta})$, with $\nabla^2\psi_{\mathrm{amb}}$ computed exactly (Gauss--Hermite quadrature in $\bx_1$ given $\bx_2$, product rule on the sphere), against $N\langle\bDelta,\Ieff^{-1}\bDelta\rangle$, $d=2$, $m\in\{2,3\}$: relative $2.3\times10^{-13}$. Annihilation $\langle\mathbf c(\mathbf A),d\btheta[X]\rangle=0$ and $H^{\mathrm e}(X,Y;\mathbf A)=-\tr(\mathbf A\dot{\bB}_X^{\top}\bS^{-1}\dot{\bB}_Y)$ by finite differences of (9) of the main text for $(d,m)\in\{(1,2),(2,3),(3,4)\}$: $10^{-8}$ and relative $5\times10^{-8}$. Likelihood ratio from the ambient likelihoods (through the closed form of $\psi_{\mathrm{amb}}$ in Section~S8 with the Fisher--Bingham normalizer by quadrature) against $2N\KL[\FB^{*}:\vMF]$, relative $3\times10^{-14}$; against the m-geodesic integral of Theorem~13(ii), relative $2\times10^{-11}$; the Pythagorean defect (26) of the main text and the exact relation on the leaf $\{\bB=\hat{\bB}\}$, $4\times10^{-15}$; Kent's statistic in case (b), relative $5\times10^{-15}$; Bingham's statistic at $\hat\kappa=0$, relative $7\times10^{-13}$; the noncentrality $\Ieff[\boldsymbol\Upsilon,\boldsymbol\Upsilon]$ against $2N\inf\KL$ at $N=10^6$, relative $6\times10^{-4}$, the size of the $O(N^{-1/2})$ remainder. The script \texttt{check\_infoloss.py} verifies Theorem~16: the squared exponential curvature computed by direct $g$-orthogonal projection of numerical second derivatives of (9) onto the normal space, with the exact ambient Fisher information, agrees with (27) for $(d,m,\kappa)\in\{(1,2,4),(2,2,1.5),(2,3,4)\}$ to relative $2\times10^{-9}$; a Monte Carlo evaluation of $\E[\mathbf Z\mathbf V^{-1}\mathbf Z]$ with $4\times10^5$ draws agrees to $10^{-3}$; and the sample covariance of the directions minus $\mathbf V(\hat{\bvartheta})$ equals $\bDelta$ to $4\times10^{-16}$.

\end{document}